\documentclass[a4paper,12pt,twoside]{book}
\usepackage[english]{babel}
\usepackage[utf8]{inputenc}
\usepackage[T1]{fontenc}
\usepackage{times}
\usepackage{amssymb,amsmath,amsfonts,amsthm}
\usepackage{hyperref}
\usepackage{graphicx}
\usepackage{epsfig}
\usepackage{textcomp}
\usepackage{listings}
\usepackage{lstbayes}
\def\be{\begin{equation}}
\def\ee{\end{equation}}
\def\bea{\begin{eqnarray}}
\def\eea{\end{eqnarray}}

\def\hsp5{\hspace{5mm}}
\def\lb{\label}
\def\bi{\bibitem}
\def\ct{\cite}

\newcommand{\R}{\textsf{R}}

\def\case#1/#2{\textstyle\frac{#1}{#2}}
\newcommand{\leftout}[1]{}
\begin{document}

\vspace*{1cm}
\begin{center}

\end{center}
\begin{center}
{\Huge\sc
										An Introduction to \\
                    Inductive Statistical Inference
\par\par}

\medskip
{\Large\sc
						from parameter estimation to decision-making
\par}

\par\vfill\vfill\vfill
              Lecture notes for a quantitative--methodological 
              module at the Master degree (M.Sc.) level
\par\vfill
\par\vfill
\par\vfill\vfill\vfill\vfill

                    {\large\sc Henk van Elst}

\par\vfill

                 August 30, 2022
\par\vfill\vfill

                 parcIT GmbH \\
								 Erftstra\ss e 15 \\
								 50672 K\"{o}ln \\
                 Germany
\par\vfill
	ORCID iD:
		\href{https://orcid.org/0000-0003-3331-9547}{0000-0003-3331-9547} \\
  E--Mail: \texttt{Henk.van.Elst@parcIT.de}
\par\vfill\vfill
	E--Print:
		\href{http://arxiv.org/abs/1808.10173}{arXiv:1808.10173{v3} 
[stat.AP]}
\par\vfill\vfill
\copyright\ 2016--2022 Henk van Elst
\end{center}
\vspace*{1cm}
\sloppy
      \renewcommand{\thepage}{}                 

\chapter*{}

%
\begin{verse}
{\Large\it Dedicated to}
\\
\vspace{4cm}
\hspace{4cm}
{\Large\it the good people}
\\
\vspace{8cm}
\hspace{8cm}
{\Large\it at Karlshochschule}
\end{verse}


     \cleardoublepage \pagenumbering{roman} \setcounter{page}{5}
\addcontentsline{toc}{chapter}{Abstract}
\chapter*{}
\vspace{-8ex}
\section*{Abstract}
{\small
These lecture notes aim at a post-Bachelor audience with a
background at an introductory level in Applied Mathematics and
Applied Statistics. They discuss the logic and methodology of the
Bayes--Laplace approach to inductive statistical inference that
places common sense and the guiding lines of the scientific method
at the heart of systematic analyses of quantitative--empirical
data. Following an exposition of exactly solvable cases of
single- and two-parameter estimation problems, the main focus is
laid on Markov Chain Monte Carlo (MCMC) simulations on the basis of
Hamiltonian Monte Carlo sampling of posterior joint
probability distributions for regression parameters occurring in
generalised linear models for a univariate outcome variable.
The modelling of fixed effects as well as of correlated
varying effects via multi-level models in non-centred
parametrisation is considered. The simulation of  posterior
predictive distributions is outlined. The assessment of a model's
relative out-of-sample posterior predictive accuracy
with information entropy-based criteria WAIC and LOOIC
and model comparison with Bayes factors are addressed. A brief
discussion on the description of the generation of stationary
time series data by means of autoregressive models is contained.
Concluding, a conceptual link to the behavioural subjective
expected utility representation of a single decision-maker's choice 
behaviour in static one-shot decision problems is established.
Vectorised codes for MCMC simulations of multi-dimensional
posterior joint  probability distributions with the Stan
probabilistic programming language implemented in the statistical
software~\R{} are provided. The lecture notes are fully
hyperlinked. They direct the reader to original scientific
research papers, online resources on inductive
statistical inference, and to pertinent biographical information.
Worked examples in statistical modelling have been compiled in
an~\R{} Markdown notebook \href{https://github.com/hve1964/stanCodes/blob/master/InductiveStatisticalInference.Rmd}{\texttt{InductiveStatisticalInference.Rmd}} and made available at the URL
\href{https://github.com/hve1964/stanCodes}{\texttt{github.com/hve1964/stanCodes}}.
}

\vspace{10mm}
\noindent
\underline{Cite as:} 
\href{http://arxiv.org/abs/1808.10173}{arXiv:1808.10173{v3} 
[stat.AP]}
\vfill

\medskip
\noindent
These lecture notes were typeset in \LaTeXe.

      \newpage \thispagestyle{empty}
      \cleardoublepage \pagenumbering{roman} \setcounter{page}{7}
\tableofcontents
      \newpage \thispagestyle{empty}
      \cleardoublepage \pagenumbering{arabic}
\addcontentsline{toc}{chapter}{Introductory remarks}
\chapter*{Introductory remarks}
Contemporaries of the 21$^{\mathrm{st}}$ Century find themselves 
exposed to flows of \textbf{information} of unprecedented current 
strengths. In an incredibly diverse spectrum of walks of life, the 
volumes of data amassed as a consequence of the steadily 
progressing digital transformation have gradually attained 
astronomically huge dimensions. Given this state of affairs, a 
legitimate question arising to an enquiring mind is to 
whether or not, and, if answered in the affirmative, to what 
extent this societal process ought to have any bearings on one's 
envisaged academic training?

\medskip
\noindent
Though in view of present-day global developments in, foremost, 
business and communication it is quite plausible to question 
the future status of the English language as the world's 
\textit{lingua franca} in, say, five decades from now, it is with 
near certainty that the minimum level of statistical literacy 
required to keep up with the demands and expectations in one's 
professional life will continue to rise. Moreover, the challenge 
of handling successfully the complexity of such pressing issues of 
humanity as planet Earth's attested climate change and the need to 
maintain its habitability by practising a sustainable economic use 
of its natural resources makes systematic processing of 
information a valuable and much-sought intellectual skill. Not 
disregarding complementary methodological tools, conscientious and 
sense-inducing communal decision-making in information-heavy 
managerial contexts is likely to benefit from a sound technical 
training in the principles of \textbf{statistical methods of data 
analysis} of citizens aiming to assume positions with a certain 
degree of responsibility attached to them.

\medskip
\noindent
In the course of social interactions experienced by the 
generations of people living during the last few centuries and up 
to now, pursuing the \textbf{scientific method} has proven beyond 
doubt to be the most reliable human approach to satisfactory 
problem-solving. That is, given at hand a practical or theoretical 
problem of some urgency, forming one's viewpoint on the basis of 
available factual information, and re-evaluating it in the light 
of relevant new evidence in order to draw conclusions as to 
reasonable consequent action, defines a systematic inductive 
procedure of compelling resilience. As this technique
constitutes a valid operationalisation of a notion of acting by 
\textbf{common sense}, it has the potential to increase both
idealistic as well as overall economic value for the
human community when transferred as a guiding principle for 
advancing matters to a wider field of socially important domains. 
There surely exists an obligation to hedge against a currently 
prevailing tendency of decision-making based on ``alternative 
facts,'' as the quality of the ensuing consequences and outcomes 
for the vast majority of people affected is self-evident.

\medskip
\noindent
Coles (2006)~\ct[p~3]{col2006} paraphrases the prime objective 
of the \textbf{scientific endeavour} by reminiscing:
%
\begin{quotation}
\noindent
``When I started doing research it gradually dawned on me that if 
science is about anything at all, it is not about being certain 
but about dealing rigorously with uncertainty.''
\end{quotation}
The immediate implication of this viewpoint is that the actual 
issue one finds oneself confronted with
when trying to make inferences from necessarily \textbf{incomplete 
information} is to have available a coherent and logically 
consistent framework for capturing and systematically processing 
fundamental \textbf{uncertainty}, which, in a scientific setting,
is interpreted in an epistemological fashion as representing a
researcher's \textbf{state of knowledge} concerning the problem of
her/his interest.

\medskip
\noindent
It comes across as a somewhat irritating piece of historical irony 
in the evolution of the empirical sciences that the principles of 
such a calculus of ``reasonable expectation'' (according to Cox 
(1946)~\ct{cox1946}) had been fully worked out halfway through the
20$^{\mathrm{th}}$ Century, but were largely ignored by the 
majority of active empirical researchers until lately. Lasting 
contributions to its methodology emerged from the predominantly 
data-driven scientific disciplines of \textbf{Physics},
\textbf{Astronomy}, \textbf{Anthropology}, \textbf{Biology},
\textbf{Economics}, \textbf{Psychology}, \textbf{Political Science}
and \textbf{Statistics}. For the sake of the uninitiated reader, a
few brief historical  comments are in order.

\medskip
\noindent
Arguably the history of the development of a framework 
of \textbf{inductive statistical inference} from past to present
can be grouped into the following four periods:
\begin{itemize}

\item \textbf{pioneering period}: the foundations of the framework 
of \textbf{inductive statistical inference} were laid independently 
during the 18$^{\mathrm{th}}$ Century by the English mathematician
and  Presbyterian minister 
\href{https://mathshistory.st-andrews.ac.uk/Biographies/Bayes/}{Thomas
Bayes (1702--1761)} and the French mathematician and astronomer
\href{https://mathshistory.st-andrews.ac.uk/Biographies/Laplace/}{Marquis
Pierre Simon de Laplace (1749--1827)}; see Bayes 
(1763)~\ct{bay1763} and Laplace (1774)~\ct{lap1774}. The latter of 
these two is credited for giving a full mathematical formulation 
of \textbf{probability theory} which presupposes \textbf{prior
information} on the plausibilities of \textbf{outcomes} in a set of 
different possibilities, and how to update these plausibilities in 
the light of relevant new evidence.

\item \textbf{conceptual period}: during the first half of the 
20$^{\mathrm{th}}$ Century the British economist 
\href{https://mathshistory.st-andrews.ac.uk/Biographies/Keynes/}{John
Maynard Keynes CB FBA (1883--1946)}, the British 
mathematician, statistician, geophysicist, and astronomer 
\href{https://mathshistory.st-andrews.ac.uk/Biographies/Jeffreys/}{Sir
Harold Jeffreys FRS (1891--1989)}, and the Italian 
probabilist statistician and actuary 
\href{https://mathshistory.st-andrews.ac.uk/Biographies/De_Finetti/}{Bruno
de Finetti (1906--1985)} argued strongly that a concept of
\textbf{probability} can only be meaningful when it relates to the
\textbf{states of knowledge} of individuals, and so inherently
bears  a certain dimension of \textbf{subjectivity}. Their views
are condensed in the classical monographs by Keynes 
(1921)~\ct{key1921} and Jeffreys (1939)~\ct{jef1939}, and the 
seminal paper by de Finetti (1937)~\ct{def1937} . The conceptual 
work of these authors was propagated in particular by the 
US-American physicist 
\href{https://en.wikipedia.org/wiki/Edwin_Thompson_Jaynes}{Edwin 
Thompson Jaynes (1922--1998)}, who supplemented it by a compelling 
interpretion of \textbf{probability theory} as extended logic; cf. 
Jaynes (2003)~\ct{jay2003}.

\item \textbf{engineering period}: this period, which roughly
started during the mid-1980ies, is characterised by the development
of powerful algorithms for numerically simulating complicated 
multi-dimensional distribution functions using \textbf{Markov Chain 
Monte Carlo (MCMC)} and \textbf{Hamiltonian Monte Carlo (HMC)} 
techniques, and their stable and efficient implementation in 
standard statistical software; see, e.g., Geman and Geman 
(1984)~\ct{gemgem1984}, Duane \textit{et al} 
(1987)~\ct{duaetal1987}, Gelfand and Smith (1990)~\ct{gelsmi1990}, 
Lunn \textit{et al} (2000)~\ct{lunetal2000}, or Plummer 
(2017)~\ct{plu2017}.

\item \textbf{big data period}: the present period, when it has 
become commonplace to process large amounts of data, often in a 
machine learning context (see, e.g., Ng (2018)~\ct{ng2018}).
Data sets, high-speed algorithms and other supporting material are 
shared by online communities such as the one active on the platform
\href{https://github.com/}{GitHub (\texttt{github.com})}, or the
\href{http://mc-stan.org/}{Stan Development Team
(\texttt{mc-stan.org})}. A central objective of many efforts in
this area is the performance of predictive analytics in a diverse
field of  applications. Associated with this focus is a continued
interest in the possibilities of artificial intelligence; cf.,
e.g., Penrose (1989)~\ct{pen1989}.\footnote{In the context of
recent developments in artificial intelligence, the video
documentation of the conversion between Sir Roger Penrose and the
advanced robot Sophia on YouTube provides some interesting insight.
URL (cited on June 22, 2018): 
\href{https://www.youtube.com/watch?v=YUo1FzZQzZ0}{www.youtube.com/watch?v=YUo1FzZQzZ0}.}

\end{itemize}
Until quite recently, the dominant methodological paradigm for 
quantitative--empirical research work has been the
\textbf{frequentist approach to data analysis and statistical
inference}, the most prominent proponent of which was the English 
statistician, evolutionary biologist, eugenicist and geneticist
\href{https://mathshistory.st-andrews.ac.uk/Biographies/Fisher/}{Sir
Ronald Aylmer Fisher FRS (1890--1962)}; cf. Fisher 
(1935)~\ct{fis1935a}. From the present perspective it appears as 
though Fisher had a rather strong influence on the sociology of the 
academic community in \textbf{Statistics} for the better part of
the first half of the 20$^{\mathrm{th}}$ Century; see, e.g., the
insightful and revealing recount of statistics training in
academic physics education by
Jaynes(2003)~\ct[Sec.~10.2]{jay2003}. Fisher, having 
been a fierce opponent to the \textbf{inductive statistical 
inference} framework advocated by his contemporaries Jeffreys and 
Keynes, is generally assigned the authorship of the (originally 
intended as derogatory) term ``Bayesian Statistics.'' At the time,
the leading figures of this framework referred to it as 
``inverse probability;'' see Jeffreys (1939)~\ct[p~28]{jef1939}
and Stigler (1986)~\ct[p~101]{sti1986}. The frequentist approach
was outlined in the lecture notes~\ct{hve2019}.

\medskip
\noindent
The realisation for a need of a systematic rethinking of standard 
practices in statistical methodology has been heavily boosted 
during the last decade by recurrent problems of successfully 
reproducing published results in the research literature, foremost 
in the \textbf{Social Sciences}. Explicit examples are given, e.g., 
in Gill (1999)~\ct{gil1999}, who discusses the abundant but 
unreliable practice of null hypothesis significance testing in 
Political Science, in an article published by \textit{The 
Economist} (2013)~\ct{eco2013}, in Nuzzo (2014)~\ct{nuz2014}, and 
in some recent blog entries by, amongst others, Vasishth 
(2017)~\ct{vas2017} or by Papineau (2018)~\ct{pap2018}. Also 
Kruschke and Liddell (2017)~\ct{krulid2017} and Briggs 
(2012)~\ct{bri2012} address this and other related conceptual 
difficulties with the frequentist approach. The worrisome fact of 
regularly failing reproduction attempts of asserted empirical 
effects has come to be known by the name of ``\textbf{replication
crisis in science}.''

\medskip
\noindent
In reflection of the massively increased interest in the
\textbf{Bayes--Laplace approach to data analysis and statistical 
inference} since the last turn of the centuries, and in 
recognition of its undeniable track record of successes in all 
areas of quantitative--empirical investigation over the last few 
decades, there exists a plethora of recently published 
state-of-the-art textbooks. In chronological order, these comprise 
Sivia and Skilling (2006)~\ct{sivski2006}, who focus on 
applications in Physics, Albert (2009)~\ct{alb2009}, Lee 
(2012)~\ct{lee2012}, Greenberg (2013)~\ct{gre2013}, who outlines 
uses in Econometrics, Gelman \textit{et al}
(2014)~\ct{geletal2014}, which the community of applied
statisticians considers to be \textit{the} authoritative monograph
in the field, Andreon and Weaver (2015)~\ct{andwea2015}, giving
explicit examples from statistical modelling in Astrophysics, Gill
(2015)~\ct{gil2015}, who presents applications in the Social and
Behavioural Sciences, Kruschke (2015)~\ct{kru2015}, with case
studies from Biology, Psychology, Sociology and Sports, and
McElreath (2020)~\ct{mce2020a}, who establishes a link to
inspiring  quantitative research problems in
Anthropology.\footnote{Trotta's
(2008)~\ct{tro2008} review discusses applications of the
Bayes--Laplace approach in the cosmological context.} Almost all of 
these textbooks provide an abundance of practical problems and 
exercises, generally in combination with fully operational codes 
implemented in the shareware statistical software packages~\R, Stan
and/or JAGS.

\medskip
\noindent
The methods presented in these lecture notes are rooted in
\textbf{Applied Statistics}. They address an audience at a
post-Bachelor academic level, with a vested interest in acquainting
themselves with standard pratices of modern \textbf{statistical
methods of data analysis}. The topics presented form a selection of
the most frequently employed tools for building data-based
\textbf{statistical models} for purposes of explanation and
prediction of observable phenomena. These comprise in particular:
\begin{itemize}

\item analytical \textbf{single-parameter estimation},

\item fitting multi-dimensional \textbf{generalised linear models}
employing MCMC simulations,

\item basic modelling of processes generating stationary
\textbf{time series data},

\item \textbf{model comparison}, and

\item elementary \textbf{decision-making} under conditions of
uncertainty.
\end{itemize}
As implicitly hinted at above, we here deliberately assume an
\textbf{interdisciplinary perspective}, being thoroughly convinced
that the chances for successfully dealing with most kinds of
modern-day problems of societal relevance will not be reasonably
improved by confining one's efforts to a possibly comfortable
though often narrow-minded intellectual niche.

\medskip
\noindent
The present lecture notes are designed to be dynamical in 
character. On the one-hand side, this means that they will be 
updated on a regular basis. On the other, that its *.pdf version 
contains interactive features such as fully 
hyperlinked references to original publications at the websites 
\href{https://doi.org}{\texttt{doi.org}},
\href{http://www.jstor.org}{\texttt{jstor.org}}, or elsewhere, and 
also many active links to biographical information on scientists 
that have been influential in the historical development of
\textbf{probability theory} and the \textbf{Bayes--Laplace approach
to data analysis and statistical inference}, hosted by the websites 
\href{https://mathshistory.st-andrews.ac.uk/}{MacTutor History of
Mathematics Archive (\texttt{mathshistory.st-andrews.ac.uk})}
and \href{http://en.wikipedia.org/wiki/Main_Page}{\texttt{en.wikipedia.org}}.

\medskip
\noindent
Opting for the application of the \textbf{Bayes--Laplace approach
to data analysis and statistical inference} entails the frequent
performance of a large number of computations and numerical 
simulations, which, to ensure reliability, need to be meticulously 
checked for potential errors. However, these computations are an 
integral part of the fun of the research activity, and they are 
enormously facilitated by the provision of taylor-made software 
packages that are distributed as shareware on the internet. A 
widespread computational tool that we, too, will employ and refer
to in the  course of these lecture notes is the statistical
software package~\R{} distributed by the R~Core Team 
(2022)~\ct{rct2022} free of charge for many different operating
systems via the website
\href{http://cran.r-project.org}{\texttt{cran.r-project.org}}.
Useful and easily accessible introductory textbooks on the
application of \R{} for purposes of statistical 
data analysis are, e.g., Dalgaard (2008)~\ct{dal2008}, or 
Hatzinger \textit{et al} (2014)~\ct{hatetal2014}. Additional
helpful information and assistance is available from the website 
\href{http://www.r-tutor.com/}{\texttt{www.r-tutor.com}}.
We strongly recommend the use of the convenient custom-made work 
environment \R{}Studio (soon to be known as
\href{https://posit.co/}{posit}) provided at
\href{https://www.rstudio.com}{\texttt{www.rstudio.com}}. Also, we 
point the reader to an overview of \R{} tools made available for 
\textbf{Bayes--Laplace statistical inference} which is maintained
by Park \textit{et al}; cf. Park \textit{et al}
(2022)~\ct{paretal2022}. All figures have been
generated in \R{} employing the advanced graphical package
\texttt{ggplot2} by Wickham (2016)~\ct{wic2016}. Notation to be
used follows the conventions of Refs.~\ct{hve2019}
and~\ct{hve2015b}.

\medskip
\noindent
Vectorised codes written in the Stan probabilistic programming
language to run MCMC simulations of multi-dimensional posterior
joint probability distributions for model parameters in both fixed
effects and  varying effects generalised linear models as well
as in stationary linear time series models are made available at
the web address \href{https://github.com/hve1964/stanCodes}{\texttt{github.com/hve1964/stanCodes}}. These Stan codes are used in worked examples
in statistical modelling that have been compiled in the
R Markdown *.html notebook
\href{https://github.com/hve1964/stanCodes/blob/master/InductiveStatisticalInference.Rmd}{\texttt{InductiveStatisticalInference.Rmd}}, which is distributed via the same address.

\chapter[Mathematical rules of probability theory]{Mathematical 
rules of probability theory}
\lb{ch1}
We begin our journey through the framework of \textbf{inductive
statistical inference} by reviewing the mathematical rules of
\textbf{probability theory}.

\section[Probability and uncertainty]{Probability and uncertainty}
\lb{sec:probunc}
Jaynes (2003)~\ct{jay2003}, in his influential monograph, 
conceptualises \textbf{probability theory} as an extension of
\textbf{Aristotelian deductive logic}.\footnote{Named after the
ancient Greek philosopher and scientist 
\href{https://mathshistory.st-andrews.ac.uk/Biographies/Aristotle/}{Aristotle
(384~BC--322~BC)}.} In the latter discipline, the objects of
investigation are \textbf{propositions}. Propositions can 
be verbal statements that relate to some observable real-world 
phenomenon of a certain practical interest, or they can be 
suppositions in the context of an academic discourse. For 
instance, the assertions
\begin{itemize}

\item[] $A$: \textit{The average travel time for human space 
missions from planet Earth to a like planet in the Andromeda 
Galaxy ranges between four and five hours.}

\item[] $B$: \textit{Prince Rogers Nelson was a US-American 
musician.}

\item[] $C$: \textit{The German team will win the next Cricket
World Cup.}

\end{itemize}
are representative of simple kinds of propositions. In
\textbf{Aristotelian deductive logic}, the truth content of a
proposition can be exclusively either \textit{true} or
\textit{false}, and so in this respect this specific logic is
inherently two-valued in nature.

\medskip
\noindent
Employing \textbf{Boolean algebra},\footnote{Named after 
the English mathematician, educator, philosopher and logician 
\href{https://mathshistory.st-andrews.ac.uk/Biographies/Boole/}{George
Boole (1815--1864)}.} two propositions $A$ and $B$ can be combined
to form a new proposition via
\begin{itemize}

\item[(i)] the \textbf{logical product} (or mutual conjunction),
\be
\lb{eq:logprod}
AB:\ \text{``\textit{both} of the propositions $A$ and $B$ are 
true''} \ ;
\ee
note that naturally $AB = BA$ applies, i.e., commutativity is a 
property of the product operation, and

\item[(ii)] the \textbf{logical sum} (or mutual disjunction),
\be
\lb{eq:logsum} 
A+B:\ \text{``\textit{at least one} of the propositions $A$ and $B$ 
is true''} \ ;
\ee
again, naturally commutativity holds true for the sum operation, 
$A+B = B+A$.

\end{itemize}
More complex propositions can be constructed by combining 
different propositions with both the logical product and the 
logical sum, and making use of bracketing sub-operations, i.e., 
inserting $(\ldots)$ where intended or needed.

\medskip
\noindent
With $\overline{A}$ denoting the \textbf{logical complement} of
some proposition~$A$ (referred to as ``not $A$''), it follows that
\bea
A\,\overline{A} & & \text{is\ always\ \textit{false}}
\qquad (\text{a\ contradiction}) \\
A + \overline{A} & & \text{is\ always\ \textit{true}}
\qquad (\text{a\ tautology}) \ .
\eea
Moreover, the logical identities
\bea
\lb{eq:logid1}
AA & = & A \\
\lb{eq:logid2}
A + A & = & A
\eea
apply. Of particular practical use are \textbf{De Morgan's 
laws},\footnote{Named after the British mathematician and logician
\href{https://mathshistory.st-andrews.ac.uk/Biographies/De_Morgan/}{Augustus
De Morgan (1806--1871)}.} which state that
\bea
\overline{AB} & = & \overline{A} + \overline{B} \\
\overline{A+B} & = & \overline{A}\,\overline{B} \ .
\eea
Negating the latter relation yields
\be
\lb{eq:equiv1}
A+B = \overline{\overline{A}\,\overline{B}} \ ,
\ee
a result that is to be used later on.

\medskip
\noindent
Jaynes' (2003)~\ct{jay2003} notion of an \textbf{extended logic} 
comes into effect by relaxing the strict demand for the binary 
truth content property of a proposition, but rather to assign to 
it a normalised \textbf{degree of plausibility}\footnote{Jeffreys
(1961)~\ct[p~5]{jef1939} coins the illustrative term ``reasonable
degree of belief.'' This might be the origin of decision theory's
standard terminology of the ``degree-of-belief'' assigned by a
rational agent to the realisation of an uncertain outcome.} which 
depends on a researcher's individual \textbf{state of knowledge} on 
the current matter of interest. This is to say that, subject to 
available background information, collectively denoted by~$I$, a 
real number $P(\ldots|I)$ from the interval $\left[0, 1\right]$ is 
assigned to a proposition which is referred to as its
\textbf{probability}.\footnote{The exploration of the psychological 
dimension underlying the assignment of probabilities to 
propositions was pioneered by Kahneman and Tversky 
(1972)~\ct{kahtve1972}.} The $I$-proviso here serves to express 
the position that, by way of conception, a \textbf{probability} is 
\textit{always} conditional on some form of \textbf{prior
information}; see Jaynes (2003)~\ct[p~87]{jay2003} and Sivia and
Skilling (2006)~\ct[p~5]{sivski2006}, or, as Keynes 
(1921)~\ct[p~102]{key1921} puts it, ``relative to given 
premisses.''

\medskip
\noindent
To link back to the three example propositions introduced 
above, one may thus assign on the basis of presently available 
understanding the (prior) probabilities
\begin{itemize}

\item[] $P(A|I_{A}) = 0$, expressing a logical resp.~practical 
impossibility,

\item[] $P(B|I_{B}) = 1$, expressing a logical resp.~practical 
certainty, and

\item[] $0 \leq P(C|I_{C}) \leq 1$, expressing a logical 
resp.~practical possibility of which the attributed degree of 
plausibility is considered limited.

\end{itemize}

\medskip
\noindent
In view of the interpretation of \textbf{probability assignments}
as a researcher's systematic way of hand\-ling practical situations
with \textbf{incomplete information} (which, typically, is more
often the case than not), it proves of little help to try to
associate any physical reality with the corresponding numerical
value from the interval $\left[0, 1\right]$. Rather, it constitutes
a specific proposal for dealing with \textbf{uncertainty} within a
logically consistent and coherent quantitative 
framework,\footnote{Philosophical viewpoints opposing the idea of 
uncertainty being amenable to a compelling treatment within a
quantitative framework have been put forward nearly a full century
ago by Knight (1921)~\ct{kni1921} and by Keynes
(1921)~\ct{key1921}.} the basic rules of which are to be described
in the following. Indeed, it comes as quite a surprise, and a
veritable manifestation of formal ellegance, that the
\textbf{probabilistic calculus} for plausible reasoning originated
by Bayes and Laplace rests on the foundation of only a few rather
simple first principles.

\section[Sum and product rules]{Sum and product rules}
\lb{sec:sumprod}
\subsection[Sum rule]{Sum rule}
For probabilities assigned to a proposition $A$ and its logical 
complement $\overline{A}$, the \textbf{sum rule} states that
\be
\lb{eq:sumrule}
\fbox{$\displaystyle
P(A|I) + P(\overline{A}|I) = 1
$}
\ee
must \textit{always} be true.

\subsection[Product rule]{Product rule}
To calculate the probability of the \textbf{logical product} of two 
propostions,  $AB$, the \textbf{product rule} holds that
\be
\lb{eq:productrule}
\fbox{$\displaystyle
P(AB|I) = P(A|BI)P(B|I) = P(B|AI)P(A|I) = P(BA|I) \ ,
$}
\ee
taking into account commutativity of the product operation, $AB = 
BA$, in the second part of this rule.

\medskip
\noindent
Re-arranging, and assuming that $P(A|I) > 0$ resp.~$P(B|I) > 0$ 
apply, alternative representations of the product rule are given by
\be
P(A|BI) = \frac{P(AB|I)}{P(B|I)} \ ,
\qquad\qquad
P(B|AI) = \frac{P(BA|I)}{P(A|I)} \ .
\ee
The first variant is generally referred to as the 
\textbf{conditional probability} for proposition~$A$ to be true,
given proposition $B$ is true and relevant background
information~$I$ is available. Analogously, the second variant
expresses the \textbf{conditional probability} for proposition~$B$
to be true, given proposition~$A$ is true and~$I$ is known.

\medskip
\noindent
In preparation of concepts of importance to be introduced in
subsequent chapters, it is fitting at this stage to briefly raise
the following point. The quantity $P(AB|I)$ [or $P(BA|I)$] occuring
in the product rule~(\ref{eq:productrule}) represents the
\textbf{joint probability} for propositions~$A$ and~$B$ to be
simultaneously true, given background information~$I$. It is
instructive to formally supplement~$A$ and~$B$ by their logical 
complements, $\overline{A}$ and~$\overline{B}$, and to represent 
the \textbf{joint probabilities} for all possible product 
combinations of these propositions, given~$I$, in the form of a 
$2 \times 2$~contingency table. By way of summation, separately 
across row and column entries, while respecting the sum 
rule~(\ref{eq:sumrule}), the concept of a \textbf{marginal 
probability} for a proposition to be true, given~$I$, is
introduced. The kind of $2 \times 2$~contingency table just
outlined is depicted in Tab.~\ref{tab:jointdistr}.
\begin{table}
\begin{center}
    \begin{tabular}[h!]{c|cc|c}
     & & \\
    \textbf{joint distribution} (2-D) & proposition~$B$ & 
    proposition~$\overline{B}$ &
    \textbf{marginal distribution} (1-D) \\
		& & \\
    \hline
     & & \\
    proposition~$A$ & $P(AB|I)$ & $P(A\overline{B}|I)$ & $P(A|I)$
		\\
     & & \\
    proposition~$\overline{A}$ & $P(\overline{A}B|I)$ & 
    $P(\overline{A}\,\overline{B}|I)$ & $P(\overline{A}|I)$ \\
     & & \\
    \hline
     & & \\
    \textbf{marginal distribution} (1-D)
    & $P(B|I)$ & $P(\overline{B}|I)$ 
    & $1$ 
    \end{tabular}
\end{center}
\caption{Representation of a discrete (prior) joint probability 
distribution and corresponding discrete marginal probability 
distributions for propositions $A$ and $B$ and their logical 
complements $\overline{A}$ and $\overline{B}$, given background 
information~$I$, in terms of a $2 \times 2$~contingency table. The 
marginal cell entries are obtained by row-wise resp.~column-wise 
summation over associated interior cell entries while respecting 
the sum rule~(\ref{eq:sumrule}).}
\lb{tab:jointdistr}
\end{table}
%

\subsection[Generalised sum rule]{Generalised sum rule}
Starting from the negation of the second De Morgan's law which was 
noted in Eq.~(\ref{eq:equiv1}), the derivation of a rule for 
calculating the probability of the \textbf{logical sum} of two 
propostions, $A+B$, is calculated. A string of algebraic 
manipulations leads to\footnote{There is a typo in the 
first line of Eq.~(2.65) in Jaynes (2003)~\ct{jay2003}. We here 
give the necessary correction in the expression following the 
second equality sign in Eq.~(\ref{eq:gensumrulederiv}).}
\bea
\lb{eq:gensumrulederiv}
P(A+B|I) & \stackrel{\text{Eq.~(\ref{eq:equiv1})}}{=} & 
P(\overline{\overline{A}\,\overline{B}}|I)
\ \stackrel{\text{Eq.~(\ref{eq:sumrule})}}{=} \ 1 - 
P(\overline{A}\,\overline{B}|I) \nonumber \\
& \stackrel{\text{Eq.~(\ref{eq:productrule})}}{=} & 1 - 
P(\overline{B}|\overline{A}I)P(\overline{A}|I)
\ \stackrel{\text{Eq.~(\ref{eq:sumrule})}}{=} \ 1 - 
[1-P(B|\overline{A}I)]P(\overline{A}|I) \nonumber \\
& \stackrel{\text{Eqs.~(\ref{eq:sumrule}), (\ref{eq:productrule})}}{=} 
& P(A|I) + P(\overline{A}B|I)
\ \stackrel{\text{Eq.~(\ref{eq:productrule})}}{=} \ P(A|I) + 
P(\overline{A}|BI)P(B|I) \nonumber \\
& \stackrel{\text{Eq.~(\ref{eq:sumrule})}}{=} & P(A|I) + 
[1-P(A|BI)]P(B|I) \ ,
\eea
so that with one final application of Eq.~(\ref{eq:productrule}) 
one obtains the \textbf{generalised sum rule}
as\footnote{Re-arranging Eq.~(\ref{eq:gensumrule}), to solve for 
$P(AB|I)$ instead, yields an alternative representation of the 
generalised sum rule, or ``conjunction 
rule,'' Eq.~(\ref{eq:productrule}). By means of their famous 
``Linda the bank teller'' example (amongst others), Tversky and 
Kahneman (1983)~\ct[p~297ff]{tvekah1983} were able to demonstrate 
the startling empirical fact that the conjunction rule is 
frequently violated in everyday (intuitive) decision-making. They 
termed this empirical phenomenon the ``conjunction fallacy.'' In 
their view, it can be explained as a consequence of 
decision-makers often resorting to a ``representativeness 
heuristic'' as an aid; see also Kahneman 
(2011)~\ct[Sec.~15]{kah2011}.}
\be
\lb{eq:gensumrule}
\fbox{$\displaystyle
P(A+B|I) = P(A|I) + P(B|I) - P(AB|I) \ .
$}
\ee

\medskip
\noindent
At this point the list of elementary mathematical rules of 
\textbf{probability theory} is complete. It comprises the
\textbf{sum rule}~(\ref{eq:sumrule}), the \textbf{product 
rule}~(\ref{eq:productrule}), and the \textbf{generalised sum 
rule}~(\ref{eq:gensumrule}). We next turn to highlight a few 
important extensions of these rules when dealing with special 
kinds of sets of propositions.

\subsection[Extensions to sets of propositions]{Extensions to sets 
of propositions}
Suppose given a finite set of $k \in \mathbb{N}$ \textbf{mutually 
exlusive and exhaustive propositions} $\{A_{1}, \ldots, A_{k}\}$, 
conditioned on some background information $I$, so that
\be 
P(A_{i}A_{j}|I) = 0 \qquad \text{for}\ i \neq j\ , \quad
i, j = 1, \ldots, k
\ee
is true. Then the sum rule~(\ref{eq:sumrule}) extends to the
\textbf{normalisation condition} of \textbf{probability theory},
namely
\be
\lb{eq:normcomplpart}
P(A_{1} + \ldots + A_{k}|I)
= P(A_{1}|I) + \ldots + P(A_{k}|I)
= \sum_{i=1}^{k}P(A_{i}|I) = 1 \ .
\ee
This states that, when assigning probabilities across a complete
set of mutually exclusive possibilities, for reasons of overall
consistency these must add up to~$1$.

\medskip
\noindent
Furthermore, on the basis of the \textbf{normalisation 
condition}~(\ref{eq:normcomplpart}) and the \textbf{product 
rule}~(\ref{eq:productrule}), it holds that
\be
\lb{eq:margrule}
P(B|I)
\stackrel{\text{Eq.~(\ref{eq:normcomplpart})}}{=}
\sum_{i=1}^{k}P(BA_{i}|I)
\stackrel{\text{Eq.~(\ref{eq:productrule})}}{=}
\sum_{i=1}^{k}P(B|A_{i}I)P(A_{i}|I) > 0 \ .
\ee
This is generally referred to as the \textbf{marginalisation rule} 
(see, e.g., Saha (2002)~\ct[p~5]{sah2002}, Sivia and Skilling 
(2006)~\ct[p~7]{sivski2006}, or Andreon and 
Weaver~\ct[p~4]{andwea2015}), and it possesses high practical 
value in the context of numerical simulations of probability 
distributions, a major topic in \textbf{inductive statistical
inference} that is to be discussed later on. An immediate simple
application of the \textbf{marginalisation rule} was illustrated in 
Tab.~\ref{tab:jointdistr} above.

\section[Bayes' theorem]{Bayes' theorem}
\lb{sec:bayes}
The core of the plausible reasoning framework developed as an 
efficient and reliable practical tool for \textbf{inductive 
statistical inference} is constituted by a result that is due to 
the English mathematician and Presbyterian minister 
\href{https://mathshistory.st-andrews.ac.uk/Biographies/Bayes/}{Thomas
Bayes (1702--1761)}; see the posthumous publication Bayes 
(1763)~\ct{bay1763}. It states that for two propositions~$A$ 
and~$B$, given background information~$I$, it is always true that
\be
\fbox{$\displaystyle
\lb{eq:bayes0}
P(A|BI) = \frac{P(B|AI)}{P(B|I)}\,P(A|I) \ ,
$}
\ee
On the face of it, \textbf{Bayes' theorem}, as it has come to be 
known for a long time, is just a convenient re-arrangement of the 
product rule~(\ref{eq:productrule}), provided $P(B|I) > 0$. 
However, its immense conceptual significance for plausible 
reasoning and \textbf{inductive statistical inference} was
glimpsed at already by Bayes himself; cf. Stigler
(1986)~\ct[pp~98--98]{sti1986}.

\medskip
\noindent
In qualitative terms \textbf{Bayes' theorem} is 
saying:\footnote{Depicting the structure of 
Bayes' theorem in this particular fashion ties in nicely with a 
famous quotation by the British economist 
\href{https://mathshistory.st-andrews.ac.uk/Biographies/Keynes/}{John
Maynard Keynes CB FBA (1883--1946)}, who is said to have once
remarked: ``When the facts change, I change my mind. What do you 
do, sir?'' See URL (cited on August 17, 2022): 
\href{https://mathshistory.st-andrews.ac.uk/Biographies/Keynes/quotations/}{mathshistory.st-andrews.ac.uk/Biographies/Keynes/quotations/}.}

\medskip
\medskip
\noindent
\fbox{
\begin{minipage}[c][2.5cm][c]{\textwidth}
\begin{center}
\textit{(prior knowledge on proposition~$A$)}
\ combined with \ 
\textit{(empirical evidence~$B$ on proposition~$A$)} \par

\medskip
yields \par

\medskip
\textit{(updated knowledge on proposition~$A$)}
\end{center}
\end{minipage}
}

\medskip
\medskip
\noindent
According to Jaynes (2003)~\ct[p~112]{jay2003}), in the clear-cut 
representation of Eq.~(\ref{eq:bayes0}), the theorem was first 
formulated by the French mathematician and astronomer
\href{https://mathshistory.st-andrews.ac.uk/Biographies/Laplace/}{Marquis Pierre Simon de Laplace (1749--1827)}; cf. Laplace 
(1774)~\ct{lap1774}.

\medskip
\noindent
To facilitate efficient communication, the different factors
featuring in \textbf{Bayes' theorem}~(\ref{eq:bayes0}) have been
given names in their own right. These are:
\begin{itemize}

\item $P(A|I)$ is referred to as the \textbf{prior probability} 
for proposition $A$ to be true, subject to background
information~$I$,\footnote{Kahneman (2011)~\ct[p~147]{kah2011}, in
his stimulating popular book, refers to $P(A|I)$ as the ``base
rate'' for proposition~$A$.}

\item $P(B|AI)$ is the \textbf{likelihood} for a proposition~$B$, 
providing potentially relevant information for proposition~$A$, 
given background information $I$,

\item $P(A|BI)$ is called the \textbf{posterior probability} for
proposition $A$ to be true in light of the information pertaining
to proposition~$B$ and background information $I$, and, lastly,

\item $P(B|I) > 0$ is usually known as the \textbf{evidence}
relating to proposition~$B$.

\end{itemize}
By means of marginalisation, and on the basis of the sum 
rule~(\ref{eq:sumrule}) and the product 
rule~(\ref{eq:productrule}), the evidence $P(B|I)$ can be 
re-expressed as
\bea
\lb{eq:margrule1}
P(B|I)
& \stackrel{\text{Eq.~(\ref{eq:sumrule})}}{=} &
P(BA|I) + P(B\overline{A}|I) \nonumber \\
& \stackrel{\text{Eq.~(\ref{eq:productrule})}}{=} & 
P(B|AI)P(A|I) + P(B|\overline{A}I)P(\overline{A}|I) \ .
\eea
In this form it is also referred to as the \textbf{
average likelihood} or \textbf{marginal likelihood};
cf. McElreath (2020)~\ct[Sec.~2.4]{mce2020a}.

\medskip
\noindent
From \textbf{Bayes' theorem}~(\ref{eq:bayes0}), one directly infers 
for the relation between the \textbf{prior probability} for
proposition~$A$ to be true and its \textbf{posterior probability}
that
\be
\begin{cases}
\ \text{if}\quad 0 < {\displaystyle\frac{P(B|AI)}{P(B|I)}} < 1
& \quad\Rightarrow\quad
P(A|BI) < P(A|I) \\ \\
\ \text{if}\quad {\displaystyle\frac{P(B|AI)}{P(B|I)}} > 1
& \quad\Rightarrow\quad
P(A|BI) > P(A|I)
\end{cases} \ ;
\ee
depending on the evidence available through proposition~$B$, the 
probability for proposition~$A$ to be true can potentially either
decrease or increase.

\medskip
\noindent
For simple practical applications with only two propositions~$A$ 
and $B$ involved, as is typically the case in situations analogous 
to drug testing, disease testing, or signal detection, it is 
helpful to rewrite \textbf{Bayes' theorem}~(\ref{eq:bayes0}) by 
making use of the marginalisation rule~(\ref{eq:margrule1}). One 
thus obtains
\be
\lb{eq:bayes0.1}
\fbox{$\displaystyle
P(A|BI) = \frac{P(B|AI)}{P(B|AI)P(A|I)
+ P(B|\overline{A}I)P(\overline{A}|I)}\,P(A|I) \ .
$}
\ee
The posterior probability for a proposition~$A$ to be true, in view
of some evidence relating to a proposition~$B$ and background 
information~$I$, can then be easily computed provided the 
following three pieces of information are available; cf. Silver 
(2012)~\ct[p~244]{sil2012}:
\begin{itemize}

\item[(i)] the prior probability for proposition~$A$ to be true in
the absence of evidence, $P(A|I)$,

\item[(ii)] the ``true positive rate,'' $P(B|AI)$, and

\item[(iii)] the ``false positive rate,'' $P(B|\overline{A}I)$.

\end{itemize}

\medskip
\noindent
This last bit of discussion generalises to the case of a set of $k 
\in \mathbb{N}$ \textbf{mutually exlusive and exhaustive 
propositions} $\{A_{1}, \ldots, A_{k}\}$ in a straightforward 
fashion. Given prior probabilities~$P(A_{i}|I)$ for each 
proposition~$A_{i}$ in the set to be true, and with the
marginalisation rule~(\ref{eq:margrule}) employed to express the
average likelihood~$P(B|I)$ for some evidential proposition~$B$,
one calculates posterior probabilities~$P(A_{i}|BI)$ for each 
proposition~$A_{i}$ in the set to be true from \textbf{Bayes' 
theorem}~(\ref{eq:bayes0}) according to
\be
\lb{eq:bayes0.2}
\fbox{$\displaystyle
P(A_{i}|BI) = \frac{P(B|A_{i}I)}{
{\displaystyle\sum_{j=1}^{k}P(B|A_{j}I)P(A_{j}|I)}}\,P(A_{i}|I) \ ,
\qquad \text{for}\ i = 1, \ldots, k \ .
$}
\ee
In this specific form, \textbf{Bayes' theorem} has high practical 
value as a computational basis for discretised numerical 
simulations of complicated high-dimensional probability 
distribution functions.

\section[Outlook on inductive data analysis and model 
building]{Outlook on inductive data analysis and model building}
\lb{sec:datamodel}
So why does the \textbf{Bayes--Laplace approach} to
\textbf{probability theory} provide such a conceptually compelling
basis for \textbf{plausible reasoning} and \textbf{inductive
statistical inference}?

\medskip
\noindent
One of a number of strong arguments in its favour is that 
\textbf{scientific objectivity} is ensured by strict adherence to
the requirements of \textbf{logical consistency} and
\textbf{fact-based reasoning}. That is to say, on the basis of the
mathematical rules of \textbf{probability theory} outlined in
Sec.~\ref{sec:sumprod}, two researching individuals that
\begin{itemize}

\item[(i)] hold the \textit{same} relevant background
information~$I$ on a specific proposition~$A$ of scientific
interest, and

\item[(ii)] have access to the \textit{same} empirical evidence 
associated with a proposition~$B$,

\end{itemize}
must assign the \textit{same} \textbf{prior probability} $P(A|I)$
to proposition~$A$, and calculate the \textit{same}
\textbf{posterior probability} $P(A|BI)$ for proposition~$A$ from
the empirical evidence available. In practice, of course,
\textit{different} individuals typically have access to
\textit{differing} amounts of relevant background information and
empirical evidence. The \textbf{Bayes--Laplace approach}, however,
exposes itself deliberately to criticism in that it requires a
researcher to state openly all of 
her/his assumptions that went into an inductive statistical 
inference process. There are no hidden agendas, which certainly 
facilitates to a novice the task of becoming acquainted with the 
specific rationale employed in this framework.\footnote{Jaynes 
(2003)~\ct[p~22]{jay2003} identifies as a dangerous pitfall for 
plausible reasoning what he refers to as the ``mind projection 
fallacy.'' He depicts this as the error of confusing 
epistemological statements (statements of knowledge of things) 
with ontological statements (statements of existence of things), 
and vice versa. Put differently, this describes a case where an 
individual confuses what they personally think exists in reality
with what \textit{actually} (and, therefore, testably) does exist
in reality.}

\medskip
\noindent
Formally, \textbf{Bayes' theorem}, for example in its 
variant~(\ref{eq:bayes0.2}), represents the fundamental principle 
according to which \textbf{inductive statistical inference} is to
be performed, given prior information and empirical data of
relevance to an actual research question. To begin with, the
following specific substitutions need to be made:
\begin{eqnarray*}
\text{proposition}~A_{i} & \longleftarrow & \text{model}(i) \ ,
\ \text{or} \ \text{hypothesis}(i) \ , \ \text{or} \ 
\ \text{set\ of\ parameter\ values}(i) \\
\text{proposition}~B & \longleftarrow & \text{data} \ ;
\end{eqnarray*}
here the concepts ``$\text{model}(i)$'' and 
``$\text{hypothesis}(i)$'' can be interpreted as being synomynous. 
In the \textbf{Bayes--Laplace approach}, data is considered 
\textit{fixed} \textbf{incomplete information}, while the model, 
hypothesis, or set of parameter values of the researcher's focus 
is the \textit{unknown} entity about the plausibility of which
inferences are to be made in light of available evidence. The
\textbf{unknown entity} is to be described probabilistically by 
assigning a \textbf{probability distribution} to  the range of
possible \textbf{outcomes} it involves. One thus obtains
\be
\fbox{$\displaystyle
\lb{eq:bayesData}
\underbrace{P(\text{model}(i)|\text{data}, I)}_{\text{posterior}}
= \frac{\overbrace{P(\text{data}|\text{model}(i),
I)}^{\text{likelihood}}}{\underbrace{P(\text{data}|
I)}_{\text{average\ likelihood}}}\,
\underbrace{P(\text{model}(i)|I)}_{\text{prior}} \ .
$}
\ee

\medskip
\noindent
The nature of \textbf{Bayes' theorem} hereby undergoes a 
qualitative change in that it transforms from a statement 
concerning four probability values (non-negative real numbers) to 
a functional relationship between entire probability 
distributions. The main statement is that the \textbf{posterior 
probability distribution} for the \textbf{unknown entity} of
interest amounts to the product between the \textbf{likelihood
function} for  the data, given the unknown entity, and the
\textbf{prior probability distribution} for the unknown entity,
divided by a normalising constant (a positive real number, as the
data is considered fixed) referred to as the \textbf{average
likelihood}. Within the \textbf{Bayes--Laplace approach}, the
\textbf{posterior  probability distribution} is viewed as a ``
compromise'' between  the background information-driven
\textbf{prior probability distribution} and the data-driven
\textbf{likelihood function}; see, e.g., Kruschke
(2015)~\ct[p~112]{kru2015}.

\medskip
\noindent
At the heart of the activity of \textbf{inductive statistical 
inference} is the proposition of a \textbf{statistical model} 
derived from transparent and comprehensible \textbf{theoretical
considerations}. The general purpose of a \textbf{scientific
theory} is to describe, explain and predict observable phenomena in
its particular field of application. A \textbf{statistical model}
is formulated in the concise language of \textbf{mathematics}. It
typically comprises a certain finite number of
\textit{unobservable} continuous \textbf{model parameters}, the
values of which are to be estimated probabilistically via
calculating a \textbf{posterior joint probability distribution}
from (i)~a discrete set of \textit{measured} \textbf{data} for a
finite number of relevant \textbf{statistical variables}, and
(ii)~a sensible~\textbf{prior joint probability distribution}
reflecting a given \textbf{state of knowledge} concerning the
possible ranges of values for the unobservable \textbf{model
parameters}. In some rather special lower-dimensional cases it is 
possible to obtain closed-form analytical (exact) solutions for 
posterior joint probability distributions. In general, however, 
posterior joint probability distributions prove to be of a
non-standard form due to inherent complexity which often features 
already at the two-parameter level. The aim of numerical
\textbf{Markov Chain Monte Carlo (MCMC)} simulations is to generate 
\textit{discretised} approximations to the continuous
high-dimensional \textbf{posterior joint probability 
distributions} for the unobservable \textbf{model parameters} to an 
accuracy that is reasonable for practical inference. Simulated
\textbf{posterior joint probability distributions}, and even more
so their associated \textbf{posterior marginal probability
distributions}, can then be summarised by standard methods such as
five number summaries, means, standard deviations and standard
errors, skewnesses, kurtoses, and further taylor-made statistics of
convenience.

\medskip
\noindent
To be recognised as meaningful and valuable by the scientific 
community, a proposed \textbf{statistical model} must cope well 
with two major challenges: (i)~\textbf{retrodiction} of observed 
(and, therefore, known to the researcher) data, and
(ii)~\textbf{prediction} of new (and, therefore, unknown to the
researcher) data. From a technical point of view this means
engineering an acceptable balance between \textbf{under-fitting}
and \textbf{over-fitting} when adapting a proposed
\textbf{statistical model} to available  empirical data. In this
process, a researcher can resort to methods of \textbf{model
comparison} by means of \textbf{information criteria}, and checks
of a model's \textbf{out-of-sample posterior predictive accuracy}.
These techniques are to be addressed in these lecture notes in
later chapters.

\medskip
\noindent
Input into the statistical model building process for explaining 
the variation of a \textbf{statistical variable}~$Y$ in dependence
on a set of $k \in \mathbb{N}$ predicting \textbf{independent 
variables}~$\{X_{1}, \ldots, X_{k}\}$, given background 
information~$I$, is a \textbf{prior joint probability distribution} 
$P(\theta_{0}, \ldots, \theta_{k}|I)$ for a set of typically
$k+1$~unknown \textbf{model parameters}~$\{\theta_{0}, \ldots, 
\theta_{k}\}$. On the basis of relevant measured
\textbf{quantitative--empirical data}~$\{y_{i}\}_{i=1, \ldots, n}$
of a finite sample size~$n$ (which is thus given and fixed, and
inherently amounts to incomplete information), the fundamental
objective is to deduce a \textbf{posterior joint probability
distribution} $P(\theta_{0}, \ldots, \theta_{k}|\{y_{i}\}_{i=1,
\ldots, n}, I)$ for these \textbf{model parameters}, employing the
logic of plausible reasoning according to the
\textbf{Bayes--Laplace approach}. This activity leads
to\footnote{To avoid cluttering of notation, here and in the
following we suppress conditioning on the (fixed) data for the
$k$~independent variables~$\{X_{1}, \ldots, X_{k}\}$.}
\be
\lb{eq:postprob}
\fbox{$\displaystyle
P(\theta_{0}, \ldots, \theta_{k}|\{y_{i}\}_{i=1, \ldots, n}, I) 
=
\frac{P(\{y_{i}\}_{i=1, \ldots, n}|\theta_{0}, \ldots, \theta_{k},
I)}{P(\{y_{i}\}_{i=1,\ldots,n}|I)}
\,P(\theta_{0}, \ldots, \theta_{k}|I) \ ,
$}
\ee
and is referred to as the \textbf{updating process};
subject-specific information available to a researcher is being
enlarged by learning from relevant quantitative--empirical data.
It is from  the \textbf{posterior joint probability distribution} 
that the researcher draws all relevant inferences concerning 
her/his research question, while, in parallel, acknowledging 
and quantifying overall \textbf{uncertainty} due to
\textbf{incomplete information}.

\medskip
\noindent
Multiple application of the product rule~(\ref{eq:productrule}) 
transforms the \textbf{prior joint probability distribution} for 
the $k+1$ unknown \textbf{model parameters}~$\{\theta_{0}, \ldots, 
\theta_{k}\}$ to the practically more convenient form
\be
P(\theta_{0}, \ldots, \theta_{k}|I)
= P(\theta_{0}|\theta_{1}, \ldots, \theta_{k}, I)
\times P(\theta_{1}|\theta_{2}, \ldots, \theta_{k}, I)
\times \ldots \times
P(\theta_{k}|I) \ .
\ee

\medskip
\noindent
Frequently, in fact  --- in particular in the context of numerical 
simulations --- the simplifying assumption of separability is 
introduced for the \textbf{prior joint probability distribution}
(reflecting the assumption of prior mutual logical independence of 
the model parameters), i.e., the product structure
\be
\lb{eq:jointpriorprod}
P(\theta_{0}, \ldots, \theta_{k}|I)
= P(\theta_{0}|I)
\times \ldots \times
P(\theta_{k}|I)
\ee
is employed. Such a choice, however, disregards potential non-zero 
bivariate \textbf{correlations} between the \textbf{model
parameters} which arise in generic situations. It can be justified, 
though, as expressing a researcher's \textbf{complete ignorance} as
to the existence and strengths of such correlations. The idea is
that exactly the quantitative--empirical data to be analysed will 
provide the clues necessary to make progress on the answer to this 
specific question.

\medskip
\noindent
We conclude this section by briefly reviewing two concepts that 
are used for assessing the quality of fit and the predictive 
accuracy of a \textbf{statistical model}. The so-called
\textbf{prior predictive probability distribution} for a single
datum~$y$, given a \textbf{prior joint probability distribution}
for a set of $k+1$ unknown continuous \textbf{model
parameters}~$\{\theta_{0}, \ldots, \theta_{k}\}$
and the relevant \textbf{single-datum likelihood function}, is
defined by (see, e.g., Gelman \textit{et al}
(2014)~\ct[Sec.~1.3]{geletal2014}, Andreon and Weaver
(2015)~\ct[Sec.~8.10]{andwea2015}, or Gill 
(2015)~\ct[Sec.~6.4]{gil2015})
\bea
\lb{eq:priorpred}
P(y|I)
& := & \int\cdots\int_{\theta_{j}\,\text{ranges}}
P(y, \theta_{0}, \ldots, \theta_{k}|I)\,
\mathrm{d}\theta_{0}\cdots\mathrm{d}\theta_{k} \nonumber \\
& = & \int\cdots\int_{\theta_{j}\,\text{ranges}}
\underbrace{P(y|\theta_{0}, \ldots, \theta_{k}, 
I)}_{\text{likelihood}}\,
\underbrace{P(\theta_{0}, \ldots, 
\theta_{k}|I)}_{\text{prior}}\,\mathrm{d}\theta_{0}\cdots
\mathrm{d}\theta_{k} \ .
\eea
Here, the single-datum likelihood function is weighted by the 
prior joint probability distribution for the model parameters and
then integrated over the entire range of the various
$\theta_{j}$-spectra. This operation amounts to \textit{averaging}
the \textbf{single-datum likelihood function} with the
\textbf{prior joint probability distribution} over the
$(k+1)$-dimensional \textbf{parameter space}.

\medskip
\noindent
The so-called \textbf{posterior predictive probability
distribution} for a new datum~$y_\mathrm{new}$, given a
\textbf{posterior joint probability distribution} for a set of
$k+1$ continuous \textbf{model parameters}~$\{\theta_{0}, \ldots,
\theta_{k}\}$ and the relevant \textbf{single-datum likelihood
function}, is defined by (see, e.g., Gelman 
\textit{et al} (2014)~\ct[Sec.~1.3]{geletal2014}, Andreon and 
Weaver (2015)~\ct[Sec.~8.10]{andwea2015}, or Gill 
(2015)~\ct[Sec.~6.4]{gil2015})
\bea
\lb{eq:postpred}
P(y_\mathrm{new}|\{y_{i}\}_{i=1, \ldots, n}\}, I)
& := & \int\cdots\int_{\theta_{j}\,\text{ranges}} P(y_\mathrm{new}, 
\theta_{0}, \ldots, \theta_{k}|\{y_{i}\}_{i=1, \ldots, n}\}, 
I)\,\mathrm{d}\theta_{0}\cdots\mathrm{d}\theta_{k} \nonumber \\
& = & \int\cdots\int_{\theta_{j}\,\text{ranges}}
P(y_\mathrm{new}|\theta_{0}, \ldots, \theta_{k}, \{y_{i}\}_{i=1,
\ldots, n}\}, I) \nonumber \\
& & \qquad\qquad\qquad \times
P(\theta_{0}, \ldots, \theta_{k}|\{y_{i}\}_{i=1, \ldots, n}, 
I)\,\mathrm{d}\theta_{0}\cdots\mathrm{d}\theta_{k} \nonumber \\
& = & \int\cdots\int_{\theta_{j}\,\text{ranges}}
\underbrace{P(y_\mathrm{new}|\theta_{0}, \ldots, \theta_{k}, 
I)}_{\text{likelihood}} \\
& & \qquad\qquad\qquad \times
\underbrace{P(\theta_{0}, \ldots, \theta_{k}|\{y_{i}\}_{i=1, 
\ldots, n}\}, I)}_{\text{posterior}}\,\mathrm{d}\theta_{0}\cdots
\mathrm{d}\theta_{k} \ , \nonumber 
\eea
assuming in the final step conditional logical independence of the 
new datum~$y_\mathrm{new}$ from the previous sample 
$\{y_{i}\}_{i=1, \ldots, n}$, given values for the model 
parameters. Here, the single-datum likelihood function is weighted 
by the posterior joint probability distribution for the model
parameters  and then integrated over the entire range of the
various $\theta_{j}$-spectra. It thus represents the expectation of
the conditional probability  $P(y_\mathrm{new}|\theta_{0}, \ldots,
\theta_{k}, I)$ over the  posterior joint probability distribution
for~$\{\theta_{0}, \ldots, \theta_{k}\}$. Alternatively, this
operation is viewed as \textit{averaging} the \textbf{single-datum
likelihood function} with the \textbf{posterior joint probability
distribution} over the $(k+1)$-dimensional \textbf{parameter
space}. Note that the \textbf{posterior predictive probability 
distribution} possesses a standard deviation that is larger than
for the \textbf{posterior joint probability distribution}, because
it joins \textbf{uncertainty} that is inherent \textit{both} in the
sampling of new \textbf{quantitative--empirical data} for~$Y$
\textit{and} in the estimation process for the unknown
\textbf{model parameters}.

\medskip
\noindent
We now turn to describe ways of capturing in formal language
different kinds of data-generating processes that are of
importance for many practical applications.

\chapter[Likelihood functions and sampling 
distributions]{Likelihood functions and sampling distributions}
\lb{ch2}
In \textbf{statistical modelling} a first fundamental assumption is 
to suppose that every single measured datum~$y_{i}$ for a
particular \textbf{statistical variable}~$Y$ of interest originates 
from a definite \textbf{data-generating process}. This
data-generating process is to be described parametrically by means
of a \textbf{single-datum likelihood function}, the specific form
of which depends on the actual nature of the statistical
variable~$Y$ in question: its scale level of measurement, and
whether its values vary discretely or continuously. Conceptually,
this \textbf{single-datum likelihood function} amounts to a
probability distribution for the single measured datum~$y_{i}$,
given fixed values for the parameters of the data-generating
process. A second fundamental assumption comprises the view that
when taking measurements with respect to~$Y$ from a total of
$n$~\textbf{sample units}, then the \textit{order} in which this
data was obtained would not matter. This second assumption
corresponds to de
Finetti's (1930)~\ct{def1930} concept of \textbf{exchangeability}.
This assumption is valid, if \textit{no} particular chronological
order is to be respected in the measurement process as, for
instance, needs to be taken care of when gathering \textbf{time
series data}. In technical language this is the requirement that
the \textbf{total-data likelihood function} for the entire data set
for~$Y$ be \textit{invariant} under permutations (re-ordering) of
the measured values for~$Y$. In practice, however, often an
even stronger assumption is built upon, namely that obtaining
\textit{one} value~$y_{i}$ for~$Y$ from some data-generating
process may be considered logically independent from obtaining a
\textit{second} value~$y_{j}$ from the same data-generating
process, and vice versa, so that in consequence (as it is referred
to) \textbf{independently and identically distributed~(iid)} data
arises; cf. Jaynes (2003)~\ct[p~62]{jay2003}, Gilboa
(2009)~\ct[p~42f]{gil2009}, 
Greenberg (2013)~\ct[p~52f]{gre2013}, Gelman \textit{et al} 
(2014)~\ct[p~104f]{geletal2014}, and Gill 
(2015)~\ct[Sec.~12.4]{gil2015}. The iid~assumption, which serves as
a practical convenience, may be justified as reflecting
\textbf{prior ignorance} on the part of the researcher as regards
potential \textbf{autocorrelations} amongst the measured
values of~$Y$;\footnote{Generally, data from convenience samples is  
plagued with a high degree of autocorrelation.} see McElreath 
(2020)~\ct[p~81]{mce2020a}.

\medskip
\noindent
We denote the \textbf{single-datum likelihood
function}\footnote{Alternatively: single-case likelihood function.}
for a \textbf{statistical variable}~$Y$ in a \textbf{statistical
model} comprising $k+1$~\textbf{model
parameters}~$\{\theta_{0}, \ldots, \theta_{k}\}$ by 
$f(y_{i}|\theta_{0}, \ldots, \theta_{k}, I)$. As it happens, 
assuming known and thus \textit{fixed} values for the
set~$\{\theta_{0}, \ldots, 
\theta_{k}\}$, for discretely varying~$y_{i}$ the single-datum
likelihood function represents a normalised probability function,
while for continuously varying~$y_{i}$ it represents a normalised
probability density function~(pdf). Viewed from this particular
angle, $f(y_{i}|\theta_{0}, \ldots, \theta_{k}, I)$ is referred to
as a \textbf{sampling distribution}. From the perspective of the
\textbf{Bayes--Laplace approach}, however, it is the
set~$\{\theta_{0}, \ldots, \theta_{k}\}$ that is considered unknown
and \textit{variable}, while the value of~$y_{i}$ is regarded as
known and \textit{fixed}. As $f(y_{i}|\theta_{0}, \ldots,
\theta_{k}, I)$ is typically \textit{not} normalised with respect
to the~$\{\theta_{0}, \ldots, \theta_{k}\}$, it is then referred to
as a \textbf{likelihood function}; cf.
Jaynes (2003)~\ct[p~89]{jay2003}, Sivia and Skilling
(2006)~\ct[p~80]{sivski2006}, or Lee (2012)~\ct[p~37]{lee2012}.

\medskip
\noindent
When the \textbf{iid assumption} appears sensible for describing a 
particular \textbf{data-generating process} for~$Y$, then for a
given total sample of measured values
$\left\{y_{i}\right\}_{i=1,\ldots,n}$ of size~$n$ an immediate 
consequence for the \textbf{total-data likelihood function} is 
proportionality to the product of $n$~single-datum likelihood 
functions~$f(y_{i}|\theta_{0}, \ldots, \theta_{k}, I)$, 
i.e.,
\be
\lb{eq:likefact}
P(\left\{y_{i}\right\}_{i=1,\ldots,n}|\theta_{0}, \ldots, 
\theta_{k}, I) \propto \prod_{i=1}^{n}f(y_{i}|\theta_{0}, \ldots, 
\theta_{k}, I) \ .
\ee

\medskip
\noindent
In the following we will review some of the standard univariate and 
multivariate \textbf{single-datum likelihood functions} for both 
discretely and continuously varying statistical variables~$Y$. Most
of the examples presented belong to the class of \textbf{maximum
entropy distributions} that reflect for a specific context, and 
conditional on some set of definite constraints, maximum
\textbf{ignorance} on the part of the researcher as to the unknown
actual data-generating process; see, e.g., McElreath 
(2020)~\ct[p~314]{mce2020a}.

\section[Univariate discrete data]{Univariate discrete data}
\lb{sec:likelidiscr}
The single-datum likelihood functions introduced in this section 
apply to univariate data~$y_{i}$ for a discrete one-dimensional 
\textbf{statistical variable}~$Y$. They depend on one or two
unobservable and therefore unknown continuously varying parameters.

\subsection[Bernoulli distributions]{Bernoulli distributions}
\lb{subsec:berndistr}
The one-parameter family of univariate \textbf{Bernoulli 
distributions},
\be
\left.y_{i}\right|\theta, I \sim \mathrm{Bern}(\theta) \ ,
\ee
was put forward by the Swiss mathematician 
\href{https://mathshistory.st-andrews.ac.uk/Biographies/Bernoulli_Jacob/}{Jakob Bernoulli (1655--1705)}. It can be used to model 
data-generating processes in which, in a single observation, 
$y_{i}$~has two possible outcomes: ``failure''~($0$) or 
``success''~($1$). This could be, for example,
\begin{itemize}

\item whether or not a student gets accepted for the degree
programme she/he had applied for,

\item whether or not a bank customer is granted the mortgage loan 
she/he had asked for,

\item whether or not it will rain tomorrow at your present
location, or

\item whether or not your favourite football team will win their
next league match.

\end{itemize}
$y_{i}$ here is a dimensionless quantity. Properties of
\textbf{Bernoulli distributions} are (see, e.g., Rinne 
(2008)~\ct[Subsec.~3.8.2]{rin2008}):

\medskip
\noindent
Spectrum of values:
\be
y_{i} \in \left\{0, 1\right\} \ .
\ee
Probability function:
\be
\lb{eq:bernprob}
\fbox{$\displaystyle
P(y_{i}|\theta, I)
= \theta^{y_{i}}\,(1-\theta)^{1-y_{i}} \ , 
\quad\text{with}\quad 0 \leq \theta \leq 1 \ ,
$}
\ee
where the dimensionless parameter~$\theta$ quantifies the 
\textbf{probability for ``success,''~($1$)}. The graph of a
Bernoulli probability function is shown in
Fig.~\ref{fig:bernoulliprob} below for four different values
for~$\theta$.

\medskip
\noindent
Expectation value and variance:
\bea
\mathrm{E}(y_{i}) & = & \theta \\
\mathrm{Var}(y_{i}) & = & \theta(1-\theta) \ .
\eea
\begin{figure}[!htb]
\begin{center}
\fbox{\includegraphics[width=14cm]{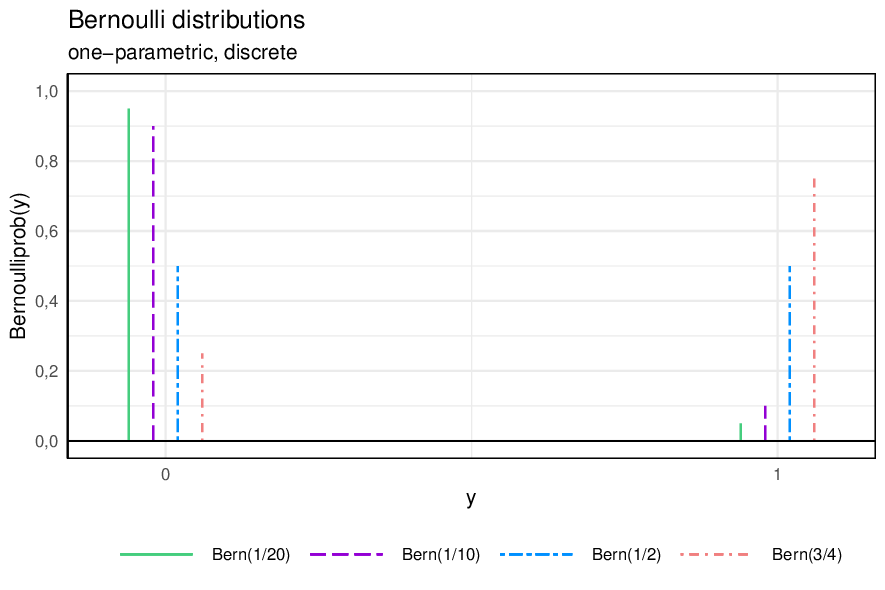}}
\end{center}
\caption{Four examples of Bernoulli sampling distributions for an 
uncertain dichotomous discrete quantity~$y$.}
\lb{fig:bernoulliprob}
\end{figure}
Note that under an exchange $\theta \leftrightarrow (1-\theta)$ 
one obtains a qualitatively identical distribution.

\medskip
\noindent
\underline{\R:}
$\texttt{dbinom}(y_{i}, 1, \theta)$,
$\texttt{pbinom}(y_{i}, 1, \theta)$,
$\texttt{qbinom}(p, 1, \theta)$,
$\texttt{rbinom}(n_{\mathrm{simulations}}, 1, \theta)$\\
\underline{Stan:} Cf. \href{https://mc-stan.org/docs/functions-reference/bernoulli-distribution.html}{Stan Functions Reference
(v2.30)} \ct{sta2022b}
\begin{itemize}
\item $\texttt{bernoulli}(\theta)$ (sampling)
\item $\texttt{bernoulli\_lpmf}( y | \theta )$ (log-sampling)
\item $\texttt{bernoulli\_rng}(\theta)$ (generating)
\end{itemize}
\underline{JAGS:}
$\texttt{dbern}(\theta)$ (sampling)

\subsection[Binomial distributions]{Binomial distributions}
\lb{subsec:binomdistr}
The natural extension of Bernoulli distributions to situations 
with a finite number of repetitions under iid~conditions of the 
underlying binary decision process was discussed by Bernoulli 
himself. He introduced the two-parameter family of univariate 
\textbf{binomial distributions},
\be
\left.y\right|n, \theta, I \sim \mathrm{Bin}(n, \theta) \ ,
\ee
where $n \in \mathbb{N}$ is the \textbf{number of iid-repetitions}, 
and~$\theta$ again denotes the \textbf{probability for 
``success,''~($1$)}. The dimensionless non-negative integer 
quantity~$y$ is by its very nature varying discretely and
represents a pure \textbf{count} with a \textit{known} finite
maximum. It could measure, for example,
\begin{itemize}

\item how many out of $n$~students get accepted for the degree 
programme they had applied for,

\item how many out of $n$~bank customers are granted the mortgage
loan they had asked for,

\item on how many out of the next $n$~days it will rain at your 
present location, or

\item how many out of $n$~upcoming matches in the league your 
favourite football team will win.

\end{itemize}
\textbf{Binomial distributions} are described by (see, e.g., Rinne 
(2008)~\ct[Subsec.~3.8.3]{rin2008}):

\medskip
\noindent
Spectrum of values:
\be
y \in \mathbb{N}_{0} \ .
\ee
Probability function:
\be
\lb{eq:binomprob}
\fbox{$\displaystyle
P(y|n, \theta, I)
= \left(
\begin{array}{c}
n \\
y
\end{array}
\right)
\theta^{y}\,(1-\theta)^{n-y} \ , 
\quad\text{with}\quad 0 \leq \theta \leq 1 \ ,
$}
\ee
wherein the \textbf{binomial coefficient} is defined by
\be
\lb{eq:bincoeff}
\left(
\begin{array}{c}
n \\
y
\end{array}
\right)
:= \frac{n!}{y!(n-y)!} \ ,
\ee
for $n \in \mathbb{N}$ and $y \leq n$. Note that the binomial
probability function is normalised with respect to the discrete 
variable~$y$ but \textit{not} with respect to the continuous 
parameter~$\theta$. Its graph is shown in Fig.~\ref{fig:binomprob} 
below for four different values for~$\theta$ and
$n = 20$.\footnote{We plot the graphs of probability functions with
connecting lines to highlight the shapes of enveloping curves. The
probability functions are, of course, discrete by nature.}

\medskip
\noindent
Expectation value and variance:
\bea
\mathrm{E}(y) & = & n\theta \\
\mathrm{Var}(y) & = & n\theta(1-\theta) \ .
\eea
\begin{figure}[!htb]
\begin{center}
\fbox{\includegraphics[width=14cm]{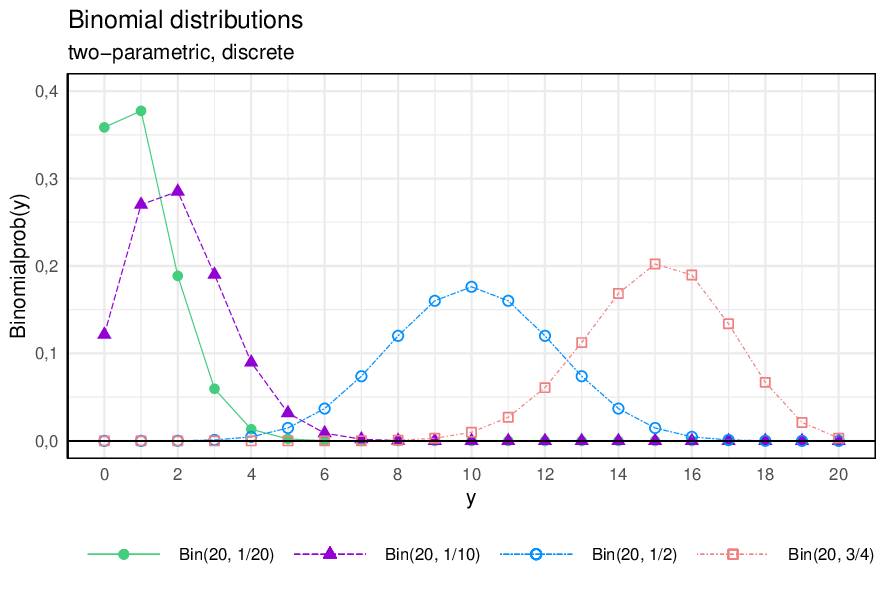}}
\end{center}
\caption{Four examples of binomial sampling distributions for an 
uncertain discrete quantity~$y$.}
\lb{fig:binomprob}
\end{figure}
A qualitatively identical distribution is obtained under the 
exchange $\theta \leftrightarrow (1-\theta)$.

\medskip
\noindent
\underline{\R:}
$\texttt{dbinom}(y, n, \theta)$,
$\texttt{pbinom}(y, n, \theta)$,
$\texttt{qbinom}(p, n, \theta)$,
$\texttt{rbinom}(n_{\mathrm{simulations}}, n, \theta)$\\
\underline{Stan:} Cf. \href{https://mc-stan.org/docs/functions-reference/binomial-distribution.html}{Stan Functions Reference (v2.30)}
\ct{sta2022b}
\begin{itemize}
\item $\texttt{binomial}( n , \theta )$ (sampling)
\item $\texttt{binomial\_lpmf}( y | n , \theta )$ (log-sampling)
\item $\texttt{binomial\_rng}( n , \theta )$ (generating)
\end{itemize}
\underline{JAGS:}
$\texttt{dbin}(\theta, n)$ (sampling)

\medskip
\noindent
It is of some practical interest that, by means of the logistic 
transformation $\theta = \exp(u) / (1 + \exp(u))$, binomial 
sampling distributions according to Eq.~(\ref{eq:binomprob}) can 
explicitly be shown to be members of the so-called
\textbf{exponential family} of sampling distributions; see
Sec.~\ref{sec:expfam} below. 

\subsection[Poisson distributions]{Poisson distributions}
\lb{subsec:poisdistr}
The one-parameter family of univariate \textbf{Poisson
distributions},
\be
\left.y_{i}\right|\theta, I \sim \mathrm{Pois}(\theta) \ ,
\ee
named after the French mathematician, engineer, and physicist 
\href{https://mathshistory.st-andrews.ac.uk/Biographies/Poisson/}{Baron
Sim\'{e}on Denis Poisson FRSFor HFRSE MIF (1781--1840)}, is 
the most important tool for modelling \textbf{count data} with an
\textit{unknown} maximum. They can be considered to arise as
special cases of binomial distributions when $n$ is very large
($n \gg 1$) and $\theta$ is very small ($0 < \theta \ll 1$)
(cf. Sivia and Skilling (2006)~\ct[Sec.~5.4]{sivski2006}), and 
so typically describe instances of data-generating proccesses 
associated with (relatively) rare events. Examples for~$y_{i}$ as 
a pure count (i.e., a dimensionless non-negative integer) are
\begin{itemize}

\item the number of automobiles sold by a car vendor,

\item the number of goals scored by a football team,

\item the number of elephants living in certain parts of eastern 
or southern Africa, or

\item the number of photons received from a faint distant luminous 
source by an astronomical telescope.

\end{itemize}
\textbf{Poisson distributions} have the properties (see, 
e.g., Rinne (2008)~\ct[Subsec.~3.9.2]{rin2008}):

\medskip
\noindent
Spectrum of values:
\be
y_{i} \in \mathbb{N}_{0} \ .
\ee
Probability function:
\be
\lb{eq:poisprob}
\fbox{$\displaystyle
P(y_{i}|\theta, I)
= \frac{\theta^{y_{i}}}{y_{i}!}\,\exp\left(-\theta\right) \ , 
\quad\text{with}\quad \theta \in \mathbb{R}_{\geq 0} \ ,
$}
\ee
and~$\theta$ is the dimensionless \textbf{rate parameter} (also 
referred to as the intensity parameter). The Poisson probability 
function is normalised with respect to the discrete 
variable~$y_{i}$ but \textit{not} with respect to the continuous 
parameter~$\theta$. Its graph is shown in Fig.~\ref{fig:poisprob}
for four different values for~$\theta$.

\medskip
\noindent
Expectation value and variance:
\bea
\mathrm{E}(y_{i}) & = & \theta \\
\mathrm{Var}(y_{i}) & = & \theta \ .
\eea
Note that for \textbf{Poisson distributions} the (dimensionless) 
expectation value and variance \textit{coincide}.
\begin{figure}[!htb]
\begin{center}
\fbox{\includegraphics[width=14cm]{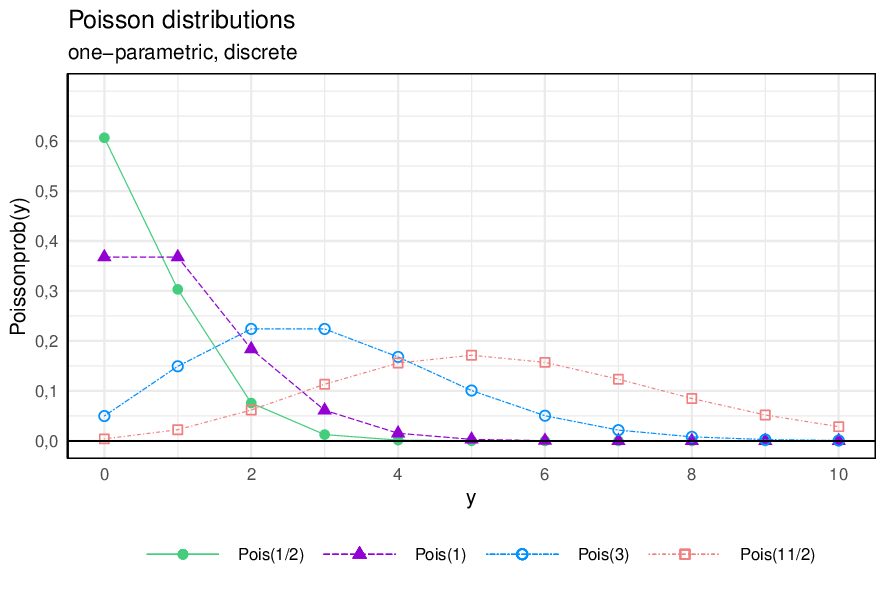}}
\end{center}
\caption{Four examples of Poisson sampling distributions for an 
uncertain discrete quantity~$y$.}
\lb{fig:poisprob}
\end{figure}

\medskip
\noindent
\underline{\R:}
$\texttt{dpois}(y_{i}, \theta)$,
$\texttt{ppois}(y_{i}, \theta)$,
$\texttt{qpois}(p, \theta)$,
$\texttt{rpois}(n_{\mathrm{simulations}}, \theta)$\\
\underline{Stan:} Cf. \href{https://mc-stan.org/docs/functions-reference/poisson.html}{Stan Functions Reference (v2.30)} \ct{sta2022b}
\begin{itemize}
\item $\texttt{poisson}(\theta)$ (sampling)
\item $\texttt{poisson\_lpmf}( y | \theta )$ (log-sampling)
\item $\texttt{poisson\_rng}(\theta)$ (generating)
\end{itemize}
\underline{JAGS:}
$\texttt{dpois}(\theta)$ (sampling)

\medskip
\noindent
In many applications one finds the expectation value
$\mathrm{E}(y_{i})$ for the count variable~$Y$ decomposed into 
a product of two dimensionful quantities,
\be
\text{mean\ count} = \text{exposure} \times \text{rate}
\qquad\Rightarrow\qquad
\theta = \tau \times \lambda \ ,
\ee
so that the \textbf{rate parameter}~$\theta$ amounts to the product 
``length/size of interval/domain of observation, $\tau$, times 
number of events per unit interval/domain, $\lambda$.'' In 
temporal contexts $\lambda$ represents counts per unit time, while 
in spatial contexts is stands for counts per unit length, counts 
per unit area,
or counts per unit volume.
In this view, the parameter $\tau$ is referred to as the
\textbf{exposure} (of a sample unit to some data-generating
influence) and carries the physical dimension of $[\text{time}]$, 
$[\text{length}]$,  $[\text{area}]$, or $[\text{volume}]$. The 
corresponding rate parameter~$\lambda$ could represent, for 
example,
\begin{itemize}

\item the average number of automobiles sold by a car vendor per 
working day,

\item the average number of goals scored by a football team per 
match,

\item the average number of elephants living per 
ten-kilometres-squared of area in the 
\href{http://www.krugerpark.co.za/}{Kruger National Park}, or

\item the average number of elliptical galaxies observed per 
megaparsec-cubed of comoving volume of space at a redshift of 
$z=0.5$.

\end{itemize}
%


\section[Univariate continuous data]{Univariate continuous data}
\lb{sec:likelidiscr}
The single-datum likelihood functions introduced in this section 
apply to univariate data~$y_{i}$ for a continuous one-dimensional 
\textbf{statistical variable}~$Y$. They depend on a certain number
of unobservable and therefore unknown continuously varying
parameters.

\subsection[Gau\ss\ distributions]{Gau\ss\ distributions}
\lb{subsec:gaussdistr}
The two-parameter family of univariate \textbf{Gau\ss
\ distributions} (or normal distributions),
\be
\left.y_{i}\right|\theta_{1}, \theta_{2}, I
\sim \mathrm{N}(\theta_{1}, \theta_{2}^{2}) \ ,
\ee
has gained its status as ranking amongst the best-known and most 
frequently applied continuous distributions foremost by the work 
of the German mathematician and astronomer 
\href{https://mathshistory.st-andrews.ac.uk/Biographies/Gauss/}{Carl
Friedrich Gau\ss\ (1777--1855)}; cf. Gau\ss\ 
(1809)~\ct{gau1809}. Examples for the usually dimensionful 
continuous metrical quantity~$y_{i}$ that can be described as 
arising from a Gau\ss\ process are
\begin{itemize}

\item the price of a $1~\mathrm{kg}$ loaf of bread in a
medium-sized town of your home country,

\item the average monthly waiting time in minutes spent by car 
users in traffic jams during the morning rush hour near an 
industrial centre,

\item the IQ of an adult female or male individual, or

\item the wavelength in nanometres of the red line in the visible
hydrogen emission spectrum.

\end{itemize}
Often one finds the natural logarithm of a strictly positive 
metrical statistical variable~$Y$, after properly normalising the 
latter via division by a convenient reference quantity of the 
same physical dimension, to be describable as approximately 
originating from a Gau\ss\ data-generating process. Note that
departures of  Gau\ss-distributed data from their common mean by
more than three  standard deviations are very rare, and by more
than six standard  deviations are practically impossible.

\medskip
\noindent
\textbf{Gau\ss\ distributions} have the properties (see, 
e.g., Rinne (2008)~\ct[Subsec.~3.10.1]{rin2008}):

\medskip
\noindent
Spectrum of values:
\be
y_{i} \in \mathbb{R} \ .
\ee
Probability density function (\texttt{pdf}):
\be
\lb{eq:gausspdf}
\fbox{$\displaystyle
f(y_{i}|\theta_{1}, \theta_{2}, I) = 
\frac{1}{\sqrt{2\pi}\,\theta_{2}}\,\exp\left[\,-\frac{1}{2}
\left(\frac{y_{i}-\theta_{1}}{\theta_{2}}\right)^{2}\,\right] \ , 
\quad\text{with}\quad
\theta_{1} \in \mathbb{R} \ , \ \theta_{2} \in \mathbb{R}_{>0} \ ;
$}
\ee
$\theta_{1}$ constitutes a \textbf{location parameter} and 
$\theta_{2}$ a \textbf{scale parameter}, both of which share the 
physical dimension of~$y_{i}$ itself. The reciprocal of the 
squared scale parameter, $1/\theta_{2}^{2}$, is conventionally 
referred to as the \textbf{precision}. Note that the Gau\ss\ 
probability density function is normalised with respect to the 
continuous variable~$y_{i}$ but \textit{not} with respect to the 
continuous parameters~$\theta_{1}$ and~$\theta_{2}$. Its graph is 
shown in Fig.~\ref{fig:gausspdf} for four different combinations 
of values for~$\theta_{1}$ and~$\theta_{2}$.

\medskip
\noindent
Expectation value and variance:
\bea
\mathrm{E}(y_{i}) & = & \theta_{1} \\
\mathrm{Var}(y_{i}) & = & \theta_{2}^{2} \ .
\eea
\begin{figure}[!htb]
\begin{center}
\fbox{\includegraphics[width=14cm]{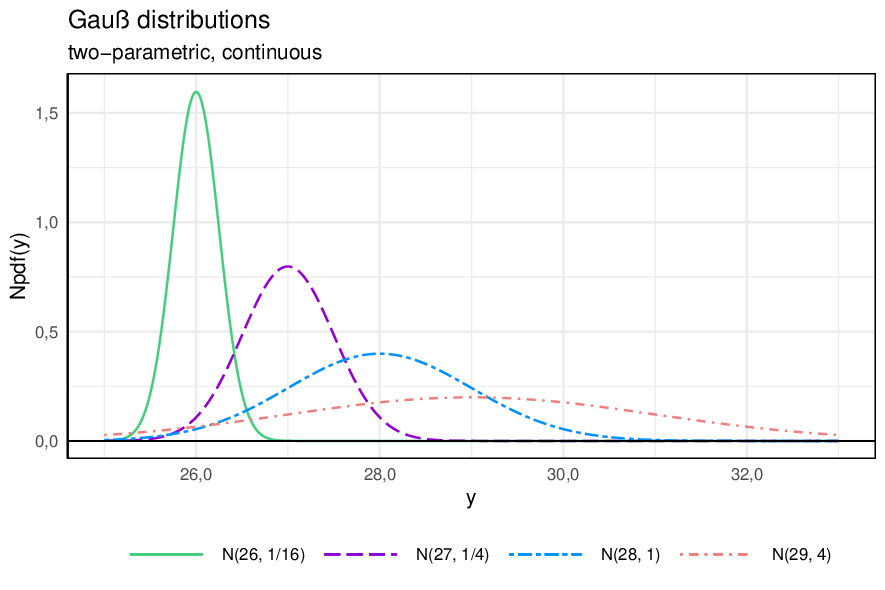}}
\end{center}
\caption{Four examples of Gau\ss\ sampling distributions for 
an uncertain continuous quantity~$y$.}
\lb{fig:gausspdf}
\end{figure}

\medskip
\noindent
\underline{\R:}
$\texttt{dnorm}(y_{i}, \theta_{1}, \theta_{2})$,
$\texttt{pnorm}(y_{i}, \theta_{1}, \theta_{2})$,
$\texttt{qnorm}(p, \theta_{1}, \theta_{2})$,
$\texttt{rnorm}(n_{\mathrm{simulations}}, \theta_{1}, \theta_{2})$
\\
\underline{Stan:} Cf. \href{https://mc-stan.org/docs/functions-reference/normal-distribution.html}{Stan Functions Reference (v2.30)}
\ct{sta2022b}
\begin{itemize}
\item $\texttt{normal}( \theta_{1} , \theta_{2} )$ (sampling)
\item $\texttt{normal\_lpdf}( y | \theta_{1} , \theta_{2} )$
(log-sampling)
\item $\texttt{normal\_rng}( \theta_{1} , \theta_{2} )$
(generating)
\end{itemize}
\underline{JAGS:}
$\texttt{dnorm}(\theta_{1}, 1/\theta_{2}^{2})$ (sampling)

\subsection[Non-central $t$--distributions]{Non-central 
$\boldsymbol{t}$--distributions}
\lb{subsec:tdistr}
The three-parameter family of \textbf{non-central}
$\boldsymbol{t}${\bf--distributions},
\be
\left.y_{i}\right|\theta_{1}, \theta_{2}, \nu, I
\sim t(\theta_{1}, \theta_{2},\nu) \ ,
\ee
constitutes a generalisation of a well-known family of continuous 
probability distributions discovered by the English statistician 
\href{https://mathshistory.st-andrews.ac.uk/Biographies/Gosset/}{William
Sealy Gosset (1876--1937)}. Profoundly confusing for the 
scientific community, he published his findings under the 
pseudonym of ``Student;'' cf. Student (1908)~\ct{stu1908}. While 
being qualitatively similar to Gau\ss\ distributions, their main 
characteristic is the larger propability weight contained in the 
``tails'' of the distributions. Therefore, \textbf{non-central}
$\boldsymbol{t}${\bf--distributions} are being employed to model 
data-generating processes for usually dimensionful 
continuous metrical quantities~$y_{i}$ which regularly produce 
\textbf{outliers}. Specific features of \textbf{non-central} 
$\boldsymbol{t}${\bf--distributions} are (see, e.g., Rinne 
(2008)~\ct[Subsec.~3.10.6]{rin2008}):

\medskip
\noindent
Spectrum of values:
\be
y_{i} \in \mathbb{R} \ .
\ee
Probability density function (\texttt{pdf}):
\be
\lb{eq:tpdf}
\fbox{$\displaystyle
f(y_{i}|\theta_{1}, \theta_{2}, \nu, I)
= \frac{\Gamma\left[(\nu+1)/2\right]}{\Gamma\left(\nu/2\right)
\sqrt{\pi\nu}\,\theta_{2}}\,\left[\,1 + 
\frac{1}{\nu}\left(\frac{y_{i}-\theta_{1}}{\theta_{2}}\right)^{2}\,
\right]^{-(\nu+1)/2} \ ,
\quad\text{with}\quad \nu \in \mathbb{R}_{\geq 1} \ ;
$}
\ee
%
$\theta_{1} \in \mathbb{R}$ represents a \textbf{location
parameter} and $\theta_{2} \in \mathbb{R}_{>0}$ a \textbf{scale
parameter}, both of which share the physical dimension of~$y_{i}$,
and $\nu \geq 1$ is the dimensionless positive \textbf{degrees of
freedom parameter}. All three parameters are continuous. The
\textbf{Gamma function} used above is defined via an Euler integral
of the second kind by (see, e.g., Rinne (2008)~\ct[p~168]{rin2008})
\be
\lb{eq:gammafct}
\Gamma(x) := \int_{0}^{\infty}t^{x-1}\,\exp(-t)\,\mathrm{d}t \ ,
\quad\text{with}\quad x \in \mathbb{R}_{\geq 0} \ .
\ee
For later application it is important to note that for positive 
integer values of $x$, i.e., $x = n \in \mathbb{N}$, it holds true 
that
\be
\lb{eq:gammaid}
\Gamma(n+1) = n! \ .
\ee
The graph of the non-central $t$--probability density function is 
shown in Fig.~\ref{fig:tpdf} for four different combinations of 
values for~$\theta_{1}$, $\theta_{2}$ and~$\nu$. Gosset's 
one-parameter family of standard $t$--distributions is contained 
in Eq.~(\ref{eq:tpdf}) for the special parameter choices 
$\theta_{1} = 0$ and $\theta_{2} = 1$.

\medskip
\noindent
Expectation value and variance (cf. Greenberg 
(2013)~\ct[p~230]{gre2013}):
\bea
\lb{eq:texpv}
\mathrm{E}(y_{i}) & = & \theta_{1} \ ,
\quad\text{if}\quad \nu > 1 \\
\lb{eq:tvar}
\mathrm{Var}(y_{i}) & = & \frac{\nu}{\nu-2}\,\theta_{2}^{2} \ ,
\quad\text{if}\quad \nu > 2 \ .
\eea
\begin{figure}[!htb]
\begin{center}
\fbox{\includegraphics[width=14cm]{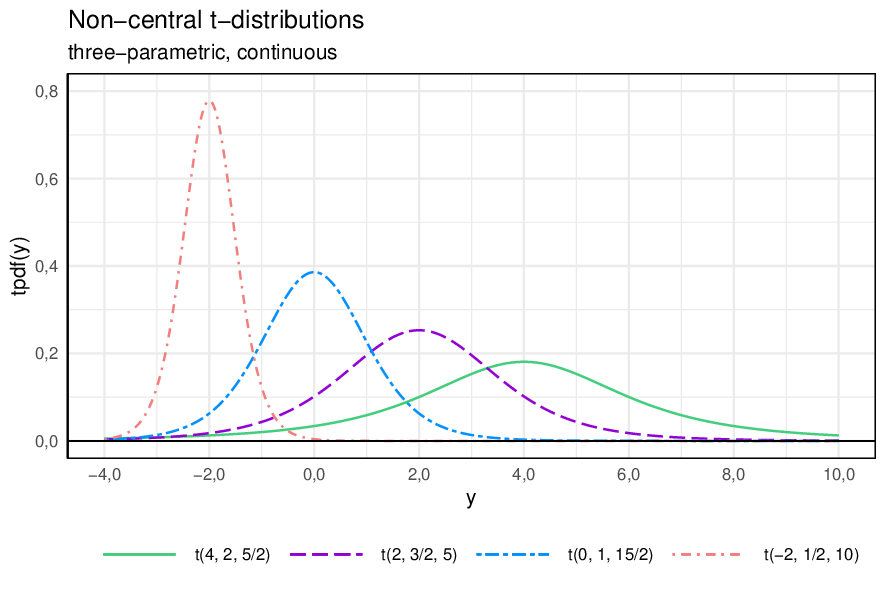}}
\end{center}
\caption{Four examples of non-central $t$--sampling distributions  
for an uncertain continuous quantity~$y$.}
\lb{fig:tpdf}
\end{figure}

\medskip
\noindent
\underline{\R:}
$(1/\theta_{2})*\texttt{dt}( (y_{i}+\theta_{1})/\theta_{2} , \nu )$,
$(1/\theta_{2})*\texttt{pt}( (y_{i}+\theta_{1})/\theta_{2} , \nu )$\\
\underline{Stan:} Cf. \href{https://mc-stan.org/docs/functions-reference/student-t-distribution.html}{Stan Functions Reference (v2.30)}
\ct{sta2022b}
\begin{itemize}
\item $\texttt{student\_t}( \nu , \theta_{1} , \theta_{2} )$
(sampling)
\item $\texttt{student\_t\_lpdf}( y | \nu , \theta_{1} ,
\theta_{2} )$ (log-sampling)
\item $\texttt{student\_t\_rng}( \nu , \theta_{1} , \theta_{2} )$
(generating)
\end{itemize}
\underline{JAGS:}
$\texttt{dt}(\theta_{1}, 1/\theta_{2}^{2}, \nu)$ (sampling)

\medskip
\noindent
In the limit $\nu \to +\infty$, non-central $t$--distributions 
asymptote towards Gau\ss\ distributions. In actual practical 
situations, differences between the two kinds of distributions 
become effectively irrelevant when $\nu \geq 50$, in which case 
Gau\ss\ distributions may be used to simplify computations.

\subsection[Exponential distributions]{Exponential distributions}
\lb{subsec:expdistr}
The one-parameter family of \textbf{exponential distributions},
\be
\left.y_{i}\right|\theta, I \sim \mathrm{Exp}(\theta) \ ,
\ee
is regularly employed in modelling data-generating processes for
waiting times or spatial distances. For example, the generically
dimensionful continuous metrical quantity~$y_{i}$ may represent
\begin{itemize}

\item the lifetime in months of a fashion hype,

\item the distance in kilometres a commuter travels from their 
home to their workplace near an industrial centre,

\item the time in minutes until the next incoming telephone call 
in a call centre, or

\item the lifetime in seconds of a rainbow.

\end{itemize}
Main properties of \textbf{exponential distributions} are (see, 
e.g., Rinne (2008)~\ct[Subsec.~3.9.3]{rin2008}):

\medskip
\noindent
Spectrum of values:
\be
y_{i} \in \mathbb{R}_{\geq 0} \ .
\ee
Probability density function (\texttt{pdf}):
\be
\lb{eq:exppdf}
\fbox{$\displaystyle
f(y_{i}|\theta, I)
= \theta\,\exp\left(-\theta y_{i}\right) \ , 
\quad\text{with}\quad \theta \in \mathbb{R}_{>0} \ ,
$}
\ee
and $\theta$ represents a \textbf{rate parameter} of physical 
dimension inverse to~$y_{i}$. Note that the exponential 
probability density function is normalised with respect to the 
continuous variable~$y_{i}$ but \textit{not} with respect to the 
continuous parameter~$\theta$. Its graph is shown in 
Fig.~\ref{fig:exppdf} for four different values for~$\theta$. 

\medskip
\noindent
Expectation value and variance:
\bea
\mathrm{E}(y_{i}) & = & \frac{1}{\theta} \\
\mathrm{Var}(y_{i}) & = & \frac{1}{\theta^{2}} \ .
\eea
\begin{figure}[!htb]
\begin{center}
\fbox{\includegraphics[width=14cm]{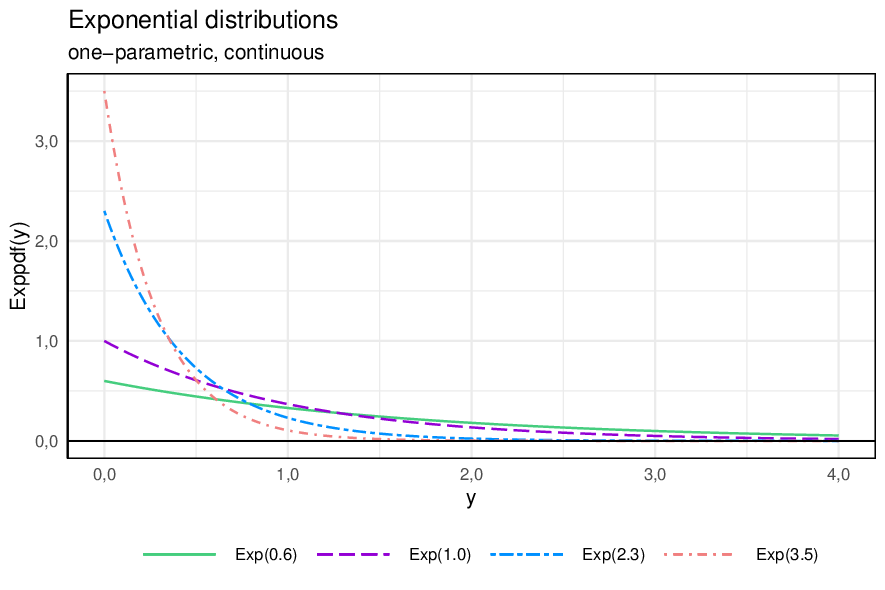}}
\end{center}
\caption{Four examples of exponential sampling distributions 
for an uncertain continuous quantity~$y$.}
\lb{fig:exppdf}
\end{figure}

\medskip
\noindent
\underline{\R:}
$\texttt{dexp}(y_{i}, \theta)$,
$\texttt{pexp}(y_{i}, \theta)$,
$\texttt{qexp}(p, \theta)$,
$\texttt{rexp}(n_{\mathrm{simulations}}, \theta)$\\
\underline{Stan:} Cf. \href{https://mc-stan.org/docs/functions-reference/exponential-distribution.html}{Stan Functions Reference (v2.30)}
\ct{sta2022b}
\begin{itemize}
\item $\texttt{exponential}(\theta)$ (sampling)
\item $\texttt{exponential\_lpdf}( y | \theta )$ (log-sampling)
\item $\texttt{exponential\_rng}(\theta)$ (generating)
\end{itemize}
\underline{JAGS:}
$\texttt{dexp}(\theta)$ (sampling)

\medskip
\noindent
Exponential distributions constitute a special case of the
two-parameter family of \textbf{Gamma distributions} (cf. Greenberg 
(2013)~\ct[p~225]{gre2013}), which will be introduced 
in Subsec.~\ref{subsec:gammadistr} below. We remark in passing
that in fixed effects and varying effects generalised linear models
(see Ch.~\ref{ch7} and Ch.~\ref{ch9}) exponential
distributions, which represent a certain type of maximum entropy
distribution, often serve as (weakly or strongly regularising)
prior distributions for scale parameters~$\theta_{2}$ of
Gau\ss\ likelihood functions, or for degree-of-freedom
parameters~$\nu$ of $t$--likelihood functions; cf. McElreath
(2020)~\ct[p~407]{mce2020a}, and Krusch\-ke 
(2015)~\ct[p~462]{kru2015}.

\subsection[Pareto distributions]{Pareto distributions}
\lb{subsec:paretodistr}
The two-parameter family of univariate \textbf{Pareto
distributions},
\be
\left.y_{i}\right|\theta, y_{\mathrm{min}}, I \sim \mathrm{Par}(
\theta, y_{\mathrm{min}}) \ ,
\ee
was introduced, to Economics in the first place, by the Italian 
engineer, sociologist, economist, political scientist 
and philosopher
\href{http://en.wikipedia.org/wiki/Vilfredo_Pareto}{Vilfredo
Federico Damaso Pareto (1848--1923)}; cf. Pareto 
(1896)~\ct{par1896}. The usually dimensionful continuous 
positive quantity~$y_{i}$ could represent, for example,
\begin{itemize}

\item the annual revenue of a company listed at the 
\href{https://www.nyse.com/index}{New York Stock Exchange},

\item the number of clicks attracted by a video on 
\href{https://www.youtube.com/}{YouTube} that was watched at least 
once,

\item the number of books sold by a writer in a given year, or

\item the mass of a galaxy cluster.

\end{itemize}
\textbf{Pareto distributions} possess the features (see, e.g.,
Rinne (2008)~\ct[Subsec.~3.11.7]{rin2008}):

\medskip
\noindent
Spectrum of values:
\be
y_{\mathrm{min}} \leq y_{i} \in \mathbb{R}_{> 0} \ .
\ee
Probability density function (\texttt{pdf}):
\be
\lb{eq:paretopdf}
\fbox{$\displaystyle
f(y_{i}|\theta, y_{\mathrm{min}}, I)
= \frac{\theta}{y_{\mathrm{min}}}
\left(\frac{y_{\mathrm{min}}}{y_{i}}\right)^{\theta+1} \ , 
\quad\text{with}\quad \theta \in \mathbb{R}_{>0} \ ;
$}
\ee
$\theta$ constitutes a dimensionless \textbf{scale parameter} and
$y_{\mathrm{min}}$ a \textbf{location parameter} of the same
physical dimension as~$y_{i}$. Note that the Pareto probability 
density function is normalised with respect to the continuous 
variable~$y_{i}$ but \textit{not} with respect to the continuous 
parameters~$\theta$ and $y_{\mathrm{min}}$. Its graph is shown in 
Fig.~\ref{fig:Paretopdf} for four different combinations of values 
for~$\theta$ and~$y_{\mathrm{min}}$.

\medskip
\noindent
Expectation value and variance:
\bea
\mathrm{E}(y_{i}) & = & \frac{\theta}{\theta-1}\,y_{\mathrm{min}}
\qquad\text{for}\quad \theta > 1 \\
\mathrm{Var}(y_{i}) & = & \frac{\theta}{(\theta-1)^{2}(\theta-2)}\,
y_{\mathrm{min}}^{2}
\qquad\text{for}\quad \theta > 2 \ .
\eea
\begin{figure}[!htb]
\begin{center}
\fbox{\includegraphics[width=14cm]{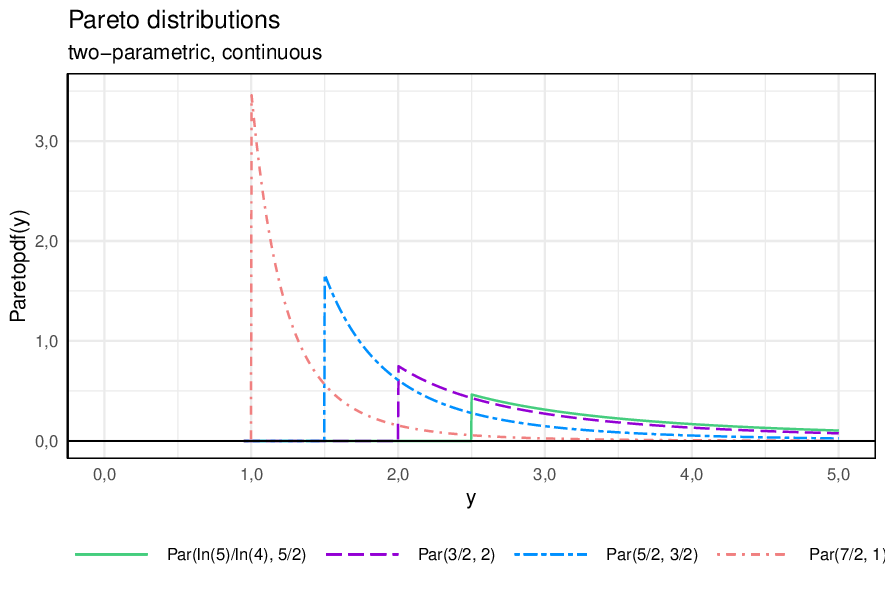}}
\end{center}
\caption{Four examples of Pareto sampling distributions 
for an uncertain continuous quantity~$y$.}
\lb{fig:Paretopdf}
\end{figure}

\medskip
\noindent
\underline{\R:}
$\texttt{dpareto}(y_{i}, \theta, y_{\mathrm{min}})$,
$\texttt{ppareto}(y_{i}, \theta, y_{\mathrm{min}})$,
$\texttt{qpareto}(p, \theta, y_{\mathrm{min}})$,
$\texttt{rpareto}(n_{\mathrm{simulations}}, \theta,
y_{\mathrm{min}})$ (\texttt{extraDistr} package, by Wolodzko
(2020)~\ct{wol2020})\\
\underline{Stan:} Cf. \href{https://mc-stan.org/docs/functions-reference/pareto-distribution.html}{Stan Functions Reference (v2.30)}
\ct{sta2022b}
\begin{itemize}
\item $\texttt{pareto}( y_{\mathrm{min}} , \theta )$ (sampling)
\item $\texttt{pareto\_lpdf}( y | y_{\mathrm{min}} , \theta )$
(log-sampling)
\item $\texttt{pareto\_rng}( y_{\mathrm{min}} , \theta )$
(generating)
\end{itemize}
\underline{JAGS:}
$\texttt{dpar}(\theta, y_{\mathrm{min}})$ (sampling)


\section[Multivariate data]{Multivariate data}
\lb{sec:likelimulti}
The single-datum likelihood functions introduced in this section 
apply to multivariate data~$\boldsymbol{y}$ from a continuous 
vector-valued, $m$-dimensional \textbf{statistical 
variable}~$\boldsymbol{Y}$. They depend on unobservable and 
therefore unknown continuously varying scalar-, vector- and 
matrix-valued parameters. We will here briefly review only the two 
most frequently used multivariate single-datum likelihood 
functions for vector-valued continuously varying data.

\subsection[Multivariate Gau\ss\ distributions]{Multivariate 
Gau\ss\ distributions}
\lb{subsec:multigaussdistr}
Multivariate Gau\ss\ processes are described by a single-datum 
likelihood function for a vector-valued~$\boldsymbol{y} \in 
\mathbb{R}^{m \times 1}$ given by (see, e.g., 
Rinne (2008)~\ct[Subsec.~3.10.4]{rin2008}, Gelman \textit{et al} 
(2014)~\ct[Sec.~3.5]{geletal2014}, or Gill 
(2015)~\ct[Sec.~3.5]{gil2015})
\be
\lb{eq:mvgausspdf}
\fbox{$\displaystyle
f(\boldsymbol{y}|\boldsymbol{\mu}, \boldsymbol{\Sigma}, I)
= \frac{1}{\sqrt{(2\pi)^{m}\det(\boldsymbol{\Sigma})}}\,
\exp\left[\,-\frac{1}{2}\,(\boldsymbol{y}-\boldsymbol{\mu})^{T}
\boldsymbol{\Sigma}^{-1}(\boldsymbol{y}-\boldsymbol{\mu})\,\right]
\ ,
$}
\ee
wherein $\boldsymbol{\mu} \in \mathbb{R}^{m \times 1}$ represents 
a \textbf{mean vector} of the same physical dimension
as~$\boldsymbol{y}$, and $\boldsymbol{\Sigma} \in \mathbb{R}^{m 
\times m}$ a regular \textbf{covariance matrix} of the squared
physical dimension of~$\boldsymbol{y}$ which is always
symmetric and positive semi-definite.

\medskip
\noindent
\underline{\R:}
$\texttt{dmvnorm}(\boldsymbol{y}, \boldsymbol{\mu}, 
\boldsymbol{\Sigma})$,
$\texttt{rmvnorm}(n_{\mathrm{simulations}}, \boldsymbol{\mu}, 
\boldsymbol{\Sigma})$ (\texttt{mvtnorm} package, by Genz \textit{et
al} (2021)~\ct{genetal2021})\\
\underline{Stan:} Cf. \href{https://mc-stan.org/docs/functions-reference/multivariate-normal-distribution.html}{Stan Functions Reference
(v2.30)} \ct{sta2022b}
\begin{itemize}
\item $\texttt{multi\_normal}( \boldsymbol{\mu} ,
\boldsymbol{\Sigma} )$ (sampling)
\item $\texttt{multi\_normal\_lpdf}( \boldsymbol{y} |
\boldsymbol{\mu} , \boldsymbol{\Sigma} )$ (log-sampling)
\item $\texttt{multi\_normal\_rng}( \boldsymbol{\mu} ,
\boldsymbol{\Sigma} )$ (generating)
\end{itemize}
\underline{JAGS:}
$\texttt{dmnorm(mu[1:m], Omega[1:m, 1:m])}$ (precision matrix:
\texttt{Omega}) (sampling)

\subsection[Multivariate non-central 
$t$--distributions]{Multivariate non-central
$t$--distributions}
\lb{subsec:multitdistr}
The generalisation of the three-parameter non-central 
$t$--distribution discussed in Subsec.~\ref{subsec:tdistr} to the 
multivariate case is given by the single-datum likelihood function 
for a vector-valued~$\boldsymbol{y} \in \mathbb{R}^{m \times 1}$ 
(see, e.g., Gelman \textit{et al} (2014)~\ct[Tab.~A.1]{geletal2014})
\be
\lb{eq:mvtpdf}
\fbox{$\displaystyle
f(\boldsymbol{y}|\boldsymbol{\mu}, \boldsymbol{\Sigma}, \nu, I)
= \frac{\Gamma\left[(\nu+m)/2\right]}{\Gamma\left(\nu/2\right)
(\pi\nu)^{m/2}\sqrt{\det(\boldsymbol{\Sigma})}}\,\left[\,1 + 
\frac{1}{\nu}\,(\boldsymbol{y}-\boldsymbol{\mu})^{T}
\boldsymbol{\Sigma}^{-1}(\boldsymbol{y}-\boldsymbol{\mu})\,
\right]^{-(\nu+m)/2} \ ,
$}
\ee
where $\boldsymbol{\mu} \in \mathbb{R}^{m \times 1}$ is a
\textbf{mean vector} of the same physical dimension
as~$\boldsymbol{y}$, $\boldsymbol{\Sigma} \in \mathbb{R}^{m \times
m}$ is a regular symmetric and positive semi-definite
\textbf{covariance matrix} of the squared physical dimension
of~$\boldsymbol{y}$, and $\nu \geq 1$ is the positive dimensionless
\textbf{degrees of freedom parameter}.

\medskip
\noindent
\underline{\R:}
$\texttt{dmvt}(\textit{(data vector)}, \textit{(ncp vector)}, 
\textit{(scale matrix)}, \nu)$, \\
$\texttt{rmvt}(n_\mathrm{simulations}, \textit{(ncp vector)}, 
\textit{(scale matrix)}, \nu)$ (\texttt{mvtnorm} package, by Genz
\textit{et al} (2021)~\ct{genetal2021})\\
\underline{Stan:} Cf. \href{https://mc-stan.org/docs/functions-reference/multivariate-student-t-distribution.html}{Stan Functions Reference
(v2.30)} \ct{sta2022b}
\begin{itemize}
\item $\texttt{multi\_student\_t}( \nu , \boldsymbol{\mu} ,
\boldsymbol{\Sigma} )$ (sampling)
\item $\texttt{multi\_student\_t\_lpdf}( \boldsymbol{y} |
\nu , \boldsymbol{\mu} , \boldsymbol{\Sigma} )$ (log-sampling)
\item $\texttt{multi\_student\_t\_rng}( \nu , \boldsymbol{\mu} ,
\boldsymbol{\Sigma} )$ (generating)
\end{itemize}
\underline{JAGS:}
$\texttt{dmt(mu[1:m], Omega[1:m, 1:m], nu)}$ (precision matrix:
\texttt{Omega}) (sampling)

\section[Exponential family]{Exponential family}
\lb{sec:expfam}
It is of some practical interest to realise that each of the 
binomial, Poisson, Gau\ss\ and exponential distributions belong to 
a larger class of probability distributions referred to as the 
\textbf{exponential family}; this was first discussed by Fisher 
(1935)~\ct{fis1935a}. These are particularly important as they 
can be used to quantitatively model data-generating processes for 
a wide spectrum of observable natural phenomena in a comprehensive 
fashion. It can be shown that each member of this family 
constitutes a \textbf{maximum entropy probability distribution},
given specific constraints corresponding to available information
in the different contexts wherein they appear; cf. McElreath 
(2020)~\ct[p~7]{mce2020a}, and Sec.~\ref{sec:maxent} below.

\medskip
\noindent
In a \textbf{statistical model} which aims to capture the
distributional features of a univariate \textbf{statistical
variable}~$Y$ by employing a set of $k+1$~\textbf{model
parameters}~$\{\theta_{0}, \ldots, \theta_{k}\}$ , the
\textbf{total-data likelihood function} for members of the
\textbf{exponential family} exhibits the general structure 
(cf. Lee (2012)~\ct[Sec.~2.11]{lee2012}, Gelman \textit{et al} 
(2014)~\ct[Sec.~2.4]{geletal2014}, Gill 
(2015)~\ct[Subsec.~4.3.2]{gil2015}, or McElreath 
(2020)~\ct[Sec.~9.2.]{mce2020a})
\bea
\lb{eq:expfam}
P(\{y_{i}\}_{i=1,\ldots,n}|\theta_{0}, \ldots, \theta_{k}, I)
& = & \left[\,\prod_{i=1}^{n}r(y_{i})\,\right]
s^{n}(\theta_{0}, \ldots, \theta_{k}) \nonumber \\
& & \qquad\qquad\times
\exp\left[\,\boldsymbol{u}^{T}(\theta_{0}, \ldots, 
\theta_{k})\cdot\sum_{i=1}^{n}\boldsymbol{t}(y_{i})\,\right] \ ,
\eea
with, in general, vector-valued factors $\boldsymbol{u} \in 
\mathbb{R}^{(k+1) \times 1}$ and $\boldsymbol{t} \in 
\mathbb{R}^{(k+1) \times 1}$. In the special one-parameter case, 
$k=0$, both of these reduce to scalars. The vector-valued quantity 
$\displaystyle\sum_{i=1}^{n}\boldsymbol{t}(y_{i}) \in 
\mathbb{R}^{(k+1) \times 1}$ is referred to as a \textbf{sufficient 
statistic} for the set of model parameters $\{\theta_{0}, \ldots, 
\theta_{k}\}$, as in the \textbf{total-data likelihood function}
the latter interact with the quantitative-empirical data
$\{y_{i}\}_{i=1,\ldots,n}$ only 
via the former. If a \textbf{prior probability distribution}
possesses the \textit{same} structure as the \textbf{total-data
likelihood function} given in Eq.~(\ref{eq:expfam}) and so will
generate a \textbf{posterior probability distribution} belonging
to its own family, then it is said to be of the \textbf{conjugate}
type; cf. Sec.~\ref{sec:conjpriors} below.

\medskip
\noindent
We now turn to discuss in the next chapter prior probability 
distributions, which serve to model initial states of knowledge of 
a researcher concerning the range of plausible values of a single
parameter in specific empirical situations of enquiry.

\chapter[Prior probability distributions]{Prior probability 
distributions}
\lb{ch3}
It lies at the very heart of the methodological philosophy
of the \textbf{Bayes--Laplace approach to data analysis and
statistical inference} that unknown quantities such as
\textbf{parameters} in \textbf{statistical models} are treated probabilistically by assigning to them probability distributions
that represent a \textbf{state of knowledge} on the part of the
researcher as to their plausible ranges of values. Therefore, there
is an immediate necessity in \textbf{statistical modelling} to
specify a \textbf{prior joint
probability distribution} for all \textit{unknown} \textbf{model
parameters}. This mode of action is to be viewed as a
mathematical formalisation of including \textit{all} available
background information~$I$ on a matter of interest, such as
obtained from related past data analyses, scientific discourse, or
even from personal prejudices; cf. Coles
(2006)~\ct[p~61]{col2006}. The latter option provides the
psychological basis for many people to associate with the
\textbf{Bayes--Laplace approach} the notion of ``subjective
probabilities,'' although such a view neglects some deep
epistemological issues.

\medskip
\noindent
\textbf{Prior probability distributions} represent a researcher's
\textbf{state of knowledge} \textit{before} gaining access to
relevant observational or experimental data on the problem under 
investigation. They can be broadly classified into one of three 
qualitative categories, ranked according to information content:
\begin{itemize}
\item uninformative prior probability distributions,

\item weakly informative prior probability distributions, and

\item sceptical prior probability distributions, resp. regularising
prior probability distributions.

\end{itemize}

\medskip
\noindent
For practical reasons, and as an expression of typical
\textbf{ignorance} of a researcher of parameter correlations prior
to data analysis, it is often assumed  that a \textbf{prior joint
probability distribution} for multiple \textbf{model parameters} 
factorises into a product of \textbf{single-parameter prior
probability distributions};\footnote{Single-parameter prior
probability distributions treat a single parameter in a
model-building process probabilistically prior to data analysis.
These distributions themselves depend generically on further
parameters, which may be specified as fixed, or as adaptive to 
additional information input.} cf. Eq.~(\ref{eq:jointpriorprod})
and the remarks made in Sec.~\ref{sec:datamodel}.

\medskip
\noindent
In the following we will review the cases of
\textbf{single-parameter prior probability distributions} that are
most important for actual practical model-building, and how some of
them can be motivated
conceptually. The different options that will be outlined offer
a sufficient amount of flexibility in that they let a researcher
express a diverse range of prior \textbf{states of knowledge}, from
uninformed to sceptical, by tuning accordingly the free parameters
in the various probability distributions employed. We will begin by
addressing formal ways of specifying a \textbf{state of complete
ignorance} as a \textbf{reference point} for prior probability
distributions.

\section[Principle of indifference]{Principle of indifference}
\lb{sec:indif}
Suppose given a set of $k \in \mathbb{N}$ \textbf{mutually
exclusive and exhaustive propositions} $\{A_{1}, \ldots, A_{k}\}$, 
conditioned on background information $I$, so that 
$P(A_{i}A_{j}|I) = 0$ for $i \neq j$, and $i, j = 1, \ldots, k$.
If $I$ provides \textit{no} reason to assign a higher plausibility
to any one proposition in the set than to any other, thus
expressing a \textbf{state of complete ignorance}, then Keynes
(1921)~\ct[p~41]{key1921} suggested the only consequential
probability assignment could be
\be
\lb{eq:discreteUniformprob}
\fbox{$\displaystyle
P(A_{i}|I) = \frac{1}{k} \ , \qquad i = 1, \ldots, k \ .
$}
\ee
Originally, this approach was introduced by the Swiss mathematician
\href{https://mathshistory.st-andrews.ac.uk/Biographies/Bernoulli_Jacob/}{Jakob Bernoulli (1655--1705)}, who referred to it as the
``principle of non-sufficient reason;'' Keynes
(1921)~\ct[p~41]{key1921} himself preferred to call it the
\textbf{principle of indifference}, which is the term that spread
in the literature.

\medskip
\noindent
The assignment~(\ref{eq:discreteUniformprob}) yields univariate
\textbf{discrete uniform distributions} for sets of
propositions~$\{A_{1}, \ldots, A_{k}\}$ that are properly
normalised, in line with Eq.~(\ref{eq:normcomplpart}). The graph of
the probability function is shown in
Fig.~\ref{fig:discreteUniformprob} below for four different 
values for~$k$.
\begin{figure}[!htb]
\begin{center}
\fbox{\includegraphics[width=14cm]{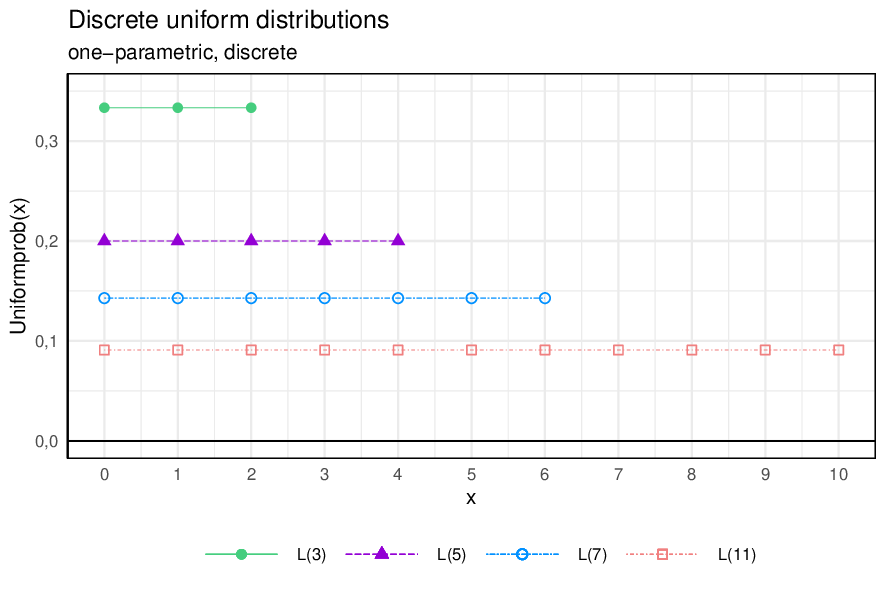}}
\end{center}
\caption{Four examples of discrete uniform distributions for an 
uncertain integer quantity~$x$.}
\lb{fig:discreteUniformprob}
\end{figure}

\medskip
\noindent
\underline{\R:} $\texttt{ddunif}(x, 1, k)$,
$\texttt{pdunif}(x, 1, k)$, 
$\texttt{qdunif}(\alpha, 1, k)$,
$\texttt{rdunif}(n_{\mathrm{simulations}}, 1, k)$ (package:
{\tt extraDistr}, by Wolodzko (2020)~\ct{wol2020})

\section[Transformation invariance]{Transformation invariance}
\lb{sec:transfinv}
A different method for establishing in mathematical terms a
\textbf{state of complete ignorance} has been elucidated by Jaynes
(2003)~\ct[Subsec.~12.4.1]{jay2003}, and by Sivia and Skilling
(2006)~\ct[Subsec.~5.1.2]{sivski2006}. Here the requirement imposed
on probability distributions for model parameters is that they
remain invariant under \textbf{transformations} of the model
parameters. We will now address the two simplest examples of
transformation-invariant single-parameter distributions:
\begin{itemize}

\item[(i)] Let $\rho \in \mathbb{R}$ be a continuous
\textbf{location parameter}. When invariance is demanded under a
shift of the parameter's origin, i.e., a \textbf{translation}
$\rho \mapsto \rho + a$, for a constant $a \in \mathbb{R}$, then
the condition
\be
P(\rho|I)\,\mathrm{d}\rho \stackrel{!}{=}
P(\rho+a|I)\,\mathrm{d}(\rho+a)
\quad\Rightarrow\quad
P(\rho|I)\,\mathrm{d}\rho \stackrel{!}{=}
P(\rho+a|I)\,\mathrm{d}\rho
\ee
needs to be solved to determine an adequate form for~$P(\rho|I)$.
The general solution is given by
\be
\lb{eq:flatprior1}
\fbox{$\displaystyle
P(\rho|I) = \text{constant} \ ,
$}
\ee
which expresses uniformity of~$P(\rho|I)$, irrespective of the
value of~$\rho$. To obtain a properly normalised \textbf{continuous
uniform distribution} satisfying Eq.~(\ref{eq:normcont}) below,
additional information as to the range of $\rho \in \left[a,
b\right] \subset \mathbb{R}$ needs to be injected, provided it is
available. In that case one obtains
\be
\lb{eq:continuousUniformpdf}
\left.\rho\right|a, b, I \sim \mathrm{U}(a, b) \ ,
\qquad\qquad
P(\rho|a,b,I) = \frac{1}{(b-a)} \ .
\ee
The graph of this probability density function is shown in
Fig.~\ref{fig:continuousUniformpdf} for four different 
combinations of values for~$a$ and~$b$.

\begin{figure}[!htb]
\begin{center}
\fbox{\includegraphics[width=14cm]{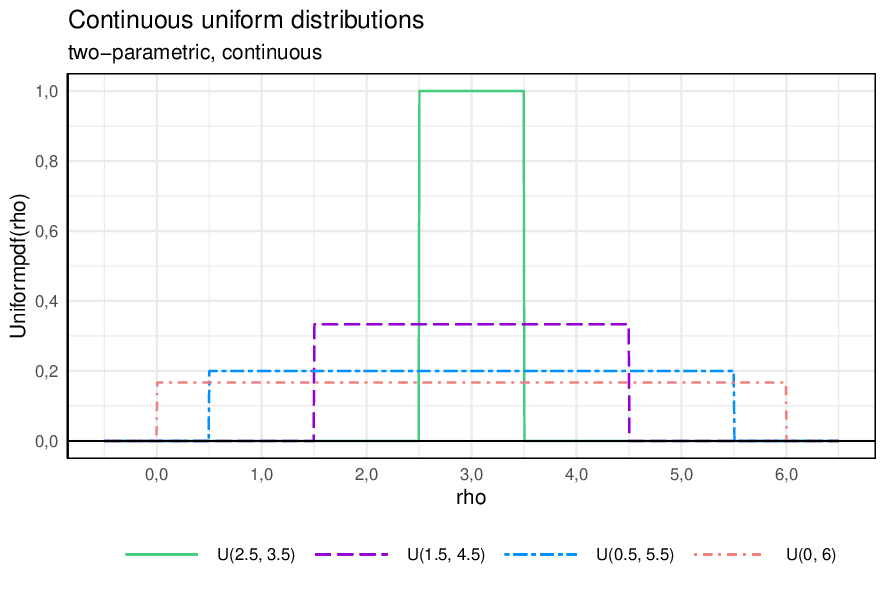}}
\end{center}
\caption{Four examples of uniform distributions for a
continuous location parameter~$\rho$.}
\lb{fig:continuousUniformpdf}
\end{figure}

\medskip
\noindent
\underline{\R:}
$\texttt{dunif}(\rho, a, b)$,
$\texttt{punif}(\rho, a, b)$,
$\texttt{qunif}(\alpha, a, b)$,
$\texttt{runif}(n_{\mathrm{simulations}}, a, b)$\\
\underline{Stan:} Cf. \href{https://mc-stan.org/docs/functions-reference/uniform-distribution.html}{Stan Functions Reference (v2.30)} \ct{sta2022b}
\begin{itemize}
\item $\texttt{uniform}( a , b )$ (sampling)
\item $\texttt{uniform\_lpdf}( \rho | a , b )$ (log-sampling)
\item $\texttt{uniform\_rng}( a , b )$ (generating)
\end{itemize}
\underline{JAGS:}
$\texttt{dunif}( a , b )$ (sampling)

\item[(ii)] Let $\ell \in \mathbb{R}_{>0}$ be a continuous
positive \textbf{scale parameter}. When invariance is demanded
under a change of the parameter's size, i.e., a \textbf{re-scaling}
$\ell \mapsto \beta\ell$, for a positive constant $\beta \in
\mathbb{R}_{>0}$, then the condition
\be
P(\ell|I)\,\mathrm{d}\ell \stackrel{!}{=}
P(\beta\ell|I)\,\mathrm{d}(\beta\ell)
\quad\Rightarrow\quad
P(\ell|I)\,\mathrm{d}\ell \stackrel{!}{=}
P(\beta\ell|I)\,\beta\mathrm{d}\ell
\ee
needs to be solved to determine an adequate form for~$P(\ell|I)$.
The general solution is given by
\be
\lb{eq:flatprior2}
\fbox{$\displaystyle
P(\ell|I) = \frac{\text{constant}}{\ell} \ ,
$}
\ee
which is generally referred to as a \textbf{Jeffreys prior}; cf.
Jeffreys (1961)~\ct[pp~117--122]{jef1939}, Sivia and Skilling
(2006~\ct[p~109]{sivski2006}, or Gill
(2015)~\ct[Subsec.~4.4.2]{gil2015}. To obtain a properly normalised
probability distribution satisfying Eq.~(\ref{eq:normcont}) below,
additional information as to the range of $\ell \in \left[a,
b\right] \subset \mathbb{R}_{> 0}$ needs to be injected, provided
it is available. In that case one obtains a \textbf{truncated
Jeffreys distribution} given by
\be
\lb{eq:Jeffreyspdf}
\left.\ell\right|a, b, I \sim \mathrm{Jeff}(a, b) \ ,
\qquad\qquad
P(\ell|a,b,I) = \frac{1}{\ln(b/a)}\left(\frac{1}{\ell}\right) \ .
\ee
The graph of this probability density function is shown in
Fig.~\ref{fig:Jeffreyspdf} for four different combinations of
values for~$a$ and~$b$.

\begin{figure}[!htb]
\begin{center}
\fbox{\includegraphics[width=14cm]{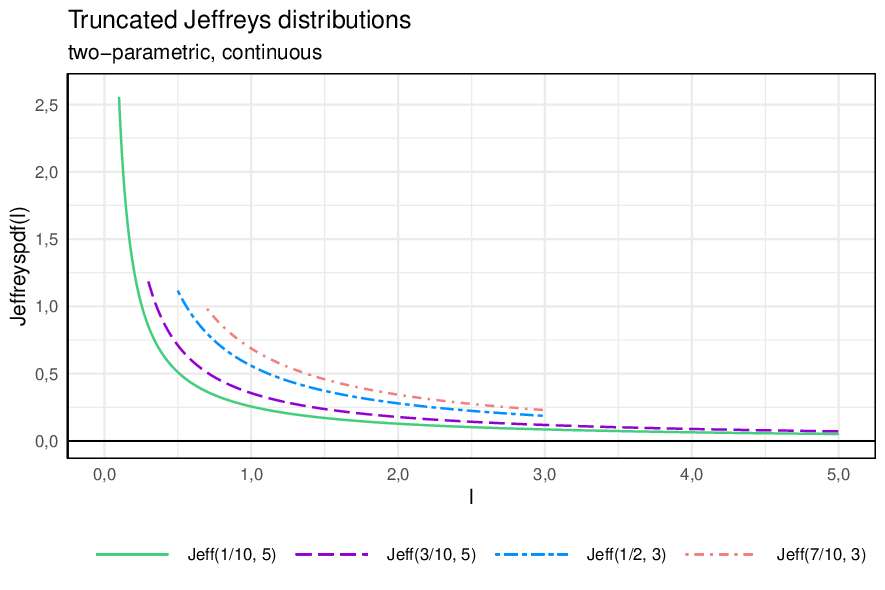}}
\end{center}
\caption{Four examples of truncated Jeffreys distributions for a
continuous positive scale parameter~$\ell$.}
\lb{fig:Jeffreyspdf}
\end{figure}

\end{itemize}
%

\section[Principle of maximum entropy]{Principle of maximum 
entropy}
\lb{sec:maxent}
The most sophisticated technical procedure for systematically
converting relevant background information~$I$ into usable
specific \textbf{prior probability distributions}
has been proposed by Jaynes (1957)~\ct{jay1957} through his
\textbf{principle of maximum entropy}; see also Jaynes
(2003)~\ct[Ch.~11]{jay2003}.

\medskip
\noindent
In this procedure he employs the notion of an \textbf{information
entropy} associated with a specific \textbf{probability
distribution} for a set of $k \in \mathbb{N}$ mutually exclusive
and exhaustive propositions~$\{A_{1}, \ldots, A_{k}\}$, conditioned
on background information $I$, that was developed by the
US-American mathematician, electrical engineer, and
cryptographer
\href{https://mathshistory.st-andrews.ac.uk/Biographies/Shannon/}{Claude
Elwood Shannon (1916--2001)}; see Shannon (1948)~\ct{sha1948}. This
is defined by\footnote{The minus sign preceding the expression on
the right-hand side of Eq.~(\ref{eq:entropy}) ensures for the
information entropy~$S$ a spectrum of non-negative values.}
\be
\lb{eq:entropy}
S := -\sum_{i=1}^{k}P(A_{i}|I)\,\ln\left(
\frac{P(A_{i}|I)}{m_{i}}\right) \ .
\ee
As Jaynes (2003)~\ct[p~358]{jay2003} suggests, it may be
interpreted as a measure of the ``amount of \textbf{uncertainty}''
represented by a \textbf{probability distribution}. His
modification of the information entropy
formula~(\ref{eq:entropy}) by a normalised \textbf{Lebesgue
measure}\footnote{Named after the French mathematician
\href{https://mathshistory.st-andrews.ac.uk/Biographies/Lebesgue/}{Henri
L\'{e}on Lebesgue (1875--1941)}.} $m_{i}$, $i = 1, \ldots, k$,
keeps this non-negative quantity invariant under
re-parametrisations of the set of propositions~$\{A_{1}, \ldots,
A_{k}\}$; cf. Jaynes (2003)~\ct[Sec.~12.3]{jay2003} and Sivia and
Skilling (2006)~\ct[p~116]{sivski2006}.

\medskip
\noindent
To maximise the \textbf{information entropy} of the probability
assignment for a given set of propositions~$\{A_{1}, \ldots,
A_{k}\}$ and pertinent background information~$I$, Jaynes devises a
\textbf{variational principle} for a scalar-valued \textbf{Lagrange
function}\footnote{Named after the an Italian mathematician and
astronomer \href{https://mathshistory.st-andrews.ac.uk/Biographies/Lagrange/}{Joseph--Louis Lagrange (1736--1813)}.} that is a
linear combination of the \textbf{information
entropy}~(\ref{eq:entropy}) itself, the \textbf{normalisation
condition}~(\ref{eq:normcomplpart}), and a set of
$l \in \mathbb{N}$ further \textbf{constraints} $0 = C_{i}$,
$i = 1, \ldots, l$, each of which represents testable pertinent
information~$I$. These constraints usually depend on the unknown
probabilities~$P(A_{i}|I)$. Setting $p_{i} := P(A_{i}|I)$ to
simplify notation, the \textbf{Lagrange function} is given
by\footnote{Here the signs of the second and third terms are
motivated by computational convenience.}
\be
L = -\sum_{i=1}^{k}p_{i}\ln\left(\frac{p_{i}}{m_{i}}\right)
- \lambda_{0}\left(\sum_{i=1}^{k}p_{i}-1\right)
- \sum_{i=1}^{l}\lambda_{i}\,C_{i} \ ;
\ee
the unknown coefficients $\lambda_{0}$ and $\lambda_{i}$ are
referred to as \textbf{Lagrange multipliers}.

\medskip
\noindent
To attain a maximum for the \textbf{information
entropy}~(\ref{eq:entropy}), the unknowns $p_{i}$, $\lambda_{0}$
and $\lambda_{i}$ must necessarily satisfy the system of $k+l+1$
non-linear algebraic equations given by
\bea
\lb{eq:maxentcond1}
0 & \stackrel{!}{=} & \frac{\partial L}{\partial p_{j}}
\ = \ -\,1 - \ln\left(\frac{p_{j}}{m_{j}}\right) - \lambda_{0}
-  \sum_{i=1}^{l}\lambda_{i}\,\frac{\partial 
C_{i}}{\partial p_{j}} \ ,
\qquad j = 1, \ldots, k \ , \\
\lb{eq:maxentcond2}
0 & \stackrel{!}{=} & \frac{\partial L}{\partial\lambda_{0}}
\ = \ \sum_{i=1}^{k}p_{i}-1 \\
\lb{eq:maxentcond3}
0 & \stackrel{!}{=} & \frac{\partial L}{\partial\lambda_{j}}
\ = \ C_{j} \ ,
\qquad j = 1, \ldots, l \ .
\eea
%
%
The general solution to condition~(\ref{eq:maxentcond1}) is given
by
\be
p_{j} = m_{j}\,e^{-(1+\lambda_{0})}\,\exp\left[\,-\,\sum_{i=1}^{l}
\lambda_{i}\,\frac{\partial C_{i}}{\partial p_{j}}\,\right] \ ,
\qquad j = 1, \ldots, k \ ,
\ee
while Eqs.~(\ref{eq:maxentcond2}) and~(\ref{eq:maxentcond3}) 
serve to enforce the normalisation condition and the 
$l$~constraints on the~$p_{i}$. Viewed from a qualitative
perspective, it turns out that, amongst all competitors, those
\textbf{probability distributions} attain maximum
\textbf{information entropy} which spread out probability as evenly
as possible between the given propositions, while fully
incorporating the available background information by respecting
all the given constraints. The extremisation procedure outlined
aims at rendering a probability distribution as uniform as
possible, in the sense of the principle of indifference. However,
the more testable information is available, the more non-uniform
the resultant probability distribution will become.

\medskip
\noindent
We point the interested reader to Sivia and Skilling 
(2006)~\ct[Sec.~5.3]{sivski2006} for specific applications of the
\textbf{principle of maximum entropy}. Representing for a discrete
resp.~continuous statistical variable~$Y$ its \textbf{expectation
value} and \textbf{variance} as constraints by
\bea
\lb{eq:expectconstr}
0 = C_{1} = \sum_{i=1}^{k}y_{i}p_{i} - \mu \ ,
\ & \text{resp.} & \ \mbox{}
0 = C_{1} = \int_{-\infty}^{+\infty}y f(y)\,\mathrm{d}y - \mu \ ,
\\
\lb{eq:varconstr}
0 = C_{2} = \sum_{i=1}^{k}(y_{i}-\mu)^{2}p_{i} - \sigma^{2} \ ,
\ & \text{resp.} & \ \mbox{}
0 = C_{2} = \int_{-\infty}^{+\infty}(y-\mu)^{2} f(y)\,\mathrm{d}y
- \sigma^{2} \ ,
\eea
and giving the \textbf{normalisation condition} and the
\textbf{information entropy} for the continuous case as
\bea
\lb{eq:normcont}
0 & = & \int_{-\infty}^{+\infty}f(y)\,\mathrm{d}y - 1 \\
\lb{eq:entropycont}
S & = & -\int_{-\infty}^{+\infty}f(y)\,\ln\left(
\frac{f(y)}{m(y)}\right)\mathrm{d}y \ ,
\eea
these authors demonstrate how some standard \textbf{probability
distributions} for discrete and continuous~$Y$ arise as
\textbf{maximum entropy distributions}. In particular, combining
the \textbf{information entropy}~(\ref{eq:entropy})
or~(\ref{eq:entropycont}) with
\begin{itemize}

\item[(i)] the normalisation conditions~(\ref{eq:normcomplpart})
or~(\ref{eq:normcont}) and a uniform Lebesgue
measure, the \textbf{discrete} or \textbf{continuous uniform
distributions} discussed in Subsecs.~\ref{sec:indif}
and~\ref{sec:transfinv} can be derived;

\item[(ii)] the normalisation condition~(\ref{eq:normcont}),
the expectation value constraint~(\ref{eq:expectconstr})
and a uniform Lebesgue measure, while restricting the range of~$Y$
to $[0, +\infty)$, the \textbf{exponential distributions} discussed
in Subsec.~\ref{subsec:expdistr} are obtained;

\item[(iii)] the normalisation condition~(\ref{eq:normcont}),
the expectation value constraint~(\ref{eq:expectconstr}), the
variance constraint~(\ref{eq:varconstr}) and a uniform Lebesgue
measure, for a range of~$Y$ given by $(-\infty, +\infty)$,
the \textbf{Gau\ss\ distributions} discussed in
Subsec.~\ref{subsec:gaussdistr} arise; and, lastly,

\item[(iv)] the \textbf{binomial} and \textbf{Poisson
distributions} discussed in Subsecs.~(\ref{subsec:binomdistr})
and~(\ref{subsec:poisdistr}) can be obtained when employing
\textit{non-uniform} Lebesgue measures.

\end{itemize}

\leftout{
\textbf{partition function} $Z$, concept from Statistical Physics, 
cf. Jaynes (2003)~\ct[Eq.~(11.45)]{jay2003}, redefining 
$\lambda_{0} := 1+\lambda$,
\be
Z(\lambda_{1}, \ldots, \lambda_{l})
:= \sum_{i=1}^{k}\exp\left[\,-\,\sum_{j=1}^{l}\lambda_{j}\,
\frac{\partial C_{j}}{\partial p_{i}}\,\right]
\ee
The normalisation condition then reduces to
\be
\lambda_{0} = \ln\left[\,Z(\lambda_{1}, \ldots, 
\lambda_{l})\,\right] \ ,
\ee
and, formally,
\be
F_{i} := -\frac{\partial\ln\left[\,Z(\lambda_{1}, \ldots, 
\lambda_{l})\,\right]}{\partial\lambda_{i}} \ ,
\qquad i = 1, \ldots, l \ ,
\ee
where the $F_{k}$ represent $l$ expectation values of \ldots, One 
can then prove that in the present case the maximum for the 
information entropy is attained by (see Jaynes 
(2003)~\ct[]{jay2003})
\be
S_\mathrm{max} = \lambda_{0} + \sum_{i=1}^{l}\lambda_{i}\,F_{i} \ .
\ee
}

\section[Conjugate prior probability distributions]{Conjugate 
prior probability distributions}
\lb{sec:conjpriors}
In the first place, \textbf{conjugate prior probability 
distributions} for single unknown model parameters constitute a
welcome computational convenience. But the choice of a conjugate
prior in actual data analysis is by no means compulsory. The
ultimate selection depends on the quality of the information~$I$ 
available to a researcher prior to gaining access to relevant
quantitative--empirical data.

\medskip
\noindent
\textbf{Conjugate prior probability distributions} are
characterised by their property that, in combination with
\textbf{total-data likelihood functions}, they generate
\textbf{posterior probability distributions} that belong to the
very \textit{same} family of distributions as the priors one
started from. In particular, for \textbf{total-data likelihood
functions} from the \textbf{exponential family}, discussed in
Sec.~\ref{sec:expfam} before, it is straightforward to specify
related \textbf{conjugate prior probability
distributions}; see, e.g., Gelman \textit{et al}
(2014)~\ct[Sec.~2.4]{geletal2014}, or Gill
(2015)~\ct[Sec.~4.3]{gil2015} and Tab.~4.1 therein.

\medskip
\noindent
In the following, we will discuss the most frequently encountered
\textbf{conjugate prior probability distributions} used to describe
single unknown model parameters probabilistically.

\subsection[Beta distributions]{Beta distributions}
\lb{subsec:betadistr}
The two-parameter family of univariate \textbf{Beta distributions},
\be
\left.x\right|\alpha, \beta, I \sim \mathrm{Be}(\alpha,\beta) \ ,
\ee
is signified by the properties (see, e.g., Greenberg
(2013)~\ct[p~226]{gre2013}):

\medskip
\noindent
Spectrum of values:
\be
x \in \left[0,1\right] \ .
\ee
Probability density function (\texttt{pdf}):
\be
\lb{eq:betapdf}
\fbox{$\displaystyle
f(x|\alpha, \beta, I)
= \frac{1}{B(\alpha,\beta)}\,x^{\alpha-1}\,(1-x)^{\beta-1}
\ , \quad\text{with}\quad \alpha, \beta \in 
\mathbb{R}_{>0} \ ,
$}
\ee
where $\alpha$ and $\beta$ are dimensionless \textbf{shape
parameters}. The \texttt{pdf}-normalising \textbf{Beta function} is
defined via an Euler integral of the first kind by (see, e.g.,
Rinne (2008)~\ct[p~340]{rin2008})
\be
\lb{eq:betafct}
B(\alpha,\beta) := \int_{0}^{1}x^{\alpha-1}\,(1-x)^{\beta-1}
\,\mathrm{d}x
= \frac{\Gamma(\alpha)\,\Gamma(\beta)}{\Gamma(\alpha+\beta)} \ ;
\ee
the definition of the Gamma function was given in 
Eq.~(\ref{eq:gammafct}). The graph of the probability density 
function is shown in Fig.~\ref{fig:betapdf} for four different 
combinations of values for~$\alpha$ and~$\beta$. Note that the 
continuous uniform distribution on $[0,1]$ is contained as the 
special case $\mathrm{Be}(1,1) = U(0,1)$.

\medskip
\noindent
Expectation value and variance:
\bea
\lb{eq:betaexp}
\mathrm{E}(x) & = & \frac{\alpha}{\alpha+\beta} \\
\lb{eq:betavar}
\mathrm{Var}(x) & = & 
\frac{\alpha\beta}{(\alpha+\beta)^{2}(\alpha+\beta+1)} \ .
\eea
\begin{figure}[!htb]
\begin{center}
\fbox{\includegraphics[width=14cm]{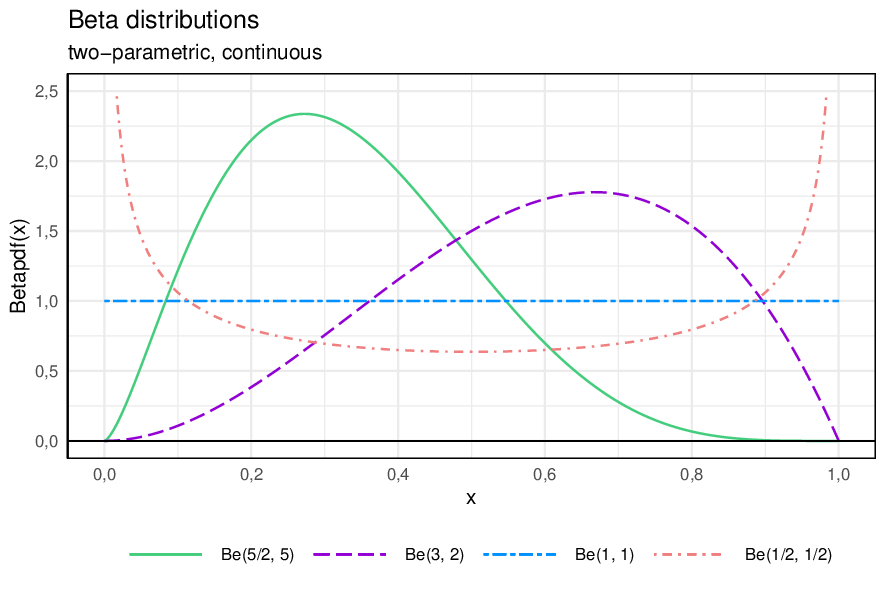}}
\end{center}
\caption{Four examples of \texttt{pdf}s of Beta distributions for
an uncertain quantity $x$.}
\lb{fig:betapdf}
\end{figure}

\medskip
\noindent
\underline{\R:}
$\texttt{dbeta}(x, \alpha, \beta)$,
$\texttt{pbeta}(x, \alpha, \beta)$,
$\texttt{qbeta}(p, \alpha, \beta)$,
$\texttt{rbeta}(n_{\mathrm{simulations}}, \alpha, \beta)$\\
\underline{Stan:} Cf. \href{https://mc-stan.org/docs/functions-reference/beta-distribution.html}{Stan Functions Reference (v2.30)}
\ct{sta2022b}
\begin{itemize}
\item $\texttt{beta}( \alpha , \beta )$ (sampling)
\item $\texttt{beta\_lpdf}( y | \alpha , \beta )$
(log-sampling)
\item $\texttt{beta\_rng}( \alpha , \beta )$
(generating)
\end{itemize}
\underline{JAGS:}
$\texttt{dbeta}(\alpha, \beta)$ (sampling)

\medskip
\noindent
Prior Beta distributions for a probability for ``success''
parameter lead to posterior Beta distributions for a probability
for ``success''when they are combined with the binomial likelihood
functions introduced in Subsec.~\ref{subsec:binomdistr}.

\subsection[Gamma distributions]{Gamma distributions}
\lb{subsec:gammadistr}
The two-parameter family of univariate \textbf{Gamma
distributions},
\be
\left.x\right|\alpha, \beta, I \sim \mathrm{Ga}(\alpha,\beta) \ ,
\ee
has characteristic features (see, e.g., Greenberg
(2013)~\ct[p~225]{gre2013}):

\medskip
\noindent
Spectrum of values:
\be
x \in \mathbb{R}_{\geq 0} \ .
\ee
Probability density function (\texttt{pdf}):
\be
\lb{eq:gammapdf}
\fbox{$\displaystyle
f(x|\alpha, \beta, I)
= \frac{\beta^{\alpha}}{\Gamma(\alpha)}\,x^{\alpha-1}\,
\exp\left(-\beta x\right) \ ,
\quad\text{with}\quad \alpha, \beta \in \mathbb{R}_{>0} \ ;
$}
\ee
$\alpha$ constitutes a dimensionless \textbf{shape parameter},
while $\beta$ is a \textbf{rate parameter} of physical dimension
inverse to~$x$. The Gamma function was defined in 
Eq.~(\ref{eq:gammafct}). The graph of the probability density 
function is shown in Fig.~\ref{fig:gammapdf} for four different 
combinations of values for~$\alpha$ and~$\beta$.

\medskip
\noindent
Expectation value and variance:
\bea
\lb{eq:gammaexp}
\mathrm{E}(x) & = & \frac{\alpha}{\beta} \\
\lb{eq:gammavar}
\mathrm{Var}(x) & = & \frac{\alpha}{\beta^{2}} \ .
\eea
\begin{figure}[!htb]
\begin{center}
\fbox{\includegraphics[width=14cm]{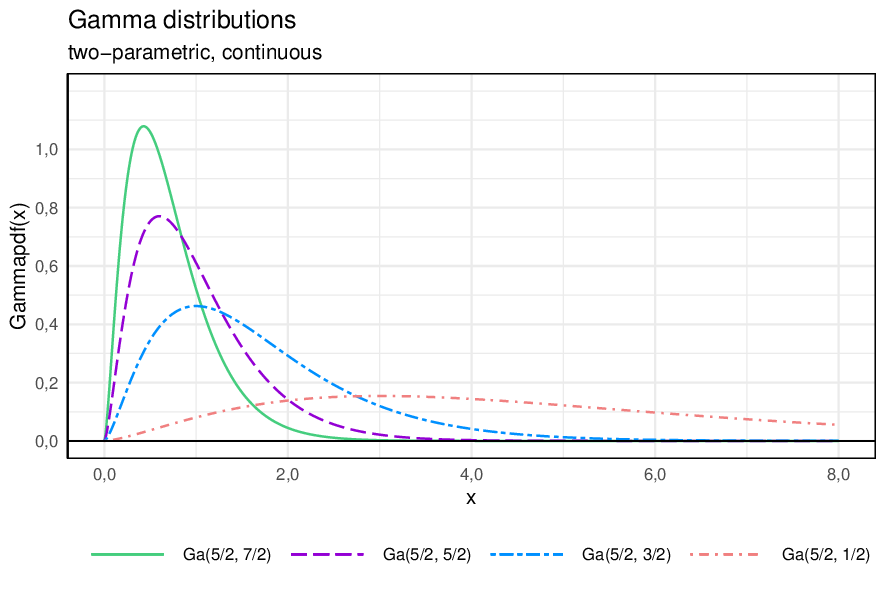}}
\end{center}
\caption{Four examples of \texttt{pdf}s of Gamma distributions for
an uncertain quantity $x$.}
\lb{fig:gammapdf}
\end{figure}

\medskip
\noindent
\underline{\R:}
$\texttt{dgamma}(x, \alpha, \beta)$,
$\texttt{pgamma}(x, \alpha, \beta)$,
$\texttt{qgamma}(p, \alpha, \beta)$,
$\texttt{rgamma}(n_{\mathrm{simulations}}, \alpha, \beta)$\\
\underline{Stan:} Cf. \href{https://mc-stan.org/docs/functions-reference/gamma-distribution.html}{Stan Functions Reference (v2.30)}
\ct{sta2022b}
\begin{itemize}
\item $\texttt{gamma}( \alpha , \beta )$ (sampling)
\item $\texttt{gamma\_lpdf}( y | \alpha , \beta )$
(log-sampling)
\item $\texttt{gamma\_rng}( \alpha , \beta )$
(generating)
\end{itemize}
\underline{JAGS:}
$\texttt{dgamma}(\alpha, \beta)$ (sampling)

\medskip
\noindent
Notice that for the particular choice of parameters $\alpha = 
\nu/2$, $\beta = 1/2$, Gamma distributions contain the 
one-parameter family of
$\boldsymbol{\chi^{2}}$\textbf{--distributions 
with} $\boldsymbol{\nu}$ \textbf{degrees of freedom} as a special 
case. Similarly, for $\alpha = 1$, $\beta = \beta$, one obtains the 
one-parameter family of \textbf{exponential distributions}
considered  in Subsec.~\ref{subsec:expdistr}; cf. Greenberg 
(2013)~\ct[p~225]{gre2013}.

\medskip
\noindent
Prior Gamma distributions for a precision parameter lead to
posterior Gamma distributions for a precision parameter when they
are combined with the Gau\ss\ likelihood functions introduced in
Subsec.~\ref{subsec:gaussdistr}. Also, prior Gamma distributions
for a rate parameter lead to posterior Gamma distributions for a
rate parameter when, for discrete count data, they are combined
with the Poisson likelihood functions introduced
in Subsec.~\ref{subsec:poisdistr}, or, for continuous interval
data, when they are combined with the exponential likelihood
functions discussed in Subsec.~\ref{subsec:expdistr}.

\subsection[Inverse Gamma distributions]{Inverse Gamma 
distributions}
\lb{subsec:invgammadistr}
The two-parameter family of univariate \textbf{inverse Gamma 
distributions},
\be
\left.x\right|\alpha, \beta, I \sim \mathrm{IG}(\alpha,\beta) \ ,
\ee
is related to Gamma distribution by a simple inversion
transformation of the independent variable, namely $x \to 1/x$; 
see, e.g., Greenberg (2013)~\ct[p~225f]{gre2013}. They have the
properties:

\medskip
\noindent
Spectrum of values:
\be
x \in \mathbb{R}_{> 0} \ .
\ee
Probability density function (\texttt{pdf}):
\be
\lb{eq:invgammapdf}
\fbox{$\displaystyle
f(x|\alpha, \beta, I)
= \frac{\beta^{\alpha}}{\Gamma(\alpha)}\,\frac{1}{x^{\alpha+1}}\,
\exp\left(-\beta/x\right) \ , \quad\text{with}\quad \alpha, \beta 
\in \mathbb{R}_{>0} \ ,
$}
\ee
where $\alpha$ is a dimensionless \textbf{shape parameter} and
$\beta$ a \textbf{rate parameter} of the same physical dimension
as~$X$. The graph of the probability density function is shown 
in Fig.~\ref{fig:invgammapdf} for four different combinations of 
values for~$\alpha$ and~$\beta$.

\medskip
\noindent
Expectation value and variance:
\bea
\mathrm{E}(x) & = & \frac{\beta}{\alpha-1} \ ,
\quad\text{if}\quad \alpha > 1 \\
\mathrm{Var}(x) & = & \frac{\beta^{2}}{(\alpha-1)^{2}(\alpha-2)}
\ , \quad\text{if}\quad \alpha > 2 \ .
\eea
\begin{figure}[!htb]
\begin{center}
\fbox{\includegraphics[width=14cm]{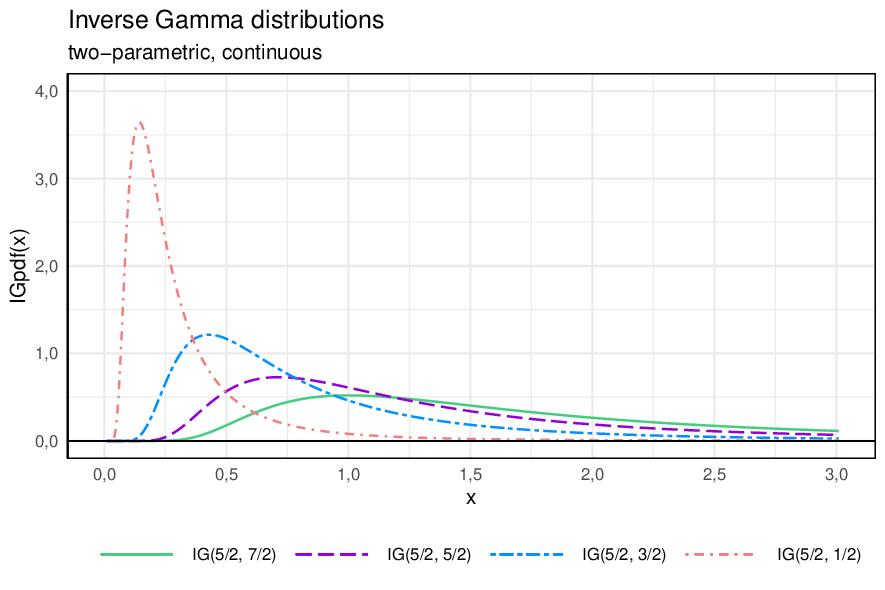}}
\end{center}
\caption{Four examples of \texttt{pdf}s of inverse Gamma 
distributions for an uncertain quantity $x$.}
\lb{fig:invgammapdf}
\end{figure}

\medskip
\noindent
\underline{\R:}
$\texttt{dinvgamma}(x, \alpha, \beta)$,
$\texttt{pinvgamma}(x, \alpha, \beta)$,
$\texttt{qinvgamma}(p, \alpha, \beta)$,
$\texttt{rinvgamma}(n_{\mathrm{simulations}}, \alpha, \beta)$
(\texttt{invgamma} package, by Kahle and Stamey
(2017)~\ct{kahsta2017})\\
\underline{Stan:} Cf. \href{https://mc-stan.org/docs/functions-reference/inverse-gamma-distribution.html}{Stan Functions Reference (v2.30)}
\ct{sta2022b}
\begin{itemize}
\item $\texttt{inv\_gamma}( \alpha , \beta )$ (sampling)
\item $\texttt{inv\_gamma\_lpdf}( y | \alpha , \beta )$
(log-sampling)
\item $\texttt{inv\_gamma\_rng}( \alpha , \beta )$
(generating)
\end{itemize}

\medskip
\noindent
Prior inverse Gamma distributions for a scale parameter lead to
posterior inverse Gamma distributions for a scale parameter when
they are combined with the Gau\ss\ likelihood functions introduced
in Subsec.~\ref{subsec:gaussdistr}.

\subsection[Gau\ss\ distributions]{Gau\ss\ distributions}
For the two-parameter family of \textbf{Gau\ss\ distributions},
\be
\lb{eq:gausspriorpdf}
\fbox{$\displaystyle
f(x|m_{0}, \tau_{0}, I)
= \frac{1}{\sqrt{2\pi}\,\tau_{0}}\,\exp\left[\,-\frac{1}{2}
\left(\frac{x-m_{0}}{\tau_{0}}\right)^{2}\,\right] \ , 
\quad\text{with}\quad m_{0} \in \mathbb{R} \ , \ \tau_{0} \in 
\mathbb{R}_{>0} \ ,
$}
\ee
the \textbf{location parameter}~$m_{0}$ and the \textbf{scale
parameter}~$\tau_{0}$ both have the physical dimension of~$x$.
Prior Gau\ss\ distributions for a location parameter lead to
posterior Gau\ss\ distributions for a location parameter when they
are combined with the Gau\ss\ likelihood functions introduced in
Subsec.~\ref{subsec:gaussdistr}; in that particular case this
family is conjugate to itself.

\section[Other prior probability distributions]{Other prior 
probability distributions}
\lb{sec:otherpriors}
Lastly, we introduce three more families of probability
distributions that are also often used as \textbf{prior probability distributions} for different kinds of single unknown model
parameters.

\subsection[Cauchy distributions]{Cauchy distributions}
\lb{subsec:cauchydistr}
The two-parameter family of univariate \textbf{Cauchy
distributions},
\be
\left.x\right|x_{0}, \gamma, I \sim \mathrm{Ca}(x_{0},\gamma) \ ,
\ee
put forward by the French mathematician, engineer and physicist
\href{https://mathshistory.st-andrews.ac.uk/Biographies/Cauchy/}{Augustin--Louis
Cauchy (1789--1857)}, is given by (see, e.g., Rinne
(2008)~\ct[Subsec.~3.11.2]{rin2008}):

\medskip
\noindent
Spectrum of values:
\be
x \in \mathbb{R} \ .
\ee
Probability density function (\texttt{pdf}):
\be
\lb{eq:cauchypdf}
\fbox{$\displaystyle
f(x|x_{0}, \gamma, I) =
\frac{1}{\pi\gamma}\,\frac{1}{1+\left({\displaystyle
\frac{x-x_{0}}{\gamma}}\right)^{2}} \ ,
\qquad\text{with}\quad x_{0} \in \mathbb{R} \ ,
\ \gamma \in \mathbb{R}_{>0},\  \ ;
$}
\ee
the \textbf{location parameter}~$x_{0}$ and the \textbf{scale 
parameter}~$\gamma$ both carry the physical dimension of~$x$. The
graph of the probability density function is shown in
Fig.~\ref{fig:Cauchypdf} for four different combinations of 
values for~$x_{0}$ and~$\gamma$. Formally, as follows from 
Eq.~(\ref{eq:tpdf}), Cauchy distributions correspond to 
non-central $t$--distributions with just one degree of freedom,
$\nu = 1$.

\medskip
\noindent
Expectation value and variance:
\bea
\mathrm{E}(x) & = & \text{does NOT exist due to a
diverging integral} \\
\mathrm{Var}(x) & = & \text{does NOT exist due to a
diverging integral} \ .
\eea
\begin{figure}[!htb]
\begin{center}
\fbox{\includegraphics[width=14cm]{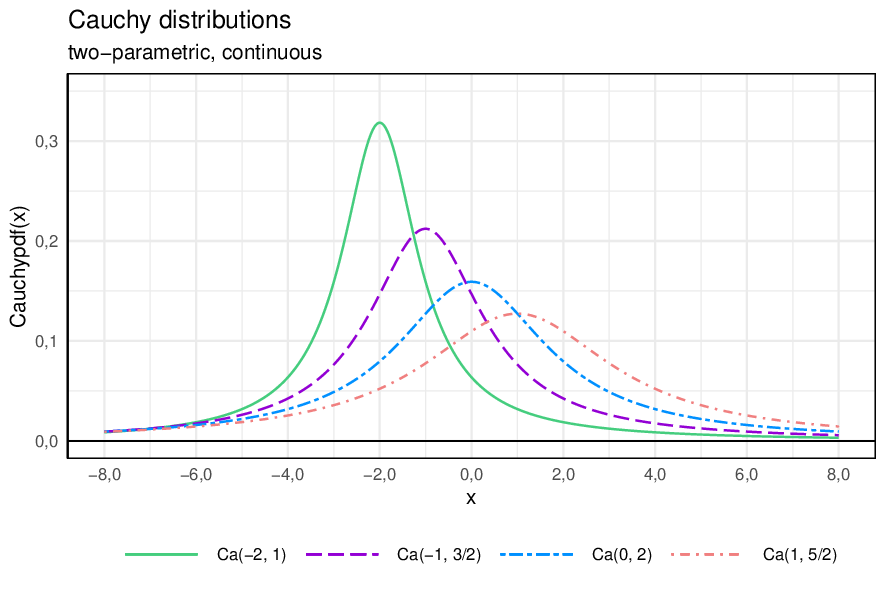}}
\end{center}
\caption{Four examples of \texttt{pdf}s of Cauchy distributions for 
an uncertain quantity $x$.}
\lb{fig:Cauchypdf}
\end{figure}

\medskip
\noindent
\underline{\R:}
$\texttt{dcauchy}(x, x_{0}, \gamma)$,
$\texttt{pcauchy}(x, x_{0}, \gamma)$,
$\texttt{qcauchy}(p, x_{0}, \gamma)$,
$\texttt{rcauchy}(n_{\mathrm{simulations}}, x_{0}, \gamma)$\\
\underline{Stan:} Cf. \href{https://mc-stan.org/docs/functions-reference/cauchy-distribution.html}{Stan Functions Reference (v2.30)}
\ct{sta2022b}
\begin{itemize}
\item $\texttt{cauchy}( x_{0} , \gamma )$ (sampling)
\item $\texttt{cauchy\_lpdf}( y | x_{0} , \gamma )$
(log-sampling)
\item $\texttt{cauchy\_rng}( x_{0} , \gamma )$
(generating)
\end{itemize}
\underline{JAGS:}
$\texttt{dt}(x_{0}, 1/\gamma^{2}, 1)$ (sampling)

\medskip
\noindent
Half-Cauchy distributions with $x_{0} = 0$ (meaning the half to the
right of $x_{0} = 0$) have become a standard in modern data
analysis as weakly regularising prior probability distributions for
unknown scale parameters such as standard deviations: e.g., in
fixed-prior models and, in particular, in adaptive-prior,
multi-level models, in which they control the degree of shrinkage
of model parameters for data obtained from different but related
groups; cf. Gelman (2006)~\ct{gel2006}, Gill
(2015)~\ct[p~178]{gil2015} and Kruschke (2015)~\ct[p~558]{kru2015}.

\subsection[Exponential distributions]{Exponential distributions}
Members from the one-parameter family of \textbf{exponential
distributions}, introduced before in Subsec.~\ref{subsec:expdistr},
i.e.,
\be
\lb{eq:exppriorpdf}
\fbox{$\displaystyle
f(x|\beta_{0}, I)
= \beta_{0}\,\exp\left(-\beta_{0}x\right) \ , 
\quad\text{with}\quad \beta_{0} \in \mathbb{R}_{>0} \ ,
$}
\ee
with dimensionful \textbf{rate parameter}~$\beta_{0}$, are likewise 
frequently employed as (weakly or strongly regularising) prior 
distributions for unknown scale parameters in fixed-prior models
and in adaptive-prior multi-level models;
cf. McElreath (2020)~\ct[p~407]{mce2020a}.

\subsection[Laplace distributions]{Laplace distributions}
The two-parameter family of univariate \textbf{Laplace
distributions},
\be
\left.x\right|\mu, b, I  \sim \mathrm{Lp}(\mu, b) \ ,
\ee
has the properties (see, e.g., Rinne
(2008)~\ct[Subsec.~3.11.5]{rin2008}):

\medskip
\noindent
Spectrum of values:
\be
x \in D \subseteq \mathbb{R} \ .
\ee
Probability density function (\texttt{pdf}):
\be
\lb{eq:laplacepdf}
\fbox{$\displaystyle
f(x|\mu, b, I) = 
\frac{1}{2b}\,\exp\left[-\frac{|x-\mu|}{b}\right] \ , 
\quad\text{with}\quad
\mu \in \mathbb{R} \ , \ b \in \mathbb{R}_{>0} \ ;
$}
\ee
$\mu$ constitutes a \textbf{location parameter}, $b$ a
\textbf{scale parameter}, and both carry the physical dimension
of~$x$. The graph of the probability density function is shown in
Fig.~\ref{fig:laplacepdf} for four different combinations of 
values for~$\mu$ and~$b$. 

\medskip
\noindent
Expectation value and variance:
\bea
\mathrm{E}(x) & = & \mu \\
\mathrm{Var}(x) & = & 2b^{2} \ .
\eea
\begin{figure}[!htb]
\begin{center}
\fbox{\includegraphics[width=14cm]{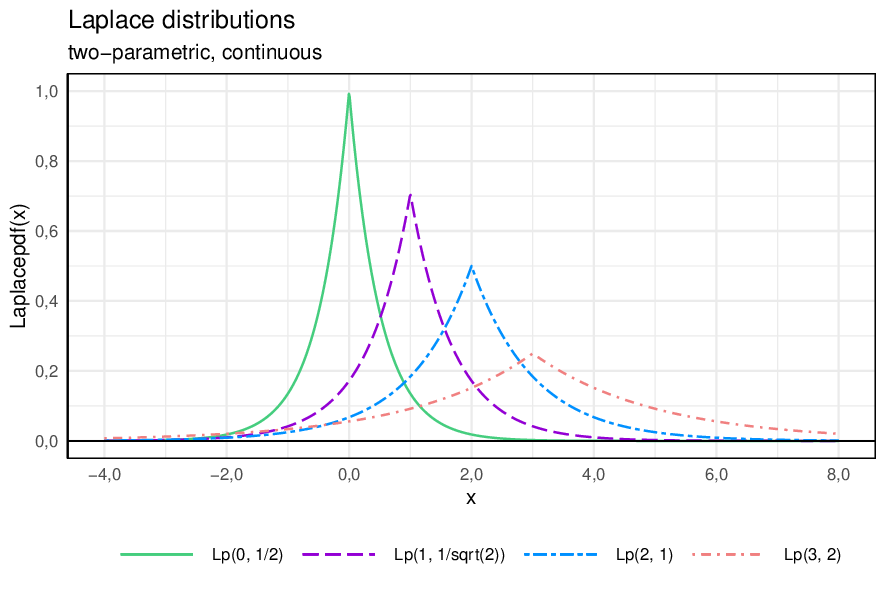}}
\end{center}
\caption{Four examples of \texttt{pdf}s of Laplace distributions
for an uncertain quantity~$x$.}
\lb{fig:laplacepdf}
\end{figure}

\medskip
\noindent
\underline{\R:}
$\texttt{dlaplace}(x, \mu, b)$,
$\texttt{plaplace}(x, \gamma, b)$,
$\texttt{qlaplace}(p, \gamma, b)$,
$\texttt{rlaplace}(n_\mathrm{simulations}, \mu, b)$
(\texttt{extraDistr} package, by Wolodzko (2020)~\ct{wol2020})\\
\underline{Stan:} Cf. \href{https://mc-stan.org/docs/functions-reference/double-exponential-laplace-distribution.html}{Stan Functions
Reference (v2.30)} \ct{sta2022b}
\begin{itemize}
\item $\texttt{double\_exponential}( \mu , b )$ (sampling)
\item $\texttt{double\_exponential\_lpdf}( y | \mu , b )$
(log-sampling)
\item $\texttt{double\_exponential\_rng}( \mu , b )$
(generating)
\end{itemize}
\underline{JAGS:}
$\texttt{ddexp}(\mu, 1/b)$ (sampling)

\medskip
\noindent
We will now turn to highlight in the next two chapters a few
of the prominent but very rare cases of single- and two-parameter
estimations for which the posterior (joint) probability
distribution can be obtained analytically.

\chapter[Single-parameter estimation]{Single-parameter estimation}
\lb{ch4}
In the special case of \textbf{single-parameter estimation} in
\textbf{inductive statistical inference}, a few analytical
solutions for the \textbf{posterior probability distribution} for
the unknown \textbf{model parameter} in question are
available, and well-known. Typically this is possible in those
cases where either \textbf{uniform} or \textbf{conjugate prior
probability distributions} are employed to express a researcher's
\textbf{state of knowledge} as to the range of values of the 
unobservable model parameter of interest, before seeing relevant
quantitative--empirical data.

\medskip
\noindent
In the single-parameter instance, \textbf{Bayes' theorem}, in its
model-building variant of Eq.~(\ref{eq:postprob}), reduces to
\be
\lb{eq:postprob1p}
\fbox{$\displaystyle
P(\theta|\{y_{i}\}_{i=1,\ldots,n}, I) = 
\frac{P(\{y_{i}\}_{i=1,\ldots,n}|\theta,
I)}{P(\{y_{i}\}_{i=1,\ldots,n}|I)}\,P(\theta|I) \ .
$}
\ee
This is the relationship for which some exact solutions can be
derived.

\medskip
\noindent
The known analytical solutions for \textbf{single-parameter
posterior probability distributions} possess a certain pedagogical
merit. Therefore, some prominent examples will be reviewed in the
next few sections.

\leftout{
Jaynes (2003)~\ct[p~172]{jay2003}, optimal point estimate by
\textbf{mean squared error criterion} of Gau\ss\ and Legendre,
quadratic ``loss function'' Greenberg (2013)~\ct[p~31]{gre2013}
\textbf{posterior mean}
\be
\theta^{\ast} = \mathrm{E}(\theta|\{\boldsymbol{x}\}I)
= \int \theta P(\theta|\{\boldsymbol{x}\}I)\,\mathrm{d}\theta
\ee

Jaynes (2003)~\ct[p~174]{jay2003} \textbf{posterior median}

\medskip
Greenberg (2013)~\ct[p~27]{gre2013}: large $n$ approximation for 
posterior distributions
}

\section[Binomial-distributed univariate discrete 
data]{Binomial-distributed univariate discrete data}
\lb{sec:bernpara}
Suppose given $n$~exchangeable and logically independent
repetitions of a \textbf{Bernoulli experiment}, with unknown but
\textit{constant} probability for ``success,'' $0 \leq \theta \leq
1$. Let the observed total number of ``successes'' in this
experiment be
\be
\lb{eq:successes}
y := \sum_{i=1}^{n}y_{i} \ ,
\ee
with $y_{i} \in \{0, 1\}$ and so $0 \leq y \leq n$. Then, as
discussed in Subsec.~(\ref{subsec:binomdistr}), the full
data-generating process is described by
\be
\left.y\right|\theta, n, I \sim \mathrm{Bin}(n, y) \ ,
\ee
with \textbf{total-data likelihood function}
\be
\lb{eq:binomprob2}
P(y|n, \theta, I)
= \left(
\begin{array}{c}
n \\
y
\end{array}
\right)
\theta^{y}\,(1-\theta)^{n-y} \ .
\ee
This is to be viewed as a function of $\theta$, for \textit{fixed}
data~$n$ and~$y$. Note that, by Eq.~(\ref{eq:expfam}),
$y =: n\bar{y}$ constitutes a sufficient statistic for~$\theta$.
The \textbf{sample mean} of a univariate metrically scaled data set
is defined by
\be
\lb{eq:samplemean}
\bar{y} := \frac{1}{n}\,\sum_{i=1}^{n}y_{i} \ .
\ee
%

\subsection[Uniform prior]{Uniform prior}
Adopting for~$\theta$ a \textbf{uniform prior probability
distribution} [cf. Eq.~(\ref{eq:continuousUniformpdf})],
\be
\left.\theta\right|n, I \sim \mathrm{U}(0, 1) \ ,
\ee
and forming the product with~(\ref{eq:binomprob2}),
then, after some algebra that involves the definition of the
binomial coefficient in Eq.~(\ref{eq:bincoeff}) and the Gamma
function identity of Eq.~(\ref{eq:gammaid}), one is led to the
normalised \textbf{posterior probability distribution} for $\theta$
given by
\be
P(\theta|y, n, I)
= \frac{\Gamma(n+2)}{\Gamma(y+1)\,\Gamma(n-y+1)}\,\theta^{y}\,
(1-\theta)^{n-y} \ ,
\ee
i.e., a \textbf{Beta distribution}
\be
\lb{eq:postbinomunif}
\fbox{$\displaystyle
\left.\theta\right|y, n, I
\sim \mathrm{Be}(y+1,n-y+1) \ .
$}
\ee

\medskip
\noindent
In the present case, it is also easy to obtain the \textbf{prior
predictive probability distribution} for $y$, when no observations
have yet been made. With Eq.~(\ref{eq:priorpred}) one finds that
\bea
P(y|n, I) & = & \int_{0}^{1}P(y|n, \theta, I)\,P(\theta|n, 
I)\,\mathrm{d}\theta \nonumber \\
& = & \int_{0}^{1}\underbrace{\left(\begin{array}{c}
n \\
y
\end{array}\right)
\theta^{y}\,(1-\theta)^{n-y}}_{\text{likelihood}}\times
\underbrace{1}_{\text{prior}}\,\mathrm{d}\theta \nonumber \\
& = & \frac{1}{n+1}\times\int_{0}^{1}\frac{(n+1)!}{y!(n-y)!}\,
\theta^{y}\,(1-\theta)^{n-y}\,\mathrm{d}\theta \nonumber \\
& = & \frac{1}{n+1}\times\underbrace{\int_{0}^{1}
\frac{\Gamma(n+2)}{\Gamma(y+1)\,\Gamma(n-y+1)}\,\theta^{y}
\,(1-\theta)^{n-y}\,\mathrm{d}\theta}_{=1,\ \text{by\ 
Eq.~(\ref{eq:betafct})}} \nonumber \\
& = & \frac{1}{n+1} \ ,
\eea
where, in the final step, the normalisation condition for Beta
distributions according to Eq.~(\ref{eq:betafct}) was used.
Clearly, this result represents an initial discrete uniform
probability distribution over the set of possible outcomes,
$y \in \{0, \ldots, n\}$.

\medskip
\noindent
Similarly, starting from Eq.~(\ref{eq:postpred}) to calculate
the \textbf{posterior predictive probability distribution} for a
single new observation~$y_{\mathrm{new}}$ to be a ``success,''
given the information that~$y$ ``successes'' were observed in~$n$
previous iid-repetitions of the Bernoulli experiment, one arrives
at
\bea
\lb{eq:binompostpredunif}
P(y_\mathrm{new}=1|y, n, I) & = & \int_{0}^{1}
P(y_\mathrm{new}=1|y, n, \theta, I)\,P(\theta|y, n, I)\,
\mathrm{d}\theta \nonumber \\
& = & \int_{0}^{1}
\underbrace{\theta}_{\text{likelihood}}\times
\underbrace{\frac{\Gamma(n+2)}{\Gamma(y+1)\,
\Gamma(n-y+1)}\,\theta^{y}(1-\theta)^{n-y}}_{\text{posterior}}
\,\mathrm{d}\theta \nonumber \\
& = & \frac{y+1}{n+2}\times\underbrace{\int_{0}^{1}
\frac{\Gamma(n+3)}{\Gamma(y+2)\,\Gamma(n-y+1)}\,
\theta^{y+1}(1-\theta)^{n-y}\,\mathrm{d}\theta}_{=1,\ \text{by\ 
Eq.~(\ref{eq:betafct})}} \nonumber \\
& = & \frac{y+1}{n+2} \ ;
\eea
again, use was made of the normalisation condition for 
Beta distributions given in Eq.~(\ref{eq:betafct}). This last
result constitutes Laplace's famous ``\textbf{rule of
succession};'' cf. Keynes (1921)~\ct[p~372]{key1921},
Cox (1946)~\ct[p~11]{cox1946}, Jaynes (2003)~\ct[pp~155, 
165]{jay2003}, and Gelman \textit{et al}
(2014)~\ct[p~32]{geletal2014}. In the present context, this may
actually be interpreted as the posterior expectation value
for~$\theta$, viz.
\be
\frac{y+1}{n+2}
= \frac{2}{n+2} \times 
\underbrace{\frac{1}{2}}_{\text{prior mean}}
+ \frac{n}{n+2} \times 
\underbrace{\frac{y}{n}}_{\text{sample mean}}
= \mathrm{E}(\theta|y, n, I) \ .
\ee
It exemplifies the notion of the posterior probability distribution
being  a ``compromise'' between the prior probability distribution
and the likelihood  function; see, e.g., Gill
(2015)~\ct[p~42]{gil2015} or Kruschke (2015)~\ct[p~112]{kru2015}.
In particular, in this representation it becomes apparent that with
increasing sample size~$n$ the weight is being pushed away from
prior information towards data information, which is a very
compelling and elegant feature.

\medskip
\noindent
From Eq.~(\ref{eq:betavar}), the posterior variance for~$\theta$
amounts to
\be
\lb{eq:postvarbinom}
\mathrm{Var}(\theta|y, n, I) = \frac{(y+1)(n-y+1)}{(n+2)^{2}(n+3)}
\ .
\ee
Further summary statistics can be easily computed with 
pre-programmed \R-functions.

\subsection[Conjugate prior]{Conjugate prior}
\lb{subsec:betabinom}
Alternatively, selecting for~$\theta$ instead a conjugate
\textbf{Beta prior probability distribution}
[cf. Eq.~(\ref{eq:betapdf})],
\be
\left.\theta\right| \alpha, \beta, n, I \sim \mathrm{
Be}(\alpha,\beta)
\ee
and forming the product with~(\ref{eq:binomprob2}), then, again,
algebra involving Eqs.~(\ref{eq:bincoeff}) and~(\ref{eq:gammaid})
yields a normalised \textbf{posterior probability distribution}
for~$\theta$ given by
\be
P(\theta|y, \alpha, \beta, n, I) =
\frac{\Gamma(\alpha+\beta+n)}{\Gamma(\alpha+y)\,\Gamma(\beta+n-y)}
\,\theta^{\alpha+y-1}\,(1-\theta)^{\beta+n-y-1} \ ,
\ee
i.e., a \textbf{Beta distribution}
\be
\fbox{$\displaystyle
\lb{eq:postbinombeta}
\left.\theta\right|y, \alpha, \beta, n, I
\sim \mathrm{Be}(\alpha+y,\beta+n-y) \ ;
$}
\ee
see, e.g., Lee (2012)~\ct[Subsec.~3.1.1]{lee2012}, Gelman
\textit{et al} (2014)~\ct[Sec.~2.4]{geletal2014}, Gill
(2015)~\ct[Subsec.~2.3.4]{gil2015}, or Kruschke 
(2015)~\ct[Sec.~6.3]{kru2015}.

\medskip
\noindent
The \textbf{posterior predictive probability distribution} for a
single new observation~$y_{\mathrm{new}}$ to be a ``success,''
given the information that~$y$ ``successes'' were observed in~$n$
previous iid-repetitions of the Bernoulli experiment, is presently
obtained by analogy to the algebraic steps taken to derive
Eq.~(\ref{eq:binompostpredunif}). This leads to\footnote{To obtain
the full posterior predictive probability distribution for an
\textit{arbitrary} single new observation~$y_{\mathrm{new}}$
according to Eq.~(\ref{eq:postpred}) proves more demanding.}
\be
\lb{eq:binompostpredconj}
P(y_\mathrm{new}=1|y, n, I) = \frac{\alpha+y}{\alpha+\beta+n} \ ,
\ee
and, again, from Eq.~(\ref{eq:betaexp}) it amounts to the posterior 
expectation value for~$\theta$ given by
\bea
\lb{eq:postmeanbinom}
\mathrm{E}(\theta|y, \alpha, \beta, n, I)
& = & \frac{\alpha+y}{\alpha+y+\beta+n-y} \nonumber \\
& = & \frac{\alpha+\beta}{\alpha+\beta+n} \times 
\underbrace{\frac{\alpha}{\alpha+\beta}}_{\text{prior mean}}
+ \frac{n}{\alpha+\beta+n} \times 
\underbrace{\frac{y}{n}}_{\text{sample mean}} \ .
\eea
It thus can be interpreted as weighted average of the prior 
mean and the sample mean, where the numerators of the respective 
weighting factors represent a \textbf{prior effective sample size} 
($\alpha+\beta$) and the actual data \textbf{sample size} ($n$),
respectively. Note that for $n \gg \alpha+\beta$ the sample mean
will dominate.

\medskip
\noindent
From Eq.~(\ref{eq:betavar}), the posterior variance for~$\theta$
amounts to
\be
\lb{eq:postvarbinom}
\mathrm{Var}(\theta|y, \alpha, \beta, n, I)
= \frac{(\alpha+y)(\beta+n-y)}{(\alpha+\beta+n)^{2}(\alpha+\beta+n
+1)} \ .
\ee
Sampling from the single-parameter posterior
distribution~(\ref{eq:postbinombeta}) and calculating further
summary statistics can be easily accomplished by use of
pre-programmed \R-functions; cf. Subsec.~\ref{subsec:betadistr}.

\section[Poisson-distributed univariate discrete 
data]{Poisson-distributed univariate discrete data}
\lb{sec:poispara}
Suppose given, from a \textbf{Poisson process}, measurements of~$n$
exchangeable and logically independent counts~$y_{i}$,
\be
\left.\{y_{i}\}_{i=1,\ldots,n}\right|\theta, I
\stackrel{\text{iid}}{\sim} \mathrm{Pois}(\theta) \ ,
\ee
so that the \textbf{total-data likelihood function} is
\be
\lb{eq:poisprob2}
P(\{y_{i}\}_{i=1,\ldots,n}|\theta, I)
= \frac{\displaystyle\theta^{\sum_{i=1}^{n}y_{i}}}{\prod_{i=1}^{n}
y_{i}!}\,\exp\left(-n\theta\right) \ ,
\ee
with unknown rate parameter~$\theta \in \mathbb{R}_{>0}$. By
Eq.~(\ref{eq:expfam}), $\sum_{i=1}^{n}y_{i} = n\bar{y}$ is a
sufficient statistic for~$\theta$.

\medskip
\noindent
Assuming for~$\theta$ a conjugate \textbf{Gamma prior probability
distribution} [cf. Eq.~(\ref{eq:gammapdf})],
\be
\left.\theta\right| \alpha, \beta, I \sim \mathrm{Ga}(\alpha,\beta)
\ ,
\ee
and forming the product with~(\ref{eq:poisprob2}), then a few
simple algebraic manipulations yield a \textbf{posterior
probability distribution} for~$\theta$ proportional to
\be
P(\theta|\{y_{i}\}_{i=1,\ldots,n}, \alpha, \beta, I) \propto
\theta^{(\alpha+\sum_{i=1}^{n}y_{i})-1}\,\exp\left[\,-(\beta+n)
\theta\,\right] \ ,
\ee
i.e., upon normalisation, a \textbf{Gamma distribution}
\be
\fbox{$\displaystyle
\lb{eq:postpois}
\left.\theta\right|\{y_{i}\}_{i=1,\ldots,n}, \alpha, \beta, I
\sim \mathrm{Ga}(\alpha+\sum_{i=1}^{n}y_{i}, \beta+n) \ ;
$}
\ee
cf. Lee (2012)~\ct[Subsec.~3.4.1]{lee2012}, or Gelman \textit{et
al} (2014)~\ct[Sec.~2.6]{geletal2014}. From
Eq.~(\ref{eq:gammaexp}), the decomposed posterior expectation value
for~$\theta$ is
\bea
\lb{eq:postmeanpois}
\mathrm{E}(\theta|\{y_{i}\}_{i=1,\ldots,n}, \alpha, \beta, I)
& = & \frac{\alpha+\sum_{i=1}^{n}y_{i}}{\beta+n} \nonumber \\
& = & \frac{\beta}{\beta+n} \times 
\underbrace{\frac{\alpha}{\beta}}_{\text{prior mean}}
+ \frac{n}{\beta+n} \times 
\underbrace{\frac{\sum_{i=1}^{n}y_{i}}{n}}_{\text{sample mean}} \ ;
\eea
the prior effective sample size thus amounts to $\beta$. For
$n \gg \beta$ the sample mean will dominate.

\medskip
\noindent
From Eq.~(\ref{eq:gammavar}), the posterior variance for~$\theta$
amounts to
\be
\lb{eq:postvarpois}
\mathrm{Var}(\theta|\{y_{i}\}_{i=1,\ldots,n}, \alpha, \beta, I)
= \frac{\alpha+\sum_{i=1}^{n}y_{i}}{(\beta+n)^{2}} \ .
\ee
Sampling from the single-parameter posterior
distribution~(\ref{eq:postpois}) and calculating further
summary statistics can be easily accomplished by use of
pre-programmed \R-functions; cf. Subsec.~\ref{subsec:gammadistr}.

\section[Gau\ss-distributed univariate continuous 
data]{Gau\ss-distributed univariate continuous data}
\lb{sec:gausspara}
In this section we will suppose given~$n$ exchangeable and
logically independent measurements~$y_{i}$ from a
\textbf{Gau\ss\ process},
\be
\left.\{y_{i}\}_{i=1,\ldots,n}\right|\theta_{1}, \theta_{2}, I 
\stackrel{\text{iid}}{\sim} \mathrm{N}(\theta_{1}, \theta_{2}^{2})
\ .
\ee
To simplify entailing algebra for demonstrational purposes, an
assumption that is usually unrealistic in practice will be imposed,
viz.~that \textit{one} of the two parameters in the
\textbf{Gau\ss\ total-data likelihood}~(\ref{eq:gausspdf}) is known
and, therefore, \textit{fixed}.

\subsection[Known variance]{Known variance}
When the scale parameter~$\theta_{2} = \sigma_{0}$ is
fixed, the \textbf{total-data likelihood function} is given by
\be
\lb{eq:gausspdf2}
P(\{y_{i}\}_{i=1,\ldots,n}|\theta_{1}, \sigma_{0}, I)
= \left(\frac{1}{\sqrt{2\pi}\,\sigma_{0}}\right)^{n}
\exp\left[\,-\frac{1}{2\sigma_{0}^{2}}\sum_{i=1}^{n}
\left(y_{i}-\theta_{1}\right)^{2}\,\right] \ ,
\ee
which is to be viewed as a function of an unknown location
parameter~$\theta_{1}$. Upon evaluating the $(y_{i}-
\theta_{1})^{2}$-term in the exponent, one finds that
$\sum_{i=1}^{n}y_{i} = n\bar{y}$ is a sufficient statistic
for~$\theta_{1}$ according to Eq.~(\ref{eq:expfam}).

\medskip
\noindent
Choosing for~$\theta_{1}$ a conjugate \textbf{Gau\ss\ prior
probability distribution} [cf.~Eq.~(\ref{eq:gausspriorpdf})],
\be
\left.\theta_{1}\right| m_{0}, s_{0}, I \sim
\mathrm{N}(m_{0}, s_{0}^{2}) \ ,
\ee
and forming the product with~(\ref{eq:gausspdf2}), then a few
algebraic steps lead to a \textbf{posterior probability
distribution} for~$\theta_{1}$ proportional to
\be
P(\theta_{1}|\{y_{i}\}_{i=1, \ldots, n}, m_{0}, s_{0}, 
\sigma_{0}, I) 
\propto \exp\left[\,-\frac{1}{2}
\left(\frac{\theta_{1}-\mu_{1}}{\sigma_{1}}\right)^{2}\,\right]
\ ,
\ee
with posterior expectation value and posterior variance 
for~$\theta_{1}$ given by
\be
\mu_{1} := 
\left(\frac{1}{s_{0}^{2}}
+ \frac{n}{\sigma_{0}^{2}}\right)^{-1}
\times
\left(\frac{1}{s_{0}^{2}}\,m_{0}
+ \frac{n}{\sigma_{0}^{2}}\,\bar{y}\right) \ ,
\qquad
\sigma_{1}^{2} := \left(\frac{1}{s_{0}^{2}}
+ \frac{n}{\sigma_{0}^{2}}\right)^{-1} \ .
\ee
Upon normalisation, this yields a \textbf{Gau\ss\ distribution}
\be
\fbox{$\displaystyle
\lb{eq:postgauss}
\left.\theta_{1}\right|\{y_{i}\}_{i=1, \ldots, n}, m_{0}, s_{0}, 
\sigma_{0}, I
\sim \mathrm{N}(\mu_{1}, \sigma_{1}^{2}) \ ;
$}
\ee
see, e.g., Lee (2012)~\ct[Subsec.~2.3.1]{lee2012},
Gelman \textit{et al} (2014)~\ct[Sec.~2.5]{geletal2014}, or Gill 
(2015)~\ct[Sec.~3.2]{gil2015}. In a fashion identical to 
previous cases, the posterior expectation value may be 
decomposed so that
\bea
\lb{eq:postmeangauss1}
\mathrm{E}(\theta_{1}|\{y_{i}\}_{i=1, \ldots, n}, m_{0}, s_{0},
\sigma_{0}, I)
& = & \mu_{1} \nonumber \\
& = & \frac{1/s_{0}^{2}}{1/s_{0}^{2}+n/\sigma_{0}^{2}} \times 
\underbrace{m_{0}}_{\text{prior mean}}
+ \frac{n/\sigma_{0}^{2}}{1/s_{0}^{2}+n/\sigma_{0}^{2}} \times 
\underbrace{\bar{y}}_{\text{sample mean}}
\eea
the prior effective sample size,~$1$, and the actual sample
size,~$n$, are weighted by the precisions $1/s_{0}^{2}$ and
$1/\sigma_{0}^{2}$, respectively. Depending on the values of these
two precisions, the sample mean will usually dominate
for~$n \gg 1$. Sampling from the single-parameter posterior
distribution~(\ref{eq:postgauss}) and calculating further
summary statistics can be easily accomplished by use of
pre-programmed \R-functions; cf. Subsec.~\ref{subsec:gaussdistr}.

\subsection[Known mean]{Known mean}
When the location parameter~$\theta_{1} = \mu_{0}$ is
fixed, the \textbf{total-data likelihood function} is given by
\be
\lb{eq:gausspdf3}
P(\{y_{i}\}_{i=1,\ldots,n}|\mu_{0}, \theta_{2}^{2}, I)
= \frac{1}{(\sqrt{2\pi})^{n}}\,\frac{1}{(\theta_{2}^{2})^{n/2}}\,
\exp\left[\,-\frac{1}{2\theta_{2}^{2}}\sum_{i=1}^{n}
\left(y_{i}-\mu_{0}\right)^{2}\,\right] \ ,
\ee
which is to be viewed as a function of an unknown \textit{squared}
scale parameter~$\theta_{2}^{2}$. By inspection, one finds that
$\sum_{i=1}^{n}(y_{i}-\mu_{0})^{2} =: nv$ is a sufficient statistic
for~$\theta_{2}^{2}$ according to Eq.~(\ref{eq:expfam}).

\medskip
\noindent
Now choosing for~$\theta_{2}^{2}$ a conjugate \textbf{inverse Gamma
prior probability distribution} [cf.~Eq.~(\ref{eq:invgammapdf})],
\be
\left.\theta_{2}^{2}\right| \alpha_{0}, \beta_{0}, I \sim
\mathrm{IG}(\alpha_{0}, \beta_{0}) \ ,
\ee
and forming the product with~(\ref{eq:gausspdf3}), then a little
algebra yields a \textbf{posterior probability distribution}
for~$\theta_{2}^{2}$ proportional to
\be
P(\theta_{2}^{2}|\{y_{i}\}_{i=1, \ldots, n}, \mu_{0}, \alpha_{0}, 
\beta_{0}, I) 
\propto \frac{1}{(\theta_{2}^{2})^{\alpha_{0}+(n/2)+1}}\,
\exp\left[\,-\left(\beta_{0}+\frac{n}{2}\,v\right)/\theta_{2}^{2}
\,\right] \ ,
\ee
i.e., upon normalisation, an \textbf{inverse Gamma distribution} 
\be
\fbox{$\displaystyle
\lb{eq:postgausssig}
\left.\theta_{2}^{2}\right|\{y_{i}\}_{i=1, \ldots, n}, \mu_{0}, 
\alpha_{0}, \beta_{0}, I
\sim \mathrm{IG}\left(\alpha_{0}+\frac{n}{2}, 
\beta_{0}+\frac{n}{2}\,v\right) \ ;
$}
\ee
see, e.g., Lee (2012)~\ct[Subsec.~2.7.1]{lee2012}, Gelman
\textit{et al} (2014)~\ct[Sec.~2.6]{geletal2014}, or Gill
(2015)~\ct[Sec.~3.2]{gil2015}. The posterior expectation value
and posterior variance for~$\theta_{2}^{2}$ are given by
\bea
\mathrm{E}(\theta_{2}^{2}|\{y_{i}\}_{i=1, \ldots, n}, \mu_{0},
\alpha_{0}, \beta_{0}, I)
& = &
\frac{\beta_{0}+\frac{n}{2}\,v}{\alpha_{0}+\frac{n}{2}-1} \\
\mathrm{Var}(\theta_{2}^{2}|\{y_{i}\}_{i=1, \ldots, n}, \mu_{0},
\alpha_{0}, \beta_{0}, I)
& = &
\frac{\left(\beta_{0}+\frac{n}{2}\,v\right)^{2}}{\left(\alpha_{0}
+\frac{n}{2}-1\right)^{2}\left(\alpha_{0}+\frac{n}{2}-2\right)}
\ ,
\eea
respectively, both of which are well-defined for
$\alpha_{0}+(n/2)-2 > 0$. Sampling from the single-parameter
posterior distribution~(\ref{eq:postgausssig}) and calculating
further summary statistics can be easily accomplished by use of
pre-programmed \R-functions; cf.
Subsec.~\ref{subsec:invgammadistr}.

\section[Exponentially distributed univariate continuous
data]{Exponentially distributed univariate continuous data}
\lb{sec:exppara}
Lastly, when there are given from an \textbf{exponential process}
measurements of $n$~exchangeable and logically independent interval
lengths~$y_{i}$,
\be
\left.\{y_{i}\}_{i=1,\ldots,n}\right|\theta, I 
\stackrel{\text{iid}}{\sim} \mathrm{Exp}(\theta) \ ,
\ee
then the \textbf{total-data likelihood function} is
\be
\lb{eq:exppdf2}
P(\{y_{i}\}_{i=1,\ldots,n}|\theta, I)
= \theta^{n}\,\exp\left(-\,\theta\,\sum_{i=1}^{n}y_{i}\right) \ ,
\ee
with unknown rate parameter~$\theta \in \mathbb{R}_{>0}$. By
Eq.~(\ref{eq:expfam}), $\sum_{i=1}^{n}y_{i} = n\bar{y}$ is a
sufficient statistic for~$\theta$.

\medskip
\noindent
Introducing for~$\theta$ a conjugate \textbf{Gamma prior
probability distribution} [cf. Eq.~(\ref{eq:gammapdf})],
\be
\left.\theta\right| \alpha, \beta, I \sim \mathrm{Ga}(\alpha,\beta)
\ ,
\ee
and forming the product with~(\ref{eq:exppdf2}), then a few
re-arrangements lead to a \textbf{posterior probability
distribution} for~$\theta$ proportional to
\be
P(\theta|\{y_{i}\}_{i=1,\ldots,n}, \alpha, \beta, I) \propto
\theta^{(\alpha+n)-1}\,\exp\left[\,
-\left(\beta+\sum_{i=1}^{n}y_{i}\right)\theta\,\right] \ ,
\ee
i.e., upon normalisation, a \textbf{Gamma distribution}
\be
\fbox{$\displaystyle
\lb{eq:postexp}
\left.\theta\right|\{y_{i}\}_{i=1,\ldots,n}, \alpha, \beta, I
\sim \mathrm{Ga}(\alpha+n,\beta+\sum_{i=1}^{n}y_{i}) \ ;
$}
\ee
cf. Gelman \textit{et al} (2014)~\ct[Sec.~2.6]{geletal2014}.
From Eq.~(\ref{eq:gammaexp}), the decomposed posterior
expectation value for~$\theta$ is
\bea
\lb{eq:postmeanexp}
\mathrm{E}(\theta|\{y_{i}\}_{i=1,\ldots,n}, \alpha, \beta, I)
& = & \frac{\alpha+n}{\beta+\sum_{i=1}^{n}y_{i}} \nonumber \\
& = & \frac{\beta}{\beta+\sum_{i=1}^{n}y_{i}} \times 
\underbrace{\frac{\alpha}{\beta}}_{\text{prior mean}}
+ \frac{\sum_{i=1}^{n}y_{i}}{\beta+\sum_{i=1}^{n}y_{i}} \times 
\underbrace{\frac{n}{\sum_{i=1}^{n}y_{i}}}_{\text{sample
mean}} \ .
\eea
For $\sum_{i=1}^{n}y_{i} \gg \beta$, the sample mean for~$\theta$,
which is $n/\sum_{i=1}^{n}y_{i} = 1/\bar{y}$, will dominate over
its prior mean, $\alpha/\beta$. 

\medskip
\noindent
From Eq.~(\ref{eq:gammavar}), the posterior variance for~$\theta$
amounts to
\be
\lb{eq:postvarexp}
\mathrm{Var}(\theta|\{y_{i}\}_{i=1,\ldots,n}, \alpha, \beta, I)
= \frac{\alpha+n}{(\beta+\sum_{i=1}^{n}y_{i})^{2}} \ .
\ee
Sampling from the single-parameter posterior
distribution~(\ref{eq:postexp}) and calculating further summary
statistics can be easily accomplished by use of pre-programmed
\R-functions; cf. Subsec.~\ref{subsec:gammadistr}.

\medskip
\noindent
We now turn to discuss some exactly solvable two-parameter
estimation problems in the next chapter.

\chapter[Joint two-parameter estimation]{Joint two-parameter
estimation for univariate Gau\ss\ processes}
\lb{ch5}
The principles remain the same, but the entailing computations
become considerably more complex quite quickly, when the
\textbf{posterior joint probability distribution} for two or more
unknown \textbf{model parameters} in a model-building project is to
be estimated from background information and available
quantitative--empirical data. In fact, there are not many known
cases in which this goal can be achieved
by means of closed-form analytical solutions. To provide a
taste of the technical complexities involved in the model-building
process based on a higher-dimensional \textbf{parameter space}, we
will outline in this chapter the derivation of the
\textbf{posterior joint probability distribution} and its
associated \textbf{posterior marginal probability distributions} in
two dimensions for univariate continuous data originating from a
\textbf{Gau\ss\ process}. In consequence, the discussion to follow 
proves a wee bit formula-(integration)-heavy, though this should
not deter the favourably inclined reader from continuing the
exciting journey through modern techniques of \textbf{inductive 
statistical inference}.

\medskip
\noindent
Suppose given a sample of size~$n$ of iid-measurements
$\{y_{i}\}_{i=1,\ldots,n}$ gained from a univariate
\textbf{Gau\ss\ process}. Then the \textbf{total-data likelihood
function} is constructed as the product of $n$~copies of the
single-datum likelihood function given in Eq.~(\ref{eq:gausspdf}).
In the following we will employ standard notation for the
\textbf{model parameters} in a \textbf{Gau\ss\ process} context and
set $\theta_{1} = \mu$ for the \textbf{location parameter}
and~$\theta_{2}^{2} = \sigma^{2}$ for the \textbf{scale
parameter}. The starting point is thus given by
\be
P(\{y_{i}\}_{i=1,\ldots,n}|\mu, \sigma^{2}, I)
= \left(\frac{1}{\sqrt{2\pi}\,\sigma}\right)^{n}
\exp\left[\,-\frac{1}{2\sigma^{2}}\sum_{i=1}^{n}
\left(y_{i}-\mu\right)^{2}\,\right] \ .
\ee
Then, evaluating first the squared term in the exponent of the
exponential function, re-ordering resultant terms, and
compactifying again by completing two convenient squares upon
adding in a zero via the identity~$0=n\bar{y}^{2}-n\bar{y}^{2}$
involving the sample mean, one arrives at
\bea
\lb{eq:gausspdf4}
P(\{y_{i}\}_{i=1,\ldots,n}|\mu, \sigma^{2}, I)
& = & \frac{1}{(\sqrt{2\pi})^{n}}\,\frac{1}{(\sigma^{2})^{n/2}}
\nonumber \\
& & \qquad \times
\exp\left[\,-\frac{1}{2\sigma^{2}}\left(\sum_{i=1}^{n}y_{i}^{2}
-2\mu\sum_{i=1}^{n}y_{i}+\mu^{2}\sum_{i=1}^{n}1\right)\right]
\nonumber \\
& = & \frac{1}{(\sqrt{2\pi})^{n}}\,\frac{1}{(\sigma^{2})^{n/2}}\,
\exp\left[\,-\frac{1}{2\sigma^{2}}\left(\sum_{i=1}^{n}y_{i}^{2}
-2n\bar{y}\mu+n\mu^{2}\right)\right]
\nonumber \\
& \stackrel{0=n\bar{y}^{2}-n\bar{y}^{2}}{=} &
\frac{1}{(\sqrt{2\pi})^{n}}\,\frac{1}{(\sigma^{2})^{1/2}}\,
\exp\left[\,-\frac{1}{2}\left(
\frac{\mu-\bar{y}}{\sigma/\sqrt{n}}\right)^{2}\,\right]
\nonumber \\
&  & \qquad \times
\frac{1}{(\sigma^{2})^{(n-1)/2}}\,
\exp\left[\,-\frac{1}{2}\,\sum_{i=1}^{n}\left(y_{i}
-\bar{y}\right)^{2}/\sigma^{2}\,\right] \ .
\eea
Altogether, we recognise in this result the product between the
kernels of a univariate \textbf{Gau\ss\ distribution} on the
one-hand side, and of a univariate \textbf{inverse Gamma
distribution} on the other, viz.,
\bea
\lb{eq:gausspdf5}
P(\{y_{i}\}_{i=1,\ldots,n}|\mu, \sigma^{2}, I)
& = &
\frac{1}{(\sqrt{2\pi})^{n}}\,\frac{1}{(\sigma^{2})^{1/2}}\,
\exp\left[\,-\frac{1}{2}\left(
\frac{\mu-\bar{y}}{\sigma/\sqrt{n}}\right)^{2}\,\right]
\nonumber \\
&  & \qquad\qquad\qquad \times
\frac{1}{(\sigma^{2})^{(n-1)/2}}\,
\exp\left(-\frac{1}{2}\,\text{TSS}/\sigma^{2}\right) \ .
\eea
In this last expression we defined a \textbf{total sum of squared
deviations} of the univariate data from their common sample mean by
\be
\lb{eq:tss}
\text{TSS} := \sum_{i=1}^{n}\left(y_{i}-\bar{y}\right)^{2} \ .
\ee

\medskip
\noindent
By inspection, it becomes apparent that, according to
Eq.~(\ref{eq:expfam}), the quantity
$\sum_{i=1}^{n}y_{i} = n\bar{y}$, which is proportional to the
\textbf{sample mean}, constitutes a sufficient statistic for the
\textbf{location parameter}~$\mu$. Likewise, the quantity
$\text{TSS}$, in terms of which one defines the \textbf{sample
variance}, is a sufficient statistic for the \textbf{scale
parameter}~$\sigma^{2}$.

\medskip
\noindent
Let us now supply the \textbf{total-data likelihood
function}~(\ref{eq:gausspdf5}) with a bivariate \textbf{prior joint
probability distribution} for the \textbf{model parameters}~$\mu$
and~$\sigma^{2}$. We will factorise it by making use of the product
rule~(\ref{eq:productrule}), so that
\be
P(\mu, \sigma^{2}|I) = P(\mu|\sigma^{2}, I) \times P(\sigma^{2}|I)
\ee
obtains. In the next two sections we will consider specific choices
for $P(\mu, \sigma^{2}|I)$ that are motivated by computational
convenience and the fact that they lead to closed form solutions
for the bivariate \textbf{posterior joint probability
distribution}. Of course, both examples possess high practical
relevance, too.

\pagebreak
\section[Uniform joint prior]{Uniform joint prior}
\lb{sec:unijointprior}
The simplest choice is that of an \textit{improper}, non-normalised
but transformation-invariant uniform \textbf{prior joint
probability distribution} for~$\mu$ and~$\sigma^{2}$ according to
Eqs.~(\ref{eq:flatprior1}) and~(\ref{eq:flatprior2}), which takes
the form
\be
P(\mu, \sigma^{2}|I) =
\underbrace{c_{1}}_{\propto
\ P(\mu|\sigma^{2}, I)}
\times
\underbrace{\frac{c_{2}}{\sigma^{2}}}_{\propto
\ P(\sigma^{2}| I)} \ ,
\ee
with $c_{1} > 0$, $c_{2} > 0$. This choice expresses the assumption
of initial logical independence between~$\mu$ and~$\sigma^{2}$.

\medskip
\noindent
Presently the derivation of the \textbf{posterior joint probability
distribution} for~$\mu$ and~$\sigma^{2}$, which is obtained from
multiplying the total-data likelihood function by the prior joint
probability distribution, does not require a lot of computational
effort. The result readily exhibits the product structure (see,
e.g., Lee (2012)~\ct[Sec.~2.12]{lee2012}, or Gelman \textit{et
al} (2014)~\ct[Sec.~3.2]{geletal2014})
\bea
P(\mu, \sigma^{2}|\{y_{i}\}_{i=1,\ldots,n}, I)
& \propto &
\underbrace{\frac{1}{(\sigma^{2})^{1/2}}\,
\exp\left[\,-\frac{1}{2}\left(
\frac{\mu-\bar{y}}{\sigma/\sqrt{n}}\right)^{2}\,\right]}_{\propto
\ \mathrm{N}(\bar{y}, \sigma^{2}/n)} \nonumber \\
& &  \qquad\qquad\qquad \times 
\underbrace{\frac{1}{(\sigma^{2})^{(n-1)/2+1}}\,
\exp\left(-\frac{1}{2}\,\text{TSS}/\sigma^{2}\right)}_{\propto
\ \mathrm{IG}((n-1)/2, \text{TSS}/2)} \ .
\eea
Upon normalisation, this gives a bivariate \textbf{Gau\ss--inverse
Gamma model} to describe the \textbf{uncertainty} inherent in the
joint estimation of~$\mu$ and~$\sigma^{2}$, i.e.,\footnote{Sivia
and Skilling (2006)~\ct[Sec.~3.3]{sivski2006} discuss the case of
estimating~$\mu$ and~$\sigma^{2}$ for a univariate Gau\ss\ process
with improper prior probability distributions that are constants
for \textit{both} parameters.}
\be
\fbox{$\displaystyle
\left.\mu, \sigma^{2}\right|\{y_{i}\}_{i=1,\ldots,n}, I
\sim
\mathrm{N}(\bar{y}, \sigma^{2}/n) \times
\mathrm{IG}((n-1)/2, \text{TSS}/2) \ .
$}
\ee

\medskip
\noindent
In practice, one is often primarily interested in the
\textbf{posterior probability distribution} for a \textit{single}
model parameter, which can be derived from the \textbf{posterior
joint probability distribution} by way of \textbf{marginalisation}
with respect to either the scale parameter~$\sigma^{2}$ or the
location parameter~$\mu$.

\medskip
\noindent
Averaging the posterior joint probability distribution over the
full range of the scale parameter~$\sigma^{2}$, and applying
the substitution method to compactify the exponent of the
exponential function, one finds that the \textbf{posterior marginal
probability distribution} for~$\mu$ is proportional to
\bea
P(\mu|\{y_{i}\}_{i=1,\ldots,n}, I)
& = &
\int_{0}^{\infty}P(\mu, \sigma^{2}|\{y_{i}\}_{i=1,\ldots,n}, I)
\,\mathrm{d}\sigma^{2} \nonumber \\
& \propto & \int_{0}^{\infty}\frac{1}{(\sigma^{2})^{(n-1)/2
+3/2}}\,\exp\left[\,-\frac{1}{2\sigma^{2}}\left(\text{TSS}
+n\left(\mu-\bar{y}\right)^{2}\right)\right]\,\mathrm{d}\sigma^{2}
\nonumber \\
& \stackrel{\text{substitution}}{\propto} &
\left[\,\text{TSS}+n\left(\mu-\bar{y}\right)^{2}\,
\right]^{-n/2} \times
\underbrace{\int_{0}^{\infty}t^{(n/2)-1}\,e^{-t}\,
\mathrm{d}t}_{\text{constant}} \nonumber \\
& \propto & \left[\,\text{TSS}+n\left(\mu-\bar{y}\right)^{2}\,
\right]^{-n/2} \nonumber \\
& \propto & \left[\,1+\frac{1}{n-1}\left(
\frac{\mu-\bar{y}}{\sqrt{\text{TSS}/(n-1)n}}\right)^{2}
\,\right]^{-n/2} \ ,
\eea
so that, upon proper normalisation, this yields a univariate
\textbf{non-central $\boldsymbol{t}$--distribution},
\be
\left.\mu\right|\{y_{i}\}_{i=1,\ldots,n}, I
\sim
t(\bar{y}, \sqrt{\text{TSS}/(n-1)n}, n-1) \ .
\ee
Sampling from this non-central $t$--distribution may be realised
by use of the pre-programmed \R-function  given in
Subsec.~\ref{subsec:tdistr}.

\medskip
\noindent
Analogously, averaging the posterior joint probability distribution
over the full range of the location parameter~$\mu$ leads to the
\textbf{posterior marginal probability distribution}
for~$\sigma^{2}$ being proportional to
\bea
P(\sigma^{2}|\{y_{i}\}_{i=1,\ldots,n}, I)
& = &
\int_{-\infty}^{+\infty}P(\mu, \sigma^{2}|\{y_{i}\}_{i=1,\ldots,n},
I)\,\mathrm{d}\mu \nonumber \\
& \propto & \frac{1}{(\sigma^{2})^{(n-1)/2+1}}\,
\exp\left(-\frac{1}{2}\,\text{TSS}/\sigma^{2}\right) \nonumber \\
& & \times
\underbrace{\int_{-\infty}^{+\infty}\frac{1}{(\sigma^{2})^{1/2}}\,
\exp\left[\,-\frac{1}{2}\left(
\frac{\mu-\bar{y}}{\sigma/\sqrt{n}}\right)^{2}\,\right]\,
\mathrm{d}\mu}_{\text{constant}} \nonumber \\
& \propto & \frac{1}{(\sigma^{2})^{(n-1)/2+1}}\,
\exp\left(-\frac{1}{2}\,\text{TSS}/\sigma^{2}\right) \ .
\eea
Normalisation gives a univariate \textbf{inverse Gamma
distribution},
\be
\left.\sigma^{2}\right|\{y_{i}\}_{i=1,\ldots,n}, I
\sim
\mathrm{IG}((n-1)/2, \text{TSS}/2) \ .
\ee
Sampling from this inverse Gamma distribution may be realised
by use of the pre-programmed \R-function  given in
Subsec.~\ref{subsec:invgammadistr}.

\section[Conditionally conjugate joint prior]{Conditionally
conjugate joint prior}
\lb{sec:condconjjointprior}
More flexibility for practical applications offers the choice of a
conditionally conjugate \textbf{prior joint probability
distribution} for~$\mu$ and~$\sigma^{2}$. This is expressed by
\be
P(\mu, \sigma^{2}|I) \propto
\underbrace{\frac{1}{(\sigma^{2})^{1/2}}\,
\exp\left[\,-\frac{1}{2}\left(
\frac{\mu-m_{0}}{\sigma}\right)^{2}\,\right]}_{\propto
\ P(\mu|\sigma^{2}, I):\ \text{Gau\ss}}
\times
\underbrace{\frac{1}{(\sigma^{2})^{\alpha_{0}+1}}\,
\exp\left(-\beta_{0}/\sigma^{2}\right)}_{\propto
\ P(\sigma^{2}| I):\ \text{inverse Gamma}} \ ;
\ee
the prior probability distribution for~$\mu$ here is conditioned
on the value of~$\sigma^{2}$.

\medskip
\noindent
Next, a number of algebraic manipulations that involve the
completion of squares in the exponents of the exponential
functions yield the \textbf{posterior joint probability
distribution} for~$\mu$ and~$\sigma^{2}$ in the product structure
given by (see, e.g., Lee (2012)~\ct[Sec.~2.13]{lee2012}, Greenberg
(2013)~\ct[Sec.~4.3]{gre2013}, Gelman \textit{et al}
(2014)~\ct[Sec.~3.3]{geletal2014}, or Gill
(2015)~\ct[Sec.~3.4]{gil2015})
\bea
P(\mu, \sigma^{2}|\{y_{i}\}_{i=1,\ldots,n}, I)
& \propto &
\underbrace{\frac{1}{(\sigma^{2})^{1/2}}\,
\exp\left[\,-\frac{1}{2}\left(
\frac{\mu-\mu_{n}}{\sigma/\sqrt{n+1}}\right)^{2}\,\right]}_{\propto
\ \mathrm{N}(\mu_{n}, \sigma^{2}/(n+1))} \nonumber \\
& &  \qquad\qquad\qquad \times 
\underbrace{\frac{1}{(\sigma^{2})^{\alpha_{n}+1}}\,
\exp\left(-\beta_{n}/\sigma^{2}\right)}_{\propto
\ \mathrm{IG}(\alpha_{n}, \beta_{n})} \ ,
\eea
where we defined
\bea
\mu_{n} & := & \frac{1}{n+1}\,m_{0} + \frac{n}{n+1}\,\bar{y} \\
\alpha_{n} & := & \alpha_{0} + \frac{n}{2} \\
\beta_{n} & := & \beta_{0} + \frac{\text{TSS}}{2}
+ \frac{1}{2}\,\frac{n}{n+1}\,(m_{0}-\bar{y})^{2} \ .
\eea
Note that the parameter~$\mu_{n}$ is a weighted average of the
prior mean,~$m_{0}$, and the sample mean,~$\bar{y}$. Normalisation
again obtains a bivariate \textbf{Gau\ss--inverse Gamma model} for
the joint estimation of~$\mu$ and~$\sigma^{2}$, viz.,
\be
\fbox{$\displaystyle
\left.\mu, \sigma^{2}\right|\{y_{i}\}_{i=1,\ldots,n}, I
\sim
\mathrm{N}(\mu_{n}, \sigma^{2}/(n+1)) \times 
\mathrm{IG}(\alpha_{n}, \beta_{n}) \ .
$}
\ee

\medskip
\noindent
\textbf{Marginalisation} to find the corresponding univariate
\textbf{posterior probability distribution} for~$\mu$ or
for~$\sigma^{2}$ proceeds along the same lines as outlined in the
previous section. Hence, averaging the posterior joint probability
distribution over the full range of the scale
parameter~$\sigma^{2}$, and applying the substitution method to
compactify the exponent of the exponential function, the {\bf
posterior marginal probability distribution} for~$\mu$ is
proportional to
\bea
P(\mu|\{y_{i}\}_{i=1,\ldots,n}, I)
& = &
\int_{0}^{\infty}P(\mu, \sigma^{2}|\{y_{i}\}_{i=1,\ldots,n}, I)
\,\mathrm{d}\sigma^{2} \nonumber \\
& \propto & \int_{0}^{\infty}\frac{1}{(\sigma^{2})^{\alpha_{n}
+3/2}}\,\exp\left[\,-\frac{1}{2\sigma^{2}}\left(2\beta_{n}
+(n+1)\left(\mu-\mu_{n}\right)^{2}\right)\right]\,
\mathrm{d}\sigma^{2} \nonumber \\
& \stackrel{\text{substitution}}{\propto} &
\left[\,2\beta_{n}+(n+1)\left(\mu-\mu_{n}\right)^{2}\,
\right]^{-(\alpha_{n}+1/2)} \times
\underbrace{\int_{0}^{\infty}t^{(\alpha_{n}+1/2)-1}\,e^{-t}\,
\mathrm{d}t}_{\text{constant}} \nonumber \\
& \propto & \left[\,2\beta_{n}+(n+1)\left(\mu-\mu_{n}\right)^{2}\,
\right]^{-(\alpha_{n}+1/2)} \nonumber \\
& \propto & \left[\,1+\frac{1}{2\alpha_{n}}\left(
\frac{\mu-\mu_{n}}{\sqrt{(\beta_{n}/\alpha_{n})/(n+1)}}\right)^{2}
\,\right]^{-(2\alpha_{n}+1)/2} \ ,
\eea
so that, after normalisation, a univariate \textbf{non-central
$t$--distribution} arises,
\be
\left.\mu\right|\{y_{i}\}_{i=1,\ldots,n}, I
\sim
t(\mu_{n}, \sqrt{(\beta_{n}/\alpha_{n})/(n+1)}, 2\alpha_{n})
\ .
\ee
Again, sampling from this non-central $t$--distribution may be
realised by use of the pre-programmed \R-function  given in
Subsec.~\ref{subsec:tdistr}.

\medskip
\noindent
Lastly, averaging the posterior joint probability distribution over
the full range of the location parameter~$\mu$, the kernel of
the \textbf{posterior marginal probability distribution}
for~$\sigma^{2}$ is given by
\bea
P(\sigma^{2}|\{y_{i}\}_{i=1,\ldots,n}, I)
& = &
\int_{-\infty}^{+\infty}P(\mu, \sigma^{2}|\{y_{i}\}_{i=1,\ldots,n},
I)\,\mathrm{d}\mu \nonumber \\
& \propto & \frac{1}{(\sigma^{2})^{\alpha_{n}+1}}\,
\exp\left(-\beta_{n}/\sigma^{2}\right) \nonumber \\
& & \times
\underbrace{\int_{-\infty}^{+\infty}\frac{1}{(\sigma^{2})^{1/2}}\,
\exp\left[\,-\frac{1}{2}\left(
\frac{\mu-\mu_{n}}{\sigma/\sqrt{n+1}}\right)^{2}\,\right]\,
\mathrm{d}\mu}_{\text{constant}} \nonumber \\
& \propto & \frac{1}{(\sigma^{2})^{\alpha_{n}+1}}\,
\exp\left(-\beta_{n}/\sigma^{2}\right) \ ,
\eea
and normalisation converts this into a univariate \textbf{inverse
Gamma distribution},
\be
\left.\sigma^{2}\right|\{y_{i}\}_{i=1,\ldots,n}, I
\sim
\mathrm{IG}(\alpha_{n}, \beta_{n}) \ .
\ee

\medskip
\noindent
Once more, sampling from this inverse Gamma distribution may be
realised by use of the pre-programmed \R-function  given in
Subsec.~\ref{subsec:invgammadistr}.

\medskip
\noindent
The extension of the parameter estimation procedure discussed in
this section to cases of quantitative--empirical data from
Gau\ss\ processes with more than two model parameters is
conceptually (though not computationally) straightforward. In the
next chapter we will describe in some detail in the context of 
generalised linear models how posterior joint probability
distributions over multi-dimensional parameter spaces and their
marginal accomplices can be simulated numerically by resorting to 
techniques that make use of Markov Chain Monte Carlo iteration
codes. This particular methodology provides them with a broad basis
for many practical applications in inductive statistical inference.

\chapter[Fitting and assessing generalised linear models]{Fitting 
and assessing generalised linear models}
\lb{ch6}
So how, in the context of a research problem of interest, does one
pursue the building of a concrete \textbf{statistical model} from
relevant background information and direct observational or
experimental evidence, when the model's related \textbf{parameter
space} becomes high-dimensional due to problem-inherent complexity,
and closed-form analytical solutions are no longer possible?

\medskip
\noindent
\textbf{Regression analysis} of quantitative--empirical data has
long been the workhorse of \textbf{inductive statistical
inference}. Its prime objective is the construction of an
empirically validated \textbf{statistical model} which is to be
viewed as a representation of a specific \textbf{scientific theory}
in the realm of one's research activities. In the
\textbf{Bayes--Laplace approach} the construction of a
\textbf{statistical model} means foremost determining a
\textbf{posterior joint probability distribution} for a certain
finite number of unknown \textbf{model parameters} from a suitable 
\textbf{joint prior probability distribution} for these
\textbf{model parameters} and \textbf{quantitative--empirical data} obtained from sample measurements for the various
\textbf{statistical variables} the researcher included in her/his
portfolio on the grounds of intensive theoretical considerations.
Making valuable progress in the task of finding the
\textbf{posterior joint probability distribution} for the
\textbf{model parameters} has, by now, been possible for a few
decades via employing one of the many well-distributed powerful and
efficient \textbf{numerical algorithms} that yield
discrete approximations of an accuracy sufficient for reliable
inference. For example, in this way hitherto unknown
\textbf{bivariate correlations} between essential \textbf{model
parameters} can be learned, and also the \textbf{posterior marginal
probability distributions} for single \textbf{model parameters}
that are of central interest can be computed. Any
\textbf{statistical model}, whether obtained by analytical means or
via numerical simulation, needs to be checked for its
\textbf{sensitivity} to the \textbf{prior assumptions}
injected in the model-building process, and it also has to be
assessed for its \textbf{out-of-sample posterior predictive
accuracy}.

\medskip
\noindent
We will outline in the following the main steps of
\textbf{regression analysis} within the \textbf{Bayes--Laplace
approach} for different types of exchangeable
\textbf{quantitative--empirical data} in
the context of \textbf{generalised linear models} (see Nelder and
Wedderburn (1972)~\ct{nelwed1972}). We will describe how the
numerical approximation of high-dimensional \textbf{posterior joint
probability distributions} for unknown \textbf{model parameters}
can be performed employing the Stan probabilistic programming
language that makes available one of the currently most reliable
and efficient numerical algorithms; cf. Stan Development Team
(2022a)~\ct{sta2022a} and Carpenter \textit{et al}
(2017)~\ct{caretal2017}.

\section[Generalised linear models]{Generalised linear models}
\lb{sec:glm}
Let us first introduce some compact notation that is to be used in
the applications presented in subsequent sections. The
\textbf{quantitative--empirical data},
$\{x_{ij}\}_{i=1,\ldots,n}^{j=1,\ldots,k}$, from a sample of
size~$n$ for a set of $k \in
\mathbb{N}$ \textbf{independent variables}, is to be collected in a
so-called \textbf{design matrix} (or model matrix), 
\be
\lb{eq:designmat}
\boldsymbol{X} :=
\left(\begin{array}{ccccc}
1 & x_{11} & x_{12} & \ldots & x_{1k} \\
1 & x_{21} & x_{22} & \ldots & x_{2k} \\
\vdots & \vdots & \vdots & \ddots & \vdots \\
1 & x_{n1} & x_{n2} & \ldots & x_{nk}
\end{array}\right) \in \mathbb{R}^{n \times (k+1)} \ ;
\ee
herein the data is conventionally augmented by a column of ones,
$x_{i0} = 1$, for reasons that will become apparent shortly. The
data contained in the design matrix~$\boldsymbol{X}$ can be either
\textbf{metrically scaled}, or of a binary nature as arising from
\textbf{indicator variables} that take values in the set~$\{0,
1\}$.

\medskip
\noindent
In the model-building process, the data for the
\textbf{independent variables} will be employed as
\textbf{predictors} (or explanatory variables), and, typically,
\textit{no} assumptions are made concerning their distributional
origin, or as to the accuracy of their measurement. Potential
problems are associated with the metrical data in the design
matrix~$\boldsymbol{X}$. When
\textbf{multi-collinearity} abounds, the \textbf{out-of-sample
posterior predictive accuracy} of a statistical model is 
weakened from the
outset; see, e.g., McElreath (2020)~\ct[Sec.~6.1.]{mce2020a}.
Multi-collinearity amounts to redundant information in the
$\boldsymbol{X}$-data that arises when strong bivariate
correlations exist between some of the metrically scaled
``independent variables.'' It has the effect that \textbf{posterior
joint probability distributions} for unknown \textbf{model
parameters} become spread out more strongly in \textbf{parameter
space}, being synonymous with
an increase in \textbf{uncertainty}. The inclusion of redundant
information in the model-building process should thus be avoided. 
In the following we will assume that bivariate correlations in the
metrical part of the $\boldsymbol{X}$-data are negligibly small.

\medskip
\noindent
A vector of $k+1$~ real-valued \textbf{model parameters}
(or regression coefficients) is introduced next by
\be
\boldsymbol{\beta} :=
\left(\begin{array}{c}
\beta_{0} \\
\beta_{1} \\
\vdots \\
\beta_{k}
\end{array}\right) \in \mathbb{R}^{(k+1) \times 1} \ ,
\ee
so that arbitrary linear combinations of the data for the
independent variables can be represented by the \textbf{linear
form}
\be
\lb{eq:lin}
\boldsymbol{X}\boldsymbol{\beta} \ .
\ee

\medskip
\noindent
The~$n$ measured values $\{y_{i}\}_{i=1,\ldots,n}$ for the single
\textbf{dependent variable}~$Y$ are assembled in a vector
\be
\boldsymbol{y} :=
\left(\begin{array}{c}
y_{1} \\
y_{2} \\
\vdots \\
y_{n}
\end{array}\right) \in \mathbb{R}^{n \times 1} \ .
\ee
The $\boldsymbol{y}$-data may vary discretely or continuously, it
may be binary in nature, represent counts, or take any real value
from a pre-specified range. The data for the dependent
variable~$\boldsymbol{y}$ is to be predicted from the data for the
independent variables contained in~$\boldsymbol{X}$ via the
\textbf{statistical model} one seeks to construct. In describing
the model-building process, we will limit considerations to the
discussion of additive \textbf{main effects} of the independent
variables, and point to the literature for the numerous
possibilities of including non-additive \textbf{interaction
effects}. The latter are to be seen as an option for devising
\textbf{statistical models} of a higher degree of
flexibility, though at the price of introducing a higher number of
unknown \textbf{model parameters} and being more difficult to
interpret.

\medskip
\noindent
In \textbf{generalised linear models (GLM)}, the relationship
between the dependent variable~$\boldsymbol{y}$ and the independent
variables contained in~$\boldsymbol{X}$ need no longer be linear,
nor does the dispersion of the data have to be of the Gau\ss ian
type; see Nelder and Wedderburn (1972)~\ct{nelwed1972}, Lee 
(2012)~\ct[Sec.~6.7]{lee2012}, Gelman \textit{et al} 
(2014)~\ct[Ch.~16]{geletal2014}, Krusch\-ke 
(2015)~\ct[Sec.~15.4]{kru2015}, Gill 
(2015)~\ct[App.~A]{gil2015}, or McElreath 
(2020)~\ct[Sec.~10.2.]{mce2020a}. \textbf{GLM} exhibit a generic
two-level structure, comprising both a \textit{deterministic} and a
\textit{probabilistic} component. These are given respectively by
\bea
\lb{eq:glm1}
f(\theta) & = & \boldsymbol{X}\boldsymbol{\beta} \\
%
\lb{eq:glm2}
\left.\boldsymbol{y}\right|\boldsymbol{X}, \boldsymbol{\beta},
\text{other\ parameter(s)}, I
& \stackrel{\text{ind}}{\sim} & \text{pdf}\left(\theta, 
\text{other\ parameter(s)}, I\right) \ ;
\eea
%
Eq.~(\ref{eq:glm1}), the \textbf{linear model}, relates a
\textbf{parameter}~$\theta$ of a
suitable \textbf{single-datum likelihood function} for the
\textbf{dependent variable}~$\boldsymbol{y}$ to the deterministic
\textbf{linear form}~$\boldsymbol{X}\boldsymbol{\beta}$ via a
continuously differentiable and invertible scalar-valued
\textbf{link function}, $f$, while Eq.~(\ref{eq:glm2}) represents
the chosen \textbf{single-datum likelihood function} for
the~$\boldsymbol{y}$-data-generating process itself, which, besides
$\theta$, may also depend on some other parameter(s). In some cases
of practical interest the \textbf{parameter}~$\theta$ is chosen to
be the \textbf{expectation value} $\mathrm{E}(\boldsymbol{y}|
\boldsymbol{X}, \boldsymbol{\beta}, \text{other\ parameter(s)}, I)
=: \mu$ for the \textbf{dependent variable}~$\boldsymbol{y}$.
Examples will be given in Ch.~\ref{ch7} and Ch.~\ref{ch9} below.

\medskip
\noindent
In preparation of subsequent discussions on the application of
iterative numerical simulations for the building of a
\textbf{statistical model}, we draw the reader's attention to the
empirical fact that \textbf{standardisation} of the metrically
scaled components in the $\boldsymbol{y}$- and
$\boldsymbol{X}$-data, and, consequently, of the related
\textbf{model parameters}~$\boldsymbol{\beta}$, renders iterative 
numerical simulations more efficient by reducing autocorrelation in 
the sampling outcomes. This immediately improves the \textbf{mixing
properties} of the sampling outcomes, and, ultimately, the overall 
\textbf{numerical stability} of the approximative solutions for
\textbf{posterior joint probability distributions}; cf. Krusch\-ke
(2015)~\ct[Sec.~17.2]{kru2015} and McElreath
(2020)~\ct[p~111]{mce2020a}. \textbf{Standardisation} amounts to a
homogenisation of \textbf{measurement scales}; cf.
Ref.~\ct[Subsec.~3.2.6]{hve2019}. It proves to be a 
straightforward algebraical exercise to transform back variables
and model parameters from standardised measurement scales to
original measurement scales, once simulations have been completed.

\section[Monte Carlo sampling algorithms]{Monte Carlo sampling
algorithms}
\lb{sec:MCsampling}
The Polish--US-American mathematician and nuclear physicist
\href{https://mathshistory.st-andrews.ac.uk/Biographies/Ulam/}{Stanislaw
Marcin Ulam (1909--1984)} and the Hungarian--US-American
mathematician, physicist and computer scientist
\href{https://mathshistory.st-andrews.ac.uk/Biographies/Von_Neumann/}{John von Neumann (1903--1957)}
pioneered the development of a
family of algorithmic techniques that have come to be known across
the empirical scientific disciplines as \textbf{Monte Carlo
simulations}. For their numerical experiments these researchers
employed the first generation of computers. The specific term
``Monte Carlo'' (MC) was coined as a code name for undisclosed 
activities in a joint paper by Metropolis and Ulam
(1949)~\ct{metula1949}. It is a historical fact that the first
powerful simulation algorithms were a spin-off of intense
conceptual research work at \href{https://lanl.gov}{Los Alamos
National Laboratory}, NM, USA during the 1940ies and 1950ies which
was invested with the aim of acquiring nuclear fission and fusion
bombs.

\medskip
\noindent
The developments on the algorithmic front were followed from the
1990ies onwards by revolutionising technological advances 
in the hardware sector that triggered an incredible boost of 
computing power on standard household notebooks and similar
computing devices.\footnote{To put this into perspective: today,
for example, every average smartphone outperforms by a few
orders of magnitude the gigantic computing machines that were
available to  NASA when landing human beings on the Moon during the
late 1960ies and early 1970ies. See URL (cited on August 7, 2018): 
\href{https://www.zmescience.com/research/technology/smartphone-power-compared-to-apollo-432/}{www.zmescience.com/research/technology/smartphone-power-compared-to-apollo-432/}.}
With boundary conditions so hospitable to transformation,
the ensuing \textbf{Markov chain Monte Carlo (MCMC)} simulation
techniques and their integration into the \textbf{Bayes--Laplace
approach to data analysis and statistical inference} were offered a
real chance to excel. The term ``Markov Chain'' refers to the
property of the algorithms driving the simulations that an
iteration step in a multi-dimensional \textbf{parameter space} to a
new position depends only on the present position, and \textit{not}
on any earlier positions. The targeted high-dimensional, stationary
\textbf{posterior joint probability distribution} for unknown
\textbf{model parameters} used in \textbf{inductive statistical
inference} is gradually built over typically thousands of iteration
steps.

\medskip
\noindent
There are three types of MCMC sampling algorithms that find
widespread use in \textbf{statistical model-building}:
\begin{enumerate}

\item The \textbf{Metropolis--Hastings (MH) sampling algorithm} was
originally put forward in the paper by Metropolis \textit{et al}
(1953)~\ct{metetal1953}, and significantly upgraded by Hastings 
(1970)~\ct{has1970} nearly two decades later. The basic principles
of MH~sampling are nicely explained and motivated with simple
simulations by Krusch\-ke (2015)~\ct[Sec.~7.2]{kru2015}
and by McElreath (2020)~\ct[Sec.~8.2.]{mce2020a}. Full details of
the MH~sampling algorithm are given by Greenberg
(2013)~\ct[Sec.~7.2]{gre2013} and by Gill 
(2015)~\ct[Sec.~10.4]{gil2015}.

\medskip
\noindent
Some MH~routines are contained in the \R{} package
\texttt{MCMCpack} by Martin \textit{et al} (2011)~\ct{maretal2011}.
Given how the core proposal distribution, the acceptance ratio and
the decision rule are designed to operate, the MH~sampling
algorithm does \textit{not} necessarily update simulated parameter
values and their associated posterior joint probability
distribution in every iteration step; it so proves computationally
less efficient.

\item The \textbf{Gibbs sampling algorithm},\footnote{Named after
the US-American scientist
\href{https://mathshistory.st-andrews.ac.uk/Biographies/Gibbs/}{Josiah
Willard Gibbs (1839--1903)}.} a special
case of the MH sampling algorithm, was developed by Geman and Geman
(1984)~\ct{gemgem1984}, and popularised through an influential
review paper by Gelfand and Smith (1990)~\ct{gelsmi1990}. This
method requires as input a complete set of analytically expressible
full conditional probability distributions for all the model
parameters involved. The simulated parameter values and their
associated conditional probability distributions are being
updated in cyclical order, one in every iteration step, while
holding the remaining ones fixed. Consequently, the targeted
posterior joint probability distribution is improved in every
iteration step and no computing time is squandered. The Gibbs
sampling algorithm is one of the most frequently employed MCMC
simulation methods. Full details of its structure are given by
Greenberg (2013)~\ct[Sec.~7.1]{gre2013} and by Gill
(2015)~\ct[Sec.~10.3]{gil2015}.

\medskip
\noindent
The MRC Biostatistics Unit at the University of Cambridge, UK
spearheaded the dissemination of the Gibbs sampling algorithm with
their BUGS (``Bayesian inference Using Gibbs Sampling'')
project. The BUGS code is freely available from the website 
\href{https://www.mrc-bsu.cam.ac.uk/software/bugs/}{\texttt{www.mrc-bsu.cam.ac.uk/software/bugs/}},
and its use is described
by Lunn \textit{et al} (2000)~\ct{lunetal2000}. A closely related
product is the GNU-licensed software package JAGS (``Just Another
Gibbs  Sampler'') developed by Martyn Plummer that is available
from the website  \href{https://mcmc-jags.sourceforge.io/}{\texttt{mcmc-jags.sourceforge.io/}};
see Plummer (2017, 
2019)~\ct{plu2017, plu2019}. JAGS can be operated in an \R{} 
environment upon loading the packages \texttt{rjags} and
\texttt{coda}; cf. Plummer (2019)~\ct{plu2019} and Plummer
\textit{et al} (2019)~\ct{pluetal2019}. The \R{} package
\texttt{runjags} by Denwood (2016)~\ct{den2016} offers the
possibility for parallel MCMC~generation with JAGS by activating
more than one processor on the computing device one uses for the
simulation.

\item The \textbf{Hamiltonian Monte Carlo (HMC) sampling algorithm}
was devised by Duane \textit{et al} (1987)~\ct{duaetal1987} for
simulating the quantum dynamic motion of nuclear particles
that are subjected to the strong nuclear force. In broad terms, the
method models a trapped massive quantum point particle that is
moving frictionless under the influence of an attractive external
potential. The
acceleration experienced by such a particle is proportional to the
local spatial gradient (``slope'') of the external potential. 
When adapted to the type of simulations needed in statistical
modelling, one finds that the HMC sampling algorithm traverses
a high-dimensional parameter space and scales a posterior joint
probability distribution much more efficiently than either of
the HM and Gibbs variants. This is of great advantage especially
when simulating very complex, multi-level models.
Quantitative details of the HMC sampling algorithm are described in
the renowned review by Betancourt (2018)~\ct{bet2018},
qualitative details in the textbooks by Krusch\-ke
(2015)~\ct[Sec.~14.1]{kru2015} and by McElreath
(2020)~\ct[Sec.~9.3.]{mce2020a}.

\medskip
\noindent
The leading implementation of the HMC sampling algorithm is
provided by the Stan probabilistic programming language distributed
freely by the \href{https://mc-stan.org/}{Stan
Development Team (\texttt{mc-stan.org})}; cf. Stan Development Team
(2022a)~\ct{sta2022a}. Its operation in an \R{} environment
requires installation of the package \texttt{rstan}, also
programmed by the Stan Development Team (2022d)~\ct{sta2022d}. In
these lecture notes we will provide an introduction to performing
numerical HMC simulations of posterior joint probability
distributions for unknown model parameters employing the \R{}
package \texttt{rstan}.

\end{enumerate}
On a technical note it is worthwhile pointing out that in order
to maintain numerical stability and accuracy the implementations
of all three types of MCMC sampling algorithms operate with the
\textbf{natural logarithms} of each of \textbf{likelihood
functions}, \textbf{prior} and \textbf{posterior probability
distributions}. In
this way it is possible to handle successfully extremely tiny
probability values which are commonplace when probability needs to
be spread out across multiple directions in a high-dimensional
\textbf{parameter space}. Moreover, upon acting with the natural
logarithm upon both sides of \textbf{Bayes' theorem} in its
data analysis focussed variant of Eq.~(\ref{eq:postprob}), the
right-hand side transforms into a \textit{sum} of logarithmic terms
and so facilitates computation. This becomes particularly
convenient when the \textbf{quantitative--empirical data} to be
analysed is \textbf{exchangeable}, implying that the
\textbf{total-data likelihood function} factorises into a
product of $n$ \textbf{single-datum likelihood functions}
according to Eq.~(\ref{eq:likefact}). Then the natural logarithm of
the \textbf{total-data likelihood function} converts into a
\textit{sum} of logarithmic terms itself.

\medskip
\noindent
The application of any one of the three MC sampling algorithms
named in the list above, e.g., when operating their
implementation in an \R{} environment, yields a discretised 
approximation to the targeted \textbf{posterior joint probability 
distribution} for a usually large set of unknown \textbf{model
parameters}. Conceptually these discretised approximations to a
\textbf{posterior joint probability distribution}, determined by
Eq.~(\ref{eq:postprob}), constitute higher-dimensional
generalisations of the very structure of the contingency table
displayed in Tab.~\ref{tab:jointdistr}.

\section[MCMC simulations using Stan]{MCMC simulations using Stan}
\lb{sec:stansim}
In this section we will now describe how to construct some standard
\textbf{GLMs} from \textbf{quantitative--empirical data} and
relevant \textbf{background information} by means of MCMC
simulations based on the Stan probabilistic programming language;
cf. Stan Development Team
(2022a)~\ct{sta2022a}. Stan is available for different operating
systems and for different statistical software packages. Here we
will employ the \R{} package \texttt{rstan}, also
programmed by the Stan Development Team (2022d)~\ct{sta2022d}.
Before we proceed to discuss the details of MCMC simulations using
Stan, we draw the reader's attention to three \R{}
packages that operate as high-level interfaces to Stan in order to 
reduce the amount of programming required on the part of the user.
These packages are
\texttt{brms} (``Bayesian Regression Models using Stan'') by
B\"{u}rkner (2017)~\ct{bue2017}, \texttt{rstanarm} (``Bayesian
Applied Regression Modeling via Stan'') by Goodrich
\textit{et al} (2022)~\ct{gooetal2022}, and \texttt{rethinking} by
McElreath (2021)~\ct{mce2021}. The interested reader is strongly
encouraged to gain experience in the applications of any of these
three packages for her/him-self. Very useful, too, proves the
carefully drafted tutorial on using Stan by Sorensen
\textit{et al} (2016)~\ct{soretal2016}, which primarily addresses
psychologists, linguists, and cognitive scientists, but goes along
well also with a much broader audience interested in inductive
statistical inference.

\medskip
\noindent
The process of building a \textbf{statistical model} from MCMC
simulations comprises four main steps:
\begin{enumerate}

\item model specification

\item model fitting

\item model assessment

\item model application

\end{enumerate}
We will focus on the first three in the following.

\medskip
\noindent
Before looking into the details, we briefly comment on the way
the frequently occuring process of definite integration can be 
numerically approximated using the discretised output of MCMC 
simulations. One-dimensional probability-weighted integrals of 
continuous functions $f$ of a single \textbf{model
parameter}~$\theta$ are approximated by (see Gill
(2015)~\ct[Eq.~(9.3)]{gil2015} or
Krusch\-ke (2015)~\ct[Eq.~(10.7)]{kru2015})
\be
\lb{eq:intdiscrete}
\fbox{$\displaystyle
\int_{\theta\,\text{range}}f(\theta)\,P(\theta|I)\,\mathrm{d}\theta
\approx \frac{1}{n}\sum_{\theta_{i} \sim P(\theta|I)}^{n} 
f(\theta_{i}) \ ,
$}
\ee
where $n$~denotes the total number of values~$\theta_{i}$ sampled
from the numerically simulated distribution $P(\theta|I)$. The
principle underlying this particular approximation technique can be
transferred to obtaining from the output of MCMC simulations
discretised versions of the higher-dimensional integrals over the
multivariate \textbf{posterior joint probability distribution} that
express the univariate \textbf{posterior marginal probability
distributions} for each of $k+1$ supposed \textbf{model
parameters}~$\beta_{i} \in \{\beta_{0}, \ldots, \beta_{k}\}$. This
is given by
\bea
P(\beta_{i}|\boldsymbol{y}, \boldsymbol{X}, I)
& = & \underbrace{\int\cdots\int_{\beta_{j}\neq\beta_{i}}
P(\beta_{0}, \ldots, \beta_{k}|\boldsymbol{y},
\boldsymbol{X}, I)\,\mathrm{d}\beta_{0}\cdots\mathrm{d}\beta_{k}}_{
k\ \beta-\text{integrations over posterior joint distribution,
excluding}\ \beta_{i}} \\
& \approx &
\underbrace{
\sum_{\beta_{0} \sim P(\beta|y, X, I)}^{n}
\ldots
\sum_{\beta_{k} \sim P(\beta|y, X, I)}^{n}
P(\beta_{0}, \ldots, \beta_{k}|\boldsymbol{y}, \boldsymbol{X}, I)
}_{k\ \beta-\text{summations
over posterior joint distribution, excluding}\ \beta_{i}} \ ,
\eea
which, conceptually, corresponds to a direct application of 
the \textbf{marginalisation rule} that was illustrated in
Tab.~\ref{tab:jointdistr} of Ch.~\ref{ch1}.
$P(\beta_{0}, \ldots, \beta_{k}|\boldsymbol{y},
\boldsymbol{X}, I)$~denotes the numerically simulated
\textbf{posterior joint probability distribution}.

\medskip
\noindent
So let us now address the main elements of MCMC simulations of
\textbf{posterior probability distributions} with Stan, and how the
discretised output so obtained can be checked for reliability.
Single-line comments in Stan code are to be preceded by a double
forward slash ``//,'' while comments extending across two lines or
more need to be enclosed in a ``/* \ldots */'' bracket, a feature
familiar to some readers from writing or reading C++ code. Any Stan
code must adhere to the following programme block structure; cf.
\href{https://mc-stan.org/docs/functions-reference/index.html}{Stan
Functions Reference (v2.30)}~\ct{sta2022b}:
\begin{lstlisting}[language = Stan, caption = {Basic structure of
a Stan code for defining a statistical model and for generating
numerical HMC simulations of the posterior joint probability
function for the unknown model parameters. The order of the
programme blocks is compulsory.},
captionpos = b, label = {lst:stanstruc}]
functions {
  /* Declaration of functions (optional) */
  ...
}

data {
  /* Declaration of dimensions */
  ...
	
  /* Declaration of observed variables */
  ...
}

transformed data {
  /* Transformations of observed variables (optional),
     e.g. standardisation */
  ...
}

parameters {
  /* Declaration of unobserved variables */
  ...
}

transformed parameters {
  /* Transformations of unobserved variables (optional) */
  ...
}

model {
  /* Declaration of (log-)prior distributions */
  ...
	
  /* Declaration of (log-)likelihood function */
  ...
}

generated quantities {
  /* Calculation of posterior predictive distribution (optional) */
  ...

  /* Calculation of pointwise log-likelihood function (optional) */
  ...
}
\end{lstlisting}
Note that the order amongst the different programme blocks is
compulsory. The \texttt{model} block specifies for a
\textbf{statistical model} to be fitted to given
\textbf{quantitative--empirical data} the \textbf{prior probability
distribution} for every single \textbf{model parameter} and the
\textbf{single-datum likelihood function} for the observed
\textbf{dependent variable}~$Y$. Note that the \textbf{prior
probability distributions} enter the MCMC simulations of the
multivariate \textbf{posterior joint probability distribution} only
\textit{once}, while the \textbf{quantitative--empirical data} as a 
whole enters via $n$ factors of the \textbf{single-datum likelihood
function} a total of $n$~\textit{times}.

\medskip
\noindent
Specific examples of Stan code are to be given in
Ch.~\ref{ch7} and Ch.~\ref{ch9} below and are available from
\href{https://github.com/hve1964/stanCodes}{\texttt{github.com/hve1964/stanCodes}}.
A Stan code is to be saved in a file with extension
*.stan. Now we will look at how a Stan code for generating
numerical MCMC simulations of a \textbf{posterior joint probability 
distribution} is to be operated from inside an \R{} environment
using the \R{} package \texttt{rstan}; cf. Stan Development Team
(2022d)~\ct{sta2022d}.

\subsection[Specification of quantitative--empirical
data]{Specification of quantitative--empirical data}
\lb{subsec:loaddata}
We begin by loading into \R{} the \textbf{quantitative--empirical
data} to be analysed, assuming that it does \textit{not} contain
any missing values and that it is available in a data file of the
*.csv-format or of the *.RData-format. If the data file is in a 
different format, the \R{} Data Import/Export manual at the website
\href{https://cran.r-project.org/doc/manuals/r-release/R-data.html}{\texttt{cran.r-project.org/doc/manuals/r-release/R-data.html}}
advises the reader on how to proceed. The
\textbf{quantitative--empirical data} typically comprises
measurements for a single \textbf{dependent
variable}~$Y$ and for $k \in \mathbb{N}$
\textbf{independent variables} (or predictor or explanatory
variables)~$\{X_{1}, \ldots, X_{k}\}$. The \textbf{sample size}
be~$n \in \mathbb{N}$. 

\medskip
\noindent
We follow the general recommendation of subjecting the data input
for the independent variables $\{X_{1}, \ldots, X_{k}\}$ to
\textbf{standardisation} for reasons given in Sec.~\ref{sec:glm}.
The standardised independent data is then to be gathered in a
\textbf{design matrix}~$\boldsymbol{X} \in
\mathbb{R}^{n \times (k+1)}$ according to Eq.~(\ref{eq:designmat}).
The data input for the dependent variable~$Y$ is subjected to 
standardisation only when it is metrically scaled, unless it
represents waiting times or spatial distances in which case we
leave the data on its original scale to maintain zero as the lower
boundary of its values. The dependent data is then to be gathered
in a \textbf{vector}~$\boldsymbol{y} \in \mathbb{R}^{n \times 1}$.

\medskip
\noindent
For the performance of MCMC simulations with Stan, we first have to
load the \R{} package~\texttt{rstan} and then make available the
\textbf{quantitative--empirical data} to be analysed in a specific
list format requested by Stan. All of this information is
communicated to \R{} via the code below, which includes the number
of independent groups from which data was obtained, should this
option apply to the case at hand.
\begin{lstlisting}[language = R, caption = {HMC sampling with Stan
in \R{}: specification of quantitative--empirical data for
fixed effects generalised linear models.},
captionpos = b, label = {lst:stanload}]
# Load R package rstan
library(rstan)

# Load specific data set from *.csv file, creating a R data.frame
dataSet <- utils::read.csv(file = "<filename>.csv", header = TRUE)
# load("<filename>.RData")  # alternative for *.RData file
# attach(<filename>)
# dataSet <- <filename>

# Standardisation of data for X1, ..., Xk
Z <- scale(
  x = dataSet[, c("X1", ..., "Xk")],
  center = TRUE,
  scale = TRUE
) # {base}

# Construction of design matrix
X <- unname(
  stats::model.matrix(
    object = dataSet$y ~ 1 + X1 + ... + Xk,
    data = as.data.frame(Z)
  )
)
attr( X , "assign" ) <- NULL

# Data list for Stan, including declaration of dimensions and
# of observed variables
dataList <- list(
  N = nrow(X),  # sample size
  M = ncol(X),  # no. of independent variables plus 1
  K = length( unique(dataSet$gpVar) ),  # no. of groups (if relevant)
  X = X,  # design matrix of standardised independent variables
  y = dataSet$y,  # dependent variable
  gp = as.integer(dataSet$gpVar)  # group variable (if relevant)
)
\end{lstlisting}
%

\subsection[HMC sampling]{HMC sampling}
The core structure of the performance of MCMC simulations with Stan
in an \R{} environment follows. This command block employs the
function \texttt{rstan::stan()} through which
central aspects of the MCMC simulations an the basis of the HMC
algorithm are specified:
\begin{itemize}
\item the *.stan file which defines the statistical model of
interest in the Stan probabilistic programming language (argument
\texttt{file}),

\item the list that provides the dimensions of and the data for
the observed variables (argument \texttt{data}),

\item the number of Markov chains to be generated (argument
\texttt{chains}),

\item the total number of iterations per Markov chain (argument
\texttt{iter}),

\item the number of warmup iterations per Markov chain
(argument \texttt{warmup}) during which the HMC sampler explores
properties of the geometry of the log-posterior joint probability
distribution in a high-dimensional parameter space; these
properties are relevant to determining optimal values for the
HMC sampler's step size and related parameters,

\item the method for setting initial values for the unknown
model parameters in the high-dimensional parameter space
(argument \texttt{init}),

\item the particular sampling algorithm to be used by Stan
(argument \texttt{algorithm}),

\item the number of cores to use when generating Markov chains in
parallel (argument \texttt{cores}).

\end{itemize}
\begin{lstlisting}[language = R, caption = {HMC sampling with Stan
in \R{}: performance of HMC sampling.},
captionpos = b, label = {lst:stansample}]
# HMC sampling instruction
modelStan <- rstan::stan(
  file = "<filename>.stan",
  data = dataList,
  chains = 4,
  iter = 5000,
  warmup = 1000,
  thin = 1,
  init = "random",
  algorithm = "NUTS",
  control = list(adapt_delta = 0.99,
                 max_treedepth = 15),
  cores = 3
)
\end{lstlisting}
The actual number of simulated samples per Markov chain generated
to be used for inference amounts to the difference of the 
values for the arguments \texttt{iter} and \texttt{warmup}. Via
the argument \texttt{control}, properties of the HMC sampler may
be adapted to cope with such issues as determination of the mode
values of unimodal and multimodal \textbf{posterior joint
probability distributions}.

\subsection[Analysis of convergence properties]{Analysis of
convergence properties of Markov chains}
Computational and graphical tools for the analysis of the
convergence properties of the MCMC simulations generated with Stan
are provided by the \R{} package \texttt{rstan} itself. These tools 
comprise for each unknown \textbf{model parameter} the calculation
of the value of the benchmark \textbf{Gelman--Rubin convergence
diagnostic}~$\hat{R}$ (which should be as close to
the value~$1$ as possible), the \textbf{effective MC sample
size}~$n_{\mathrm{eff}}$, the summaries of important \textbf{HMC
diagnostics}, and a visualisation of the convergence and mixing
properties of the Markov chains by means of a  \textbf{trace plot}.
\begin{lstlisting}[language = R, caption = {HMC sampling with Stan
in \R{}: convergence diagnostics.},
captionpos = b, label = {lst:standiagn}]
# Gelman-Rubin diagnostic Rhat (should be very close to 1) and
# effective MC sample size n_eff (the larger, the better) for
# parameter simulations
print(
  x = modelStan,
  pars = c("parameter1", "parameter2", ..., "lp__"),
  probs = c(0.015, 0.25, 0.50, 0.75, 0.985)
)

# HMC diagnostics
rstan::check_hmc_diagnostics(object = modelStan)

# Trace plots of MC chains for parameter simulations
rstan::stan_trace(
  object = modelStan ,
  pars = c("parameter1", "parameter2", ..., "lp__") ,
  inc_warmup = TRUE
)
\end{lstlisting}
%

\subsection[Posterior marginal probability distributions]{
Description of posterior marginal probability distributions}
Finally, properties of the MCMC sample representation of the
univariate \textbf{posterior marginal probability distribution} for
each unknown \textbf{model parameter} and of potential
\textbf{posterior bivariate correlations} between these
\textbf{model parameters} are made explicit. This comprises
information on the values of the mean, the standard deviation, and
of some selected quantiles, a plot of a \textbf{compatibility
interval} with pre-specified confidence levels,\footnote{McElreath
(2020)~\ct[p~56]{mce2020a} reminds the reader emphatically that
the $(95 \%, 5 \%)$- and $(99 \%, 1 \%)$-splits known from
frequentist null hypothesis significance testing are a mere
convention and have anything but an intrinsic meaning for drawing 
inferences.} and a plot of the \texttt{pdf}. Moreover, mutual
dependencies between \textbf{model  parameters} are visualised via
a \textbf{scatter plot matrix}. In addition, MCMC samples for the
unknown \textbf{model parameters} generated via the HMC sampling
algorithm can be extracted with the function
\texttt{rstan::extract()}.
\begin{lstlisting}[language = R, caption = {HMC sampling with Stan
in \R{}: extraction of HMC samples and description of posterior
marginal probability distributions and of bivariate correlations.},
captionpos = b, label = {lst:stanpostmarg}]
# Extract HMC samples
 HMCsamples <- rstan::extract(object = modelStan)

# Summaries of posterior marginal probability distributions for
# parameter simulations
print(
  x = modelStan,
  pars = c("parameter1", "parameter2", ..., "lp__"),
  probs = c(0.015, 0.25, 0.50, 0.75, 0.985)
)

# Plots of compatibility intervals for parameter simulations
rstan::stan_plot(
  object = modelStan,
  pars = c("parameter1", "parameter2", ...),
  ci_level = 0.89,
  outer_level = 0.97
)

# Densities of posterior marginal probability distributions
# for parameter simulations
rstan::stan_dens(
  object = modelStan,
  pars = c("parameter1", "parameter2", ..., "lp__")
)

# Pairwise scatter plots for parameter simulations
rstan::pairs(
  x = modelStan,
  pars = c("parameter1", "parameter2", ...)
)
\end{lstlisting}
We will now turn to demonstrate setting up a few standard examples
of fixed effects generalised linear models in the Stan
probabilistic programming language.

\chapter[Fixed effects models]{Fixed effects generalised linear
models}
\lb{ch7}
In this chapter we will discuss how to set up \textbf{generalised
linear models} with \textbf{fixed effects} in the Stan
probabilistic programming language. The \textbf{fixed effects}
method implies that for every unknown \textbf{model parameter} one
\textit{fixed} \textbf{prior probability distribution} will be
assumed. The \textbf{quantitative--empirical data} to be analysed
will be treated in an undifferentiated fashion as arising from a
single \textbf{sample}. In technical language this is referred to
as \textbf{complete pooling} of \textbf{information}. Throughout
this chapter we will also assume that the
\textbf{quantitative--empirical data} to be analysed is
\textbf{exchangeable}.

\section[Linear regression]{Linear regression}
\lb{sec:linreg}
As a first specific application of MCMC simulations with Stan
in~\R{} we turn to address \textbf{linear regression}. This serves
to model a linear relationship between data~$\boldsymbol{y}$ for
a single metrically scaled dependent variable and data
for~$k \in \mathbb{N}$ independent variables contained
in~$\boldsymbol{X}$, which may either be metrically scaled or
binary indicators. We assume that the metrically scaled data
available for \textit{both}~$\boldsymbol{y}$ and~$\boldsymbol{X}$
has been standardised prior to analysis. Many examples of
\textbf{linear regression} with one or more
independent variables can be found in the discussions given by
Sivia and Skilling (2006)~\ct[Subsec.~3.5.1]{sivski2006}, Albert 
(2009)~\ct[Sec.~9.2]{alb2009}, Krusch\-ke \textit{et al} 
(2012)~\ct{kruetal2012}, Lee (2012)~\ct[Sec.~6.3]{lee2012},
Greenberg~\ct[Sec.~8.1]{gre2013}, Gelman \textit{et al}
(2014)~\ct[Ch.~14]{geletal2014}, Andreon and Weaver
(2015)~\ct[Ch.~8]{andwea2015}, Krusch\-ke 
(2015)~\ct[Ch.~18]{kru2015}, or
McElreath (2020)~\ct[Ch.~5]{mce2020a}.

\medskip
\noindent
One of the most frequently encountered approaches to devising a
\textbf{linear regression model} is to start from a maximum entropy
perspective and choose a \textbf{Gau\ss\ single-datum likelihood
function} according to Eq.~(\ref{eq:gausspdf}), with a
\textit{homogeneous} variance, and supplement it with a
\textbf{Gau\ss--exponential prior joint
probability distribution} for the unknown \textbf{model
parameters}~$\boldsymbol{\beta}$ and~$\sigma^{2}$. The assumption
of a homogeneous variance across observations for~$\boldsymbol{y}$
is commonly known as \textbf{homoscedasticity}. We thus obtain
\bea
\lb{eq:likelihoodfctlinreg2}
\text{likelihood:} \qquad
\left.\boldsymbol{y}\right| \boldsymbol{X}, \boldsymbol{\beta},
\sigma^{2}, I
& \stackrel{\mathrm{ind}}{\sim} & \mathrm{N}(\mu, \sigma^{2}) \\
\lb{eq:linklinreg}
\text{linear model:} \qquad
\mu & = & \boldsymbol{X}\boldsymbol{\beta} \\
\lb{eq:priorlinregbeta}
\text{priors:} \qquad
\left.\boldsymbol{\beta} \right| \sigma^{2}, I
& \sim & \mathrm{N}(0, \sigma_{0}^{2}) \\
\lb{eq:priorlinregsigma}
\left.\sigma\right| I
& \sim & \mathrm{Exp}(\beta_{0}) \ ,
\eea
where $\sigma_{0}^{2}$ and $\beta_{0}$ denote fixed hyperparameters
of the prior probability distributions for
$\boldsymbol{\beta}$ and $\sigma$. Note that in
Eq.~(\ref{eq:priorlinregbeta})
zero-centred Gau\ss\ prior probability distributions were specified
for the unknown model parameters~$\boldsymbol{\beta}$. This
choice is to represent scepticism as to the presence of any of
these model parameters in a best-fit model.

\medskip
\noindent
Presently the \textbf{identity map}
\be
\lb{eq:identity}
y = z
\ee
is selected as the \textbf{link function} according to
Eq.~(\ref{eq:glm1}) for the \textbf{model
parameter} $\mu$ (the generally dimensionful location parameter),
i.e., $f\left(\mu\right) = \mu = \boldsymbol{X}
\boldsymbol{\beta}$.\footnote{The inverse link function in this
case is given by $\mu = f^{-1}\left(\boldsymbol{X}
\boldsymbol{\beta}\right) = \boldsymbol{X}\boldsymbol{\beta}$.}
This map is depicted in Fig.~\ref{fig:identity}.
\begin{figure}[!htb]
\begin{center}
\fbox{\includegraphics[width=14cm]{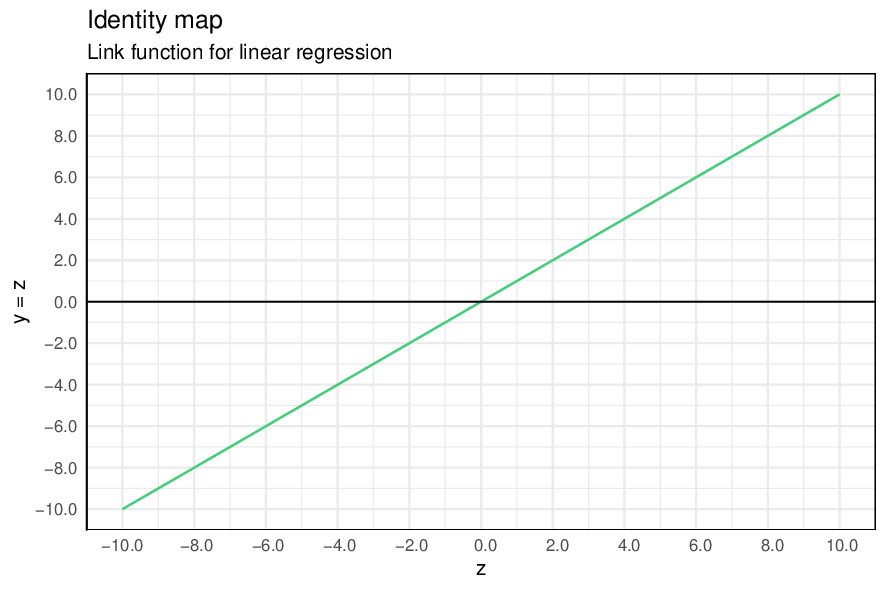}}
\end{center}
\caption{Plot of the identity map~(\ref{eq:identity}).}
\lb{fig:identity}
\end{figure}

\medskip
\noindent
In the following we give the vectorised code for the specification
in Stan of a \textbf{linear regression model}, employing
standardised metrically scaled variables, a homogeneous
variance, and fixed prior probability distributions for the
\textbf{model parameters}. We confine our consideration to the
inclusion of additive \textbf{main effects} of the independent 
variables, but point to the numerous possibilities of rendering the 
model more flexible by including also multiplicative
\textbf{interaction terms} for the independent variables. The code
\href{https://github.com/hve1964/stanCodes/blob/master/linRegNormFixed.stan}{\texttt{linRegNormFixed.stan}} and corresponding sample data are
available from
\href{https://github.com/hve1964/stanCodes}{\texttt{github.com/hve1964/stanCodes}}.\footnote{In this and all following examples of Stan
model codes the number of model parameters (or regression
coefficients), $k+1$, is specified as \texttt{M}.}

%
\begin{lstlisting}[language = Stan, caption = {Vectorised Stan
model code for fixed effects multiple linear regression. This code
also computes the posterior predictive probability distribution
re-using available predictor data. The two occurring dimensionless
fixed hyperparameters have been given the values
$\sigma_{0} = 1$ and $\beta_{0} = 1$.},
captionpos = b, label = {lst:linregfixed}]
data {
  /* Dimensions */
  int<lower=1> N;   // number of sampling units
  int<lower=1> M;   // number of predictors plus one

  /* Observed variables */
  matrix[N, M] X;   // design matrix: predictors
  vector[N] y;      // outcome
}

parameters {
  /* Unobserved variables */
  vector[M] beta;
  real<lower=0> sigma;
}

model {
  /* Fixed log-priors (regularising) */
  target += normal_lpdf ( beta | 0 , 1 );
  target += exponential_lpdf( sigma | 1 );

  /* Gauss log-likelihood w/ identity link */
  target += normal_lpdf( y | X * beta , sigma );
}

generated quantities {
  /* Posterior predictive distribution (re-using predictor data) */
  vector[N] yrep;

  for ( i in 1:N ) {
    yrep[i] = normal_rng( (X * beta)[i] , sigma );
  }
}
\end{lstlisting}

\medskip
\noindent
When there is a need to deal with \textbf{outliers} in the
$\boldsymbol{y}$-data, one might resort to a non-central
$t$--single-datum likelihood function according to
Eq.~(\ref{eq:tpdf}) by replacing in the Stan model
code~\ref{lst:linregfixed}
\texttt{normal\_lpdf( y | X * beta , sigma )} by
\texttt{student\_t\_lpdf( y | nu , X * beta , sigma )}, and then
specify an additional fixed prior
probability distribution for the degrees of freedom
parameter~$\nu \geq 1$; e.g., an exponential distribution from the
family defined by Eq.~(\ref{eq:exppdf}).


\section[ANOVA-like regression]{ANOVA-like regression}
\lb{sec:anova}
Now we describe coding MCMC simulations with Stan in~\R{} for
\textbf{ANOVA-like regression}. This serves to model a linear
relationship between data~$\boldsymbol{y}$ for a single metrically
scaled dependent variable, and data~$[gp]$ for a
qualitative variable which can take values in~$k \in \mathbb{N}$
non-ordered categories. This method is useful when the researcher's
objective is to compare distributional features of one and the same
metrically scaled variable between several independent groups.
Examples of \textbf{ANOVA-like regression} can be found in the 
discussions given, e.g., by Lee
(2012)~\ct[Sec.~6.5]{lee2012}, Gelman \textit{et al}
(2014)~\ct[Sec.~15.6]{geletal2014}, and Krusch\-ke
(2015)~\ct[Ch.~19]{kru2015}.

\medskip
\noindent
Again, taking the maximum entropy perspective, one chooses a
\textbf{Gau\ss\ single-datum likelihood function} according to
Eq.~(\ref{eq:gausspdf}), with a \textit{homogeneous} variance, and
specifies a \textbf{Gau\ss--exponential prior
joint probability distribution} for the unknown \textbf{location
parameters}~$\boldsymbol{\mu}[gp]$ and the \textbf{scale
parameter}~$\sigma^{2}$. Adopting
the cell means view of ANOVA-like regression, this gives
\bea
\lb{eq:likelihoodfctanova2}
\text{likelihood:} \qquad
\left.\boldsymbol{y}\right| \boldsymbol{\mu}, \sigma^{2}, [gp], I
&\stackrel{\mathrm{ind}}{\sim} &
\mathrm{N}(\boldsymbol{\mu}[gp], \sigma^{2}) \\
\lb{eq:priorsanova}
\text{priors:} \qquad
\left.\boldsymbol{\mu}[gp]\right| \sigma^{2}, I
& \sim & \mathrm{N}(\mu_{0}, \sigma_{0}^{2}) \\
\left.\sigma\right| I
& \sim & \mathrm{Exp}(\beta_{0}) \ ,
\eea
where $\mu_{0}$, $\sigma_{0}$ and $\beta_{0}$ denote fixed
hyperparameters of the prior probability distributions.

\medskip
\noindent
In the vectorised code for the specification in Stan of an
\textbf{ANOVA-like regression model} in the cell means view, we
employ the \textbf{homoscedasticity} assumption and fixed prior 
probability distributions for the \textbf{model parameters}.
Experience shows that for the present kind of statistical models
the Markov chains generated by the HMC algorithm are usually not
plagued with autocorrelation, so there is no need for standardising
the metrically scaled $\boldsymbol{y}$-data. The Stan code
\href{https://github.com/hve1964/stanCodes/blob/master/anovaRegNormFixed.stan}{\texttt{anovaRegNormFixed.stan}} and corresponding sample data
are available from
\href{https://github.com/hve1964/stanCodes}{\texttt{github.com/hve1964/stanCodes}}.

%
\begin{lstlisting}[language = Stan, caption = {Vectorised Stan
model code for fixed effects homoscedastic ANOVA-like regression.
This code also computes the posterior predictive probability
distribution re-using available predictor data.
The three occurring fixed hyperparameters have been given the
values $\mu_{0} = 90~\text{(units)}$,
$\sigma_{0} = 2~\text{(units)}$ and
$\beta_{0} = 1~\text{(units)}$.},
captionpos = b, label = {lst:anovaregfixed}]
data {
  /* Dimensions */
  int<lower=1> N;    // number of sampling units
  int<lower=1> K;    // number of groups

  /* Observed variables */
  int<lower=1,upper=K> gp[N];    // group indicator
  vector[N] y;                   // outcome
}

parameters {
  /* Unobserved variables */
  vector[K] mu;           // group means
  real<lower=0> sigma;    // common stdev (homogeneous)
}

model {
  /* Fixed log-priors */
  target += normal_lpdf( mu | 90 , 2 );
  target += exponential_lpdf( sigma | 1 );

  /* Gauss log-likelihood */
  target += normal_lpdf( y | mu[gp] , sigma );
}

generated quantities {
  /* Posterior predictive distribution (re-using predictor data) */
  vector[N] yrep;

  for ( i in 1:N ) {
    yrep[i] = normal_rng( mu[gp[i]] , sigma );
  }
}
\end{lstlisting}

\medskip
\noindent
If the group-specific $\boldsymbol{y}$-data contains
\textbf{outliers}, the single-datum likelihood function may be
replaced by a non-central $t$--distribution according to
Eq.~(\ref{eq:tpdf}) by replacing in the Stan model
code~\ref{lst:anovaregfixed}
\texttt{normal\_lpdf( y | mu[gp] , sigma )} by
\texttt{student\_t\_lpdf( y | nu , mu[gp] , sigma )}, and then
specify an additional fixed prior
probability distribution for the degrees of freedom
parameter~$\nu \geq 1$; e.g., an exponential distribution from the
family defined by Eq.~(\ref{eq:exppdf}). Also, the available
$\boldsymbol{y}$-data
may suggest that an assumption of \textbf{heteroscedasticity} is
more realistic. In this case, one specifies an adaptive prior
probability distribution for the scale parameter~$\sigma^{2}$. This
case will be addressed in Sec.~\ref{sec:hieranova}.

\medskip
\noindent
As this fits the present discussion, we remark that the
Bayes--Laplace analogue of the frequentist Student's independent
samples $t$--test has been developed by G\"{o}nen \textit{et al}
(2005)~\ct{goeetal2005}. Further applications and implementations
in~\R{} are given by Krusch\-ke (2013)~\ct{kru2013} and
Kruschke (2015)~\ct[Sec.~16.3]{kru2015}.

\section[Logistic regression]{Logistic regression}
\lb{sec:logitreg}
When the research objective is to explain the dependency of count
data with a \textit{known} finite maximum of~$n \in \mathbb{N}$
on a set of $k \in \mathbb{N}$ either metrically scaled or binary
indicator independent variables, \textbf{logistic regression} is
the standard tool for model-building. We will here sketch the way
of integrating this technique into MCMC simulations with Stan in
\R{}. We assume that the metrically scaled data available
for~$\boldsymbol{X}$ has been standardised prior to
analysis. Many interesting examples of applications of
\textbf{logistic regression} can be found in
Albert (2009)~\ct[Sec.~4.4]{alb2009}, Lee
(2012)~\ct[Subec.~9.8.1]{lee2012},
Greenberg~\ct[Subsec.~8.2.3]{gre2013},
Gelman \textit{et al} (2014)~\ct[Sec.16.3]{geletal2014},
Krusch\-ke (2015)~\ct[Ch.~21]{kru2015}, or
McElreath (2020)~\ct[Sec.~11.1.]{mce2020a}.

\medskip
\noindent
The maximum entropy perspective suggests to capture the
individual instant of whether a count was observed or not by a
\textbf{Bernoulli single-datum likelihood function} according to
Eq.~(\ref{eq:bernprob}), so that, overall, a binomial total-data
likelihood function according to Eq.~(\ref{eq:binomprob}) abounds
as describing the data-generating process for the total count.
Contrary to linear regression and ANOVA-like regression discussed
in Secs.~\ref{sec:linreg} and~\ref{sec:anova} above, these
likelihood functions have no explicit dependence on a dispersion
parameter. In addition, it is meaningful to assign zero-centred
\textbf{Gau\ss\ prior probability distributions} for the unknown
\textbf{model parameters}~$\boldsymbol{\beta}$, a choice we will
comment on shortly. The standard set-up for \textbf{logistic
regression} is given by
\bea
\lb{eq:likelihoodfctlogreg}
\text{likelihood:} \qquad
\left.\boldsymbol{y}\right| \boldsymbol{X}, \boldsymbol{\beta}, I 
& \stackrel{\mathrm{ind}}{\sim} &
\mathrm{Bern}(p) \\
\text{linear model:} \qquad
\ln\left(\frac{p}{1-p}\right)
& = & \boldsymbol{X}\boldsymbol{\beta} \\
\lb{eq:priorlogregbeta}
\text{priors:} \qquad
\left.\boldsymbol{\beta}\right| I
& \sim & \mathrm{N}(0, \sigma_{0}^{2}) \ ,
\eea
where $\sigma_{0}^{2}$ denotes a fixed hyperparameter of the
Gau\ss\ prior probability distributions.

\medskip
\noindent
In \textbf{logistic regression} the \textbf{logit function},
defined by
\be
\lb{eq:logit}
y = \text{logit}(z) := \ln\left(\frac{z}{1-z}\right) \ ,
\ee
serves as the \textbf{link function} according to
Eq.~(\ref{eq:glm1}) for the \textbf{model
parameter}~$p$ (the dimensionless probability for ``success''),
i.e., $f\left(p\right) = \text{logit}(p)
= \boldsymbol{X}\boldsymbol{\beta}$.\footnote{The inverse link
function in this case is given by the standard logistic function,
$p = f^{-1}\left(\boldsymbol{X}\boldsymbol{\beta}\right)
= \text{logistic}\left(\boldsymbol{X}\boldsymbol{\beta}\right)
:= \left[\,1+\exp\left(-\boldsymbol{X}\boldsymbol{\beta}\right)
\,\right]^{-1}$.} This map is depicted in Fig.~\ref{fig:logit}.
\begin{figure}[!htb]
\begin{center}
\fbox{\includegraphics[width=14cm]{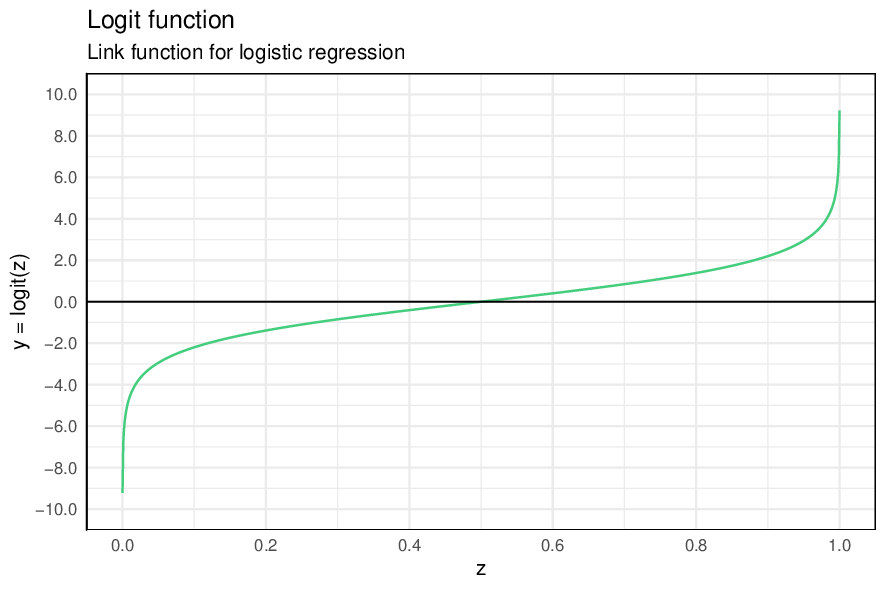}}
\end{center}
\caption{Plot of the logit function~(\ref{eq:logit}).}
\lb{fig:logit}
\end{figure}
The logit function maps the bounded interval~$\left[0,1\right]$
to the entire real line, the range of values of the 
unbounded real-valued linear form $\boldsymbol{X}
\boldsymbol{\beta}$. With the ranges of possible values thus
coinciding, the linear form $\boldsymbol{X}\boldsymbol{\beta}$
can be related directly to the transformed dimensionless
probability for ``success'' parameter~$p$. A particular
feature of logistic regression is the general choice of a ``zero
log-odds'' reference baseline,
\be
0 \stackrel{!}{=} \ln\left(\frac{p}{1-p}\right)
= \boldsymbol{X}\boldsymbol{\beta}
\qquad\Leftrightarrow\qquad
p = \frac{1}{2}\ ,
\ee
i.e., equal probabilities for the binary outcomes ``failure'' or
``success'' according to Bernoulli's ``principle of non-sufficient
reason'' of Sec.~\ref{sec:indif}. For non-zero
data, $\boldsymbol{X} \neq \boldsymbol{0}$, the implication is
$\boldsymbol{\beta} = \boldsymbol{0}$, explaining the
zero-centering of the Gau\ss\ prior probability distributions.

\medskip
\noindent
The code for the specification in Stan of a \textbf{logistic
regression model} with fixed prior probability distributions for
the \textbf{model parameters} is given in the following. This code,
\href{https://github.com/hve1964/stanCodes/blob/master/logistRegBernFixed.R}{\texttt{logistRegBernFixed.stan}},
is available from
\href{https://github.com/hve1964/stanCodes}{\texttt{github.com/hve1964/stanCodes}}.

%
\begin{lstlisting}[language = Stan, caption = {Vectorised Stan
model code for fixed effects logistic regression. This code
also computes the posterior predictive probability distribution
re-using available predictor data. The occurring single
dimensionless fixed hyperparameter has been given
the value $\sigma_{0} = 1$.},
captionpos = b, label = {lst:logitregfixed}]
data {
  /* Dimensions */
  int<lower=1> N;   // number of sampling units
  int<lower=1> M;   // number of predictors plus one

  /* Observed variables */
  matrix[N, M] X;               // design matrix: predictors
  int<lower=0,upper=1> y[N];    // outcome
}

parameters {
  /* Unobserved variables */
  vector[M] beta;
}

model {
  /* Fixed log-prior (regularising) */
  target += normal_lpdf ( beta | 0 , 1 );
  
  /* Bernoulli log-likelihood w/ logit link */
  target += bernoulli_logit_lpmf( y | X * beta );
}

generated quantities {
  /* Posterior predictive distribution (re-using predictor data) */
  int<lower=0,upper=1> yrep[N];

  for ( i in 1:N ) {
    yrep[i] = bernoulli_logit_rng( (X * beta)[i] );
  }
}
\end{lstlisting}
%


\section[Poisson regression]{Poisson regression}
\lb{sec:poisreg}
Qualitatively different to the previous case is the situation when
the \textbf{quantitative--empirical data} to be explained is
count data with an \textit{unknown} maximum. To model the
respective data-generating
process in dependence of a set of $k \in \mathbb{N}$ either
metrically scaled or binary indicator independent variables, it is
best practice to employ \textbf{Poisson regression}. The
implementation of corresponding MCMC simulations with Stan in \R{}
will be the topic of this section. We assume that the metrically
scaled data available for~$\boldsymbol{X}$ has been standardised
prior to analysis. Many practical applications of \textbf{Poisson 
regression} to interesting research problems are outlined by
Sivia and Skilling (2006)~\ct[Sec.~3.1]{sivski2006},
Albert (2009)~\ct[Sec.~11.4]{alb2009},
Gelman \textit{et al} (2014)~\ct[Sec.~16.4]{geletal2014},
Andreon and Weaver (2015)~\ct[Subsec.~6.1.2]{andwea2015}, 
or McElreath (2020)~\ct[Sec.~11.2.]{mce2020a}. Krusch\-ke
(2015)~\ct[Ch.~24]{kru2015} discusses a count data case with two 
nominally scaled independent variables, which amounts to the
analysis of data from a contingency table. Methodologically, this
case can be broadly likened to a frequentist
$\chi^{2}$--test of independence.

\medskip
\noindent
Like all the methods of regression analysis discussed in this
chapter, \textbf{Poisson regression} takes a prime motivation from
a maximum entropy perspective. Hence the choice of a
\textbf{Poisson single-datum likelihood function} according to
Eq.~(\ref{eq:poisprob}), which represents a particular state of
ignorance. It has no explicit dependence on a dispersion parameter.
Again, we select zero-centred \textbf{Gau\ss\ prior probability
distributions} for the unknown \textbf{model
parameters}~$\boldsymbol{\beta}$ to express scepticism as to
their presence in a best-fit model. A resultant common set-up for
\textbf{Poisson regression} is given by
\bea
\lb{eq:likelihoodfctpoisreg}
\text{likelihood:} \qquad
\left.\boldsymbol{y}\right|\boldsymbol{X}, \boldsymbol{\beta}, I 
& \stackrel{\mathrm{ind}}{\sim} & \text{Pois}(\theta) \\
\text{linear model:} \qquad
\ln(\theta) & = & \boldsymbol{X}\boldsymbol{\beta} \\
\lb{eq:priorpoisregbeta}
\text{priors:} \qquad
\left.\boldsymbol{\beta}\right| I
& \sim & \mathrm{N}(0, \sigma_{0}^{2}) \ ,
\eea
where $\sigma_{0}^{2}$ is a fixed hyperparameter of the
Gau\ss\ prior probability distributions.

\medskip
\noindent
In \textbf{Poisson regression} the \textbf{natural logarithmic
function}
\be
\lb{eq:natLog}
y = \ln(z)
\ee
serves as the \textbf{link function} according to
Eq.~(\ref{eq:glm1}) for the positive \textbf{model
parameter}~$\theta$ (the dimensionless rate parameter), i.e.,
$f\left(\theta\right) = \ln(\theta)
= \boldsymbol{X}\boldsymbol{\beta}$.\footnote{The inverse link
function in this case is given by the natural exponential function,
$\theta = f^{-1}\left(\boldsymbol{X}\boldsymbol{\beta}\right)
= \exp\left(\boldsymbol{X}\boldsymbol{\beta}\right)$.} This map is
depicted in Fig.~\ref{fig:natLog}.
\begin{figure}[!htb]
\begin{center}
\fbox{\includegraphics[width=14cm]{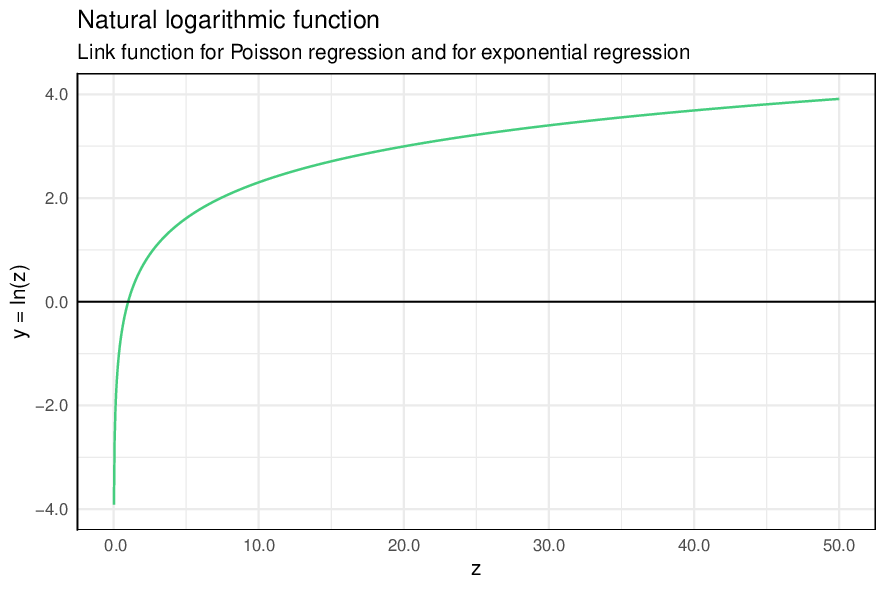}}
\end{center}
\caption{Plot of the natural logarithmic
function~(\ref{eq:natLog}).}
\lb{fig:natLog}
\end{figure}
The natural logarithmic function maps the open interval~$\left(0,
\infty\right)$ to the entire real line, the range of values of the 
unbounded real-valued linear form $\boldsymbol{X}
\boldsymbol{\beta}$. The linear form $\boldsymbol{X}
\boldsymbol{\beta}$ can thus be directly related to the transformed
non-negative dimensionless rate
parameter~$\theta$. The choice of zero-centred Gau\ss\ prior
probability distributions for the~$\boldsymbol{\beta}$ implies
via the link function:
\be
\boldsymbol{\beta} = \boldsymbol{0} \ ,
\ \boldsymbol{X} \neq \boldsymbol{0}
\qquad\Leftrightarrow\qquad \ln(\theta) = 0
\ \Rightarrow\ \theta = 1 \ ;
\ee
the order-of-magnitute of~$\theta$ is the information that
is most relevant in Poisson data-generating processes.

\medskip
\noindent
Often in practical applications the \textbf{model
parameter}~$\theta$ is given a product structure $\theta
= \tau\lambda$, with the understanding that the dimensionful
\textbf{exposure}~$\tau$ amounts to available empirical data,
while~$\lambda$ is the \textit{unknown} dimensionful
\textbf{rate parameter}. The linear model then becomes
$\ln\left(\lambda/\lambda_{0}\right)
= -\ln\left(\tau/\tau_{0}\right) + \boldsymbol{X}
\boldsymbol{\beta}$, where $\lambda_{0}$ and $\tau_{0}$ denote
corresponding normalising units with $\lambda_{0}
= \left(1/\tau_{0}\right)$.

\medskip
\noindent
The code for the specification in Stan of a \textbf{Poisson
regression model} with fixed prior probability distributions for
the \textbf{model parameters} is to follow. This code,
\href{https://github.com/hve1964/stanCodes/blob/master/poisRegFixed.stan}{\texttt{poisRegFixed.stan}}, and corresponding sample data are
available from
\href{https://github.com/hve1964/stanCodes}{\texttt{github.com/hve1964/stanCodes}}.\footnote{Note
that in the generated quantities block
of this Stan code the upper limit of $30\ln(2) \approx 20.79$ for
the dimensionless logarithmic rate parameter  in the
\texttt{poisson\_log\_rng()} function has been respected in order
to avoid a crash when compiling this code;
cf. \href{https://mc-stan.org/docs/functions-reference/poisson-distribution-log-parameterization.html}{Stan Functions Reference
(v2.30)} \ct{sta2022b}.}

%
\begin{lstlisting}[language = Stan, caption = {Vectorised Stan
model code for fixed effects Poisson regression. This code
also computes the posterior predictive probability distribution
re-using available predictor data. The occurring single
dimensionless fixed hyperparameter has been given
the value $\sigma_{0} = 1$.},
captionpos = b, label = {lst:poisregfixed}]
data {
  /* Dimensions */
  int<lower=1> N;   // number of sampling units
  int<lower=1> M;   // number of predictors plus one
      
  /* Observed variables */
  matrix[N, M] X;       // design matrix: predictors
  int<lower=0> y[N];    // outcome
}

parameters {
  /* Unobserved variables */
  vector[M] beta;
}

model {
  /* Fixed log-priors (regularising) */
  target += normal_lpdf ( beta | 0 , 1 );
    
  /* Poisson log-likelihood w/ log link */
  target += poisson_log_lpmf( y | X * beta );
}

generated quantities {
  /* Posterior predictive distribution (re-using predictor data) */
  int<lower=0> yrep[N];
    
  for ( i in 1:N ) {
    if ( (X * beta)[i] > 20 ) {
      yrep[i] = poisson_log_rng( 20 );
    } else {
        yrep[i] = poisson_log_rng( (X * beta)[i] );
    }
  }
}
\end{lstlisting}
%


\medskip
\noindent
Should one find that the empirical count data that one analysed is
over-dispersed, which manifests itself by the property
$\mathrm{E}(\boldsymbol{y}|\boldsymbol{X}, \boldsymbol{\beta}, I)
\ll \mathrm{Var}(\boldsymbol{y}|\boldsymbol{X}, \boldsymbol{\beta},
I)$ (when near equality for these two quantities was expected
initially), then a
\textbf{Gamma--Poisson mixture model} is suggested as a viable 
alternative; cf., e.g., Gelman \textit{et al}
(2014)~\ct[p~437f]{geletal2014},
or McElreath (2020)~\ct[Sec.~12.1.]{mce2020a}. The
\textbf{Gamma--Poisson probability distribution} is also referred
to as negative binomial distribution.

\medskip
\noindent
\underline{\R:}
$\texttt{dnbinom}(y_{i}, n, \theta)$,
$\texttt{pnbinom}(y_{i}, n, \theta)$,
$\texttt{qnbinom}(p, n, \theta)$,
$\texttt{rnbinom}(n_\mathrm{simulations}, n, \theta)$\\
\underline{Stan:} Cf. \href{https://mc-stan.org/docs/functions-reference/negative-binomial-distribution.html}{Stan Functions
Reference (v2.30)}~\ct{sta2022b}
\begin{itemize}
\item $\texttt{neg\_binomial}( n , \theta/(1-\theta) )$ (sampling)
\item $\texttt{neg\_binomial\_lpmf}( y | n , \theta/(1-\theta) )$
(log-sampling)
\item $\texttt{neg\_binomial\_rng}( n , \theta/(1-\theta) )$
(generating)
\end{itemize}
\underline{JAGS:}
$\texttt{dnegbin}(\theta, n)$ (sampling)

\section[Exponential regression]{Exponential regression}
\lb{sec:expreg}
The last type of regression analysis we want to introduce in this
chapter is \textbf{exponential regression}. This can be employed to
explain data~$\boldsymbol{y}$ for the lengths of continuous
temporal or spatial intervals, i.e., waiting times or spatial
distances, in terms of a set of $k \in \mathbb{N}$ either
metrically scaled or binary indicator independent variables
contained in~$\boldsymbol{X}$. We assume that the metrically
scaled data available for~$\boldsymbol{X}$ has been standardised
prior to analysis. An example of \textbf{exponential
regression} can be found in Gill (2015)~\ct[Sec.~12.4]{gil2015}.

\medskip
\noindent
As outlined before in Sec.~\ref{sec:maxent}, also the
\textbf{exponential single-datum likelihood function} is of a
certain maximum entropy kind, thus expressing a corresponding
specific state of ignorance. It, too, features no explicit
dependence on a dispersion parameter. Selecting zero-centred
\textbf{Gau\ss\ prior probability distributions} for the unknown
\textbf{model parameters}~$\boldsymbol{\beta}$ to express
scepticism as to their presence in a best-fit model, the present
set-up for \textbf{exponential regression} is given by
\bea
\lb{eq:likelihoodfctexpreg}
\text{likelihood:} \qquad
\left.\boldsymbol{y}\right|\boldsymbol{X}, \boldsymbol{\beta}, I 
& \stackrel{\mathrm{ind}}{\sim} & \text{Exp}(\theta) \\
\text{linear model:} \qquad
\ln(\theta/\theta_{0})
& = & \boldsymbol{X}\boldsymbol{\beta} \\
\lb{eq:priorexpregbeta}
\text{priors:} \qquad
\left.\boldsymbol{\beta}\right| I
& \sim & \mathrm{N}(0, \sigma_{0}^{2}) \ ,
\eea
where $\theta_{0}$ is the unit of the dimensionful \textbf{model
parameter}~$\theta$ and $\sigma_{0}^{2}$ is a fixed hyperparameter
of the Gau\ss\ prior probability distributions. In analogy
to Poisson regression, we have here chosen the \textbf{natural
logarithmic function} of Eq.~(\ref{eq:natLog}) as the
\textbf{link function} according to Eq.~(\ref{eq:glm1}) for the
positive \textbf{model parameter}~$\theta$ (the generally
dimensionful rate parameter).

\medskip
\noindent
We now give the code for the specification in Stan of an
\textbf{exponential regression model} with fixed prior probability
distributions for the \textbf{model parameters}. This code,
\href{https://github.com/hve1964/stanCodes/blob/master/expRegFixed.stan}{\texttt{expRegFixed.stan}}, and corresponding sample data are
available from
\href{https://github.com/hve1964/stanCodes}{\texttt{github.com/hve1964/stanCodes}}.

%
\begin{lstlisting}[language = Stan, caption = {Vectorised Stan
model code for fixed effects exponential regression. This code
also computes the posterior predictive probability distribution
re-using available predictor data. The occurring single
dimensionless fixed hyperparameter has been given
the value $\sigma_{0} = 1$.},
captionpos = b, label = {lst:exoregfixed}]
data {
  /* Dimensions */
  int<lower=1> N;   // number of sampling units
  int<lower=1> M;   // number of predictors plus one
      
  /* Observed variables */
  matrix[N, M] X;        // design matrix: predictors
  real<lower=0> y[N];    // outcome
}

parameters {
  /* Unobserved variables */
  vector[M] beta;
}

model {
  /* Fixed log-priors (regularising) */
  target += normal_lpdf ( beta | 0 , 1 );
    
  /* Exponential log-likelihood w/ exponential inverse link */
  target += exponential_lpdf( y | exp(X * beta) );
}

generated quantities {
  /* Posterior predictive distribution (re-using predictor data) */
  real<lower=0> yrep[N];
    
  for ( i in 1:N ) {
    yrep[i] = exponential_rng( exp((X * beta)[i]) );
  }
}
\end{lstlisting}
%

\section[Simulating posterior predictive distributions]{Simulating
posterior predictive probability distributions}
\lb{sec:samppostpred}
The researcher's actual inferential work begins only when the model
fitting process described in the previous sections has been
finalised, implying that the \textbf{statistical model} to be
employed for explanatory and predictive purposes has been validated
to a sufficient degree. At the centre of attention at the
\textbf{validation} stage of \textbf{inductive statistical
inference} are two approaches to model assessment: the
evaluation for a fitted \textbf{statistical model} of (i)~the
\textbf{posterior predictive probability distribution} for a
\textbf{dependent variable}~$Y$ according to
Eq.~(\ref{eq:postpred}) and (ii)~(one of) the \textbf{information
citeria} to be introduced in Sec.~\ref{sec:infcrit} both serve as
safeguards with respect to the threats of \textbf{under-fitting}
and \textbf{over-fitting}. The special status of the
\textbf{posterior predictive probability
distribution} arises out of the dual role the \textbf{single-datum 
likelihood function} takes from operating in \textit{two}
directions: from \textbf{quantitative--empirical data} to values
for unknown \textbf{model parameters}, and from known (simulated)
values for \textbf{model parameters} back to yet unobserved
\textbf{quantitative--empirical data}; cf. McElreath
(2020)~\ct[p~62]{mce2020a}. In order to evaluate the
multi-dimensional integral of the \textbf{single-datum likelihood 
function} of a specific \textbf{data-generating process}, weighted
by the \textbf{posterior joint probability distribution} of
the proposed \textbf{statistical model}, over a $(k+1)$-dimensional
\textbf{parameter space} on the basis of approximative MCMC
simulations, direct application of the discretised integration over
a continuous parameter according to Eq.~(\ref{eq:intdiscrete}) is 
imminent. By taking this action one \textit{averages} the
\textbf{single-datum likelihood function} over the
\textbf{parameter space} after relevant information from
available \textbf{quantitative--empirical data} has been accounted
for. A \textbf{posterior predictive probability distribution}
combines \textbf{observation uncertainty} with \textbf{parameter 
uncertainty} and so provides a fairly conservative representation
of a researcher's \textbf{state of knowledge} on the given problem
they subjected to statistical data analysis.

\medskip
\noindent
The main steps to be taken to gain MCMC simulations of a
\textbf{posterior predictive probability distribution} for a
\textbf{GLM} are as follows:
\begin{enumerate}

\item Evaluate the linear form~$\boldsymbol{X}\boldsymbol{\beta}$
from a researcher-specified design
matrix~$\boldsymbol{X}_\mathrm{new}$ for $n_{\mathrm{new}}$~cases
on $k$~independent variables and the MCMC simulated $k+1$ model
parameters~$\boldsymbol{\beta}$.

\item Apply the inverse link function~$f^{-1}$ to the linear
form~$\boldsymbol{X}_\mathrm{new}\boldsymbol{\beta}$ to obtain
a value~$\theta_{\mathrm{new}}$ for the parameter~$\theta$ of
the single-datum likelihood function according to
Eq.~(\ref{eq:glm1}).

\item Sample repeatedly $n_{\mathrm{new}}$~new
values~$y_{\mathrm{new}}$ for~$Y$
from the single-datum likelihood function with the MCMC posterior
settings for~$\theta_{\mathrm{new}}$ and possible additional
parameters to obtain simulated dependent data according to
Eq.~(\ref{eq:glm2}).

\item Visualise and summarise specific features of the MCMC
simulated posterior predictive probability distributions for~$Y$.

\end{enumerate}
To trigger MCMC simulations that yield distributions of
predictions~$\boldsymbol{y}_{\mathrm{new}}$ for the
\textbf{dependent variable}~$Y$ from new data
for the \textbf{independent variables} gathered in a design
matrix~$\boldsymbol{X}_\mathrm{new}$, the following
supplementations to the \texttt{data} and \texttt{generated
quantities} blocks of a Stan model code are necessary, here
given for the example of fixed effects multiple linear regression
discussed in Sec.~\ref{sec:linreg}:
\begin{lstlisting}[language = Stan, caption = {Supplementations
to vectorised Stan model code for fixed effects multiple linear regression to enable the computation of the posterior predictive probability distribution for new predictor data.},
captionpos = b, label = {lst:postprednew}]
data {
  /* Dimensions */
  ...
  int<lower=1> Nnew;   // number of new sampling units

  /* Observed variables */
  ...
  matrix[Nnew, M] Xnew;   // design matrix: new values
}

...

generated quantities {
  /* Posterior predictive distribution (new predictor data) */
  vector[Nnew] ynew;

  for ( i in 1:Nnew ) {
    ynew[i] = normal_rng( (Xnew * beta)[i] , sigma );
  }
}
\end{lstlisting}
For the remaining fixed effects GLMs in this
chapter the necessary supplementations follow by analogy.

\medskip
\noindent
Alternatively, posterior predictive checks may also be performed
by re-using the available data for the \textbf{independent
variables} in the design matrix~$\boldsymbol{X}$, thus generating 
distributions of replications~$\boldsymbol{y}_{\mathrm{rep}}$
for the \textbf{dependent variable}~$Y$ from the fitted
\textbf{statistical model}. Corresponding commands are contained
in the \texttt{generated quantities} blocks of the Stan model
codes~\ref{lst:linregfixed} to~\ref{lst:exoregfixed}.

\medskip
\noindent
Concerning the visualisation of particular features of the
\textbf{posterior predictive probability distribution} for a
\textbf{dependent variable}~$Y$ in an \R{} environment,
Gabry and Mahr (2022)~\ct{gabmah2022} make available the
user-friendly package \texttt{bayesplot}. The general value of the
integration of visualisations in the workflow of \textbf{inductive
statistical inference} has been emphasised strongly in the paper by 
Gabry \textit{et al} (2019)~\ct{gabetal2019}. A selection
of functions from \texttt{bayesplot} for visual posterior
predictive checks is presented in the \R{} code
block~\ref{lst:stanppc}. They generate, amongst others, graphs of
the \textbf{empirical data distribution} for~$Y$ overlaid with
HMC samples of the \textbf{posterior predictive distribution}
for~$Y$, or plots of the \textbf{data points} for~$Y$ overlaid with corresponding \textbf{posterior predictive compatibility intervals}
for~$Y$.
\begin{lstlisting}[language = R, caption = {HMC sampling with Stan
in \R{}: visual posterior predictive checks using
\texttt{bayesplot}.},
captionpos = b, label = {lst:stanppc}]
# Load R package bayesplot
library(bayesplot)

# Extract posterior predictive samples
draws <- as.matrix(
  modelStan,
  pars = "yrep"
)

# Selection of posterior predictive checks
bayesplot::ppc_dens_overlay(
  y = dataList$y,
  yrep = draws[1:100,]
)
bayesplot::ppc_hist(
  y = dataList$y,
  yrep = draws[11:15,],
  binwidth = 1
)
bayesplot::ppc_stat(
  y = dataList$y,
  yrep = draws,
  stat = "mean",
  binwidth = 0.1
)
bayesplot::ppc_intervals(
  y = dataList$y,
  yrep = draws, 
  prob = 0.5,
  prob_outer = 0.89,
  size = 1,
  fatten = 3
)

# For Poisson regression only
bayesplot::ppc_rootogram(
  y = dataList$y,
  yrep = draws,
  style = "standing",
  prob = 0.89,
  size = 1
)
\end{lstlisting}
Some of the graphs these functions produce for visual posterior
predictive checks by re-using available predictor data are given in
Figs.~\ref{fig:linRegNormFixedPPC1} to~\ref{fig:poisRegFixedPPC2}
for the cases of linear regression and Poisson regression.
\begin{figure}[!htb]
\begin{center}
\fbox{\includegraphics[width=14cm]{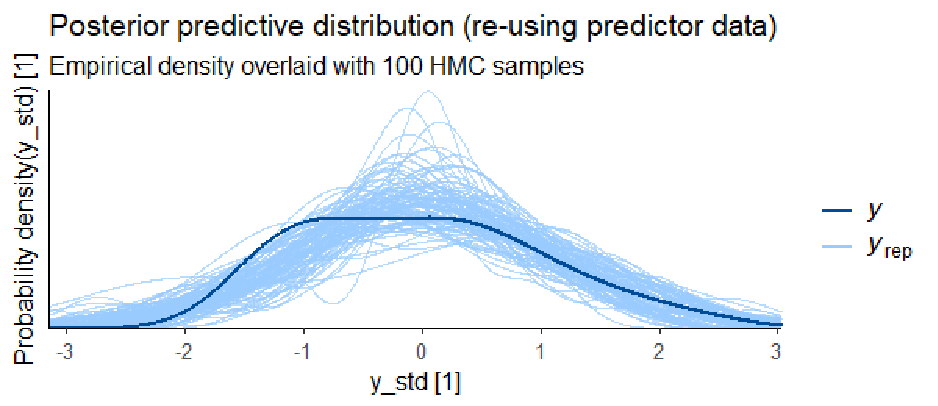}}
\end{center}
\caption{Posterior predictive distribution for fixed effects
linear regression re-using available predictor data, generated with
the function \texttt{bayesplot::ppc\_dens\_overlay()}.}
\lb{fig:linRegNormFixedPPC1}
\end{figure}
\begin{figure}[!htb]
\begin{center}
\fbox{\includegraphics[width=14cm]{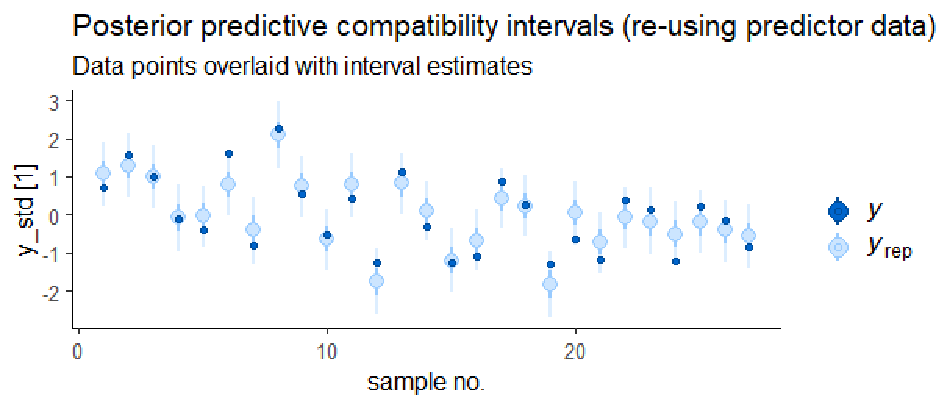}}
\end{center}
\caption{Posterior predictive compatibility intervals for fixed
effects linear regression re-using available predictor data,
generated with the function \texttt{bayesplot::ppc\_intervals()}.}
\lb{fig:linRegNormFixedPPC2}
\end{figure}
\begin{figure}[!htb]
\begin{center}
\fbox{\includegraphics[width=14cm]{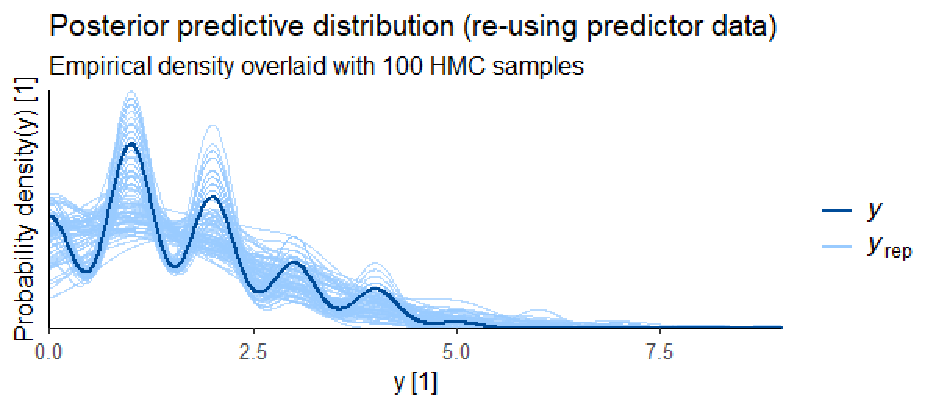}}
\end{center}
\caption{Posterior predictive distribution for fixed effects
Poisson regression re-using available predictor data, generated
with the function \texttt{bayesplot::ppc\_dens\_overlay()}.}
\lb{fig:poisRegFixedPPC1}
\end{figure}
\begin{figure}[!htb]
\begin{center}
\fbox{\includegraphics[width=14cm]{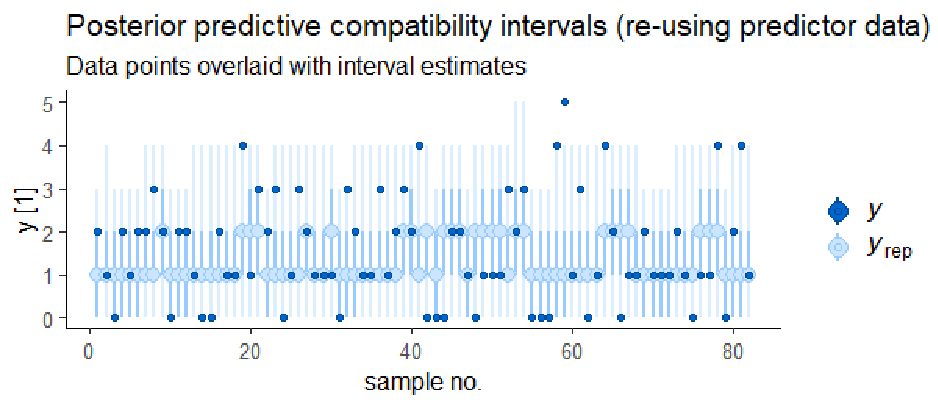}}
\end{center}
\caption{Posterior predictive compatibility intervals for fixed
effects Poisson regression re-using available predictor data,
generated with the function \texttt{bayesplot::ppc\_intervals()}.}
\lb{fig:poisRegFixedPPC2}
\end{figure}

\medskip
\noindent
If the set of \textbf{quantitative--empirical data} available to
the researcher is sufficiently large in size, it might be worth
improving the \textbf{out-of-sample posterior predictive accuracy}
of a fitted \textbf{statistical model} by following the standard 
practice of supervised machine learning and split the \textbf{data
set} into a \textbf{training set} (ca. 60\% of samples), a
\textbf{cross-validation set} (ca. 20\% of samples), and
\textbf{test set} (ca. 20\% of samples). Here, \textbf{parameter 
estimation} is performed with the training set, while
\textbf{out-of-sample posterior predictive accuracy} is gauged with
the remaining two data sets. We point the reader to Gill 
(2015)~\ct[Sec.~6.4]{gil2015}, Krusch\-ke
(2015)~\ct[Sec.~17.5]{kru2015} and McElreath
(2020)~\ct[Sec.~3.3]{mce2020a} for further information on
simulating and interpreting \textbf{posterior predictive
probability distributions}.

\medskip
\noindent
In the next chapter we will look at ways of discriminating between
competing statistical models as regards their performance in
fitting quantitative--empirical data and predicting yet unobserved
new data.

\chapter[Model comparison and hypothesis testing]{Model comparison 
and hypothesis testing}
\lb{ch8}
The comparison of \textbf{statistical models} that compete for an
optimal fit to the \textit{same} given set of
\textbf{quantitative--empirical data} is an essential issue of
reliability assessment which needs to be addressed in
\textbf{inductive statistical inference}. Concerning
their main objectives, some parallels can be drawn between
\textbf{model comparison} in the \textbf{Bayes--Laplace approach}
on the one-hand side, and frequentist null hypothesis significance
testing on the other, but the former operation is based on a
conceptually sound and transparent methodology, as we intend to
describe in this chapter.

\medskip
\noindent
In the \textbf{Bayes--Laplace approach} there have been developed
two different kinds of frameworks to capture the \textit{relative}
performance of two or more competing \textbf{statistical models} as 
regards quality of \textbf{model-fit} attained, and their ensuing
\textbf{out-of-sample posterior predictive accuracy}. The aim is to 
reduce potential \textbf{over-} or \textbf{under-fitting} of a
candidate \textbf{statistical model} to given
\textbf{quantitative--empirical data} as much as possible. As, for
a specific research problem, \textit{no} \textbf{statistical model}
that was built from relevant prior information and available
empirical evidence will ever be ``correct'' in the ontological
sense, i.e., be identical to the true \textbf{data-generating
process}, it is worthwhile considering the option of drawing
inferences from a statistical model that was obtained from
averaging over two or more of the competing \textbf{statistical
models}, should the fit of any of the statistical models involved
in the averaging process not be entirely unacceptable. The weight
factors necessary for \textbf{model averaging} are obtained from
normalising the differences in the quantified \textit{relative}
\textbf{out-of-sample posterior predictive accuracy} of the
different \textbf{statistical models} considered.

\medskip
\noindent
A widely accepted guiding principle in statistical model-building
is \textbf{parameter parsimony}. This is motivated by a particular
view that dates back to the Middle Ages and proved highly
influential in the history of Science. It was voiced by the English 
Franciscan friar, scholastic philosopher and theologian 
\href{https://mathshistory.st-andrews.ac.uk/Biographies/Ockham/}{William
of Ockham (1288--1348)}, who asserted:
\begin{quotation}
``Frustra fit per plura, quod potest fieri per 
pauciora.''\footnote{English translation: ``It is vain to do with
more, what can be done with less.''  See URL (cited on August 17,
2022): \href{https://mathshistory.st-andrews.ac.uk/Biographies/Ockham/quotations/}{mathshistory.st-andrews.ac.uk/Biographies/Ockham/quotations/}, and Sivia and Skilling (2006)~\ct[p~81]{sivski2006}.}
\end{quotation}
In the various scientific communities of these modern days, this
heuristic principle is known as \textbf{Ockham's razor}, and it is
often used as a justification for preferring one specific
\textbf{statistical model} as representing a proposed
\textbf{theoretical framework} in place of a competing
\textbf{statistical model} of comparable performance properties
which, however, contains a larger number of \textbf{model
parameters}.

\section[Information criteria and posterior predictive 
accuracy]{Information criteria and posterior predictive accuracy}
\lb{sec:infcrit}
Shannon's \textbf{information entropy}, as defined in
Eq.~(\ref{eq:entropy}) for discrete cases and in
Eq.~(\ref{eq:entropycont}) for continuous cases, is the
universally approved unique measure of the amount of
\textbf{uncertainty} represented by a given \textbf{probability
distribution}. Qualitatively speaking, \textbf{information entropy}
is growing the more probability (the total of which must sum
to~$1$) is spread out across viable possibilities. In this respect,
the number of dimensions of the parameter space in which
probability is  distributed plays a particularly important role.
\textbf{Information entropy} is also
the central pillar of the currently most widely applied techniques
for \textbf{model comparison}, for which \textit{relative}
\textbf{out-of-sample posterior predictive accuracy} is the
criterion for evaluating the performance of competing
\textbf{statistical models}. We will describe the details in this
section.

\medskip
\noindent
The non-symmetric, directed \textbf{Kullback--Leibler divergence},
defined in terms of the \textbf{information entropy} in the
discrete case of Eq.~(\ref{eq:entropy}) by\footnote{Here we assume
a uniform Lebesgue measure~$m_{i}$. In the continuous case, the
Kullback--Leibler divergence becomes
${\displaystyle D_\mathrm{KL}(P,Q) = \int_{-\infty}^{+\infty}
p(y)\,\ln\left(\frac{p(y)}{q(y)}\right)\mathrm{d}y}$.}
\be
\lb{eq:kulleidiv}
D_\mathrm{KL}(P,Q) := S(P,Q) - S(P) =
\sum_{i=1}^{k}p_{i}\ln\left(\frac{p_{i}}{q_{i}}\right) \ ,
\ee
measures the \textit{additional} \textbf{information entropy}
generated when approximating one discrete probability
distribution, $P = \{p_{i}\}_{i=1,\ldots,k}$, by a second discrete
probability distribution, $Q = \{q_{i}\}_{i=1,\ldots,k}$;
see Kullback and Leibler (1951)~\ct{kullei1951}. Note that,
by construction, it holds that $D_\mathrm{KL}(P,P) \equiv 0$.

\medskip
\noindent
Suppose given two candidate probability distributions,
$Q^{(1)}$ and $Q^{(2)}$, that are competing to approximate an
\textit{unknown} \textbf{target probability distribution}, $P$.
Then, when measured with the \textbf{Kullback--Leibler divergence},
the \textit{relative} distance of $Q^{(1)}$ and $Q^{(2)}$ from the
unknown target~$P$ amounts to
\bea
\lb{eq:kldiff}
D_\mathrm{KL}(P,Q^{(1)}) -  D_\mathrm{KL}(P,Q^{(2)})
& = & \left[\,S(P,Q^{(1)}) - S(P)\,\right]
- \left[\,S(P,Q^{(2)}) - S(P)\,\right] \nonumber \\
& = & S(P,Q^{(1)}) - S(P,Q^{(2)})
\ = \ 
-\sum_{i=1}^{k}p_{i}\ln\left(\frac{q^{(1)}_{i}}{q^{(2)}_{i}}\right)
 \nonumber \\
& = & -\sum_{i=1}^{k}p_{i}\,\left[\,\ln\left(q^{(1)}_{i}\right)
-\ln\left(q^{(2)}_{i}\right)\,\right] \ ;
\eea
this is just the $P$-average of the difference in
\textbf{logarithmic probability} between $Q^{(1)}$ and $Q^{(2)}$.
When this difference comes out
negative, $Q^{(1)}$ is closer to the unknown target~$P$ than
$Q^{(2)}$, and vice versa when the difference comes out positive.
The snag is that the $P$-measure used for the averaging is
\textit{not known}, and so, to continue the comparison of
$Q^{(1)}$ and $Q^{(2)}$, one needs to resort to a reliable
approximative evaluation of the $P$-averaged difference in
Eq.~(\ref{eq:kldiff}). A tried and tested estimation procedure
for this purpose that has  become commonplace employs the
\textbf{deviance} of a probability distribution~$Q$, defined by
\be
\lb{eq:dev}
D(Q) := -2\sum_{i=1}^{n}\ln(q_{i}) \ .
\ee
In terms of the \textbf{deviance}, the difference of {\bf
Kullback--Leibler divergences} in Eq.~(\ref{eq:kldiff}) can now be
estimated by
\bea
\lb{eq:kldiff2}
D_\mathrm{KL}(P,Q^{(1)}) -  D_\mathrm{KL}(P,Q^{(2)})
& \approx & D(Q^{(1)}) - D(Q^{(2)}) \nonumber \\
& = & -2\sum_{i=1}^{k}\left[\,\ln\left(q^{(1)}_{i}\right)
-\ln\left(q^{(2)}_{i}\right)\,\right] \ .
\eea

\medskip
\noindent
All of these considerations provide the foundation for a strategy
that seeks to safeguard against \textbf{over-fitting} when
approximating \textit{unknown} \textbf{posterior joint probability
distributions} by means of MCMC simulations. The objective of
the model-building process is to compare competing
\textbf{statistical models} on the basis of an estimation of their 
anticipated \textbf{out-of-sample posterior predictive accuracy},
or, in more technical terms, their expected \textbf{out-of-sample 
deviance}. We will now introduce two of the most widely accepted
estimators of a fitted statistical model's expected
\textbf{out-of-sample deviance},
\textbf{information criteria} that take different aspects
of the \textbf{posterior average} (or posterior expectation) of the 
natural logarithm of a model's \textbf{likelihood function} as
major building-blocks for valuing \textbf{out-of-sample posterior 
predictive accuracy}. This type of \textbf{logarithmic score}
corresponds to the unique local and proper scoring rule and is
commonly employed for assessing probabilitic predictions; cf.
Gelman \textit{et al} (2014)~\ct[p~167]{geletal2014}.

\subsection[Watanabe--Akaike information 
criterion]{Watanabe--Akaike information criterion}
\lb{subsec:waic}
The \textbf{Watanabe--Akaike information criterion (WAIC)},
suggested by Watanabe (2010)~\ct[Eq.~(6)]{wat2010}, provides a
cumulative pointwise estimate of a fitted statistical model's
expected \textbf{out-of-sample deviance}. Specifically it can be
applied for assessing \textbf{out-of-sample posterior predictive
accuracy} of highly skewed \textbf{posterior joint probability 
distributions} for \textbf{model parameters}, i.e., it does
\textit{not} assume that they be multivariate
Gau\ss\ distributions. It also allows for arbitrary \textbf{prior 
probability distributions}.

\medskip
\noindent
WAIC is built from two additive components. One component is
the sum of the natural logarithms
of the posterior-averaged single-datum likelihood functions for
every \textit{observed} datum, $y_{i}$, --- this is referred to as
the \textbf{log-pointwise posterior predictive density (lpd)} and
is given by
\be
\lb{eq:lpd}
\text{lpd} := \sum_{i=1}^{n}\,\underbrace{\ln\left(
\int_{\boldsymbol{\theta}\,\text{range}}
P(y_{i}|\boldsymbol{\theta}, I)
P(\boldsymbol{\theta}|\boldsymbol{y}, I)\,
\mathrm{d}\boldsymbol{\theta}
\right)}_{\text{log-posterior-average-single-datum-likelihood}} \ .
\ee
Using already \textit{observed} quantitative--empirical data,
the $\text{lpd}$ represents an over-estimate of the unknown
\textbf{expected log-pointwise posterior predictive density (elpd)}
for a \textit{new} data set; cf. Vehtari \textit{et al}
(2017)~\ct{vehetal2017}. The unknown \textbf{elpd} constitutes the
actual object of interest and needs to be estimated from
observed quantitative--empirical data. The expected value for the
$\text{lpd}$ defined in Eq.~(\ref{eq:lpd}) is estimated from MCMC 
simulations of size~$S$ by
\be
\lb{eq:lpdcomp}
\widehat{\text{lpd}} \approx \sum_{i=1}^{n}\,\ln\left(
\frac{1}{S}\,\sum_{s=1}^{S}\,P(y_{i}|\boldsymbol{\theta}^{s}, I)
\right) \ .
\ee
The second component is a term representing an
\textbf{effective number of parameters} that penalises model 
complexity and the fact that $\text{lpd}$ is an over-estimate
of $\text{elpd}$. It can be defined by
\be
p_\mathrm{WAIC} := -2\sum_{i=1}^{n}\mathrm{E}_{\mathrm{post}}
\left(\,
\ln[\,P(y_{i}|\boldsymbol{\theta}, I)\,]\,\right)
+ 2\sum_{i=1}^{n}\ln\left(\,\mathrm{E}_{\mathrm{post}}[P(y_{i}|
\boldsymbol{\theta}, I)]\,\right) \ .
\ee
The notation $\mathrm{E}_{\mathrm{post}}(\ldots)$ indicates
averaging a quantity over the parameter space with the
posterior joint probability distribution~$P(\boldsymbol{\theta}|
\boldsymbol{y}, I)$. Altogether, the cumulative pointwise
WAIC-estimate for a fitted \textbf{statistical model}'s expected
\textbf{out-of-sample deviance} is thus given by
\be
\lb{eq:waic}
\text{WAIC} := \underbrace{-2\,\text{lpd}}_{\text{measure\ of\ 
model\ fit}}
+ \underbrace{2\,p_\mathrm{WAIC}}_{\text{complexity\ penalty}} \ ,
\ee
where the particular re-scaling proposed by Gelman \textit{et al} 
(2014)~\ct[p~174]{geletal2014} is employed in order to comply with
the usual deviance-based structure of other information criteria;
see also McElreath (2020)~\ct[Sec.~7.4]{mce2020a}. Fitted
statistical models with smaller WAIC scores are preferred, but it
is generally advisable to retain weaker-performing competing
fitted statistical models for further reference. It needs to be
emphasised that model comparison by means of the WAIC score is
meaningful only between fitted statistical models that are built on
the \textit{same} single-datum likelihood function.

\subsection[PSIS leave-one-out cross-validation]{Pareto-smoothed 
importance-sampling leave-one-out cross-validation}
\lb{subsec:loo}
The philosophy underlying \textbf{cross-validation} is to split a
sample of size~$n$ of observed \textbf{quantitative--empirical
data} into a \textbf{training set} and a \textbf{test set} and to
fit a canditate \textbf{statistical model} to the
\textbf{training set} while assessing its \textbf{out-of-sample posterior predictive accuracy} on the \textbf{test set}.
\textbf{Leave-one-out cross-validation (LOO-CV)} opts for the
most extreme interpretation of this perspective in that it
puts $n-1$ of the available samples into the \textbf{training
set} and only a \textit{single} sample into the \textbf{test set},
while repeating the fitting procedure $n$ times, with every
observed datum~$y_{i}$ allocated to the \textbf{test set}
\textit{once}.

\medskip
\noindent
The formal cumulative pointwise LOO-CV estimate of
the unknown \textbf{expected log-pointwise posterior predictive
density (elpd)} for a \textit{new} data set using the
\textit{available} data set can be defined as
\be
\lb{eq:elpdloo}
\text{elpd}_{\mathrm{LOO-CV}} := \sum_{i=1}^{n}\,
\underbrace{\ln\left(
\int_{\boldsymbol{\theta}\,\text{range}}
P(y_{i}|\boldsymbol{\theta}, I)
P(\boldsymbol{\theta}|\boldsymbol{y}_{(-i)}, I)\,
\mathrm{d}\boldsymbol{\theta}
\right)}_{\text{log-posterior-average-single-datum-likelihood}} \ ,
\ee
where $P(\boldsymbol{\theta}|\boldsymbol{y}_{(-i)}, I)$
denotes the \textbf{posterior joint probability distribution}
for the \textbf{model parameters}~$\boldsymbol{\theta}$ obtained
when omitting the single observed datum $y_{i}$ from the 
fitting procedure. The expected value for
$\text{elpd}_{\mathrm{LOO-CV}}$ is computed from MCMC simulations
of size~$S$ by
\be
\lb{eq:elpdloocomp}
\widehat{\text{elpd}}_{\mathrm{LOO-CV}} \approx \sum_{i=1}^{n}\,\ln
\left(\frac{1}{S}\,\sum_{s=1}^{S}\,P(y_{i}|
\boldsymbol{\theta}^{s}_{(-i)}, I)
\right) \ .
\ee
As this approach requires the performance of a total of $n$ fitting 
procedures in order to determine $n$ \textbf{posterior joint
probability distributions} for the \textbf{model
parameters}~$\boldsymbol{\theta}$ from observed
\textbf{quantitative--empirical data} and so can be computationally
very expensive, a very elegant and efficient approximation
technique to overcome this problem was devised by Vehtari
\textit{et al} (2017)~\ct{vehetal2017}. They refer to their
approximation technique for evaluating
$\text{elpd}_{\mathrm{LOO-CV}}$ as \textbf{Pareto-smoothed
importance sampling (PSIS)}. The efficient approximate PSIS-LOO-CV 
cumulative pointwise estimate of the unknown \textbf{expected log-
pointwise posterior predictive density (elpd)} they propose is
given by
\be
\lb{eq:elpdloocomp}
\widehat{\text{elpd}}_{\mathrm{PSIS-LOO-CV}}
\approx \sum_{i=1}^{n}\,
\ln\left(
\frac{\sum_{s=1}^{S}w_{i}^{s}P(y_{i}|
\boldsymbol{\theta}^{s}, I)}{\sum_{s=1}^{S}w_{i}^{s}}\,
\right) \ ,
\ee
with Pareto-smoothed importance weights
$\{w_{i}^{s}\}_{i=1, \ldots, n}^{s = 1, \ldots, S}$. The
approximative PSIS approach requires the performance of only a
\textit{single} fitting  procedure to determine the
\textbf{posterior joint probability  distribution} for the
\textbf{model parameters}~$\boldsymbol{\theta}$ from observed
\textbf{quantitative--empirical data}. It integrates a
number of quite demanding information-theoretical considerations of
\textbf{statistical methods of data analysis} and of
\textbf{model-fitting}; for further details the interested reader
is referred to the original publication by Vehtari \textit{et al}
(2017)~\ct{vehetal2017}.

\medskip
\noindent
In analogy with other deviance-based information criteria, the
\textbf{LOO information criterion (LOOIC)} is defined by
\be
\lb{eq:looic}
\text{LOOIC} :=
\underbrace{-2\,\widehat{
\text{elpd}}_{\mathrm{PSIS-LOO-CV}}}_{\text{measure\ of
\ model\ fit}} \ .
\ee
Fitted statistical models with smaller LOOIC scores are preferred,
but it is generally advisable to retain weaker-performing competing
fitted statistical models for further reference. It needs to be
emphasised that model comparison by means of the LOOIC score is
meaningful only between fitted statistical models that are built on
the \textit{same} single-datum likelihood function.
In recent years, the LOOIC score has acquired in the pertinent
research literature the status of the prime tool for comparing
competing fitted \textbf{statistical models}
with respect to properties of \textbf{over-fitting} and
\textbf{out-of-sample posterior predictive accuracy}.
Exhaustive information on LOOIC and its possibilities of
application in \textbf{Applied Statistics} is available from the
website~\href{https://mc-stan.org/loo/}{mc-stan.org/loo/}.

\medskip
\noindent
To enable the computation for a fitted \textbf{statistical model}
of both the WAIC and LOOIC scores from observed
\textbf{quantitative--empirical data}, the following
supplementations to the \texttt{generated
quantities} block of a Stan model code are necessary to obtain
MCMC simulations of the
log-posterior-average-single-datum-likelihood, here given for the 
example of fixed effects multiple linear regression discussed in
Sec.~\ref{sec:linreg}:
\begin{lstlisting}[language = Stan, caption = {Supplementations
to vectorised Stan model code for fixed effects multiple linear
regression to enable the computation of the pointwise
log-likelihood for observed data.},
captionpos = b, label = {lst:pwloglike}]
generated quantities {
  ...
  /* Calculation of pointwise log-likelihood
     (re-using observed data) */
  vector[N] log_lik;
  
  for ( i in 1:N ) {
    log_lik[i] = normal_lpdf( y[i] | (X * beta)[i] , sigma );
  }
}
\end{lstlisting}
For the remaining fixed effects GLMs in Ch.~\ref{ch7} the necessary 
supplementations follow by analogy.

\medskip
\noindent
The actual computation for a fitted \textbf{statistical model}
of its WAIC and LOOIC scores in an \R{} environment employs the
functions \texttt{waic()} and \texttt{loo()} from the
\texttt{loo}~package by Vehtari \textit{et al}
(2020)~\ct{vehetal2020}. This is demonstrated in the following
\R{} code block~\ref{lst:stanloo}.
\begin{lstlisting}[language = R, caption = {HMC sampling with Stan
in \R{}: computation of WAIC and LOOIC scores using \texttt{loo}.},
captionpos = b, label = {lst:stanloo}]
# Load R package loo
library(loo)

# Extract pointwise log-likelihood samples
pwLogLik <- as.matrix(
  x = modelStan,
  pars = "log_lik"
)

# Compute waic
waicModelStan <- loo::waic(x = pwLogLik)

# Print value of waic
print(x = waicModelStan)


# Compute elpd_loo and looic
looModelStan <- loo::loo(
  x = modelStan,
  pars = "log_lik"
)

# Print values of elpd_loo and looic
print(x = looModelStan)

# Generate PSIS diagnostic plot
plot(
  x = looModelStan,
  label_points = TRUE
)
\end{lstlisting}

\medskip
\noindent
For historical reasons we now turn to address \textbf{model
comparison} by means of \textbf{Bayes factors}.

\section[Bayes factors]{Bayes factors}
\lb{sec:bayesf}
The idea behind the concept of \textbf{Bayes factors} as a
practical tool for \textbf{model comparison} is a simple one. Start
from \textbf{Bayes' theorem} in its variant of
Eq.~(\ref{eq:bayesData}) given in Ch.~\ref{ch1}, and apply it, for
a given fixed set of \textbf{quantitative--empirical data}, to both
a ``$\text{model}(i)$'' and a ``$\text{model}(j)$.'' Upon forming
the \textbf{posterior odds}, i.e.,
the ratio of the posterior probability for ``$\text{model}(i)$''
and the posterior probability for ``$\text{model}(j)$,'' one
obtains\footnote{In the present formulation of the posterior odds
and their relation to the prior odds, the terms
``$\text{model}(i)$'' and ``$\text{hypothesis}(i)$'' may be
perceived to be synonymous.}
\be
\lb{eq:postodds}
\underbrace{\frac{P(\text{model}(i)|\text{data}, 
I)}{P(\text{model}(j)|\text{data}, I)}}_{\text{posterior\ odds}}
= \frac{P(\text{data}|\text{model}(i), 
I)}{P(\text{data}|\text{model}(j), I)} \times
\underbrace{\frac{P(\text{model}(i)|I)}{P(\text{model}(j)|I)}}_{
\text{prior\ odds}} \ ,
\ee
where a common divisor of $P(\text{data}|I)$ cancelled out along
the way. Conventionally one defines the ratio multiplying the
\textbf{prior odds} on the right-hand side of
Eq.~(\ref{eq:postodds}), i.e.,
\be
\lb{eq:bayesfactor1}
B_{ij} := \frac{P(\text{data}|\text{model}(i), 
I)}{P(\text{data}|\text{model}(j), I)}
\ee
as the \textbf{Bayes factor}. In this form it gives the ratio of
the \textbf{average likelihoods} for ``$\text{model}(i)$'' and
``$\text{model}(j)$.'' By re-arranging Eq.~(\ref{eq:postodds}),
one find that this is equal to
\be
\lb{eq:bayesfactor2}
B_{ij} = \frac{\displaystyle\frac{P(\text{model}(i)|\text{data}, 
I)}{P(\text{model}(j)|\text{data}, 
I)}}{\displaystyle\frac{P(\text{model}(i)|I)}
{P(\text{model}(j)|I)}} \ ,
\ee
i.e., the ratio of the \textbf{posterior odds} and the
\textbf{prior odds} for ``$\text{model}(i)$'' and
``$\text{model}(j)$.'' The \textbf{Bayes factor} provides an
immediate manifestation of the very fact that within the
\textbf{Bayes--Laplace approach} only the values of
\textit{relative measures} contain tangible information. In this
framework it is often not possible to define absolute values
in any sensible way.

\medskip
\noindent
Suppose ``$\text{model}(i)$'' contains a set of $k+1$
\textbf{model parameters},
$\{\theta_{0}^{(i)}, \ldots, \theta_{k}^{(i)}\}$. Then its
associated \textbf{average likelihood} is calculated by
averaging the \textbf{total-data likelihood function} in
$(k+1)$-dimensional \textbf{parameter space} with the \textbf{prior
joint probability distribution},
\bea
P(\text{data}|\text{model}(i), I)
& = & \int\cdots\int_{\theta_{j}^{(i)}\,\text{ranges}}
\underbrace{P(\text{data}|\theta_{0}^{(i)}, \ldots, 
\theta_{k}^{(i)}, \text{model}(i), I)}_{\text{likelihood}}
\nonumber \\
& & \qquad \times
\underbrace{P(\theta_{0}^{(i)}, \ldots, 
\theta_{k}^{(i)}|\text{model}(i), I)}_{\text{prior}}\,
\mathrm{d}\theta_{0}^{(i)} \cdots \mathrm{d}\theta_{k}^{(i)} \ .
\eea
This calculation is to be repeated in an analogous fashion for
``$\text{model}(j)$,'' which, however, usually contains a number
of \textbf{model parameters} different from~$k+1$. The point is
that \textbf{statistical models} with a higher number of
\textbf{model parameters} need to spread out \textbf{prior joint probability density} (which, of course, needs to integrate to~$1$)
over a larger number of dimensions in \textbf{parameter space} than 
\textbf{statistical models} with a smaller number of
\textbf{model parameters}. A larger number of dimensions
of \textbf{parameter space} amounts to a larger hyper-volume to be
covered by the \textbf{prior joint probability density}. In this
respect, \textbf{statistical models} with a higher number of
\textbf{model parameters} automatically get
penalised by the present procedure. Only if there is a sufficient
amount of supporting evidence in the
\textbf{quantitative--empirical data} for the presence of
additional \textbf{model parameters} (which will factor into the
procedure via the total-data likelihood function) can the
penalty for a dimension-inflated hyper-volume be compensated.

\medskip
\noindent
\textbf{Bayes factors} can be calculated analytically for all the
single-parameter estimation examples with exact solutions for the
parameter's \textbf{posterior probability distribution} that were
discussed in Ch.~\ref{ch4}. The reason is that in those cases only
the \textbf{prior probability distribution} can be varied between
competing models, while their \textbf{total-data likelihood
functions} are identical.\footnote{For once, they have no
$\boldsymbol{X}$-data for independent variables blended in via an
(inverse) link function like in GLMs.} Here we present the explicit
\textbf{Bayes factor} solution for the \textbf{Beta--binomial
model} of Subsec.~\ref{subsec:betabinom}. Suppose given
\textbf{quantitative--empirical data}~$\{y, n\}$, and introduce
a ``$\text{model}(1)$'' and a ``$\text{model}(2)$'' as competing to 
explain the underlying \textbf{data-generating process}.
The binomial total-data likelihood function has the same structure
for both cases; only the parameter values of the two Beta prior
probability distributions, $\{\alpha_{1}, \beta_{1}\}$ and
$\{\alpha_{2}, \beta_{2}\}$, will be different. Then the ratio of
\textbf{posterior model odds} and \textbf{prior model odds} amounts
to
\be
\lb{eq:bayesfbetabinom}
B_{12} = \frac{\displaystyle\frac{\mathrm{
Be}(\alpha_{1}+y,\beta_{1}+n-y)}{\mathrm{
Be}(\alpha_{2}+y,\beta_{2}+n-y)
}}{\displaystyle\frac{\mathrm{Be}(\alpha_{1},\beta_{1})}
{\mathrm{Be}(\alpha_{2},\beta_{2})}}
=
\frac{\mathrm{Be}(\alpha_{1}+y,\beta_{1}+n-y)}{\mathrm{Be}(
\alpha_{2}+y,\beta_{2}+n-y)} \times
\frac{\mathrm{Be}(\alpha_{2},\beta_{2})}{\mathrm{Be}(
\alpha_{1},\beta_{1})} \ .
\ee

\medskip
\medskip
\noindent
Jeffreys (1939)~\ct[App.~B]{jef1939} devised a heuristic scale for
interpreting the values of \textbf{Bayes factors} when only
\textit{two} competing \textbf{statistical models} are considered.
According to this scale, one classifies the explanatory power of
the two models under investigation as

\medskip
\noindent
\underline{\textbf{Jeffreys' scale for comparison of two competing
models:}}\\
$B_{12} > 1$: $\text{model}(1)$ supported\\
$1 > B_{12} > 10^{-1/2}$: weak evidence against $\text{model}(1)$\\
$10^{-1/2} > B_{12} > 10^{-1}$: substantial evidence against
$\text{model}(1)$\\
$10^{-1} > B_{12} > 10^{-3/2}$: strong evidence against
$\text{model}(1)$\\
$10^{-3/2} > B_{12} > 10^{-2}$: very strong evidence against
$\text{model}(1)$\\
$10^{-2} > B_{12}$: decisive evidence against $\text{model}(1)$;

\medskip
\noindent
see also Gill (2015)~\ct[p~217]{gil2015}. Kass and
Raftery (1995)~\ct[p~777]{kasraf1995}, and Jaynes
(2003)~\ct[p~91]{jay2003}, transform Jeffreys' scale to an
orders-of-magnitude emphasising, base-$10$ logarithmic scale,
which appears closer to intuition. In this case it holds that
\be
\log_{10}(B_{12})
= \log_{10}\left(\frac{P(\text{model}(1)|\text{data}, 
I)}{P(\text{model}(2)|\text{data}, I)}\right)
- \log_{10}\left(\frac{P(\text{model}(1)|I)}{P(\text{model}(2)|I)}
\right) \ ,
\ee
which gives the difference between the posterior decadic
log-odds and the prior decadic log-odds; see also Greenberg
(2013)~\ct[p~36]{gre2013}.

\medskip
\noindent
We draw the reader's attention to the lively review by Jefferys
and Berger (1992)~\ct{jefber1992} on the concept of \textbf{Bayes
factors}. For illustrative purposes they relate their discussion
to the prominent historical example from the early
20$^{\mathrm{th}}$ Century of the two competing theories of 
gravitational interactions that were trying to explain the
phenomenon of the advance of the perihelion of planet Mercury on
its orbit around the Sun. This observation had been puzzling
astronomers ever since the French astronomer and mathematician
\href{https://en.wikipedia.org/wiki/Urbain_Le_Verrier}{Urbain Jean
Joseph Le Verrier (1811--1877)} had reported on this problem for
gravitational theory to the French Academy of Sciences on September
12, 1859.

\medskip
\noindent
The present exposition of \textbf{model comparison} elucidates that
for \textbf{hypothesis testing} in the \textbf{Bayes--Laplace
approach} one requires at least one proper, testable alternative 
hypothesis to a given ``\text{hypothesis}(1),'' where the former,
too, can be assigned a well-defined \textbf{total-data likelihood 
function} so that decision-making as to the data-favoured
hypothesis becomes possible. Simply making the choice
\[
\text{hypothesis}(2) := \overline{\text{hypothesis}(1)}
\]
will lead into a non-constructive dead end. There is no way to
devise a meaningful \textbf{total-data likelihood
function}~$P(\text{data}| \overline{\text{hypothesis}(1)}, I)$;
see Sivia and Skilling (2006)~\ct[p~84]{sivski2006}, and Trotta
(2008)~\ct[Sec.~4.1]{tro2008}.

\medskip
\noindent
A computational challenge that had long plagued researchers was the
evaluation of the \textbf{average likelihood} for multi-parameter
models, where domain integrations have to be performed in a
high-dimensional \textbf{parameter space}. Some of the first
algorithms that obtain this information by means of numerical
simulation were given by Chib (1995)~\ct{chi1995} and by Carlin and
Chib (1995)~\ct{carchi1995}; see also Greenberg
(2013)~\ct[Subsec.~7.1.2]{gre2013}. Specific routines for
facilitating the practical task of computing \textbf{Bayes factors}
have been made available within an \R{} environment in the
\texttt{MCMCpack}~package by Martin \textit{et al}
(2011)~\ct{maretal2011}

\medskip
\noindent
\underline{\R:}
$\texttt{BayesFactor}(\textit{MCMCpack output})$ (\texttt{MCMCpack} 
package),

\medskip
\noindent
and in the \texttt{BayesFactor}~package by Morey and Rouder 
(2018)~\ct{morrou2018}. Further information on conceptual aspects
of \textbf{model comparison} with \textbf{Bayes factors} is given
in the helpful practical tutorial by Lodewyckx \textit{et al}
(2011)~\ct{lodetal2011}, and in Gelman \textit{et al} 
(2014)~\ct[Sec.~7.4]{geletal2014}, Gill 
(2015)~\ct[Ch.~7]{gil2015}, or in Kruschke
(2015)~\ct[Ch.~10]{kru2015}. 

\medskip
\noindent
For astrophysical and cosmological problems with only a small
amount of available observational information, Trotta
(2008)~\ct[Sec.~4.7]{tro2008} advocates the \textbf{Bayes factor}
method for \textbf{model comparisons}. Gelman and Rubin
(1995)~\ct{gelrub1995} and Gelman \textit{et al} 
(2014)~\ct[Sec.~7.4]{geletal2014}, on the other hand, generally
advise against the use of \textbf{Bayes factors} as a selection
criterion for \textbf{statistical models} due to their inherent
\textbf{sensitivity} to the choice of \textbf{prior probability
distributions}.

\medskip
\noindent
Now we turn to put into perspective in the next chapter various
possibilities of increasing the flexibility of a statistical
model by adapting it to a researcher's state of knowledge
when this exhibits a more complex structure than assumed in the
examples discussed before in Ch.~\ref{ch7}.

\chapter[Varying effects models]{Varying effects generalised linear 
models}
\lb{ch9}
The simple cases of \textbf{regression analysis} we
discussed in Ch.~\ref{ch7} in the context of \textbf{generalised
linear models} confined their methodological considerations to
constructing \textbf{fixed effects models}. That is to say, the
different families of \textbf{statistical models} that were
presented are built on the implicit assumption that supposes
effects to be the same for all sample units in the \textbf{target 
population} of the researcher's investigation. In practice,
however, one often possesses \textbf{information} on some kind of 
intrinsic structure within the \textbf{target population}
manifested in some qualitative dimension, such as sample units
being  members of exclusive \textbf{groups} or \textbf{clusters}
(which are assumed to be exchangeable), with the possibility of
sample units belonging to the same group to appear more similar to
one another on the quantitative characteristic features of the
researcher's interest
than across groups. If this kind of \textbf{information} is
available, one might as well make use of it in the
\textbf{model-building process} by integrating it into the
calculation resp. simulation of \textbf{posterior joint probability
distributions} for unknown \textbf{model parameters}.
\textbf{Multi-level models}, as they are known within the
\textbf{Bayes--Laplace approach to data analysis and statistical 
inference}, constitute the present state-of-the art in
\textbf{Applied Statistics} for devising flexible frameworks that
can handle \textbf{quantitative--empirical data} obtained from 
performing measurements on complex systems
or/and dynamical processes. Multi-level data arises, e.g., from
\textbf{stratified samples} or from \textbf{repeated measurements}
of the same statistical variables on the same sample units. 

\medskip
\noindent
Technically speaking, \textbf{varying effects models} are obtained
by introducing for some unknown \textbf{model parameters}
\textit{adaptive} \textbf{prior probability distributions} as
opposed to the fixed ones that were employed in Ch.~\ref{ch7}. The 
rationale behind this procedure is the view that the specific
properties of some \textbf{model parameters} are captured more 
accurately when one describes them as arising from an entire 
distribution of possibilities across groups which have fixed
\textbf{prior probability distributions} within these groups. Such
an approach has the consequence that \textbf{information} on the
values of these \textbf{model parameters} obtained from analysing
\textbf{quantitative--empirical data} is partially shared between
groups --- known as \textbf{partial pooling} of information ---
which leads in combination with \textbf{adaptive regularisation}
to more robust \textbf{statistical models} that are less vulnerable
to the threat of \textbf{over-fitting}. This in turn will generally
improve a \textbf{statistical model}'s overall
\textbf{out-of-sample posterior predictive accuracy}: while
within-group model-fits and predictions will get worse due to
\textbf{partial pooling}, out-of-sample predictions will become
more reliable; see, e.g., Gelman \textit{et al}
(2014)~\ct[Ch.~5]{geletal2014}, Gill (2015)~\ct[Ch.~12]{gil2015},
Krusch\-ke (2015)~\ct[Ch.~9]{kru2015}, or McElreath
(2020)~\ct[Chs.~13, 14]{mce2020a}. With \textbf{partial pooling}
of information is associated the regularising phenomenon of
\textbf{shrinkage} of estimates for \textbf{model parameters},
these thus becoming less susceptible to the influence of
\textbf{outliers} in the given data. Because of \textbf{adaptive
regularisation} induced by the information exchanged between
groups, estimates for \textbf{model parameters} on the level of the
individual groups will ``shrink'' towards their total sample mean.
The size of \textbf{shrinkage} is influenced by the amount of
data available within the groups and the amount of variation
between the groups. In the empirical sciences this statistical 
phenomenon has long been known as \textbf{regression towards the
mean}. It was made popular through the work of the English
empiricist \href{https://mathshistory.st-andrews.ac.uk/Biographies/Galton/}{Sir
Francis Galton FRS (1822--1911)}, who discovered it following years
of intense  research during the late $19^{\mathrm{th}}$ Century;
see Galton  (1886)~\ct{gal1886}, and also Kahneman
(2011)~\ct[Ch.~17]{kah2011}. Contrasting with the \textbf{partial
pooling} approach are the \textbf{complete pooling} approach that
the \textbf{fixed effects GLMs} of Ch.~\ref{ch7} take, i.e.,
fitting a single statistical model to the total of
quantitative--empirical data available, and the \textbf{no pooling} 
approach that fits a separate statistical model to the
quantitative--empirical data available for each individual group.

\medskip
\noindent
There are basically two kinds of \textbf{varying effects}
extensions of the \textbf{fixed effects GLMs}  
introduced in Ch.~\ref{ch7} one may consider: (i)~the reasonably
straightforward \textbf{varying intercept models} that suppose
variation across groups of the mean of the \textbf{dependent
variable}~$Y$; this possibility is reflected in the variation
across groups of the \textbf{model parameter}~$\beta_{0}$, and
(ii)~the technically more demanding \textbf{correlated varying
intercept and slopes models} that require in particular the
modelling of \textbf{prior correlations} amongst the \textbf{model
parameters}~$\boldsymbol{\beta}$; here not only does the
mean of the \textbf{dependent variable}~$Y$ vary across groups,
but also the influence on~$Y$ of the \textbf{independent variables}
contained in the design matrix~$\boldsymbol{X}$. The second option
leads naturally to a significant increase in a \textbf{statistical 
model}'s complexity, though not in the overall logic of making 
inferences in the \textbf{Bayes--Laplace approach}. This second
option is what we will be focussing on in the present chapter.

\section[Non-centred parametrisation]{Non-centred decomposition
of model parameters}
\lb{sec:hiercholesky}
Mathematically the idea of \textbf{partial pooling} of information
across exclusive and exchangeable groups may be represented by
assuming that the vector-valued \textbf{model
parameter}~$\boldsymbol{\beta}
\in \mathbb{R}^{(k+1) \times 1}$ in the \textbf{GLMs} introduced
in Ch.~\ref{ch7} is generated by a multivariate
\textbf{Gau\ss\ distribution}, i.e.,
$\left.\boldsymbol{\beta}\right| I
\sim \mathrm{N}(\boldsymbol{\mu}_{\boldsymbol{\beta}},
\boldsymbol{\Sigma}_{\boldsymbol{\beta}})$, with
\textbf{mean vector}~$\boldsymbol{\mu}_{\boldsymbol{\beta}}
\in \mathbb{R}^{(k+1) \times 1}$ and positive semi-definite
and quadratic \textbf{covariance
matrix}~$\boldsymbol{\Sigma}_{\boldsymbol{\beta}} \in
\mathbb{R}^{(k+1) \times (k+1)}$. The latter encodes
potential interdependencies between the components
of~$\boldsymbol{\beta} \in \mathbb{R}^{(k+1) \times 1}$.
For each group in question, a separate \textbf{model
parameter}~$\boldsymbol{\beta} \in \mathbb{R}^{(k+1) \times 1}$ is
drawn from $\mathrm{N}(\boldsymbol{\mu}_{\boldsymbol{\beta}},
\boldsymbol{\Sigma}_{\boldsymbol{\beta}})$,
whereby~$\boldsymbol{\mu}_{\boldsymbol{\beta}}
\in \mathbb{R}^{(k+1) \times 1}$
and~$\boldsymbol{\Sigma}_{\boldsymbol{\beta}} \in
\mathbb{R}^{(k+1) \times (k+1)}$ are determined from the available
\textbf{quantitative--empirical data} by means of
\textbf{adaptive learning}.

\medskip
\noindent
It is a surprising but for practical purposes very convenient fact
that the property of a \textbf{model parameter} originating from a
\textbf{Gau\ss\ distribution} can be technically achieved by
resorting to a simple trick. This is what we describe next.
\begin{enumerate}

\item Suppose given a univariate \textbf{model parameter}~$\beta
\in \mathbb{R}$, that, by assumption, is Gau\ss\ distributed, i.e.,
$\left.\beta\right| I \sim \mathrm{N}(\mu_{\beta},
\sigma_{\beta}^{2})$. This \textbf{model parameter} can be
decomposed while maintaining its Gau\ss\ distributed property as
\be
\lb{eq:ncuni}
\beta = \mu_{\beta} + \sigma_{\beta}\,z_{\beta} \ ,
\ee
with scalar-valued \textbf{mean}~$\mu_{\beta} \in \mathbb{R}$,
scalar-valued \textbf{standard deviation}~$\sigma_{\beta} \in
\mathbb{R}_{\geq 0}$, and where $z_{\beta} \in \mathbb{R}$ follows
a standard normal distribution, i.e., $\left.z_{\beta}\right| I
\sim \mathrm{N}(0, 1)$. This decomposition is referred to as a
\textbf{non-centred parametrisation} of $\left.\beta\right| I
\sim \mathrm{N}(\mu_{\beta}, \sigma_{\beta}^{2})$. Beginning
with a $\boldsymbol{Z}$ \textbf{score} and then building the
\textbf{model parameter}~$\beta$ from it, the non-centred
decomposition amounts to the reverse transformation of
\textbf{standardisation}; cf. Ref.~\ct[Subsec.~3.2.6]{hve2019}.

\item The generalisation of this kind of decomposition to the case
of a multivariate \textbf{model parameter}~$\boldsymbol{\beta}
\in \mathbb{R}^{(k+1) \times 1}$, where
$\left.\boldsymbol{\beta}\right| I
\sim \mathrm{N}(\boldsymbol{\mu}_{\boldsymbol{\beta}},
\boldsymbol{\Sigma}_{\boldsymbol{\beta}})$, with
\textbf{mean vector}~$\boldsymbol{\mu}_{\boldsymbol{\beta}}
\in \mathbb{R}^{(k+1) \times 1}$ and positive semi-definite
and quadratic \textbf{covariance
matrix}~$\boldsymbol{\Sigma}_{\boldsymbol{\beta}} \in
\mathbb{R}^{(k+1) \times (k+1}$, follows the same logic while
entailing an additional level of complexity due to potential
interdependencies between the components
of~$\boldsymbol{\beta} \in \mathbb{R}^{(k+1) \times 1}$. The
Polish--French military officer and mathematician
\href{https://mathshistory.st-andrews.ac.uk/Biographies/Cholesky/}{Andr\'{e}--Louis
Cholesky (1875--1918)} realised that \textbf{quadratic matrices}
such as $\boldsymbol{\Sigma}_{\boldsymbol{\beta}} \in
\mathbb{R}^{(k+1) \times (k+1)}$ may be factorised as
\be
\lb{eq:decompSigma1}
\boldsymbol{\Sigma}_{\boldsymbol{\beta}}
= \boldsymbol{L}_{\boldsymbol{\Sigma}_{\boldsymbol{\beta}}}
\boldsymbol{L}_{\boldsymbol{\Sigma}_{\boldsymbol{\beta}}
}^{\top} \ ,
\ee
wherein the \textbf{Cholesky factor
matrix}~$\boldsymbol{L}_{\boldsymbol{\Sigma}_{\boldsymbol{\beta}}}
\in \mathbb{R}^{(k+1) \times (k+1)}$ is lower-triangular; cf.,
e.g., Rinne (2008)~\ct[Subsec.~3.10.4]{rin2008}.
Employing such a \textbf{Cholesky decompostion} of
$\boldsymbol{\Sigma}_{\boldsymbol{\beta}} \in
\mathbb{R}^{(k+1) \times (k+1)}$, the multivariate analogue of
Eq.~(\ref{eq:ncuni}) becomes
\be
\lb{eq:ncmulti1}
\boldsymbol{\beta}
= \boldsymbol{\mu}_{\boldsymbol{\beta}}
+ \boldsymbol{L}_{\boldsymbol{\Sigma}_{\boldsymbol{\beta}}}\,
\boldsymbol{z}_{\boldsymbol{\beta}} \ ,
\ee
with $\boldsymbol{z}_{\boldsymbol{\beta}}
\in \mathbb{R}^{(k+1)\times 1}$ and
$\left.\boldsymbol{z}_{\boldsymbol{\beta}}\right| I
\sim \mathrm{N}(\boldsymbol{0}, \boldsymbol{1})$.

\medskip
\noindent
Taking in turn into account that
$\boldsymbol{\Sigma}_{\boldsymbol{\beta}} \in
\mathbb{R}^{(k+1) \times (k+1)}$ may likewise be factorised as
\be
\boldsymbol{\Sigma}_{\boldsymbol{\beta}}
= \text{diag}(\boldsymbol{\sigma}_{\boldsymbol{\beta}})
\boldsymbol{R}_{\boldsymbol{\beta}}\,
\text{diag}(\boldsymbol{\sigma}_{\boldsymbol{\beta}})^{\top} \ ,
\ee
with $\text{diag}(\boldsymbol{\sigma}_{\boldsymbol{\beta}})
\in \mathbb{R}^{(k+1) \times (k+1)}$ a diagonal matrix
of $k+1$ typically independent scalar-valued \textbf{standard
deviations}, and where the associated \textbf{correlation
matrix}~$\boldsymbol{R}_{\boldsymbol{\beta}} \in
\mathbb{R}^{(k+1) \times (k+1)}$ possesses a \textbf{Cholesky
decompostion} given by
\be
\boldsymbol{R}_{\boldsymbol{\beta}}
= \boldsymbol{L}_{\boldsymbol{R}_{\boldsymbol{\beta}}}
\boldsymbol{L}_{\boldsymbol{R}_{\boldsymbol{\beta}}}^{\top} \ ,
\ee
with $\boldsymbol{L}_{\boldsymbol{R}_{\boldsymbol{\beta}}}
\in \mathbb{R}^{(k+1) \times (k+1)}$ lower-triangular, one
arrives at
\be
\lb{eq:ncmulti2}
\boldsymbol{\beta}
= \boldsymbol{\mu}_{\boldsymbol{\beta}}
+ \text{diag}(\boldsymbol{\sigma}_{\boldsymbol{\beta}})\,
\boldsymbol{L}_{\boldsymbol{R}}\,
\boldsymbol{z}_{\boldsymbol{\beta}} \ .
\ee

\medskip
\noindent
Lastly, $\boldsymbol{\mu}_{\boldsymbol{\beta}}
\in \mathbb{R}^{(k+1) \times 1}$ may be decomposed as
$\boldsymbol{\mu}_{\boldsymbol{\beta}}
= \boldsymbol{u}\boldsymbol{\gamma}_{\boldsymbol{\beta}}$, with
$\boldsymbol{u} \in \mathbb{R}^{(k+1) \times l}$ a matrix-valued
\textbf{group predictor} and
$\boldsymbol{\gamma}_{\boldsymbol{\beta}}
\in \mathbb{R}^{l\times 1}$ corresponding \textbf{group
coefficients}, where the latter can be determined by means of
regression on the former; cf.
\href{https://mc-stan.org/docs/stan-users-guide/multivariate-hierarchical-priors.html}{Stan 
User's Guide (v2.30)} \ct[Sec.~1.13]{sta2022c}. Thus, the final
\textbf{non-centred parametrisation} of
$\left.\boldsymbol{\beta}\right| I
\sim \mathrm{N}(\boldsymbol{\mu}_{\boldsymbol{\beta}},
\boldsymbol{\Sigma}_{\boldsymbol{\beta}})$ is given by
\be
\lb{eq:ncmulti3}
\boldsymbol{\beta}
= \boldsymbol{u}\boldsymbol{\gamma}_{\boldsymbol{\beta}}
+ \text{diag}(\boldsymbol{\sigma}_{\boldsymbol{\beta}})\,
\boldsymbol{L}_{\boldsymbol{R}_{\boldsymbol{\beta}}}\,
\boldsymbol{z}_{\boldsymbol{\beta}} \ .
\ee

\end{enumerate}
To complete the procedure and prepare for \textbf{partial pooling}
of information and \textbf{adaptive regularisation},
\textbf{prior probability distributions} need to
be specified for each of the \textbf{group
coefficients}~$\boldsymbol{\gamma}_{\boldsymbol{\beta}}$, the
diagonal matrix of \textbf{standard
deviations}~$\text{diag}(\boldsymbol{\sigma}_{\boldsymbol{\beta}})$,
and the \textbf{Cholesky factor
matrix}~$\boldsymbol{L}_{\boldsymbol{R}_{\boldsymbol{\beta}}}$ of
the \textbf{correlation matrix}. We will follow general custom and
choose
\bea
\left.\boldsymbol{\gamma}_{\boldsymbol{\beta}}\right| I
& \sim & \mathrm{N}(\boldsymbol{0},
\text{diag}(\boldsymbol{\sigma}_{0})) \\
\left.\boldsymbol{\sigma}_{\boldsymbol{\beta}}\right| I
& \sim & \mathrm{Exp}(\gamma_{0}) \\
\lb{eq:LKJprior}
\left.\boldsymbol{L}_{\boldsymbol{R}_{\boldsymbol{\beta}}}\right| I
& \sim & \mathrm{LKJ}(\eta_{0}) \ .
\eea
Herein, $\mathrm{LKJ}(\eta_{0})$ denotes a so-called
\textbf{LKJ prior} according to Lewandowski \textit{et al}
(2009)~\ct{lkj2009}. When the single positive parameter
$\eta_{0} = 1$, it amounts to a continuous uniform distribution
over the interval~$\left[-1, 1\right]$, the full range of a
correlation coefficient. Otherwise, the more $\eta_{0} > 1$, the
more the \textbf{LKJ prior} expresses scepticism as to the
existence of correlations; the more $\eta_{0} < 1$, the more it
considers the existence of correlations likely.

\medskip
\noindent
Generally a \textbf{non-centred parametrisation} has a
curvature-reducing effect on the geometry of the \textbf{posterior
joint probability distribution}
in a high-dimensional \textbf{parameter space} to the
extent of improving both the numerical stability and the
convergence rate of the \textbf{HMC sampling algorithm};
cf. \href{https://mc-stan.org/docs/stan-users-guide/reparameterization.html}{Stan
User’s Guide (v2.30)}~\ct[Sec.~25.7]{sta2022c}. This will lead to
significantly fewer (or even no) divergent transitions in the
MCMC simulations using Stan.

\medskip
\noindent
In the examples to follow, we will not make use of the possibility
of introducing non-trivial group predictors and so we choose
$l = 1$ and set $\boldsymbol{u} = (1, 1, \ldots, 1)$ (with $k+1$
``ones'' in total). For the performance of MCMC simulations with
Stan for correlated varying effects models, the \R{} code that
specifies the available \textbf{quantitative--empirical data} needs
to be supplemented as shown in~\ref{lst:stanloadml} in order to
pass on to the HMC sampler the additional information given for
these cases.
\begin{lstlisting}[language = R, caption = {HMC sampling with Stan
in \R{}: specification of quantitative--empirical data for
correlated varying effects generalised linear models in
non-centred parametrisation.},
captionpos = b, label = {lst:stanloadml}]
# Data list for Stan, including declaration of dimensions and
# of observed variables
dataList <- list(
  N = nrow(X),  # sample size
  M = ncol(X),  # no. of independent variables plus 1
  K = length( unique(dataSet$gpVar) ),  # no. of groups
  L = 1L,  # no. of group predictors
  u = matrix(
    data = rep(1.0, length(unique(dataSet$gpVar))),
    nrow = length(unique(dataSet$gpVar)),
    ncol = 1L
  ),  # matrix of group predictors
  X = X,  # design matrix of standardised independent variables
  y = dataSet$y,  # dependent variable
  gp = as.integer(dataSet$gpVar)  # group variable
)
\end{lstlisting}

\medskip
\noindent
We now turn to take a look at fitting to adequate
\textbf{quantitative--empirical data} specific MCMC simulated
\textbf{correlated varying effects}
extensions of the five classes of \textbf{GLMs} we
presented in Secs.~\ref{sec:linreg} to~\ref{sec:expreg}, using
Stan in \R{} for their implementation. Again, the fitted
\textbf{statistical models} so obtained need to be subjected to 
dedicated \textbf{sensitivity analyses} as to meaningful choices of 
\textbf{prior probability distributions}, and to rigorous
\textbf{posterior predictive checks}.

\section[Multi-level linear regression]{Multi-level linear 
regression}
\lb{sec:hierlinreg}
For the \textbf{correlated varying effects linear regression
model}, we continue to use a \textbf{Gau\ss\ single-datum
likelihood function} according to Eq.~(\ref{eq:gausspdf}), though a
non-central $t$--single-datum likelihood according to
Eq.~(\ref{eq:tpdf}) is conceivable to provide the flexibility
needed for adapting to potential \textbf{outliers} in
the data for the dependent variable~$Y$. We maintain the assumption
of homogeneous variances. We assume that the metrically scaled data
available for \textit{both}~$\boldsymbol{y}$ and~$\boldsymbol{X}$
has been standardised prior to analysis. Employing a non-centred
decomposition of the group-dependent \textbf{model
parameters}~$\boldsymbol{\beta}$ according to
Eq.~(\ref{eq:ncmulti3}), the \textbf{correlated varying effects
linear regression model} is described by
\bea
\lb{eq:likelihoodfctlinreghier}
\text{likelihood:} \qquad
\left.\boldsymbol{y}\right| \boldsymbol{X}, \boldsymbol{\beta},
\sigma^{2}, [gp], I
& \stackrel{\mathrm{ind}}{\sim} & \mathrm{N}(
\boldsymbol{\mu}[gp], \sigma^{2}) \\
\lb{eq:invlinklinreghier}
\text{linear model:} \qquad
\boldsymbol{\mu}[gp]
& = & \boldsymbol{X}[gp]\,\boldsymbol{\beta}[gp] \\
\lb{eq:priorlinregbetahier}
\text{non-centred decomposition:} \qquad
\boldsymbol{\beta}[gp]
& = & \boldsymbol{u}\boldsymbol{\gamma}_{\boldsymbol{\beta}}[gp]
+ \text{diag}(\boldsymbol{\sigma}_{\boldsymbol{\beta}}[gp])\,
\boldsymbol{L}_{\boldsymbol{R}_{\boldsymbol{\beta}}}\,
\boldsymbol{z}_{\boldsymbol{\beta}}[gp] \\
\text{priors:} \qquad
\left.\boldsymbol{\gamma}_{\boldsymbol{\beta}}[gp]
\right|\sigma^{2}, I
& \sim & \mathrm{N}(0, \sigma_{0}^{2}) \\
%
\left.\boldsymbol{\sigma}_{\boldsymbol{\beta}}[gp]
\right|\sigma^{2}, I
& \sim & \mathrm{Exp}(\gamma_{0}) \\
%
\left.\boldsymbol{L}_{\boldsymbol{R}_{\boldsymbol{\beta}}}
\right|\sigma^{2}, I
& \sim & \mathrm{LKJ}(\eta_{0}) \\
\left.\boldsymbol{z}_{\boldsymbol{\beta}}[gp]\right|\sigma^{2}, I
& \sim & \mathrm{N}(0, 1) \\
\left.\sigma\right| I
& \sim & \mathrm{Exp}(\beta_{0}) \ ,
\eea
where $\sigma_{0}^{2}$, $\gamma_{0}$, $\eta_{0}$ and~$\beta_{0}$
denote fixed hyperparameters of the various prior probability 
distributions.

\medskip
\noindent
Examples of applications of \textbf{multi-level linear 
regression} can be found in Gill
(2015)~\ct[Sec.~12.7]{gil2015} and in Krusch\-ke
(2015)~\ct[Sec.~17.3]{kru2015}. Andreon and Weaver
(2015)~\ct[Secs.~8.4]{andwea2015} employ logarithmic variables
for modelling a simple linear regression relationship between
galaxies' velocity dispersions on the one hand and the masses of
their central black holes on the other, wherein the variables are
subject to measurement error and object-specific intrinsic scatter.
Sorensen \textit{et al} (2016)~\ct{soretal2016} run
a \textbf{multi-level linear regression} in the context of a
two-condition self-paced reading experiment in Linguistics, where
the outcome variable~$\boldsymbol{y}$ is given by the natural
logarithm of a reading time and is assumed to arise from a
Gau\ss\ data-generating process.

\medskip
\noindent
We now give the vectorised code for the specification in Stan of a
\textbf{correlated varying effects linear regression model}. This
code, \href{https://github.com/hve1964/stanCodes/blob/master/linRegNormVarying.stan}{\texttt{linRegNormVarying.stan}},
is available from
\href{https://github.com/hve1964/stanCodes}{\texttt{github.com/hve1964/stanCodes}}.\footnote{In this and all following examples of
varying effects Stan model codes the model parameters
$\boldsymbol{\beta}[gp]$ are defined as matrix-valued objects that
contain group-specific information in their rows. The group
coefficients $\boldsymbol{\gamma}_{\boldsymbol{\beta}}[gp]$ are
adapted to this structure.}

%
\begin{lstlisting}[language = Stan, caption = {Vectorised Stan
model code for varying effects multiple linear regression. It
employs a non-centred parametrisation based on a Cholesky
decomposition of the covariance matrix and also computes the
posterior predictive probability distribution
re-using available predictor data, as well as the pointwise
log-likelihood. The four occurring
dimensionless fixed hyperparameters have been given the
values $\sigma_{0} = 1$, $\beta_{0} = 1$, $\gamma_{0} = 1$
and $\eta_{0} = 2$ and .},
captionpos = b, label = {lst:linregvarying}]
data {
  /* Dimensions */
  int<lower=1> N;   // number of sampling units
  int<lower=1> M;   // number of predictors plus one
  int<lower=1> K;   // number of groups
  int<lower=1> L;   // number of group predictors
      
  /* Observed variables */
  matrix[N, M] X;                // design matrix: predictors
  matrix[K, L] u;                // matrix of group predictors
  int<lower=1,upper=K> gp[N];    // group indicator
  vector[N] y;                   // outcome
}

parameters {
  /* Unobserved variables */
  matrix[L, M] gamma_beta;    // group coefficients
  matrix[M, K] z_beta;
  real<lower=0> sigma;
 
  /* Cholesky decomposition of covariance matrix */
  vector<lower=0>[M] sigma_beta;
  cholesky_factor_corr[M] L_R_beta;
}

transformed parameters {
  matrix[K, M] beta;

  /* Correlated varying intercepts and slopes */
  beta = u * gamma_beta +
	( diag_pre_multiply( sigma_beta , L_R_beta ) * z_beta )';
}

model {
  /* Fixed log-priors for unobserved variables (regularising) */
  target += normal_lpdf( to_vector(gamma_beta) | 0 , 1 );
  target += normal_lpdf( to_vector(z_beta) | 0 , 1 );
  target += exponential_lpdf( sigma | 1 );

  /* Fixed log-priors for Cholesky decomposition of covariance matrix */
  target += exponential_lpdf( sigma_beta | 1 );
  target += lkj_corr_cholesky_lpdf( L_R_beta | 2 );
    
  /* Gauss log-likelihood w/ identity link */
  target += normal_lpdf( y |
    rows_dot_product( beta[gp] , X ) , sigma );
}

generated quantities {
  vector[N] yrep;
  vector[N] log_lik;
  matrix[M, M] R_beta;

  /* Reconstruction of correlation matrix */
  R_beta = multiply_lower_tri_self_transpose(L_R_beta);

  /* Posterior predictive distribution (re-using predictor data)
     and calculation of pointwise log-likelihood */
  for ( i in 1:N ) {
    yrep[i] = normal_rng(
      rows_dot_product( beta[gp] , X )[i] , sigma );
    
    log_lik[i] = normal_lpdf( y[i] |
      rows_dot_product( beta[gp] , X )[i] , sigma );
  }
}
\end{lstlisting}
%

\section[Multi-level ANOVA-like regression]{Multi-level 
ANOVA-like regression}
\lb{sec:hieranova}
Also for \textbf{multi-level ANOVA-like regression}
we stick to a \textbf{Gau\ss\ single-datum likelihood function}
according to Eq.~(\ref{eq:gausspdf}). A non-central
$t$--single-datum likelihood function according to
Eq.~(\ref{eq:tpdf}) may be used instead, should there be a need to
account for \textbf{outliers} in the data for the dependent
variable~$Y$. In the present case the specific multi-level feature
is injected via abandoning the assumption of homogeneous variances.
We express this position by specifying an \textbf{exponential
distribution} as a population distribution for the group-level
scale parameters~$\boldsymbol{\sigma}[gp]$, thus allowing for
\textbf{heteroscedasticity}. The \textbf{multi-level ANOVA-like 
regression model} is then given by
\bea
\lb{eq:likelihoodfctanovahier}
\text{likelihood:} \qquad
\left.\boldsymbol{y}\right| \boldsymbol{\mu},
\boldsymbol{\sigma}^{2}, [gp], I
&\stackrel{\mathrm{ind}}{\sim} &
\mathrm{N}(\boldsymbol{\mu}[gp],
\boldsymbol{\sigma}^{2}[gp]) \\
\lb{eq:priorsanovahier}
\text{adaptive\ priors:} \qquad
\boldsymbol{\mu}[gp]
& \sim & \mathrm{N}(\mu_{\mu}, \sigma_{\mu}^{2}) \\
\boldsymbol{\sigma}[gp]
& \sim & \mathrm{Exp}(\beta_{\sigma}) \\
\text{fixed priors:} \qquad
\mu_{\mu}
& \sim & \mathrm{N}(\mu_{0}, \sigma_{0}^{2}) \\
\sigma_{\mu}
& \sim & \mathrm{Exp}(\gamma_{0}) \\
\beta_{\sigma}
& \sim & \mathrm{Exp}(\delta_{0}) \ ,
\eea
where $\mu_{0}$, $\sigma_{0}^{2}$, $\gamma_{0}$ and $\delta_{0}$
each denote fixed hyperparameters of the prior probability 
distributions. Examples of applications are presented in Gelman
\textit{et al} (2014)~\ct[Sec.~5.3]{geletal2014} and in
Krusch\-ke (2015)~\ct[Sec.~19.5]{kru2015}.

\medskip
\noindent
The vectorised code for the specification in Stan of a
\textbf{multi-level ANOVA-like regression model} is given next.
This code,
\href{https://github.com/hve1964/stanCodes/blob/master/anovaRegNormVarying.stan}{\texttt{anovaRegNormVarying.stan}}, and corresponding sample
data are available from
\href{https://github.com/hve1964/stanCodes}{\texttt{github.com/hve1964/stanCodes}}.

%
\begin{lstlisting}[language = Stan, caption = {Vectorised Stan
model code for varying effects heteroscedastic ANOVA-like
regression. It computes the
posterior predictive probability distribution
re-using available predictor data, as well as the pointwise
log-likelihood. The four
occurring fixed hyperparameters have been given the
values $\mu_{0} = 90~\text{(units)}$,
$\sigma_{0} = 2~\text{(units)}$,
$\gamma_{0} = 1~\text{(unit)}$ and
$\delta_{0} = 1~\text{(unit)}$.}, captionpos = b,
label = {lst:anovaregvarying}]
data {
  /* Dimensions */
  int<lower=1> N;    // number of sampling units
  int<lower=1> K;    // number of groups
    
  /* Observed variables */
  int<lower=1,upper=K> gp[N];    // group indicator
  vector[N] y;                   // outcome
}

parameters {
  /* Unobserved variables */
  real mu_mu;                  // mean of distribution of group means
  real<lower=0> sigma_mu;      // stdev of distribution of group means
  vector[K] mu;                // group means
  real<lower=0> beta_sigma;    // parameter of distribution of group stdevs
  vector<lower=0>[K] sigma;    // group stdevs (inhomogeneous)
}

model {
  /* Fixed log-priors (weakly informative, regularising) */
  target += normal_lpdf( mu_mu | 90 , 2 );
  target += exponential_lpdf( sigma_mu | 1 );
  target += exponential_lpdf( beta_sigma | 1 );

  /* Adaptive log-priors */
  target += normal_lpdf( mu | mu_mu , sigma_mu );
  target += exponential_lpdf( sigma | beta_sigma );

  /* Gauss log-likelihood */
  target += normal_lpdf( y | mu[gp] , sigma[gp] );
}

generated quantities {
  vector[N] yrep;
  vector[N] log_lik;

  /* Posterior predictive distribution (re-using predictor data)
     and calculation of pointwise log-likelihood */
  for ( i in 1:N ) {
    yrep[i] = normal_rng( mu[gp[i]] , sigma[gp[i]] );
    
    log_lik[i] = normal_lpdf( y[i] | mu[gp[i]] , sigma[gp[i]] );
  }
}
\end{lstlisting}
%

\section[Multi-level logistic regression]{Multi-level logistic 
regression}
\lb{sec:hierlogitreg}
The extension to the correlated varying effects level of the next
three classes of GLMs entails no further complication. The
\textbf{correlated varying effects logistic regression model} has
at its core a \textbf{Bernoulli single-datum  
likelihood function} according to Eq.~(\ref{eq:bernprob}).
We assume that the metrically scaled data available
for~$\boldsymbol{X}$ has been standardised prior to
analysis. Employing a non-centred decomposition of the
group-dependent \textbf{model parameters}~$\boldsymbol{\beta}$
according to Eq.~(\ref{eq:ncmulti3}), the \textbf{correlated
varying effects logistic regression model} is thus described by
\bea
\lb{eq:likelihoodfctlogreghier}
\text{likelihood:} \qquad
\left.\boldsymbol{y}\right| \boldsymbol{X}, \boldsymbol{\beta},
[gp], I 
& \stackrel{\mathrm{ind}}{\sim} &
\mathrm{Bern}(p[gp]) \\
\text{linear model:} \qquad
\ln\left(\frac{p}{1-p}\right)[gp]
& = & \boldsymbol{X}[gp]\,\boldsymbol{\beta}[gp] \\
\lb{eq:priorlogregbetahier}
\text{non-centred decomposition:} \qquad
\boldsymbol{\beta}[gp]
& = & \boldsymbol{u}\boldsymbol{\gamma}_{\boldsymbol{\beta}}[gp]
+ \text{diag}(\boldsymbol{\sigma}_{\boldsymbol{\beta}}[gp])\,
\boldsymbol{L}_{\boldsymbol{R}_{\boldsymbol{\beta}}}\,
\boldsymbol{z}_{\boldsymbol{\beta}}[gp] \\
\text{priors:} \qquad
\left.\boldsymbol{\gamma}_{\boldsymbol{\beta}}[gp]\right| I
& \sim & \mathrm{N}(0, \sigma_{0}^{2}) \\
\left.\boldsymbol{\sigma}_{\boldsymbol{\beta}}[gp]\right| I
& \sim & \mathrm{Exp}(\gamma_{0}) \\
\left.\boldsymbol{L}_{\boldsymbol{R}_{\boldsymbol{\beta}}}\right| I
& \sim & \mathrm{LKJ}(\eta_{0}) \\
\left.\boldsymbol{z}_{\boldsymbol{\beta}}[gp]\right| I
& \sim & \mathrm{N}(0, 1) \ ,
\eea
where $\sigma_{0}^{2}$, $\gamma_{0}$ and $\eta_{0}$ denote fixed 
hyperparameters of the different prior
probability distributions. Illustrative examples are discussed in
Gelman \textit{et al} (2014)~\ct[Sec.~5.3]{geletal2014},
Gill (2015)~\ct[Sec.~12.8]{gil2015}, Krusch\-ke
(2015)~\ct[Sec.~21.4]{kru2015}, and
McElreath (2020)~\ct[Sec.~14.2.]{mce2020a}.

\medskip
\noindent
We now give the vectorised code for the specification in Stan of a
\textbf{correlated varying effects logistic regression model}. This
code, \href{https://github.com/hve1964/stanCodes/blob/master/logistRegBernVarying.stan}{\texttt{logistRegBernVarying.stan}},
and corresponding sample data are available from
\href{https://github.com/hve1964/stanCodes}{\texttt{github.com/hve1964/stanCodes}}.

%
\begin{lstlisting}[language = Stan, caption = {Vectorised Stan
model code for varying effects logistic regression. It
employs a non-centred parametrisation based on a Cholesky
decomposition of the covariance matrix and also computes the
posterior predictive probability distribution
re-using available predictor data, as well as the pointwise
log-likelihood. The three occurring dimensionless fixed
hyperparameter have been given the values $\sigma_{0} = 1$,
$\gamma_{0} = 1$ and $\eta_{0} = 2$.},
captionpos = b, label = {lst:logitregvarying}]
data {
  /* Dimensions */
  int<lower=1> N;    // number of sampling units
  int<lower=1> M;    // number of predictors plus one
  int<lower=1> K;    // number of groups
  int<lower=1> L;    // number of group predictors
      
  /* Observed variables */
  matrix[N, M] X;                // design matrix: predictors
  matrix[K, L] u;                // matrix of group predictors
  int<lower=1,upper=K> gp[N];    // group indicator
  int<lower=0,upper=1> y[N];     // outcome
}

parameters {
  /* Unobserved variables */
  matrix[L, M] gamma_beta;    // group coefficients
  matrix[M, K] z_beta;

  /* Cholesky decomposition of covariance matrix */
  vector<lower=0>[M] sigma_beta;
  cholesky_factor_corr[M] L_R_beta;
}

transformed parameters {
  matrix[K, M] beta;

  /* Correlated varying intercepts and slopes */
  beta = u * gamma_beta +
	( diag_pre_multiply( sigma_beta , L_R_beta ) * z_beta )';
}

model {
  /* Fixed log-priors for unobserved variables (regularising) */
  target += normal_lpdf( to_vector(gamma_beta) | 0 , 1 );
  target += normal_lpdf( to_vector(z_beta) | 0 , 1 );

  /* Fixed log-priors for Cholesky decomposition of covariance matrix */
  target += exponential_lpdf( sigma_beta | 1 );
  target += lkj_corr_cholesky_lpdf( L_R_beta | 2 ); 
    
  /* Bernoulli log-likelihood w/ logit link */
  target += bernoulli_logit_lpmf( y |
    rows_dot_product( beta[gp] , X ) );
}

generated quantities {
  int<lower=0,upper=1> yrep[N];
  vector[N] log_lik;
  matrix[M, M] R_beta;

  /* Reconstruction of correlation matrix */
  R_beta = multiply_lower_tri_self_transpose(L_R_beta);

  /* Posterior predictive distribution (re-using predictor data)
     and calculation of pointwise log-likelihood */
  for ( i in 1:N ) {
    yrep[i] = bernoulli_logit_rng(
      rows_dot_product( beta[gp] , X )[i] );
    
    log_lik[i] = bernoulli_logit_lpmf( y[i] |
      rows_dot_product( beta[gp] , X )[i] );    
  }
}
\end{lstlisting}
%

\section[Multi-level Poisson regression]{Multi-level Poisson 
regression}
\lb{sec:hierpoisreg}
Next, we delineate the structure of a \textbf{correlated varying
effects Poisson regression model}. As before, the
\textbf{single-datum likelihood function} is given by a
\textbf{Poisson distribution} according to Eq.~(\ref{eq:poisprob}). 
We assume that the metrically scaled data available
for~$\boldsymbol{X}$ has been standardised prior to analysis.
Employing a non-centred decomposition of the
group-dependent \textbf{model parameters}~$\boldsymbol{\beta}$
according to Eq.~(\ref{eq:ncmulti3}), the \textbf{correlated
varying effects Poisson regression model} is given by
\bea
\lb{eq:likelihoodfctpoisreghier}
\text{likelihood:} \qquad
\left.\boldsymbol{y}\right|\boldsymbol{X}, \boldsymbol{\beta},
[gp], I 
& \stackrel{\mathrm{ind}}{\sim} & \text{Pois}(\theta[gp]) \\
\text{linear model:} \qquad
\ln(\theta)[gp]
& = & \boldsymbol{X}[gp]\,\boldsymbol{\beta}[gp] \\
\lb{eq:priorpoisregbetahier}
\text{non-centred decomposition:} \qquad
\boldsymbol{\beta}[gp]
& = & \boldsymbol{u}\boldsymbol{\gamma}_{\boldsymbol{\beta}}[gp]
+ \text{diag}(\boldsymbol{\sigma}_{\boldsymbol{\beta}}[gp])\,
\boldsymbol{L}_{\boldsymbol{R}_{\boldsymbol{\beta}}}\,
\boldsymbol{z}_{\boldsymbol{\beta}}[gp] \\
\text{priors:} \qquad
\left.\boldsymbol{\gamma}_{\boldsymbol{\beta}}[gp]\right| I
& \sim & \mathrm{N}(0, \sigma_{0}^{2}) \\
\left.\boldsymbol{\sigma}_{\boldsymbol{\beta}}[gp]\right| I
& \sim & \mathrm{Exp}(\gamma_{0}) \\
\left.\boldsymbol{L}_{\boldsymbol{R}_{\boldsymbol{\beta}}}\right| I
& \sim & \mathrm{LKJ}(\eta_{0}) \\
\left.\boldsymbol{z}_{\boldsymbol{\beta}}[gp]\right| I
& \sim & \mathrm{N}(0, 1) \ ,
\eea
where $\sigma_{0}^{2}$, $\gamma_{0}$ and $\eta_{0}$ denote fixed 
hyperparameters of the different prior
probability distributions. Interesting applications of these models
are outlined in Gelman \textit{et al}
(2014)~\ct[Sec.~16.4]{geletal2014}, Gill 
(2015)~\ct[Sec.~12.5]{gil2015}, Krusch\-ke
(2015)~\ct[Sec.~21.4]{kru2015}, and McElreath
(2020)~\ct[Sec.~14.4.]{mce2020a}. Some advanced examples of
multi-level modelling with Poisson or binomial single-datum
likelihood functions, or mixtures thereof, in an astrophysical
context, can be found in Andreon and Weaver
(2015)~\ct[Secs.~8.5 and 8.12]{andwea2015}.

\medskip
\noindent
The vectorised code for the specification in Stan of a
\textbf{correlated varying effects Poisson regression model}
follows. This code,
\href{https://github.com/hve1964/stanCodes/blob/master/poisRegVarying.stan}{\texttt{poisRegVarying.stan}}, and corresponding sample data are available from
\href{https://github.com/hve1964/stanCodes}{\texttt{github.com/hve1964/stanCodes}}.\footnote{Note that in the generated quantities block
of this Stan code the upper limit of $30\ln(2) \approx 20.79$ for
the dimensionless logarithmic rate parameter in the
\texttt{poisson\_log\_rng()} function has been respected in order
to avoid a crash when compiling this code;
cf. \href{https://mc-stan.org/docs/functions-reference/poisson-distribution-log-parameterization.html}{Stan Functions Reference
(v2.30)} \ct{sta2022b}.}

%
\begin{lstlisting}[language = Stan, caption = {Vectorised Stan
model code for varying effects Poisson regression. It
employs a non-centred parametrisation based on a Cholesky
decomposition of the covariance matrix and also computes the
posterior predictive probability distribution
re-using available predictor data, as well as the pointwise
log-likelihood. The three occurring dimensionless fixed
hyperparameter have been given the values $\sigma_{0} = 1$,
$\gamma_{0} = 1$ and $\eta_{0} = 2$.},
captionpos = b, label = {lst:poisregvarying}]
data {
  /* Dimensions */
  int<lower=1> N;    // number of sampling units
  int<lower=1> M;    // number of predictors plus one
  int<lower=1> K;    // number of groups
  int<lower=1> L;    // number of group predictors
      
  /* Observed variables */
  matrix[N, M] X;                // design matrix: predictors
  matrix[K, L] u;                // matrix of group predictors
  int<lower=1,upper=K> gp[N];    // group indicator
  int<lower=0> y[N];             // outcome
}

parameters {
  /* Unobserved variables */
  matrix[L, M] gamma_beta;    // group coefficients
  matrix[M, K] z_beta;

  /* Cholesky decomposition of covariance matrix */
  vector<lower=0>[M] sigma_beta;
  cholesky_factor_corr[M] L_R_beta;
}

transformed parameters {
  matrix[K, M] beta;

  /* Correlated varying intercepts and slopes */
  beta = u * gamma_beta +
	( diag_pre_multiply( sigma_beta , L_R_beta ) * z_beta )';
}

model {
  /* Fixed log-priors for unobserved variables (regularising) */
  target += normal_lpdf( to_vector(gamma_beta) | 0 , 1 );
  target += normal_lpdf( to_vector(z_beta) | 0 , 1 );

  /* Fixed log-priors for Cholesky decomposition of covariance matrix */
  target += exponential_lpdf( sigma_beta | 1 );
  target += lkj_corr_cholesky_lpdf( L_R_beta | 2 );
    
  /* Poisson log-likelihood w/ log link */
  target += poisson_log_lpmf( y |
    rows_dot_product( beta[gp] , X ) );
}

generated quantities {
  int<lower=0> yrep[N];
  vector[N] log_lik;
  matrix[M, M] R_beta;

  /* Reconstruction of correlation matrix */
  R_beta = multiply_lower_tri_self_transpose(L_R_beta);

  /* Posterior predictive distribution (re-using predictor data)
     and calculation of pointwise log-likelihood */
  for ( i in 1:N ) {
    if ( rows_dot_product( beta[gp] , X )[i] > 20 ) {
      yrep[i] = poisson_log_rng( 20 );
    } else {
        yrep[i] = poisson_log_rng(
          rows_dot_product( beta[gp] , X )[i] );
    }
    
    log_lik[i] = poisson_log_lpmf( y[i] |
      rows_dot_product( beta[gp] , X )[i] );    
  }
}
\end{lstlisting}
%

\section[Multi-level exponential regression]{Multi-level
exponential regression}
\lb{sec:hierexpreg}
As a last example, we turn to specify the structure of a
\textbf{correlated varying effects exponential regression model}.
Like in the fixed effects case of Sec.~\ref{sec:expreg}, the
\textbf{single-datum likelihood function} is given by an
\textbf{exponential distribution} according to
Eq.~(\ref{eq:exppdf}). We assume that the metrically scaled data
available for~$\boldsymbol{X}$ has been standardised prior to
analysis. Employing a non-centred decomposition of the
group-dependent \textbf{model parameters}~$\boldsymbol{\beta}$
according to Eq.~(\ref{eq:ncmulti3}), the \textbf{correlated
varying effects exponential regression model} is given by
\bea
\lb{eq:likelihoodfctpoisreghier}
\text{likelihood:} \qquad
\left.\boldsymbol{y}\right|\boldsymbol{X}, \boldsymbol{\beta},
[gp], I 
& \stackrel{\mathrm{ind}}{\sim} & \text{Exp}(\theta[gp]) \\
\text{linear model:} \qquad
\ln(\theta/\theta_{0})[gp]
& = & \boldsymbol{X}[gp]\,\boldsymbol{\beta}[gp] \\
\lb{eq:priorpoisregbetahier}
\text{non-centred decomposition:} \qquad
\boldsymbol{\beta}[gp]
& = & \boldsymbol{u}\boldsymbol{\gamma}_{\boldsymbol{\beta}}[gp]
+ \text{diag}(\boldsymbol{\sigma}_{\boldsymbol{\beta}}[gp])\,
\boldsymbol{L}_{\boldsymbol{R}_{\boldsymbol{\beta}}}\,
\boldsymbol{z}_{\boldsymbol{\beta}}[gp] \\
\text{priors:} \qquad
\left.\boldsymbol{\gamma}_{\boldsymbol{\beta}}[gp]\right| I
& \sim & \mathrm{N}(0, \sigma_{0}^{2}) \\
\left.\boldsymbol{\sigma}_{\boldsymbol{\beta}}[gp]\right| I
& \sim & \mathrm{Exp}(\gamma_{0}) \\
\left.\boldsymbol{L}_{\boldsymbol{R}_{\boldsymbol{\beta}}}\right| I
& \sim & \mathrm{LKJ}(\eta_{0}) \\
\left.\boldsymbol{z}_{\boldsymbol{\beta}}[gp]\right| I
& \sim & \mathrm{N}(0, 1) \ ,
\eea
where $\theta_{0}$ is the unit of the dimensionful \textbf{model
parameter}~$\theta$ and $\sigma_{0}^{2}$, $\gamma_{0}$ and
$\eta_{0}$ denote fixed hyperparameters of the different prior
probability distributions.

\medskip
\noindent
Finally, we give the vectorised code for the specification in Stan
of a \textbf{correlated varying effects exponential regression
model}. This code,
\href{https://github.com/hve1964/stanCodes/blob/master/expRegVarying.stan}{\texttt{expRegVarying.stan}}, and corresponding 
sample data are available from
\href{https://github.com/hve1964/stanCodes}{\texttt{github.com/hve1964/stanCodes}}.

%
\begin{lstlisting}[language = Stan, caption = {Vectorised Stan
model code for varying effects exponential regression. It
employs a non-centred parametrisation based on a Cholesky
decomposition of the covariance matrix and also computes the
posterior predictive probability distribution
re-using available predictor data, as well as the pointwise
log-likelihood. The three occurring dimensionless fixed
hyperparameter have been given the values $\sigma_{0} = 1$,
$\gamma_{0} = 1$ and $\eta_{0} = 2$.},
captionpos = b, label = {lst:expregvarying}]
data {
  /* Dimensions */
  int<lower=1> N;    // number of sampling units
  int<lower=1> M;    // number of predictors plus one
  int<lower=1> K;    // number of groups
  int<lower=1> L;    // number of group predictors
      
  /* Observed variables */
  matrix[N, M] X;                // design matrix: predictors
  matrix[K, L] u;                // matrix of group predictors
  int<lower=1,upper=K> gp[N];    // group indicator
  real<lower=0> y[N];            // outcome
}

parameters {
  /* Unobserved variables */
  matrix[L, M] gamma_beta;    // group coefficients
  matrix[M, K] z_beta;

  /* Cholesky decomposition of covariance matrix */
  vector<lower=0>[M] sigma_beta;
  cholesky_factor_corr[M] L_R_beta;
}

transformed parameters {
  matrix[K, M] beta;

  /* Correlated varying intercepts and slopes */
  beta = u * gamma_beta +
	( diag_pre_multiply( sigma_beta , L_R_beta ) * z_beta )';
}

model {
  /* Fixed log-priors for unobserved variables (regularising) */
  target += normal_lpdf( to_vector(gamma_beta) | 0 , 1 );
  target += normal_lpdf( to_vector(z_beta) | 0 , 1 );

  /* Fixed log-priors for Cholesky decomposition of covariance matrix */
  target += exponential_lpdf( sigma_beta | 1 );
  target += lkj_corr_cholesky_lpdf( L_R_beta | 2 );  

  /* Exponential log-likelihood w/ exponential inverse link */
  target += exponential_lpdf( y |
    exp( rows_dot_product( beta[gp] , X ) ) );
}

generated quantities {
  real<lower=0> yrep[N];
  vector[N] log_lik;
  matrix[M, M] R_beta;

  /* Reconstruction of correlation matrix */
  R_beta = multiply_lower_tri_self_transpose(L_R_beta);

  /* Posterior predictive distribution (re-using predictor data)
     and calculation of pointwise log-likelihood */
  for ( i in 1:N ) {
    yrep[i] = exponential_rng(
      exp( rows_dot_product( beta[gp] , X )[i] ) );
    
    log_lik[i] = exponential_lpdf( y[i] |
      exp( rows_dot_product( beta[gp] , X )[i] ) );
  }
}
\end{lstlisting}

\medskip
\noindent
This concludes the present chapter. Regarding further issues
concerning varying effects generalised linear models, of
direct practical interest prove the systematic 
handling of \textbf{multi-collinearity} and of \textbf{measurement
error} within the quantitative--empirical data for the independent
variables contained in~$\boldsymbol{X}$.

\medskip
\noindent
In the next chapter, we turn to address at an elementary level
the statistical concepts underlying the description of the
generation of stationary time series data by means of
autoregressive models.

\chapter[Fixed effects time series models]{Fixed effects linear 
models for stationary time series data}
\lb{ch10}
In this chapter we will give a brief introduction, in the context
of the \textbf{Bayes–Laplace approach to data analysis and
statistical inference}, to the modelling of an
\textit{unknown} data-generating processes that yields
\textit{stationary} \textbf{time series data}. Heuristically,
\textbf{stationarity} here refers either to stationarity of the
second order (or covariance stationarity), in which case
\textit{both} the \textbf{sample mean} and the
\textbf{sample auto-covariance} of the time series data prove
invariant under time translations, or to stationarity of the first
order, in which case only the \textbf{sample mean} exhibits such
an invariance property. The stationarity property of given time
series data maybe investigated with unit root test routines
contained in the R package \texttt{urca} by
Pfaff (2008)~\ct{pfa2008}.

\medskip
\noindent
A guiding principle for statistically modelling the
data-generating process underlying the time series data for a
single metrically scaled \textbf{statistical variable}~$Y$ is the
assumption that a specific value~$y_{i}$ of~$Y$ depends
statistically at least on its previous value~$y_{i-1}$, if not
also on values earlier than~$y_{i-1}$. This assumption implies in
the very least that either the \textbf{expectation
value}~$\mathrm{E}(y_{i})$ of the data-generating process for~$Y$
or its \textbf{variance}~$\mathrm{Var}(y_{i})$ can vary as time
progresses. Statistical time series models that are built on this
assumption are referred to as \textbf{autoregressive}. 

\medskip
\noindent
The theory of \textbf{time series analysis} is discussed at length
in the monograph by Hamilton (1994)~\ct{ham1994}, while the
textbook by Cowpertwait and Metcalfe (2009)~\ct{cowmet2009}
provides many examples of \textbf{time series analysis} performed
within an \R{} environment. In a spirit similar to the latter work,
the website by Hyndman and Athanasopoulos at the URL
(cited on August 27, 2022)
\href{https://otexts.com/fpp3/}{https://otexts.com/fpp3/} focusses
on the construction of time series models for the purpose of
forecasting within the frequentist approach to data analysis and
statistical inference.

\medskip
\noindent
The two kinds of \textbf{autoregressive models} for stationary
\textbf{time series data} we will highlight in the next two
sections will be restricted to the consideration of
\textbf{fixed effects} only.

\section[Stationary linear $\text{AR}(p)$--model]{Stationary linear
$\text{AR}(p)$--model with Gau\ss\ likelihood}
\lb{sec:ARp}
The linear \textbf{$\text{AR}(p)$--model} addresses a
data-generating process for a single metrically scaled
\textbf{statistical variable}~$Y$ that displays a finite but
variable \textbf{sample mean} and a stable (homogeneous)
\textbf{sample variance}. For such an \textbf{autoregressive
process of order~$p$}, the value~$y_{i}$ of~$Y$ depends
statistically on all its previous values up to~$y_{i-p}$;
$p$ is referred to as a \textbf{lag parameter}.
Typically, a \textbf{Gau\ss\ single-datum likelihood
function} according to Eq.~(\ref{eq:gausspdf}) is employed in an
\textbf{$\text{AR}(p)$--model}, in which the \textbf{expectation
value} for~$Y$ is expressed as a linear combination of previous
values of~$Y$ up to order $p$, with \textbf{model
parameters}~$\alpha$, $\beta_{i}$ and $\sigma^{2}$; see, e.g.,
Hamilton (1994)~\ct[Sec.~5.3]{ham1994},
Cowpertwait and Metcalfe (2009)~\ct[Sec.~4.5]{cowmet2009},
Greenberg~\ct[Sec.~11.1]{gre2013} or
Rinne (2008)~\ct[Subsec.~5.1.1.3]{rin2008}. A necessary (though
not  sufficient) condition for stationarity to hold for the
\textbf{$\text{AR}(p)$--model} is given by the constraint
$-1 < \beta_{i} < 1$. The linear \textbf{$\text{AR}(p)$--model}
can thus be expressed by
\bea
\lb{eq:likelihoodfctARp}
\text{likelihood:} \qquad
\left.y_{n}\right|y_{n-i}, \alpha,
\beta_{i}, \sigma^{2}, I
& \stackrel{\mathrm{ind}}{\sim} &
\mathrm{N}\left(\mu_{n}, \sigma^{2}\right) \\
\sigma^{2}
& := & \mathrm{Var}\left(\left.y_{n}\right| I\right)
\ = \ \text{constant} \nonumber \\
\lb{eq:linkARp}
\text{linear model:} \qquad
\mu_{n}
& := & \mathrm{E}\left(\left.y_{n}\right|y_{n-i}, \alpha,
\beta_{i}, \sigma^{2}, I\right) \nonumber \\
& = & \alpha + \sum_{i=1}^{p}\beta_{i}y_{n-i} \\
\lb{eq:priorARpalpha}
\text{priors:} \qquad
\left.\alpha, \beta_{i}\right| \sigma^{2}, I
& \sim & \mathrm{N}(0, \sigma_{0}^{2}) \\
\lb{eq:priorARpsigma}
\left.\sigma\right| I
& \sim & \mathrm{Exp}(\beta_{0}) \ ,
\eea
where $\sigma_{0}^{2}$ and $\beta_{0}$ denote fixed hyperparameters
of the prior probability distributions for $\alpha$, $\beta_{i}$
and $\sigma$. Note that in Eq.~(\ref{eq:priorARpalpha})
zero-centred Gau\ss\ prior probability distributions were specified
for the unknown model parameters~$\alpha$ and $\beta_{i}$. This
choice is to represent scepticism as to the presence of any of
these model parameters in a best-fit model.

\medskip
\noindent
Next, the code for the specification in Stan of a linear
\textbf{\textbf{$\text{AR}(p)$--model}} is given, employing a
homogeneous variance and fixed prior probability distributions
for the \textbf{model parameters}. The Stan code
\href{https://github.com/hve1964/stanCodes/blob/master/AR\_p\_NormFixed.stan}{\texttt{AR\_p\_NormFixed.stan}} and corresponding sample data
are available from
\href{https://github.com/hve1964/stanCodes}{\texttt{github.com/hve1964/stanCodes}}; see also Ali (2017)~\ct{ali2017} and the
\href{https://mc-stan.org/docs/stan-users-guide/autoregressive.html}{Stan 
User's Guide (v2.30)}~\ct[Sec.~2.1]{sta2022c}.
%
\begin{lstlisting}[language = Stan, caption = {Vectorised Stan
model code for stationary linear autoregressive models
of order~$p$ with a Gau\ss\ likelihood and fixed effects . This
code also computes the posterior predictive probability
distribution. The two occurring fixed hyperparameters have been
given the values $\sigma_{0} = 1~\text{unit}$ and
$\beta_{0} = 1~\text{unit}$.},
captionpos = b, label = {lst:ARpfixed}]
data {
  /* Dimensions */
  int<lower=0> T;         // length of time series
  int<lower=1,upper=T> P; // number of lags

  /* Observed variables */
  vector[T] y;            // time series data
}

transformed data {
  /* Transform data to accommodate P-lag process */
  vector[T-P] y_trans;    // outcome of time series
  matrix[T-P, P] Ymat;    // matrix of (lagged) predictors
  
  for ( i in 1:(T-P) ) {
    y_trans[i] = y[i+P];
    
    for ( p in 1:P ) {
      Ymat[i, p] = y[(P + i) - p];
    }
  }
}

parameters {
  /* Unobserved variables */
  real alpha;                         // intercept
  vector<lower=-1,upper=1>[P] beta;   // slopes w/ stationarity constraints
  real<lower=0> sigma;                // sd of error
}

model {
  /* Fixed log-priors (regularising) */
  target += normal_lpdf( alpha | 0 , 1 );
  target += normal_lpdf( beta | 0 , 1 );
  target += exponential_lpdf( sigma | 1 );

  /* Gauss log-likelihood */
  target += normal_lpdf( y_trans | alpha + Ymat * beta , sigma );
}

generated quantities {
  /* Posterior predictive distribution */
  vector[T-P] yrep;
	
  for ( i in 1:(T-P) ) {
    yrep[i] = normal_rng( alpha + Ymat[i, ] * beta , sigma );
  }
}
\end{lstlisting}
%

\section[Stationary linear $\text{GARCH}(1,1)$--model]{Stationary
linear $\text{GARCH}(1,1)$--model with Gau\ss\ likelihood}
\lb{sec:GARCH11}
The second kind of \textbf{autoregressive model} relates to a
data-generating process for a single metrically scaled
\textbf{statistical variable}~$Y$ that shows a stable
\textbf{sample mean} but a time-varying finite \textbf{sample
variance}. The associated phenomenon of \textbf{volatility
clustering} is regularly observed in economic and financial time
series data in particular. It amounts to a clear manifestation of
\textbf{heteroscedasticity}. Bollerslev (1986)~\ct{bol1986}
attributed such behaviour to \textbf{generalised autoregressive
conditional heteroscedasticity (GARCH)} and devised
a compelling method to capture it within a linear
\textbf{$\text{GARCH}(p,q)$--model}, with order parameters~$p$
and $q$. We will confine our considerations to the special case
of the \textbf{$\text{GARCH}(1,1)$--model}, which proves to be
the most relevant one for practical purposes.

\medskip
\noindent
Again, in general a \textbf{Gau\ss\ single-datum likelihood
function} according to Eq.~(\ref{eq:gausspdf}) is employed in a
\textbf{$\text{GARCH}(p,q)$--}
resp.~\textbf{$\text{GARCH}(1,1)$--model},
in which now the
\textbf{variance} for~$Y$ is expressed as a linear combination of
its previous value and of the squared deviation of the previous
value of~$Y$ from the \textbf{expectation value} for~$Y$, with
the four \textbf{model parameters}~$\mu$, $\alpha_{0} > 0$,
$\alpha_{1} > 0$ and $\beta_{1} > 0$; see, e.g.,
Hamilton (1994)~\ct[Sec.~21.2]{ham1994},
Cowpertwait and Metcalfe (2009)~\ct[Sec.~7.4.3]{cowmet2009} or
Greenberg~\ct[Sec.~11.5.1]{gre2013}. To ensure stationarity for
the linear \textbf{$\text{GARCH}(1,1)$--model}, the constraint
$\alpha_{1} + \beta_{1} < 1$ is imposed. The linear
\textbf{$\text{GARCH}(1,1)$--model} can now be expressed by
\bea
\lb{eq:likelihoodfctGARCH11}
\text{likelihood:} \qquad
\left.y_{n}\right|y_{n-1}, \mu, \alpha_{0},
\alpha_{1}, \beta_{1}, \sigma_{n-1}^{2}, I
& \stackrel{\mathrm{ind}}{\sim} &
\mathrm{N}\left(\mu, \sigma_{n}^{2}\right) \\
\mu & := & \mathrm{E}\left(\left.y_{n}\right|I\right)
= \text{constant} \nonumber \\
\lb{eq:linkGARCH11}
\text{linear model:} \qquad
\sigma_{n}^{2}
& := & \mathrm{Var}\left(\left.y_{n}\right|y_{n-1}, \mu,
\alpha_{0}, \alpha_{1}, \beta_{1}, \sigma_{n-1}^{2}, I\right)
\nonumber \\
& = & \alpha_{0} + \alpha_{1}\left(y_{n-1}-\mu\right)^{2}
+ \beta_{1}\sigma_{n-1}^{2} \\
\lb{eq:priormuGARCH11}
\text{priors:} \qquad
\left.\mu\right| I
& \sim & \mathrm{N}(0, \sigma_{0}^{2}) \\
\lb{eq:prioralpha0GARCH11}
\left.\alpha_{0}\right| \mu, I
& \sim & \mathrm{Exp}(\beta_{0}) \\
\lb{eq:prioralphabetaGARCH11}
\left.\alpha_{1}, \beta_{1}\right| \mu, I
& \sim & \mathrm{Be}(\gamma_{0}, \delta_{0}) \
\eea
where $\sigma_{0}^{2}$, $\beta_{0}$, $\gamma_{0}$ and $\delta_{0}$
denote fixed hyperparameters of the prior probability distributions
for $\mu$, $\alpha_{0}$, $\alpha_{1}$ and $\beta_{1}$.

\medskip
\noindent
The Stan code
\href{https://github.com/hve1964/stanCodes/blob/master/GARCH\_1\_1_NormFixed.stan}{\texttt{GARCH\_1\_1\_NormFixed.stan}} for the
specification of a linear \textbf{$\text{GARCH}(1,1)$--model},
employing fixed prior probability distributions for the
\textbf{model parameters}, as well as corresponding sample data
are available from
\href{https://github.com/hve1964/stanCodes}{\texttt{github.com/hve1964/stanCodes}}; see also the
\href{https://mc-stan.org/docs/stan-users-guide/modeling-temporal-heteroscedasticity.html}{Stan 
User's Guide (v2.30)}~\ct[Sec.~2.2]{sta2022c}.
%
\begin{lstlisting}[language = Stan, caption = {Vectorised Stan
model code for stationary linear $\text{GARCH}(1,1)$--models
with a Gau\ss\ likelihood and fixed effects . This code also
computes the posterior predictive probability distribution. The
four occurring fixed hyperparameters have been given the values
$\sigma_{0} = 1.5~\text{unit}$, $\beta_{0} = 1~\text{unit}$
and $\beta_{0} = \gamma_{0} = 1$.},
captionpos = b, label = {lst:GARCH11fixed}]
data {
  /* Dimensions */
  int<lower=0> T;         // length of time series

  /* Observed variables */
  real<lower=0> sigma1;   // initial value for linear variance model
  vector[T] y;            // time series data
}

parameters {
  /* Unobserved variables */
  real mu;                              // mean of likelihood
  real<lower=0> alpha0;                 // intercept
  real<lower=0,upper=1> alpha1;         // slopes w/ stationarity constraints
  real<lower=0,upper=(1-alpha1)> beta1;
}

transformed parameters {
  real<lower=0> sigma[T];
  sigma[1] = sigma1;
  
  /* Model for time-dependent scale parameter */
  for ( t in 2:T ) {
    sigma[t] = sqrt( alpha0 + alpha1 * (y[t-1] - mu) ^ 2
                     + beta1 * (sigma[t-1]) ^ 2 );
  }
}

model {
  /* Fixed log-priors (regularising) */
  target += normal_lpdf( mu | 0 , 1.5 );    // prior for mean of likelihood
  target += exponential_lpdf( alpha0 | 1 ); // prior for positive intercept
  target += beta_lpdf( alpha1 | 1 , 1 );    // prior for positive slope1
  target += beta_lpdf( beta1 | 1 , 1 );     // prior for positive slope2

  /* Gauss log-likelihood */
  target += normal_lpdf( y | mu , sigma );
}

generated quantities {
  /* Posterior predictive distribution */
  vector[T] yrep;
	
  for ( i in 1:T ) {
    yrep[i] = normal_rng( mu , sigma[i] );
  }
}
\end{lstlisting}

\medskip
\noindent
This brings our discussion of examples in \textbf{statistical
modelling} in the context of the \textbf{Bayes–Laplace approach
to data analysis and statistical inference} to an end. We have
\textit{not} discussed in these lecture notes matters of
such relevant topics as \textbf{Gau\ss ian processes}
as a generalisation of correlated varying effects models to
continuously varying ``groups,'' \textbf{survival analysis},
\textbf{multivariate outcome variables}, or \textbf{statistical
models beyond GLMs}. For many of these topics the
\href{https://mc-stan.org/docs/stan-users-guide/index.html}{Stan
User’s Guide (v2.30)}~\ct{sta2022c} provides related
Stan model codes as well as valuable practical assistance.

\medskip
\noindent
In the final chapter, we want to sketch elementary principles of an
important field of application of probability theory, viz.
the theory of decision under conditions of uncertainty,
which forms a conceptual cornerstone of the frameworks of
Economics, Political Science and the Organisational Sciences.

\chapter[Decision-making in the state space
picture]{Decision-making in the state space picture}
\lb{ch11}
\textbf{Decision-making} is a recurrent activity everyone is
confronted with virtually on a daily basis. Economic theory in
particular has long had a vested interest in a systematic
formalisation of the principles underlying basic decision
processes. A set-theoretical state space framework of descriptive
character, developed in the middle of the 20$^{\mathrm{th}}$
Century, has laid the foundation for a \textbf{theory of decision}
under conditions of \textbf{uncertainty}. It continues to be
upgraded by integrating insights gained from experimentation, and
by adapting to pertinent new conceptual ideas. In this chapter, we
want to review the standard model of \textbf{decision theory} for
the case of static one-shot \textbf{choice problems} for a single
decision-maker in the behavioural \textbf{subjective expected
utility (SEU)} representation due to Savage (1954)~\ct{sav1954} and
Anscombe and Aumann (1963)~\ct{ansaum1963}, and outline its link to
the \textbf{Bayes--Laplace approach} to \textbf{inductive
statistical inference}. We will also briefly relate to a specific
area of ongoing research. Full discussions of the principles of
\textbf{decision theory} are given in the textbooks by Gilboa
(2009)~\ct{gil2009} and Peterson (2017)~\ct{pet2017} at an
introductory level, and in the monographs by Wald
(1950)~\ct{wal1950}, Savage (1954)~\ct{sav1954} and Berger
(1985)~\ct{ber1985} at a highly advanced technical level. Some
pedagogical examples are provided in Gelman (1998)~\ct{gel1998}.

\medskip
\noindent
The simplest decision-theoretical models are built on the premiss
of the \textbf{rational-agent paradigm} of Economics. Amongst other
items, this entails the assumption of the existence of some form of
\textbf{reasoning ability} on the part of the decision-maker, so
that she/he can give justifications for the \textbf{choices} they
made. The general set-up is as follows. A \textbf{rational
decision-maker} faces a specific \textbf{choice problem}. She/he
finds herself/himself in a certain individual \textbf{prior state
of knowledge} on the matter to be decided. In particular, she/he
takes into consideration which external \textbf{states of Nature}
(or boundary conditions) could potentially take an influence on the
\textbf{consequences} of the specific \textbf{act} the
decision-maker eventually opts for, and what \textbf{outcomes} the
decision could possibly lead to. At the end of the decision
process, all \textbf{uncertainty} as to the actually realised
momentary state of Nature and the consequences of the act
preferred by the decision-maker will be resolved. Given this
\textit{new} empirical information, a basis for \textbf{learning}
has opened on which the decision-maker attains a \textbf{posterior
state of knowledge}. It is a central objective of \textbf{decision
theory} to cast the scenario just depicted into formal language.
This aims at capturing within an axiomatic framework a rational
decision-maker's \textbf{state of knowledge} concerning
decision-relevant external states of Nature, the decision-maker’s
choice behaviour under conditions of uncertainty, and resultant
prospects for herself/himself. We will now turn to describe the
main elements of this formal language, and the
choice-specific operations defined therein.

\section[Primitives]{Primitives}
\lb{sec:prims}
The description of static one-shot \textbf{choice problems} for a
single decision-maker in the set-theoretical state space
formulation of Anscombe and Aumann (1963)~\ct{ansaum1963} takes the
following set of \textbf{primitives} as building blocks. There
exist:
\begin{itemize}

\item a finite set of $n \in \mathbb{N}$ mutually exclusive and
exhaustive consequence-relevant \textbf{external states of Nature}
$\boldsymbol{\Omega}$ that are \textit{unobservable}; different
kinds of decision-relevant \textbf{events} can be represented by
arbitrary subsets $A \subseteq \boldsymbol{\Omega}$,

\item a finite set of \textbf{outcomes}~$\boldsymbol{X}$ that are
\textit{observable},

\item \textbf{consequences} given in the form of a set of
``lotteries'' (viz., discrete probability distributions),
$\Delta(\boldsymbol{X}|I)$, over the set of
outcomes~$\boldsymbol{X}$; on this set of ``lotteries'' there is
defined a mixing operation such that for every two distributions
$p, q \in \Delta(\boldsymbol{X}|I)$, every weight factor $\alpha
\in (0,1)$, and every outcome $x \in \boldsymbol{X}$ it holds that
$[\alpha p + (1-\alpha)q](x) = \alpha p(x) + (1-\alpha)q(x)$,

\item the decision-maker’s objects of choice are elements from a
finite set of $k \in \mathbb{N}$ alternative \textbf{acts}
\be
\boldsymbol{F} :=\{f|f: \boldsymbol{\Omega} \rightarrow
\Delta(\boldsymbol{X}|I)\} \ ;
\ee
acts are formally understood as maps of consequence-relevant states
of Nature in~$\boldsymbol{\Omega}$ into the space of ``lotteries''
over outcomes, $\Delta(\boldsymbol{X}|I)$,

\item an ordinal binary \textbf{preference relation}~$\succeq$ on
$\boldsymbol{F}$ and, by extension, on $\Delta(\boldsymbol{X}|I)$,
that is \textit{observable}.\footnote{The ordinal binary preference
relation~$\succeq$ is to be read as ``preferred at least as.''}

\end{itemize}
In some formulations of \textbf{choice problems}, the space of
consequence-relevant external states of Nature is given a
logic-based fine-structure; see, e.g., Gilboa (2009)~\ct{gil2009}:
\begin{itemize}

\item \textbf{canonical states of Nature} arise as truth
assignments $\{0:~\text{false}, 1:~\text{true}\}$ to a set of $l
\in \mathbb{N}$ elementary propositions,

\item the size of the resultant \textbf{canonical state space} is
given by $\text{card}(\boldsymbol{\Omega}) = 2^{l}$,

\item in this picture, the number of distinguishable
\textbf{events} amounts to $\displaystyle {2^{2}}^{l}$; this number
can easily grow very large as the number of elementary propositions
taken into account increases.

\end{itemize}
%

\section[Decision matrix]{Decision matrix}
\lb{sec:decmat}
The primitives of static one-shot \textbf{choice problems} for a
single decision-maker may be visualised by means of a
\textbf{decision matrix}, a formal concept effectively anticipated
by the French mathematician, physicist, inventor, writer and
Catholic philosopher
\href{https://mathshistory.st-andrews.ac.uk/Biographies/Pascal/}{Blaise
Pascal (1623–1662)} in his famous reasoning that has come to be
known as \textbf{Pascal's wager}; see, e.g., Gilboa
(2009)~\ct[Sec.~5.2]{gil2009}. Figure~\ref{fig:decmat} outlines the
structure of the \textbf{decision matrix} in the behavioural
\textbf{subjective expected utility} representation of Savage
(1954)~\ct{sav1954} and
Anscombe and Aumann (1963)~\ct{ansaum1963}. This representation
makes the rather unrealistic assumptions that the decision-maker
(and the decision-theoretical modeller) has \textit{complete}
knowledge of (i)~the entire set $\boldsymbol{\Omega}$ of
consequence-relevant states of Nature, to which she/he assigns a
personal discrete \textbf{prior probability distribution}~$P \in
\boldsymbol{\Delta}(\boldsymbol{\Omega}|I)$ on the basis of
available background information~$I$ --- this represents her/his
subjective \textbf{degrees-of-belief} of the plausibility of the
different states, and of (ii)~all possible outcomes $x \in
\boldsymbol{X}$ that are contingent on these states.
\begin{figure}[!htb]
\begin{center}
\[
\begin{array}{c|cccc|c}
\text{prior distribution}
\ P \in \boldsymbol{\Delta}(\boldsymbol{\Omega}|I) &
P(\omega_{1}|I) & P(\omega_{2}|I) & \ldots & P(\omega_{n}|I) &
\sum_{i=1}^{n}P(\omega_{i}|I) = 1 \\
\hline
\text{acts}\ f_{j} \in \boldsymbol{F}\ \backslash\ \text{states}
\ \omega_{i} \in \boldsymbol{\Omega} & \omega_{1} & 
\omega_{2} & \ldots & \omega_{n} & k, n \in \mathbb{N} \\
\hline
f_{1} & p_{11} & p_{12} & \ldots & p_{1n} & \text{consequences:}
\\
f_{2} & p_{21} & p_{22} & \ldots & p_{2n} & \text{``lotteries''} \\
\vdots & \vdots & \vdots & \ddots & \vdots &
p_{ij} \in \boldsymbol{\Delta}(\boldsymbol{X}|I) \\
f_{k} & p_{k1} & p_{k2} & \ldots & p_{kn} &
\text{over\ outcomes}~x \in \boldsymbol{X} \\
\hline
f: \boldsymbol{\Omega} \rightarrow \Delta(\boldsymbol{X}|I)
& & & & & \sum_{x \in \boldsymbol{X}}p_{ij} = 1
\end{array}
\]
\end{center}
\caption{Static one-shot decision matrix of subjective
expected utility theory for a single decision-maker.}
\lb{fig:decmat}
\end{figure}

\medskip
\noindent
By \textbf{learning} from observation of the consequences of
decisions made, and of the actual realisations of specific states
of Nature, the decision-maker forms a personal discrete
\textbf{posterior probability distribution} for states
of Nature that can be an informative starting point for subsequent
decision problems.

\section[Axiomatisation]{Axiomatisation}
\lb{sec:axioms}
The next step in the formal construction of the behavioural
\textbf{subjective expected utility} representation according to
Savage (1954)~\ct{sav1954} and Anscombe and Aumann
(1963)~\ct{ansaum1963}, with earlier contributions by von
Neumann\footnote{This is the same von Neumann we already
encountered in Sec.~\ref{sec:MCsampling} in the context of
Monte Carlo simulations.} and Morgenstern (1944)~\ct{neumor1944},
is the axiomatisation of a \textbf{rational decision-maker}'s
\textbf{choice behaviour}. This leads to the (cf. Gilboa
(2009)~\ct[p~143]{gil2009})
\begin{quote}
\textbf{Representation theorem:} There exist a unique discrete
\textbf{prior probability distribution}
$P \in \boldsymbol{\Delta}(\boldsymbol{\Omega}|I)$
(synonymous with a decision-maker’s ``beliefs'') and an
interval-scaled von Neumann--Morgenstern \textbf{utility function}
 $U: \boldsymbol{X} \rightarrow \mathbb{R}$ for \textbf{outcomes}
in $\boldsymbol{X}$ (the ``moral value'' of outcomes according to
Bernoulli (1738)~\ct{ber1738}),\footnote{In place of a utility
function, many authors, in a decision-theoretical context, employ
an equivalent loss function instead; see, e.g., Jaynes
(2003)~\ct[Sec.~14.3]{jay2003}, Lee (2012)~\ct[Sec.~7.5]{lee2012},
or Gill (2015)~\ct[Sec.~8.1]{gil2015}. Kahneman and Tversky
operate with a psychological value function; see Kahneman and
Tversky (1979)~\ct{kahtve1979} and Kahneman
(2011)~\ct[p~282]{kah2011}.} provided that the ordinal binary
\textbf{preference relation} $\succeq$ on the set of alternative
\textbf{acts} $\boldsymbol{F}$ satisfies a minimal set of five
\textbf{axioms of rational choice}:
\begin{enumerate}

\item \textbf{weak order}: the ordinal binary preference relation
$\succeq$ on the set of alternative acts $\boldsymbol{F}$ is
complete and transitive,

\item \textbf{continuity}: for every three acts $f, g, h \in
\boldsymbol{F}$, if the strong preference order $f \succ g \succ h$
applies, there exist weight factors $\alpha, \beta \in (0,1)$ such
that the strong preference order $\alpha f + (1-\alpha)h \succ g
\succ \beta f + (1-\beta)h$ follows,

\item \textbf{independence}: for every three acts $f, g, h \in
\boldsymbol{F}$ and weight factor $\alpha \in (0,1)$, the weak
preference order $f \succeq g$ obtains iff the weak
preference order $\alpha f + (1-\alpha)h \succeq \alpha g
+ (1-\alpha)h$ obtains,

\item \textbf{monotonicity}: for every two acts $f, g \in
\boldsymbol{F}$, the weak preference order $f(\omega) \succeq 
g(\omega)$ for all states $\omega \in \boldsymbol{\Omega}$ implies
the general weak preference order $f \succeq  g$,

\item \textbf{non-triviality}: there exist at least two acts $f, g
\in \boldsymbol{F}$ such that the strong preference order $f \succ 
g$ is true.

\end{enumerate}
\end{quote}
%

\section[Subjective expected utility model]{Subjective expected
utility model}
\lb{sec:seu}
Lastly, the \textbf{subjective expected utility model} for
describing a \textbf{rational decision-maker}'s \textbf{choice
behaviour} in the context of static one-shot \textbf{choice
problems} is embodied by the (Anscombe and Aumann
(1963)~\ct{ansaum1963}, Gilboa (2009)~\ct[p~144]{gil2009})
\begin{quote}
\textbf{Anscombe--Aumann theorem:}
The ordinal binary \textbf{preference relation} $\succeq$ on the
set of alternative \textbf{acts} $\boldsymbol{F}$ satisfies the set
of five \textbf{axioms of rational choice} if and only if there
exists a unique discrete \textbf{prior probability distribution}
$P \in \boldsymbol{\Delta}(\boldsymbol{\Omega}|I)$ for the
\textbf{state space}~$\Omega$ and a non-constant interval-scaled
von Neumann--Morgenstern \textbf{utility function}
$U: \boldsymbol{X} \rightarrow \mathbb{R}$ for \textbf{outcomes}
in $\boldsymbol{X}$ such that, for every two \textbf{acts} $f, g
\in \boldsymbol{F}$, the weak preference order
\be
f \succeq  g
\ee
is true iff for the \textbf{expected utility} of these two acts
the condition
\be
\sum_{\omega \in \boldsymbol{\Omega}}\left(
\sum_{x \in \boldsymbol{X}}U(x)f(\omega)(x)\right)P(\omega|I)
\geq
\sum_{\omega \in \boldsymbol{\Omega}}\left(
\sum_{x \in \boldsymbol{X}}U(x)g(\omega)(x)\right)P(\omega|I)
\ee
is satisfied.
\end{quote}

\medskip
\noindent
This states that a \textbf{rational decision-maker}'s
\textbf{choice behaviour} can be interpreted \textit{as if} they
apply a personal \textbf{prior probability distribution} to express
their \textbf{uncertainty} as to ensuing consequence-relevant
states of Nature, and \textit{as if} they always maximise
\textbf{subjective expected utility}. The concept of an
\textbf{expected utility} of an outcome was introduced into
economic theory by the Swiss mathematician and physicist
\href{https://mathshistory.st-andrews.ac.uk/Biographies/Bernoulli_Daniel/}{Daniel
Bernoulli FRS (1700–1782)}; cf. Bernoulli (1738)~\ct{ber1738}.

\medskip
\noindent
Savage (1954)~\ct{sav1954} posits the possibility of
reconstructing, via the axiomatic formulation, \textit{both} a
decision-maker's \textbf{prior probability distribution} and their
\textbf{utility function} when a sufficient amount of empirical
data on her/his \textbf{choice behaviour}, and the
\textbf{preferences} so revealed, becomes available.

\section[Caveats of the SEU model]{Caveats of the SEU model}
\lb{sec:cavseu}
A number of conceptual inconsistencies have been spotted over the
years by various authors within the \textbf{subjective expected
utility model} for a \textbf{rational decision-maker}'s
\textbf{choice behaviour}, when confronted with experimental data.
Particularly well-known in this respect are the following caveats:
\begin{itemize}

\item Allais’ (1953)~\ct{all1953} paradox: the violation of some
axioms of rational choice by decision-makers in empirical tests
of the SEU model, providing strong indication for a so-called
\textbf{certainty effect}; see also Kahneman
(2011)~\ct[pp~312--314]{kah2011},

\item Ellsberg’s (1961)~\ct{ell1961} paradox: his experiments
revealed that in simple specific choice situations decision-makers
often prefer known probability distributions over unknown ones,
even when the latter promise the possibility of larger
ensuing pay-offs; this effect has been termed \textbf{uncertainty
aversion},

\item Kahneman and Tversky (1979)~\ct{kahtve1979} emphatically
criticised the lack in the SEU framework of a \textbf{reference
point} for a decision-maker's individual utility function for
outcomes; in their own work they had gathered compelling empirical
evidence that vividly suggested that outcomes acquire a
\textit{different} \textbf{psychological value} for a
decision-maker, depending on whether she/he perceives the outcome
as a gain or as a loss; they referred to this (in their view)
omission as ``Bernoulli's error,''

\item Dekel, Lipman and Rustichini's (1998)~\ct{deketal1998}
impossibility results: in theoretical work these authors
demonstrated that the standard state space formulation precludes
\textit{non-trivial} forms of \textbf{unawareness} of a
decision-maker within an SEU model; a standard state space model
is incapable of consistently incorporating the dimension of a
decision-maker’s unawareness of future contingencies. In this
respect, SEU model cannot adequately capture the concept of
\textbf{surprises}.

\end{itemize}
All in all, the works listed, as well as other less prominent
publications, hinted at the possibility that decision-makers do
\textit{not} necessarily act as though they were following the
premiss of maximising their subjective expected utility on all
occasions. In contrast, decision-makers do regularly exhibit
\textbf{bounded rationality}. One particular line of investigation
started undertaking a revision of the \textbf{rational-agent
paradigm} in theories of human decision-making by integrating in a
comprehensible fashion the complex dimension of a decision-maker's
psychological variability. This lead to the initiation of the field
of \textbf{Behavioural Economics}, which is strongly associated
with the names of the Israeli–US-American experimental
psychologists
\href{https://www.nobelprize.org/nobel_prizes/economic-sciences/laureates/2002/kahneman-facts.html}{Daniel Kahneman (born~1934)} and Amos Tversky (1937–1996), and the
US-American economist
\href{https://www.nobelprize.org/nobel_prizes/economic-sciences/laureates/2017/thaler-facts.html}{Richard
H Thaler (born~1945)}; see Kahneman (2011)~\ct[p~282f]{kah2011},
Thaler and Sunstein (2008)~\ct{thasun2008}, Taleb
(2007)~\ct{tal2007}, and also Gigerenzer (2014)~\ct{gig2014}.
For their ground-breaking work, both Kahneman
in 2002 and Thaler in 2017 were awarded the
\href{https://www.nobelprize.org/prizes/economics/}{Sveriges
Riksbank Prize in Economic Sciences in Memory of Alfred Nobel}.

\leftout{
[Citation given in Kahneman (2011) (?) 
on ``good decisions'' vs ``bad decisions'' when evaluated from the 
perspective of (i)~the outcome, (ii)~the actual decision-making 
process (degree of taking consequence-relevant facts/states of 
Nature into account.)]
}

\section[Representations of non-knowledge]{Representations of
non-knowledge}
\lb{sec:nonknow}
During the last 20~years or so, researchers in economic theory
have become strongly interested again in finding coherent ways of
including in a consistent formalisation of a decision-maker's
\textbf{choice behaviour} in the light of \textbf{uncertainty} the
decision-maker's \textbf{non-knowledge} of consequence-relevant
states of Nature, and of unknown outcomes to acts she/he is going
to pursue. Presumably unintended, this topic was brought
spectacularly to the attention of an international public audience
by Donald Rumsfeld, the former U.S. Secretary of Defense, on Feb
12, 2002, when responding to a journalist's question at a U.S.
Department of Defense news briefing~\ct{rum2002} with the following
explanation:
\begin{quotation}
\noindent
``Reports that say that something hasn't happened are always 
interesting to me, because as we know, there are known knowns; 
there are things we know we know. We also know there are known 
unknowns; that is to say we know there are some things we do not 
know. But there are also unknown unknowns -- the ones we don't 
know we don't know. And if one looks throughout the history of our 
country and other free countries, it is the latter category that 
tend to be the difficult ones.''
\end{quotation}

\medskip
\noindent
The inclusion of representations of different forms of a
decision-maker's \textbf{non-knowledge} in a comprehensive
conceptual framework proves to be a challenging theoretical task.
Starting from the decision matrix of static one-shot
\textbf{choice problems} for a single decision-maker displayed in
Fig.~\ref{fig:decmat}, the immediate points for potential
modification and extension are
(i)~the decision-maker's individual prior probability distribution
for expressing degrees-of-belief for the plausibility of different
states of Nature, and (ii)~the issue of the completeness of the
space of states of Nature itself. Some approaches that follow one
or the other of these two lines of investigation have been
reviewed by Svetlova and van Elst (2012,
2014)~\ct{svehve2012,svehve2014}. Further interesting discussions
on potential ways of advancing this intriguing topic have been
collected in the ``Handbook of Ignorance Studies'' edited by Gross
and McGoey (2015)~\ct{gromcg2015}.

\medskip
\noindent
Our introductory journey through the foundations of
\textbf{inductive statistical inference} as practised within the
\textbf{Bayes--Laplace approach to data analysis and statistical
inference} has now come to an end. We hope the reader could sense a
glimpse of the fascination induced by the simplicity and elegance
of this framework, but, even more so, picked up some very useful
and efficient practical tools for building scientifically sound
statistical models, and for providing adequate interpretations and
predictions.


\addcontentsline{toc}{chapter}{Concluding remarks}
\chapter*{Concluding remarks}
Undoubtedly, an era of incomprehensibly huge big-data reservoirs
has become a reality in present-day human societies as an immediate
consequence of the changes and innovations brought about by the
all-pervasive digital transformation. The question is what
meaningful social purpose can all the information thereby generated
and collected be delivered to?

\medskip
\noindent
I think this provides us with a great opportunity for making
sustainable progress on practical as well as intellectual issues,
though these two areas do not necessarily constitute orthogonal
dimensions. There exists a multitude of intriguing and
awe-inspiring phenomena, based, located, and rooted in both the
natural and the social domains of human experience, that are
accessible to observation and measurement. Equipped with a
naturally inherited curiosity that gets passed on from one
generation to the next, and sticking to the guidelines of the
scientific method, when given access to relevant empirical data,
we can use the analytical skills we developed to try to read the
plot behind the different kinds of natural and social interactions
and interconnections that continue to pose complex and confusing
puzzles to our everyday-life
situations. The prospect of success for creating new common values
should experience a boost when humility, courage, independent and
unconstrained thinking, a diversity of ideas, taking care of one
another, and an attitude of openness towards surprises are to be
found in the portfolio of tools for investigation. Plausible
reasoning leaves no room for ``alternative facts.''
Pseudo-argumentations based on the latter have a rather
poor track record concerning their yield of tangible communal
benefits. The US-American theoretical physicist
\href{https://www.nobelprize.org/nobel_prizes/physics/laureates/1965/feynman-facts.html}{Richard
Phillips Feynman (1918--1988)} has long been an outspoken critical
voice against populist approaches towards tackling real problems
in any kind of field of societal interest; cf. his
thought-stimulating essay ``Cargo Cult Science'' published in
Ref.~\ct[pp~ 308--317]{fey1985}. And there certainly are some
really pressing issues these days that need to be addressed by
the human community urgently, the climate change on planet Earth,
the coronavirus pandemic (ongoing since early 2020), and the
aftermath of the 2008 subprime mortgage crisis amongst them, to
name but a few; see, e.g., Helbing (2013)~\ct{hel2013}.

\medskip
\noindent
The human community anticipates in about 4~to~5~billion years
a transition of the Sun from its present nuclear
hydrogen-burning state to a then nuclear helium-buring state, in
the process of which it will drastically inflate its volume to
become a red giant, so the fate of the three innermost
planets of the Solar System is practically known already today;
see, e.g., Lesch and M\"{u}ller (2003)~\ct[p~405]{lesmue2003}.
Nevertheless, we have potentially much more time ahead of us for
realising our potentials for creativity and sharing knowledge and
understanding when compared to the period that is factually on
record as regards the past history of the human species. Up to
now, quite an impressive legacy of common goals, cultural values
and outstanding intellectual triumphs has accumulated: in music,
the visual and the performing arts, in drama, poetry and
novel-writing, in sports, the different languages, and, of course,
in view of the great scientific achievements of a very diverse
spectrum of human minds. There are many prominent examples of
seminal advances in the human understanding of different kinds of
dynamical processes that have a direct bearing on human existence,
among them
\begin{itemize}
\item the natural evolution of organic creatures according to
the English naturalist, geologist and biologist
\href{https://en.wikipedia.org/wiki/Charles_Darwin}{Charles Robert
Darwin, FRS FRGS FLS FZS (1809--1882)} (see Darwin
(1859)~\ct{dar1859}),

\item the nature of gravitational interactions as explained by
\href{https://www.nobelprize.org/prizes/physics/1921/einstein/facts/}{Albert
Einstein (1871--1955)} (see Einstein (1915)~\ct{ein1915},
Hawking and Ellis (1973)~\ct{hawell1973}, Misner \textit{et al}
(1973)~\ct{mtw1973}, Abbott \textit{et al} (2016)~\ct{abbetal2016};
and also Ref.~\ct{hve2015c}), or

\item the genetic coding mechanism underlying
the reproduction of any living organism as deciphered by the
British molecular biologist, biophysicist, and neuroscientist
\href{https://www.nobelprize.org/prizes/medicine/1962/crick/facts/}{Francis
Harry Compton Crick OM FRS (1916--2004)},
the US-American molecular biologist, geneticist and zoologist
\href{https://www.nobelprize.org/prizes/medicine/1962/watson/facts/}{James
Dewey Watson (born 1928)}, and the New Zealand physicist and
molecular biologist
\href{https://www.nobelprize.org/prizes/medicine/1962/wilkins/facts/}{Maurice
Hugh Frederick Wilkins  CBE FRS (1916--2004)} (see Watson and Crick
(1953)~\ct{watcri1953} and Wilkins \textit{et al}
(1953)~\ct{wiletal1953}).
\end{itemize}

\medskip
\noindent
The face of planet Earth has been radically changed by the
influence on Nature taken by humans over a period of only a few
hundred, possibly a few thousand years --- a far cry compared to
the minimum 10~billion years it took to forge all the different
kinds of natural resources, now available to the human community,
in various kinds of astrophysical furnaces. This leaves at best a
large two-digit figure of generations that have since been involved
in exploiting these resources for generating and sharing
material as well as immaterial goods, in the large majority for
very sensible and helpful, but regrettably also for less sensible
courses. If we manage to keep an eye on the upper limits presented
to these resources, develop a common sense for the principle of
reciprocity binding wo/man-kind to Nature, foster a
related emotional connectivity, and ultimately safeguard the
habitability of planet Earth for many generations to come, we might
stand a real chance of being perceived by fellow beings in the vast
stretches of the spacetime continuum as respectable citizens of the
Universe.


\vfill
\noindent
\textit{Acknowledgements:} I am grateful to Ariane, Vincent and
Audrey for their enduring patience with my continuing
absent-mindedness during the intense writing-up process of these
lecture notes.

\appendix
\chapter[MCMC related commands in \R]{MCMC related commands in \R}
\lb{app1}
An official 
\href{https://cran.r-project.org/doc/contrib/Short-refcard.pdf}{\R{}
reference card} can be obtained from the URL (cited on August 
17, 2022): https://cran.r-project.org/doc/contrib/Short-refcard.pdf

\medskip
\noindent
\textbf{A}\\
\texttt{as.integer(\ldots)}: converting a suitable \R{} object
into integer format\\
\texttt{as.matrix(\ldots)}: converting a suitable \R{} object
into matrix format


\medskip
\noindent
\textbf{C}\\
\texttt{check\_hmc\_diagnostics(\ldots)}: returning values of
HMC diagnostics $\{\text{rstan}\}$ \\
\texttt{colMeans(\ldots)}: computing means of column 
entries of data frame or matrix\\
\texttt{colnames(\ldots)}: listing column names of data 
frame or matrix\\
\texttt{cor(\ldots)}: computing a bivariate correlation
matrix for metrically scaled data from a data frame
$\{\text{stats}\}$

\medskip
\noindent
\textbf{D}\\
\texttt{data(\ldots)}: loading a data set from within a loaded
\R{} package

\medskip
\noindent
\textbf{E}\\
\texttt{extract(\ldots)}: extracting HMC samples from a fitted
model $\{\text{rstan}\}$ \\



\medskip
\noindent
\textbf{H}\\
\texttt{head(\ldots)}: scanning the first few rows of a 
data frame

\medskip
\noindent
\textbf{I}\\
\texttt{install.packages(\ldots)}: installing a specific 
package\\
\texttt{is.na(\ldots)}: checking for missing values 
(``NA'') in a data frame



\medskip
\noindent
\textbf{L}\\
\texttt{length(\ldots)}: sample size of data for a specific 
variable\\
\texttt{library(\ldots)}: loading a specific package\\
\texttt{load(\ldots)}: loading a data set in *.RData format\\
\texttt{loo(\ldots)}: calculating Pareto-smoothed
importance-sampling estimate for leave-one-out cross-validation 
information criterion
$\{\text{loo}\}$

\medskip
\noindent
\textbf{M}\\
\texttt{mean(\ldots)}: computing the sample mean for 
metrically scaled univariate data\\
\texttt{median(\ldots)}: computing the sample median for 
metrically scaled univariate data $\{\text{stats}\}$\\
\texttt{model.matrix(\ldots)}: constructing a design matrix
$\{\text{stats}\}$

\medskip
\noindent
\textbf{N}\\
\texttt{na.omit(\ldots)}: removing cases with missing
values (``NA'') from a data frame\\
\texttt{ncol(\ldots)}: number of columns of a data frame\\
\texttt{nrow(\ldots)}: number of rows of a data frame


\medskip
\noindent
\textbf{P}\\
\texttt{pairs(\ldots)}: generating a matrix of pairwise 
scatter plots for posterior simulations of model parameters
$\{\text{rstan}\}$\\
\texttt{ppc\_dens\_overlay(\ldots)}: plotting empirical data
distribution for an outcome overlaid with HMC samples of the
posterior predictive distribution $\{\text{bayesplot}\}$\\
\texttt{ppc\_hist(\ldots)}: plotting a histogram of the
distribution for an outcome next to histograms for HMC samples of
the posterior  predictive distribution $\{\text{bayesplot}\}$\\
\texttt{ppc\_stat(\ldots)}: plotting a histogram for HMC samples of
the posterior predictive distribution overlaid by the position of
a specific sample statistic such as the mean, median or max
$\{\text{bayesplot}\}$\\
\texttt{ppc\_intervals(\ldots)}: plotting empirical data points
for an outcome overlaid with corresponding posterior predictive 
compatibility intervals $\{\text{bayesplot}\}$\\
\texttt{ppc\_rootogram(\ldots)}: plotting a rootogram for a count
outcome $\{\text{bayesplot}\}$\\
\texttt{print(\ldots)}: printing the estimated values for model
parameters or information criteria 

\medskip
\noindent
\textbf{Q}\\
\texttt{quantile(\ldots)}: computing the 
$\alpha$--quantile for metrically scaled univariate data
$\{\text{stats}\}$

\medskip
\noindent
\textbf{R}\\
\texttt{rowMeans(\ldots)}: computing means of row entries 
of data frame or matrix\\
\texttt{rownames(\ldots)}: listing row names of data 
frame or matrix

\medskip
\noindent
\textbf{S}\\
\texttt{scale(\ldots)}: standardising univariate metrically
scaled data\\
\texttt{sd(\dots)}: computing the sample standard deviation 
for metrically scaled univariate data [$(n-1)$-convention]
$\{\text{stats}\}$\\
\texttt{set.seed(\ldots)}: initialising the random number 
generator to a specific integer\\
\texttt{stan(\ldots)}: activating a MCMC simulation with Stan
$\{\text{rstan}\}$ \\
\texttt{stan\_dens(\ldots)}: plotting the posterior marginal
probability distribution for a model parameter $\{\text{rstan}\}$
\\
\texttt{stan\_plot(\ldots)}: plotting a compatibility interval for
a model parameter $\{\text{rstan}\}$\\
\texttt{stan\_trace(\ldots)}: trace plot of Markov chain
simulations for a model parameter $\{\text{rstan}\}$

\medskip
\noindent
\textbf{T}\\
\texttt{t(\ldots)}: transposition of a matrix\\
\texttt{tail(\ldots)}: scanning the last few rows of a 
data frame


\medskip
\noindent
\textbf{V}\\
\texttt{var(\ldots)}: computing the sample variance for 
metrically scaled univariate data [$(n-1)$-convention]
$\{\text{stats}\}$

\medskip
\noindent
\textbf{W}\\
\texttt{waic(\ldots)}: calculating Watanabe--Akaike information
criterion $\{\text{loo}\}$


\chapter[List of online resources]{List of online resources on
inductive statistical inference}
\lb{app2}
In this appendix we give in a random order the hyperlinks to a few
very helpful online resources on inductive statistical inference.

\begin{itemize}

\item Bayes Centre at the University of Edinburgh.
URL (cited on August 7, 2020):
\href{https://www.ed.ac.uk/bayes}{www.ed.ac.uk/bayes}.

\item Richard McElreath's 20-episodes lecture series at
\href{https://www.eva.mpg.de}{Max Planck Institute for Evolutionary 
Anthropology}, Leipzig, Germany during Winter Semester 2021/2022.
URL (cited on August 17, 2022):
\href{https://github.com/rmcelreath/stat_rethinking_2022}{github.com/rmcelreath/stat\_rethinking\_2022}.

\item Aki Vehtari's Bayesian Data Analysis course at Aalto
University, Finland.
URL (cited on August 7, 2020):
\href{https://github.com/avehtari/BDA_course_Aalto}{github.com/avehtari/BDA\_course\_Aalto}.

\item Michael Betancourt's 2014 talk on ``Hamiltonian Monte Carlo
and Stan.''
URL (cited on August 7, 2020):
\href{https://www.youtube.com/watch?v=xWQpEAyI5s8&list=WL&index=14&t=42s}{www.youtube.com/watch?v=xWQpEAyI5s8\&list=WL\&index=14\&t=42s}.

\item Michael Betancourt's 2017 vignette on ``Diagnosing Biased
Inference with Divergences.''
URL (cited on August 7, 2020):
\href{https://mc-stan.org/users/documentation/case-studies/divergences_and_bias.html}{mc-stan.org/users/documentation/case-studies/divergences\_and\_bias.html}.

\item Jonah Gabry's and Martin Modr\'{a}k's 2022 vignette on
``Visual MCMC diagnostics using the bayesplot package.''
URL (cited on August 17, 2022):
\href{https://cran.r-project.org/web/packages/bayesplot/vignettes/visual-mcmc-diagnostics.html}{cran.r-project.org/web/packages/bayesplot/vignettes/visual-mcmc-diagnostics.html}.

\item Andrew Gelman's 2015 talk on ``But When You Call Me Bayesian,
I Know I’m Not the Only One.''
URL (cited on August 7, 2020):
\href{https://www.youtube.com/watch?v=ObS1hkOxyPA}{www.youtube.com/watch?v=ObS1hkOxyPA}.

\item Will Hipson's tutorial on ``Bayesian Varying Effects Models
in R and Stan.''
URL (cited on August 7, 2020):
\href{https://willhipson.netlify.app/post/stan-random-slopes/varying_effects_stan/}{https://willhipson.netlify.app/post/stan-random-slopes/varying\_effects\_stan/}.

\item Peter Coles' 2014 entry ``Bayes, Laplace and Bayes’ Theorem''
in his Blog
\href{https://telescoper.wordpress.com/}{In The Dark}.
URL (cited on August 7, 2020):
\href{https://telescoper.wordpress.com/2014/10/01/bayes-laplace-and-bayes-theorem/}{https://telescoper.wordpress.com/2014/10/01/bayes-laplace-and-bayes-theorem/}.

\item \textit{The New York Times} in 2014 on ``The Odds,
Continually Updated.''
URL (cited on August 7, 2020):
\href{https://www.nytimes.com/2014/09/30/science/the-odds-continually-updated.html?_r=1}{www.nytimes.com/2014/09/30/science/the-odds-continually-updated.html?\_r=1}.

\item \textit{The Guardian} in 2006 on ``How a statistical formula
won the war.'' URL (cited on August 27, 2022):
\href{https://www.theguardian.com/world/2006/jul/20/secondworldwar.tvandradio}{https://www.theguardian.com/world/2006/jul/20/secondworldwar.tvandradio}.

\end{itemize}
%

\chapter[Glossary of technical terms (GB -- D)]{Glossary of 
technical terms (GB -- D)}
\lb{app3}

\noindent
{\bf A}\\
act: Handlung\\
adaptive regularisation: adaptive Regularisierung\\
algorithm: Algorithmus, Rechenanweisung, Rechenregel\\
ANOVA (analysis of variance): Varianzanalyse\\
autocorrelation: Autokorrelation\\
autoregressive process: autoregresiver Prozess

\medskip
\noindent
{\bf B}\\
Bayes' theorem: Satz von Bayes\\
behaviour: Verhalten\\
binomial coefficient: Binomialkoeffizient\\
bivariate: bivariat, zwei variable Gr\"{o}\ss en betreffend\\
bounded rationality: begrenzte Rationalit\"{a}t, begrenzte
Vernunft

\medskip
\noindent
{\bf C}\\
choice: Wahl, Auswahl\\
choice problem: Auswahlproblem\\
Cholesky decomposition: Cholesky--Zerlegung\\
cluster: Klumpen, Anh\"{a}ufung\\
common sense: gesunder Menschenverstand\\
compatibility interval: Kompatibilit\"{a}tsintervall\\
complete ignorance: g\"{a}nzliche Unkenntnis\\
conditional probability: bedingte Wahrscheinlichkeit\\
conjunction: Konjunktion, Mengenschnitt\\
conjugate: konjugiert, abgewandelt\\
consequence: Konsequenz, Auswirkung\\
contingency table: Kontingenztafel\\
convenience sample: Gelegenheitsstichprobe\\
correlation: Korrelation\\
covariance matrix: Kovarianzmatrix

\pagebreak
\medskip
\noindent
{\bf D}\\
data: Daten\\
data-generating process: Daten generierender Prozess\\
data matrix: Datenmatrix\\
decision: Entscheidung\\
degree-of-belief: Glaubw\"{u}rdigkeitsgrad, Plausibilit\"{a}t\\
degrees of freedom: Freiheitsgrade\\
degree of plausibility: Plausibilit\"{a}tsgrad\\
dependent variable: abh\"{a}ngige Variable\\
design matrix: Regressormatrix, Modellmatrix\\ 
deviance: abweichendes Verhalten\\
deviation: Abweichung \\
disjunction: Disjunktion, Mengenvereinigung\\
dispersion: Streuung\\
distribution: Verteilung\\
divergence: Divergenz, Auseinanderstreben

\medskip
\noindent
{\bf E}\\
estimation: Sch\"{a}tzung\\
evidence: Anzeichen, Hinweis, Anhaltspunkt, Indiz\\
exchangeability: Austauschbarkeit\\
expectation value: Erwartungswert\\
exposure: Ausgesetztsein

\medskip
\noindent
{\bf F}\\
fact-based reasoning: Fakten basiertes Argumentieren\\
fallacy: Trugschluss, Fehlschluss, T\"{a}uschung

\medskip
\noindent
{\bf G}\\
group: Gruppe

\medskip
\noindent
{\bf H}\\
heteroscedasticity: Heteroskedastizit\"{a}t, inhomogene Varianz\\
homoscedasticity: Homoskedastizit\"{a}t, homogene Varianz\\
hypothesis: Hypothese, Behauptung, Vermutung

\medskip
\noindent
{\bf I}\\
ignorance: Unkenntnis\\
incomplete information: unvollst\"{a}ndige Information\\
independent variable: unabh\"{a}ngige Variable\\
indicator variable: bin\"{a}re Indikatorvariable\\
inductive method: induktive Methode\\
information: Information\\
information criterion: Informationskriterium\\
information entropy: Informationsentropie\\
interaction: Wechselwirkung\\
interaction effect: Wechselwirkungseffekt\\
intercept: Achsenabschnitt

\medskip
\noindent
{\bf J}\\
joint distribution: gemeinsame Verteilung

\medskip
\noindent
{\bf K}\\
knowledge: Wissen, Kenntnis, Erkenntnis, Wissensstand

\medskip
\noindent
{\bf L}\\
lag parameter: Verz\"{o}gerungsparameter\\
Lagrange function: Lagrange-Funktion\\
Lagrange multiplier: Lagrange-Multiplikator, integrierender
Faktor\\
linear regression analysis: lineare Regressionsanalyse\\
link function: Verkn\"{u}pfungsfunktion\\
location parameter: Lokationsparameter\\
logarithmic score: logarithmische Punktzahl\\
logical complement: logisches Gegenteil\\
loss function: Verlustfunktion

\medskip
\noindent
{\bf M}\\
main effect: Haupteffekt\\
marginal distribution: Randverteilung\\
marginal frequencies: Randh\"{a}ufigkeiten\\
marginalisation: Marginalisierung(smethode)\\
maximum entropy distribution: Verteilung maximaler Entropie\\
measurement: Messung, Datenaufnahme\\
measurement scale: Ma\ss skala\\
model comparison: Modellvergleich\\
multicollinearity: Multikollinearit\"{a}t 

\medskip
\noindent
{\bf N}\\
non-knowledge: Nichtwissen\\
normalisation condition: Normierungsbedingung\\
numerical algorithm: numerischer Algorithmus, Rechenanweisung 

\medskip
\noindent
{\bf O}\\
observable: beobachtbar, messbar\\
observation: Beobachtung\\
odds: Wettchancen\\
operationalisation: Operationalisieren, latente Variable messbar 
gestalten\\
outcome: Ergebnis, Resultat\\
outlier: Ausrei\ss er\\
over-fitting: \"{u}bergenaues Anpassen

\medskip
\noindent
{\bf P}\\
parameter: Parameter, w\"{a}hlbare Stellgr\"{o}\ss e\\
parameter space: Parameterraum\\
parsimony: Sparsamkeit\\
pooling: Zusammenlegen, gemeinsames Nutzen\\
population: Grundgesamtheit\\
precision: Pr\"{a}zisionsparameter\\
prediction: Vorhersage\\
predictor: erkl\"{a}rende Variable\\
preference: Vorliebe, Bevorzugung\\
premiss: Voraussetzung, Pr\"{a}misse\\
primitive: Grundbaustein\\
probability: Wahrscheinlichkeit\\
probability density function ({\tt pdf}): 
Wahrscheinlichkeitsdichte\\
probability function: Wahrscheinlichkeitsfunktion\\
proposition: Vorschlag, Antrag, Aussage, Behauptung, Pr\"{a}misse\\
psychological value: psychologischer Wert


\medskip
\noindent
{\bf R}\\
rare event: seltenes Ereignis\\
rate parameter: Ratenparameter\\
reference point: Bezugspunkt\\
regression analysis: Regressionsanalyse\\
regression coefficient: Regressionskoeffizient\\
regression model: Regressionsmodell\\
regression toward the mean: Regression zur Mitte\\
re-scaling: Reskalierung, Gr\"{o}\ss enordnungs\"{a}nderung\\
retrodiction: Nachersage, Rekonstruktion (von Daten)\\
risk: Risiko (berechenbar)\\
rule of succession: Regel des nachfolgenden Wertes

\medskip
\noindent
{\bf S}\\
sample: Stichprobe\\
sample auto-covariance: Stichprobenautokovarianz\\
sample mean: Stichprobenmittelwert\\
sample size: Stichprobenumfang\\
sample unit: Untersuchungseinheit\\
sample variance: Stichprobenvarianz\\
sampling distribution: Stichprobenkenngr\"{o}\ss enverteilung\\
scale-invariant: skaleninvariant\\
scale parameter: Skalenparameter\\
scientific endeavour: wissenschaftliche Bem\"{u}hung\\
scientific method: Wissenschaftliche Methode\\
shape parameter: Formparameter\\
shrinkage: Schrumpfen\\
slope: Steigung\\
standardisation: Standardisierung\\
state of Nature: Zustand der Au\ss enwelt\\
stationarity: Stationarit\"{a}t\\
statistical (in)dependence: statistische (Un)abh\"{a}ngigkeit\\
statistical model: statistisches Modell\\
statistical variable: Merkmal, Variable\\
stratified sample: geschichtete Stichprobe\\
sufficient statistic: suffizientes statistisches Ma\ss\\
surprise: \"{U}berraschung, unerwartetes Ereignis\\
survival analysis: Ereigniszeitanalyse

\medskip
\noindent
{\bf T}\\
target population: Zielgruppe\\
time series: Zeitreihe\\
transformation: Transformation, Umwandlung, Ver\"{a}nderung\\
translation: Translation, Verschiebung

\medskip
\noindent
{\bf U}\\
uncertainty: Unsicherheit (nicht berechenbar)\\
under-fitting: ungen\"{u}gendes Anpassen\\
univariate: univariat, eine einzige variable Gr\"{o}\ss e 
betreffend\\
unit: (Ma\ss-)Einheit\\
unknown entity: unbekannte Gr\"{o}\ss e\\
updating process: Aktualisierungsprozess\\
utility function: Nutzenfunktion

\medskip
\noindent
{\bf V}\\
validation: Validierung, G\"{u}ltigkeitspr\"{u}fung\\
value: Wert\\
variance: Varianz\\
variation: Variation\\
volatility clustering: Volatilit\"{a}tsklumpung




\addcontentsline{toc}{chapter}{Bibliography}


\end{document}